\documentclass[twoside,openany,12pt]{book}

\pdfoutput=1
\usepackage{afterpage}
\usepackage{algorithm}
\usepackage[noend]{algpseudocode}
\usepackage{amsmath}
\usepackage{amssymb}
\usepackage{amsthm}
\usepackage[toc]{appendix}
\usepackage{caption}
\usepackage{changepage}
\usepackage{comment}
\usepackage{enumitem}
\usepackage{epigraph}
\usepackage{epstopdf}
\usepackage{etoolbox}
\usepackage{fancyhdr}
\usepackage{fix-cm}
\usepackage[margin=1in]{geometry}
\usepackage{graphicx}
\usepackage{hieroglf}
\usepackage{ifthen}
\usepackage{lipsum}
\usepackage{mathtools}
\usepackage{nomencl}
\usepackage{pdfpages}
\usepackage{setspace}
\usepackage{tabularx}
\usepackage[raggedright]{titlesec}
\usepackage[notbib]{tocbibind}
\PassOptionsToPackage{svgnames,dvipsnames}{xcolor}
\usepackage[comma,sort&compress]{natbib}
\setcitestyle{numbers,square}
\usepackage{tikz}
\PassOptionsToPackage{colorlinks=true,allcolors={blue}}{hyperref}
\usepackage{yfonts,lettrine}
\usepackage[bottom,flushmargin]{footmisc}

\newboolean{proquest}
\setboolean{proquest}{false}

\fancypagestyle{plainstyle}{\fancyhf{}\cfoot{\thepage}}
\fancypagestyle{mainstyle}{\fancyhf{}%
\fancyhead[RO,LE]{\thepage}
\fancyhead[LO]{\leftmark}
\fancyhead[RE]{\rightmark}

}
\fancypagestyle{refstyle}{\fancyhf{}\fancyhead[RO,LE]{\thepage}}
\renewcommand{\bibname}{\Large\textbf{REFERENCES}}

\newcommand{\blankpage}{\null\thispagestyle{empty}\newpage}

\definecolor{gray75}{gray}{0.75}
\newcommand{\hsp}{\hspace{0pt}}
\newcommand\mainchapter{%
\titleformat{\chapter}[hang]{\flushright
\fontseries{b}\fontsize{80}{100}\selectfont}{\fontseries{b}\fontsize{100}{130}\selectfont \textcolor{gray75}\thechapter\hsp}{0pt}{\\[-0.8cm] \titlerule\vspace*{-0.2cm}\fontsize{30}{36}\bfseries}[\titlerule]
\titlespacing*{\chapter}{0pt}{-50pt}{40pt}
}
\newcommand\twolinechapter{%
\titleformat{\chapter}[hang]{\flushright
\fontseries{b}\fontsize{80}{100}\selectfont}{\fontseries{b}\fontsize{100}{130}\selectfont \textcolor{gray75}\thechapter\hsp}{0pt}{\\[-0.8cm] \titlerule\vspace*{0.4cm}\fontsize{30}{36}\bfseries}[\titlerule]
}

\newcommand\booleq{==}
\newcommand\pluseq{\mathrel{+}=}
\newcommand\minuseq{\mathrel{-}=}

\newcommand\plusplus{\mathrel{++}}
\newcommand\andeq{\mathrel{\&}=}
\newcommand\mathand{\mathrel{\&}}
\newcommand\mathmod{\mathrel{\%}}
\newcommand\xoreq{\mathrel{\veebar}=}

\algnewcommand{\LineComment}[1]{\State\(\triangleright\) #1}
\algnewcommand\algorithmicinput{\textbf{Input:}}
\algnewcommand\Input{\item[\algorithmicinput]}
\algnewcommand\algorithmicoutput{\textbf{Output:}}
\algnewcommand\Output{\item[\algorithmicoutput]}

\makeatletter
\newcommand{\algmargin}{\the\ALG@thistlm}
\makeatother
\newlength{\ifwidth}
\newlength{\elseifwidth}
\algdef{SE}[parIF]{parIf}{EndparIf}[1]
  {\parbox[t]{\dimexpr\linewidth-\algmargin}{%
    \hangindent\ifwidth\strut\algorithmicif\ #1\ \algorithmicthen\strut}}{\algorithmicend\ \algorithmicif}
\algdef{C}[parIF]{parIF}{parElseIf}[1]
  {\parbox[t]{\dimexpr\linewidth-\algmargin}{%
    \hangindent\elseifwidth\strut\algorithmicelse\ \algorithmicif\ #1\ \algorithmicthen\strut}}
\algdef{Ce}[parELSE]{parIF}{parElse}{EndparIf}{\algorithmicelse}

\newlength{\forwidth}
\algdef{SE}[parFOR]{parFor}{EndparFor}[1]
  {\parbox[t]{\dimexpr\linewidth-\algmargin}{%
    \hangindent\forwidth\strut\algorithmicfor\ #1\ \algorithmicdo\strut}}{\algorithmicend\ \algorithmicfor}

\algnewcommand{\algorithmicgoto}{\textbf{go to}}
\algnewcommand{\Goto}[1]{\algorithmicgoto~\ref{#1}}

\algnewcommand{\parState}[1]{\State%
  \parbox[t]{\dimexpr\linewidth-\algmargin}{\strut #1\strut}}

\algnewcommand{\parComment}[1]{\LineComment%
  \parbox[t]{\dimexpr\linewidth-\algmargin}{\strut #1\strut}}

\makeatletter
\ifthenelse{\equal{\ALG@noend}{t}}%
{\algtext*{EndparIf}\algtext*{EndparFor}}
{}%
\makeatother

\let\oldReturn\Return
\renewcommand{\Return}{\State\oldReturn}

\usetikzlibrary{shapes,arrows}
\tikzstyle{startstop} = [rectangle, rounded corners, minimum width=3cm, minimum height=1cm, text centered, draw=black]
\tikzstyle{process} = [rectangle, minimum width=3cm, minimum height=1cm, text centered, draw=black]
\tikzstyle{decision} = [diamond, minimum width=2cm, minimum height=1cm, text centered, draw=black]
\tikzstyle{arrow} = [thick,->,>=stealth]

\tikzset{font={\fontsize{8pt}{12}\selectfont}}

\let\ab\allowbreak

\DeclareMathOperator{\arcsinh}{arcsinh}
\DeclareMathOperator{\arccosh}{arccosh}
\DeclareMathOperator{\sn}{sn}
\DeclareMathOperator{\cn}{cn}
\DeclareMathOperator{\dn}{dn}
\DeclareMathOperator{\fn}{fn}
\newcommand{\munu}{{\mu\nu}}
\newcommand{\dprime}{{\prime\prime}}
\DeclareMathOperator{\arctanh}{arctanh}

\newcommand{\bZ}{\mathbb{Z}}

\def\fru{\mathfrak{u}}

\allowdisplaybreaks

\makeatletter
\let\ps@plain\ps@empty
\makeatother

\definecolor{cornellRed}{HTML}{B31B1B}

\newcommand{\muop}{{$\mu$op}}
\newcommand{\muops}{{$\mu$ops}}



\AtBeginDocument{\setlength{\DefaultFindent}{0.5em}}
\patchcmd{\thenomenclature}{%
  \chapter*{\nomname}}{}{}{}

\newcommand\Nomenclature[4][O]{\nomenclature[#1#2#4]{#3}{#4}}
\newcommand\NNomenclature[4][O]{\nomenclature[#1#2]{#3}{#4}}

\renewcommand\nomgroup[1]{%
\ifthenelse{\equal{#1}{A}}{%
\item[\textbf{Acronyms}]}{%
\ifthenelse{\equal{#1}{R}}{%
\vspace*{0.5cm}
\item[\textbf{Roman Symbols}]}{%
\ifthenelse{\equal{#1}{G}}{%
\item[\textbf{Greek Symbols}]}{%
\ifthenelse{\equal{#1}{Z}}{%
\item[\textbf{Other Symbols}]}{%
{}}}}}}

\makenomenclature

\makeatletter
\renewcommand\listoffigures{%
\@starttoc{lof}%
}
\let\l@algorithm\l@figure%
\let\listofalgorithms\listoffigures%
\let\@cftmakeloatitle\@cftmakeloftitle%
\renewcommand\listofalgorithms{%
\@starttoc{loa}%
}
\patchcmd{\@cftmakeloatitle}{\listfigurename}{\listalgorithmname}{}{}%
\makeatother

\begin{document}

%%%%%%%%%%%%
\ifthenelse{\boolean{proquest}}{}{%
\frontmatter%
}
%%%%%%%%%%%%

\ifthenelse{\boolean{proquest}}{%
\include{cover_proquest}}{%
\thispagestyle{empty}
\begin{center}
\linespread{1.0}
\bf
\LARGE
High Performance Algorithms for \\Quantum Gravity and Cosmology\\
\vspace{1.2cm}
\normalsize
A Dissertation Submitted to\\
The College of Science\\
at Northeastern University\\

\vspace{1.0cm}
By\\
\large
William Joseph Cunningham\\
\vspace{1.0cm}
\normalsize
In Partial Fulfillment of the Requirements\\
for the Degree of\\
DOCTOR OF PHILOSOPHY IN PHYSICS
\vspace{1.0cm}
\begin{figure}[!h]
\centering
\includegraphics[width=0.25\linewidth]{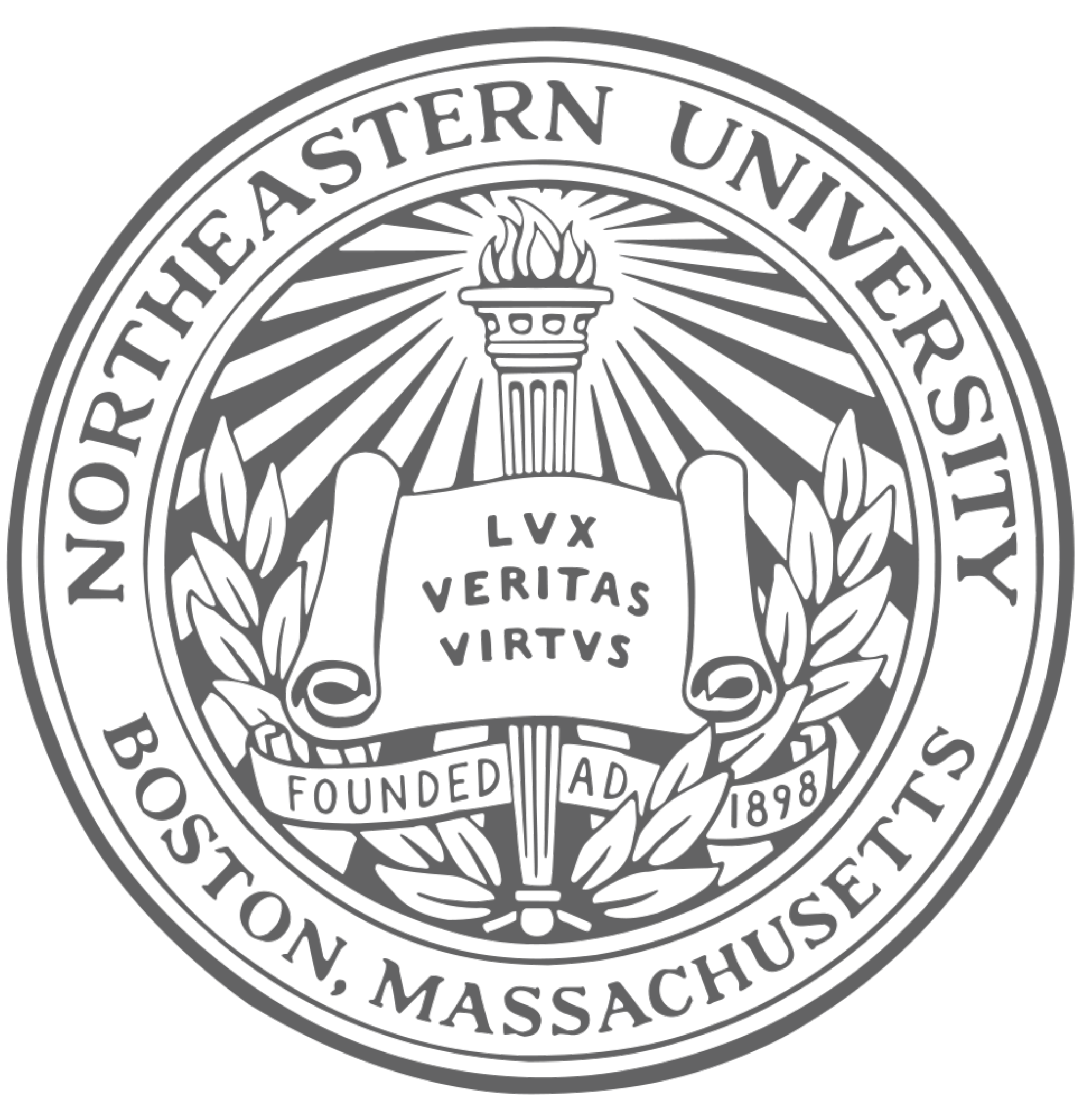}
\end{figure}

\vspace{1.0cm}
DISSERTATION COMMITTEE:\\
Dmitri Krioukov\\
James Halverson \\
Alessandro Vespignani\\
Sumati Surya \\

\vfill
May 2018\\
Boston, Massachusetts
\end{center}
\clearpage
}

\pagestyle{plainstyle}
\ifthenelse{\boolean{proquest}}{%
\thispagestyle{plainstyle}}{%
\thispagestyle{empty}}
\begin{center}
\vspace*{3.8in}
\copyright\,\,MMXVIII\\[0.2cm]
William Joseph Cunningham\\[0.2cm]
ALL RIGHTS RESERVED.
\end{center}
\clearpage

\includepdf{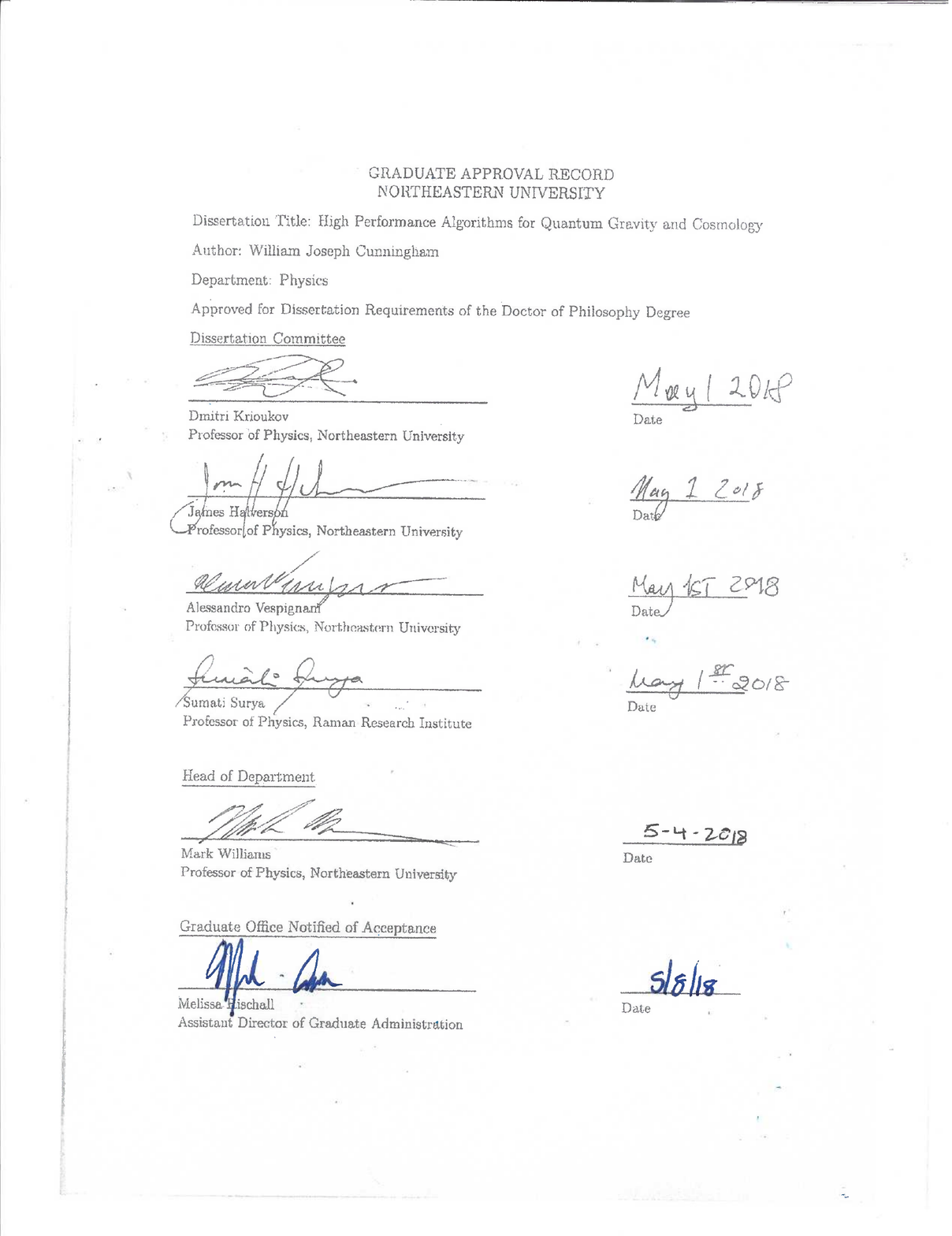}
\afterpage{\blankpage}

\doublespacing
\phantomsection
\addcontentsline{toc}{chapter}{Dedication}
\ifthenelse{\boolean{proquest}}{%
\thispagestyle{plainstyle}}{%
\thispagestyle{empty}}
\begin{center}
\ifthenelse{\boolean{proquest}}{%
Dedication\\[3.8in]}{%
\vspace*{4in}}
\textit{I dedicate this work to Daisy.}
%\textit{Dedication TBD.}
\end{center}

\phantomsection
\addcontentsline{toc}{chapter}{Acknowledgments}
\begin{center}%
\ifthenelse{\boolean{proquest}}{
Acknowledgments\\}{
\vspace*{-1.2cm}\noindent\rule{\textwidth}{1pt}
{\Large \textbf{ACKNOWLEDGMENTS}}\vspace*{-0.3cm}
\noindent\rule{\textwidth}{1pt}\\[0.2cm]}
\end{center}

I would like to extend my sincere gratitude to the following individuals for the impact they have had on me during my academic career and in my personal life. \par

First, my family has been incredibly supportive of my dream to become a physicist. My parents Scott and Michelle have always fostered my academic interests and given me the emotional support to push forward when I needed it most. Their encouragement has pushed me to be a good person and a good scientist. \par

Next, I would like to thank some of the friends I've made along the way. I'd like to acknowledge my lifelong friends in $\Phi\Sigma$K and $\Gamma$BP. Our interesting discussions and adventures have kept me from being bored during this long process. \par

I would also like to thank some of the most influential mentors I've had along the way. My undergraduate advisor Peter Persans was the first person for whom I worked as a research assistant, and he helped me better learn what science is like day-to-day. He taught me to apply myself and to be patient. I'd also like to thank Joel Giedt and Vincent Meunier for offering me my first theoretical physics research position. They challenged me and immersed me in the world of computational physics and high performance computing, where I found my deepest interests. Much of this dissertation reflects the passion for these subjects which they passed on to me. \par

In graduate school, I immediately found a friend and mentor in Dima Krioukov. To Dima, thank you for giving me the freedom to make this journey my own, and thank you for holding me to high standards. Your generosity in allowing me to travel frequently has been invaluable to my professional career, and it has helped me become a much better scientist. I would also like to thank Sumati Surya, who unexpectedly became a second mentor to me. Without knowing me at all, you invited me across the world and into a community I've come to cherish. My original childhood dream was to explore cosmology and quantum mechanics, and now with your aid I've been able to bridge the gap from network science at Northeastern to quantum gravity at PI. \par

I would also like to thank my other committee members, Jim Halverson and Alex Vespignani, for taking the time to mentor and assist me during this process. Your advice and feedback has been extremely helpful. \par

Over the last five years, I've had the pleasure of meeting many other scientists who have mentored or helped me in some way or another, and I'd like to acknowledge them here as well. Thank you to Kostia Zuev, Maksim Kitsak, Rodrigo Aldecoa, David Rideout, Michel Buck, Lisa Glaser, Cody Long, and Vania Voitalov. I would especially like to thank Pim van der Hoorn, who has been invaluable in helping me improve this dissertation. \par

Finally, I would like to thank my wonderful wife Katherine for joining me on this adventure, and for being patient and supportive. These late nights coding and writing are tough, but your love and companionship make life's problems seem trivial.\\
\begin{flushright}
\textsc{W.\ J.\ Cunningham}
\end{flushright}

\clearpage
%\ifthenelse{\boolean{proquest}}{}{%
%\afterpage{\blankpage}%}

\phantomsection
\ifthenelse{\boolean{proquest}}{%
\addcontentsline{toc}{chapter}{Abstract of Dissertation}}{%
\addcontentsline{toc}{chapter}{Abstract}}
\begin{center}
\ifthenelse{\boolean{proquest}}{%
Abstract of Dissertation\\}{%
\vspace*{-1.2cm}\noindent\rule{\textwidth}{1pt}
{\Large\textbf{ABSTRACT}}\vspace*{-0.3cm}
\noindent\rule{\textwidth}{1pt}\\[0.2cm]}
\end{center}

Large scale numerical experiments are commonplace today in theoretical physics. The high performance algorithms described herein are the most compact, efficient methods known for representing and analyzing systems modeled well by sets or graphs. After studying how these implementations maximize instruction throughput and optimize memory access patterns, we apply them to causal set quantum gravity, in which spacetime is represented by a partially ordered set. We build upon the low-level set and graph algorithms to optimize the calculation of the causal set action, and then discuss how to measure boundaries of a discrete spacetime. We then examine the broader applicability of these algorithms to greedy information routing in random geometric graphs embedded in Lorentzian manifolds, which requires us to find new closed-form solutions to the geodesic differential equations in Friedmann-Lema\^itre-Robertson-Walker spacetimes. Finally, we consider the vacuum selection problem in string theory, where we show a network-centered approach yields a dynamical mechanism for vacuum selection in the context of multiverse cosmology. These algorithms have broad applicability to many physical systems, and they improve existing methods by reducing simulation runtimes by orders of magnitude.

\singlespacing

\renewcommand{\contentsname}{Table of Contents}
\ifthenelse{\boolean{proquest}}{%
\titleformat{\chapter}[display]{\centering\normalfont\Large}{\chaptertitlename}{12pt}{}
\hypersetup{linkcolor=black}%
\hypersetup{citecolor=black}%
\hypersetup{urlcolor=black}%
}{}

\tableofcontents
\phantomsection
\addcontentsline{toc}{chapter}{List of Algorithms}
\begin{center}
\vspace*{-1cm}\noindent\rule{\textwidth}{1pt}\vspace*{0.3cm}
{\Large\textbf{LIST OF ALGORITHMS}}
\vspace*{0.2cm}\noindent\rule{\textwidth}{1pt}\\[0.2cm]
\end{center}

\listofalgorithms

\afterpage{\blankpage}
\phantomsection
\addcontentsline{toc}{chapter}{List of Figures}
\begin{center}
\vspace*{-1cm}\noindent\rule{\textwidth}{1pt}\vspace*{0.3cm}
{\Large\textbf{LIST OF FIGURES}}
\vspace*{0.2cm}\noindent\rule{\textwidth}{1pt}\\[0.2cm]
\end{center}

\listoffigures

\phantomsection
\addcontentsline{toc}{chapter}{Symbols and Abbreviations}
\begin{center}
\vspace*{-1cm}\noindent\rule{\textwidth}{1pt}\vspace*{0.3cm}
{\Large\textbf{SYMBOLS AND ABBREVIATIONS}}
\vspace*{0.2cm}\noindent\rule{\textwidth}{1pt}\\[0.2cm]
\end{center}

% Symbols

% Chapter 0
\Nomenclature[R]{N}{$N$}{Number of elements/nodes in a set or graph/network}

% Chapter 2
\Nomenclature[R]{X}{$X$}{Ordered set of elements/nodes labeled $\{0,1,\ldots,N-1\}$}
\Nomenclature[R]{i}{$i$}{Element/Node in a set or graph/network}
\Nomenclature[R]{ij}{$(i,j)$}{Pair of elements in a set, or relation/link in a graph/network}
\Nomenclature[R]{P}{$P$}{Partially ordered set (poset)}
\Nomenclature[R]{Aij}{$A_{ij}$}{Alexandroff set bounded by elements $i$ and $j$, where $i\prec j$}
\Nomenclature[R]{M}{$\mathbb{M}$}{Manifold (in general)}
\Nomenclature[R]{gmn}{$g_{\mu\nu}$}{Metric tensor}
\nomenclature[Z]{$\cap$}{Set intersection}
\nomenclature[Z]{$\cup$}{Set union}
\nomenclature[Z]{$\wedge$}{Logical \texttt{AND}}
\nomenclature[Z]{$\vee$}{Logical \texttt{OR}}
\nomenclature[Z]{$\sqcup$}{Set disjoint union}
\nomenclature[Z]{$\setminus$}{Set difference (relative complement)}
\nomenclature[Z]{$\veebar$}{Logical \texttt{XOR}}
\nomenclature[Z]{$\prec$}{Relational precedence operator}
\Nomenclature[R]{N}{$\mathbb{N}$}{Set of natural numbers}
\Nomenclature[R]{NZ}{$\mathbb{N}_0$}{Set of natural numbers including zero}
\nomenclature[Z]{$\varnothing$}{Empty set}
\Nomenclature[R]{Jpi}{$\mathcal{J}^+(i)$}{Set of elements proceeding element $i$}
\Nomenclature[R]{Jmi}{$\mathcal{J}^-(i)$}{Set of elements preceding element $i$}
\Nomenclature[R]{Ji}{$\mathcal{J}(i)$}{Set of elements related to element $i$}
\Nomenclature[R]{P}{$\mathcal{P}$}{Set of minimal elements in a causal set}
\Nomenclature[R]{F}{$\mathcal{F}$}{Set of maximal elements in a causal set}
\Nomenclature[R]{pf}{$(p,f)$}{Extremal pair, $p\in\mathcal{P}$ and $f\in\mathcal{F}$}
\Nomenclature[R]{HP}{$H_P$}{Height of a partial order (length of longest chain)}
\Nomenclature[R]{WP}{$W_P$}{Width of a partial order (size of largest antichain)}
\Nomenclature[R]{P}{$\mathbf{P}$}{Partial order matrix. i.e., adjacency matrix of corresponding graph/network}
\nomenclature[Z]{$\ll$}{Bit shift left}
\nomenclature[Z]{$\gg$}{Bit shift right}
\Nomenclature[R]{A}{$\mathcal{A}$}{Antichain}
\Nomenclature[R]{AX}{$\mathcal{A}_\Xi$}{Antichain and other candidates $\Xi$}
\Nomenclature[G]{NX}{$\Xi$}{Antichain candidates, or maximum vacuum selection strength (Chapter 8 only)}

% Chapter 3
\Nomenclature[R]{G}{$G$}{Graph/Network}
\Nomenclature[R]{d}{$d$}{Number of spatial dimensions}
\Nomenclature[R]{dp1}{$(d+1)$}{Number of spacetime dimensions}
\Nomenclature[R]{GMd}{$G_{\mathbb{M}^d}$}{Random geometric graph which converges to $d$-dimensional manifold $\mathbb{M}^d$ when the number of elements $N$ becomes infinite}
\Nomenclature[R]{R0}{$R_0$}{Connectivity threshold in Riemannian random geometric graph}
\Nomenclature[R]{xN}{$\mathbf{x}_N$}{Ordered set of coordinates $\{x_0,x_1,\ldots,x_{N-1}\}$}
\Nomenclature[R]{dxixj}{$d(x_i,x_j)$}{Geodesic distance between coordinates $x_i$ and $x_j$}
\Nomenclature[R]{L}{$\mathcal{L}$}{Lorentzian manifold}
\Nomenclature[R]{ds}{$ds$}{Geodesic line element}
\Nomenclature[R]{Rmn}{$R_{\mu\nu}$}{Ricci curvature tensor}
\Nomenclature[R]{R}{$R$}{Ricci scalar curvature}
\Nomenclature[G]{L}{$\Lambda$}{Cosmological constant}
\Nomenclature[R]{Tmn}{$T_{\mu\nu}$}{Stress-energy tensor}
\Nomenclature[R]{t}{$t$}{Cosmological time}
\Nomenclature[G]{S}{$\Sigma$}{Spatial hypersurface}
\Nomenclature[R]{at}{$a(t)$}{FLRW scale factor}
\Nomenclature[R]{K}{$\mathcal{K}$}{FLRW spatial curvature sign}
\Nomenclature[R]{g}{$g$}{FLRW matter content parameter}
\Nomenclature[R]{c}{$c$}{Constant proportional to matter density in FLRW spacetime}
\Nomenclature[G]{n}{$\nu$}{Poisson point process intensity}
\Nomenclature[G]{rx}{$\rho(x)$}{Coordinate probability distribution}
\Nomenclature[R]{A}{$\mathbf{A}$}{Adjacency matrix of a graph}
\Nomenclature[R]{E}{$\mathbf{E}$}{List of relations (edges) of a graph (network)}

% Chapter 4
\Nomenclature[R]{C}{$C$}{Causal set}
\Nomenclature[R]{CZ2D}{$\mathcal{C}_{2D}$}{Ensemble of 2D causal sets}
\Nomenclature[R]{CZ}{$\mathcal{C}$}{Canonical ensemble of causal sets}
\Nomenclature[R]{PNp}{$P_{N,p}$}{Ensemble of random partial orders}
\Nomenclature[R]{ZG}{$Z_G$}{Gravitational partition function}
\Nomenclature[R]{S}{$S$}{Action}
\Nomenclature[R]{hbar}{$\hbar$}{Dirac constant}
\Nomenclature[G]{h}{$\eta$}{Conformal time}
\Nomenclature[R]{CZMN}{$\mathcal{C}_\mathbb{M}(N)$}{Ensemble of $N$-element causal sets approximating the manifold $\mathbb{M}$}
\Nomenclature[R]{hij}{$h_{ij}$}{Induced metric tensor on subspace $\Sigma$}
\Nomenclature[R]{K}{$K$}{Extrinsic curvature of subspace $\Sigma$}
\nomenclature[Z]{$\Box^{(d+1)}$}{d'Alembertian in $(d+1)$ dimensions}
\Nomenclature[R]{Bdp1}{$B^{(d+1)}$}{Discrete d'Alembertian for causal sets in $(d+1)$ dimensions}
\Nomenclature[R]{l}{$\ell$}{Causal set discreteness scale}
\Nomenclature[G]{v}{$\phi(i)$}{Scalar field at graph element $i$}
\Nomenclature[R]{Lmi}{$L_m(i)$}{$m^{th}$ order inclusive order interval for causal set element $i$}
\NNomenclature[R]{Expectation}{$\mathbb{E}\big[$ \raisebox{-3pt}{\textpmhg{7}}$\big]$}{Expectation of~~\raisebox{-3pt}{\textpmhg{7}}}
\Nomenclature[R]{nm}{$n_m$}{Cardinality of $L_m$, i.e., the interval (IOI) abundances}
\Nomenclature[R]{lp}{$l_p$}{Planck length}
\Nomenclature[G]{e}{$\varepsilon$}{Smearing parameter for causal set action}
\Nomenclature[R]{fdp1}{$f_{d+1}(m,\varepsilon)$}{Smearing function for $(d+1)$-dimensional causal set action}
\Nomenclature[R]{T}{$T$}{Number of threads}
\Nomenclature[G]{hZ0}{$\eta_0$}{Maximum conformal time of a compact region of spacetime}
\Nomenclature[R]{V}{$V$}{Volume}

% Chapter 5
\Nomenclature[R]{Em}{$\mathcal{E}^m$}{$m$-dimensional Euclidean manifold}
\Nomenclature[R]{Rnpq}{$\mathcal{R}^n_{p,q}$}{$n$-dimensional pseudo-Riemannian manifold whose metric tensor has $p$ positive and $q$ negative eigenvalues}
\Nomenclature[R]{Emrs}{$\mathcal{E}^m_{r,s}$}{$m$-dimensional pseudo-Euclidean manifold whose metric tensor has $r$ positive and $s$ negative eigenvalues}
\Nomenclature[R]{Md}{$\mathcal{M}^{d+1}$}{$(d+1)$-dimensional Minkowski manifold}
\Nomenclature[R]{nm}{$n^\mu$}{Normal vector of a surface in an embedding space}
\Nomenclature[G]{y}{$\psi$}{Extra-dimensional coordinate used in an embedding space}
\Nomenclature[G]{Zmn}{$\Omega_{\mu\nu}$}{Embedding functions}
\Nomenclature[G]{v}{$\Phi(x^\mu,\psi)$}{Extra-dimensional metric function}
\Nomenclature[R]{L}{$L$}{Chain length}
\Nomenclature[R]{lZ0}{$l_0$}{Proper time (height) of a causal interval}
\Nomenclature[R]{lr}{$l(r)$}{Timelike geodesic distance (Chapter 5)}
\Nomenclature[R]{Sd}{$S_d$}{Volume of $d$-dimensional sphere}
\Nomenclature[G]{Gx}{$\Gamma(x)$}{Gamma function}
\Nomenclature[R]{wr}{$w(r)$}{Spacelike geodesic distance (Chapter 5)}
\Nomenclature[R]{W}{$W$}{Antichain width}
\Nomenclature[G]{nx}{$\xi$}{Fractional Alexandroff set size, $A_{ij}/N$}
\Nomenclature[R]{ai}{$\{a_i\}$}{Antichain indices $\{0,1,\ldots,H_P-1\}$}
\Nomenclature[R]{PZ0}{$P_0$}{Number of minimal elements in a causal set}
\Nomenclature[R]{FZ0}{$F_0$}{Number of maximal elements in a causal set}
\Nomenclature[R]{T}{$\mathcal{T}$}{Set of elements near a causal set's timelike boundaries}
\Nomenclature[R]{k}{$k$}{Degree (number of neighbors) of an element/node}
\Nomenclature[R]{B}{$\mathcal{B}$}{Chain which covers a timelike boundary}
\Nomenclature[R]{B}{$\mathfrak{B}$}{Set of boundary-covering chains $\{\mathcal{B}\}$}
\Nomenclature[G]{e}{$\epsilon$}{Threshold for elements near a timelike boundary}
\Nomenclature[G]{d}{$\delta$}{Threshold for chains near a timelike boundary}
\Nomenclature[R]{t0}{$t_0$}{Maximum cosmological time of a compact region of spacetime}
\Nomenclature[R]{si}{$s_i$}{Renormalized size of a chain, antichain, or Alexandroff set}
\Nomenclature[R]{ir}{$i_r$}{Renormalized index of a chain, antichain, or Alexandroff set}

% Chapter 6
\Nomenclature[G]{l}{$\lambda$}{de Sitter pseudo-radius, or temporal scale parameter in other FLRW manifolds}
\Nomenclature[G]{a}{$\alpha$}{Spatial scale parameter in FLRW manifolds with matter}
\Nomenclature[R]{dS}{$d\mathcal{S}^{(d+1)}$}{$(d+1)$-dimensional de Sitter manifold}
\Nomenclature[R]{z}{$\mathbf{z}$}{Embedding coordinates}
\Nomenclature[G]{grt}{$\Gamma_{\rho\tau}^\mu$}{Christoffel symbol}
\nomenclature[Z]{$\nabla_X$}{Covariant derivative with respect to tangent vector field $X$}
\Nomenclature[G]{z}{$\omega$}{Spatial distance}
\Nomenclature[R]{gtm}{$G(t;\mu)$}{Geodesic kernel}
\Nomenclature[G]{m}{$\mu$}{FLRW geodesic integration parameter}
\Nomenclature[R]{ds}{$D(\sigma)$}{Distance kernel}
\Nomenclature[G]{zc}{$\omega_c$}{Critical spatial separation beyond which spacelike geodesics have a turning point}
\Nomenclature[G]{mc}{$\mu_c$}{Critical geodesic parameter}
\Nomenclature[G]{zm}{$\omega_m$}{Maximum geodesic spatial separation for spacelike geodesics}
\Nomenclature[G]{hc}{$\eta_c$}{Conformal time at the turning point along a spacelike geodesic}
\Nomenclature[R]{f21}{${}_2F_1(a,b;c;z)$}{Gauss hypergeometric function}
\Nomenclature[R]{km}{$K(m)$}{Complete elliptic integral of the first kind}
\Nomenclature[R]{fpm}{$F(\phi\vert m)$}{Incomplete elliptic integral of the first kind}
\Nomenclature[R]{snphm}{$\sn(\phi\vert m)$}{Jacobi elliptic sine function}
\Nomenclature[R]{cnphm}{$\cn(\phi\vert m)$}{Jacobi elliptic cosine function}
\Nomenclature[R]{dnphm}{$\dn(\phi\vert m)$}{Jacobi delta amplitude function}
\Nomenclature[G]{Zl}{$\Omega_\Lambda$}{Fractional dark energy density in an FLRW spacetime}
\Nomenclature[G]{Zr}{$\Omega_R$}{Fractional radiation energy density in an FLRW spacetime}
\Nomenclature[G]{Zd}{$\Omega_D$}{Fractional dust matter energy density in an FLRW spacetime}
\Nomenclature[R]{h0}{$H_0$}{Hubble constant for our universe}
\Nomenclature[R]{ephk}{$E(\phi,k)$}{Incomplete elliptic integral of the second kind}
\Nomenclature[R]{fphkn}{$F(\phi,k,\nu)$}{Incomplete elliptic integral of the third kind}
\Nomenclature[G]{zt}{$\tilde\omega$}{Rescaled spatial distance, $(\alpha/\lambda)\omega$}

% Chapter 7
\Nomenclature[R]{Pk}{$P(k)$}{Degree probability distribution}
\Nomenclature[G]{g}{$\gamma$}{Power-law exponent describing $P(k)$}
\Nomenclature[G]{Zk}{$\Omega_K$}{Fractional curvature energy density in an FLRW spacetime}
\Nomenclature[G]{rc}{$\rho_c$}{Critical energy density}
\Nomenclature[G]{rZ0}{$\rho_0$}{Rescaled spatial cutoff of a compact region of spacetime}
\Nomenclature[G]{t}{$\tau$}{Rescaled cosmological time, $t/\lambda$}
\Nomenclature[R]{q}{$q$}{Rescaled sprinkling density (intensity) for a Poisson point process}
\Nomenclature[G]{tZ0}{$\tau_0$}{Rescaled temporal cutoff}
\Nomenclature[R]{kb}{$\bar{k}$}{Average degree of a graph/network}
\Nomenclature[R]{ps}{$p_s$}{Success ratio (measured for a greedy routing procedure)}
\Nomenclature[R]{s}{$s$}{Stretch (measured for a greedy routing procedure)}
\Nomenclature[R]{cb}{$\bar{c}$}{Average clustering of a graph/network}

% Chapter 8
\Nomenclature[R]{fi}{$f_i$}{Comoving volume of vacuum $i$}
\Nomenclature[G]{gij}{$\Gamma_{ij}$}{Vacuum transition matrix}
\Nomenclature[R]{hi}{$H_i$}{Hubble constant for vacuum $i$}
\Nomenclature[R]{Nij}{$N_{ij}$}{Number of vacua of type $j$ which transition to type $i$}
\Nomenclature[R]{sa}{$\mathbf{s}$}{Dominant eigenvector characterizing vacuum selection}
\Nomenclature[R]{Ni}{$N_i$}{Number of vacua of type $i$}
\Nomenclature[R]{pi}{$p_i$}{Fraction of vacua of type $i$}
\Nomenclature[G]{dc}{$\Delta^\circ$}{Reflexive polytope}
\Nomenclature[R]{X}{$X_a$}{Calabi-Yau manifold encoded in the polytope $\Delta^\circ_a$}
\Nomenclature[R]{Z}{$\mathbb{Z}$}{Set of integers}
\Nomenclature[G]{b}{$\beta$}{Transition rates among vacua}
\Nomenclature[R]{l}{$\mathbf{L}$}{Graph Laplacian}
\Nomenclature[R]{d}{$\mathbf{D}$}{Graph degree matrix}
\Nomenclature[R]{gt}{$G_T$}{Tree network}
\Nomenclature[R]{ge}{$G_E$}{Network of edge trees}
\Nomenclature[R]{gf}{$G_F$}{Network of face trees}
\Nomenclature[G]{u}{$\Upsilon$}{Typical vacuum selection strength}

% Chapter 9
\Nomenclature[R]{uv}{$u,v$}{Light cone coordinates}
\Nomenclature[R]{u}{$\mathfrak{u}$}{Uniform random variable in $[0,1)$}
\Nomenclature[R]{cx}{$C(x)$}{Cumulative probability distribution function}
\Nomenclature[G]{rxy}{$\rho(x,y)$}{Joint probability distribution}
\Nomenclature[G]{ryx}{$\rho(y|x)$}{Conditional probability distribution}
\Nomenclature[R]{w}{$W_0(x)$}{Principal branch of the Lambert function}
\Nomenclature[R]{NZ0}{$\langle N_0\rangle$}{Expected number of isolated graph elements}
\Nomenclature[R]{F11}{${}_1F_1(a;b;z)$}{Kummer hypergeometric function}

% Acronyms

% Chapter 0
\nomenclature[A]{GR}{General Relativity}
\nomenclature[A]{QFT}{Quantum Field Theory}

% Chapter 1/2
\nomenclature[A]{HPC}{High Performance Computing}
\nomenclature[A]{GPU}{Graphics Processing Unit}
\nomenclature[A]{FPGA}{Field Programmable Gate Array}
\nomenclature[A]{CPU}{Central Processing Unit}
\Nomenclature[A]{}{$\mu$op}{Micro-operation}
\nomenclature[A]{RAM}{Random Access Memory}
\nomenclature[A]{OoOE}{Out-of-Order Execution}
\nomenclature[A]{TLB}{Translation Lookaside Buffer}
\nomenclature[A]{ROB}{Re-order Buffer}
\nomenclature[A]{SMP}{Streaming Multiprocessor}
\nomenclature[A]{CUDA}{Compute Unified Device Architecture}
\nomenclature[A]{POD}{Plain Old Data}
\nomenclature[A]{AVX}{Advanced Vector Extensions}
\nomenclature[A]{SSE}{Streaming SIMD Extensions}
\nomenclature[A]{SIMD}{Single Instruction Multiple Data}
\nomenclature[A]{ISO}{International Organization for Standardization}

% Chapter 3
\nomenclature[A]{DAG}{Directed Acyclic Graph}
\nomenclature[A]{RGG}{Random Geometric Graph}
\nomenclature[A]{FLRW}{Friedmann-Lema\^itre-Robertson-Walker}
\nomenclature[A]{CSR}{Compressed Sparse Row}
\nomenclature[A]{PCIe}{Peripheral Component Interconnect Express}

% Chapter 4
\nomenclature[A]{EH}{Einstein-Hilbert}
\nomenclature[A]{GHY}{Gibbons-Hawking-York}
\nomenclature[A]{BD}{Benincasa-Dowker}
\nomenclature[A]{IOI}{Inclusive Order Interval}
\nomenclature[A]{MPI}{Message Passing Interface}

% Chapter 6
\nomenclature[A]{COBE}{Cosmic Background Explorer}
\nomenclature[A]{WMAP}{Wilkinson Microwave Anisotropy Probe}
\Nomenclature[A]{}{$\Lambda$CDM}{Lambda Cold Dark Matter}
\nomenclature[A]{AdS/CFT}{Anti de Sitter / Conformal Field Theory}
\Nomenclature[A]{dsmn}{$dS(M,N)$}{de Sitter Group}

% Chapter 7
\Nomenclature[A]{SON}{$SO(N)$}{Special Orthogonal Group}

% Chapter 8
\Nomenclature[A]{CY}{$CY_3$}{Calabi-Yau Threefold}
\Nomenclature[A]{GLNZ}{$GL(N,\mathbb{Z})$}{General Linear Group over $\mathbb{Z}$}
\Nomenclature[A]{SUN}{$SU(N)$}{Special Unitary Group}

\printnomenclature

\afterpage{\blankpage}

%%%%%%%%%%%
\ifthenelse{\boolean{proquest}}{}{%
\mainmatter
}
%%%%%%%%%%%

\doublespacing
\mainchapter
\pagestyle{mainstyle}

\setcounter{chapter}{-1}
\chapter{Introduction}
\vspace*{-1cm}
\singlespacing
\epigraph{\textit{The universe is a big place, perhaps the biggest.}}{\textsc{--- Kilgore Trout}}
\doublespacing
\vspace*{0.4cm}
\thispagestyle{empty}

Graph theory models well many of the real-world systems we encounter in our lives, from the Internet to online social networks to the microscopic biological systems within our bodies~\cite{albert2002statistical,newman2003structure,boccaletti2006complex,newman2006structure,newman2010networks,barabasi2016network}. Given the well-documented universality of the structure and function of these systems, it is no surprise that graph theory likewise describes much smaller systems proposed in theoretical quantum gravity and cosmology. The intersection of the latter fields with statistical physics has become increasingly relevant as we enter the era of exascale computing, i.e., when supercomputer power is on par with the human brain at the neuronal level, because theoretical work today requires ever larger simulations and other numerical experiments to test theories. The successes of computer scientists and engineers in developing these systems consequently requires us to design next-generation high performance algorithms intelligently. This dissertation provides an in depth analysis of efficient algorithm design for set and graph problems, with applications given here in quantum gravity, cosmology, and computer science. There is a strong emphasis on the broad applicability of these methods, thereby making them useful for countless other problems also modeled well by sets or graphs. 

\section{Quantum Gravity as Geometry}
General relativity (GR) relates a spacetime's matter content to its curvature, so one might expect a quantum theory of gravity should fundamentally reveal the deep connection between quantum matter, described by quantum field theory (QFT), and discrete geometry. Yet at the quantum gravity scale ($10^{-35}$m), which is at least a dozen orders of magnitude smaller than the quantum field scale ($10^{-18}$m), it is plausible spacetime is described by a discrete geometry rather than a purely continuous manifold. Therefore, the language of quantum gravity should extend beyond QFT's and GR's respective languages of statistical physics and differential geometry to include discrete geometry as well. \par

The most conservative approach to quantum gravity is causal set theory~\cite{bombelli1987space}, which assumes only that spacetime is discretized into ``spacetime atoms,'' and that the macroscopic causal structure, i.e., the Lorentz symmetry, is preserved at the quantum gravity scale. Spacetime itself is thus modeled by objects called causal sets. An interesting consequence of Lorentz invariant discretization is that the theory becomes non-local, meaning one must construct observables whose values possibly depend on an infinite amount of information. As radical as this sounds, there has been great progress in developing non-local expressions in recent years, even for what we consider to be extremely local quantities like the d'Alembertian, i.e., the second order differential operator in a spacetime. These results challenge our natural views of locality, and they ultimately could lead to an alternative non-local description of quantum physics. \par

There are several paths in contemporary causal set research, including kinematics of spacetime geometry, dynamics of spacetime growth and scalar fields, and even phenomenological predictions of the behavior of the cosmological constant. In this work, we restrict our discussion to kinematics and dynamics: we study algorithms which compute the causal set action, that is, the discrete analogue of the GR action, and we characterize the extrinsic geometry of some finite patch of spacetime in the context of convex hull analysis. Since the most fundamental expression in GR is the classical gravitational action, identifying and understanding the class of causal sets which extremize the causal set action is currently a top research priority. Yet, as described above, all expressions are non-local and, therefore, they depend on the entire set of information encoded in a spacetime patch. We will see later that non-locality implies algorithms have greater complexity than their counterparts in local theories. Thus, efficient algorithms are essential for the progression of this line of research. 

\section{Navigation in Information Networks}
Outside the scope of causal set theory, causal sets can also be interpreted simply as undirected random geometric graphs embedded in Lorentzian spaces. The recent work~\cite{krioukov2012network} showing de Sitter causal sets share certain universal properties with information networks (graphs) generated interest in whether Lorentzian latent geometric spaces explain the structural properties of real networks equally well as hyperbolic spaces do. In latent geometric models of information networks, information packets are routed between source-destination node pairs using a greedy algorithm, i.e., one in which local optimizations are used at each step in the path. The routing optimality of a latent space is characterized by the success ratio, which measures the fraction of pairs that reach their intended destination, and the stretch, which measures how closely each packet's greedy path is to the shortest path in the network. \par

The extension of greedy information routing to a Lorentzian space is non-trivial, since not only is it impossible to connect certain pairs of nodes in certain Lorentzian manifolds (the so-called geodesically incomplete manifolds), but it is also quite difficult to efficiently measure geodesic distances due to the complexity of the geodesic differential equations provided by general relativity. Moreover, these networks can be large, so simulating information routing across all node pairs is unfeasible if geodesic distances cannot be efficiently computed. In this dissertation, we develop new closed-form solutions to the geodesic differential equations for some of the most well-studied Lorentzian manifolds, which enables the study of these large network systems. We find that the success ratio tends to 100\% only for networks in spacetimes with dark energy, which implies that in terms of navigability, random geometric graphs in Lorentzian spacetimes are as good as random hyperbolic graphs.

\section{Vacuum Selection in String Theory}
String theory is another approach to quantum gravity which is older, and therefore more developed than causal set theory. One of the theoretical consequences of F-theory (one branch of string theory) is a prediction of at least $10^{755}$ possible spacetime vacua, i.e., universes, each with its own set of physical constants and other geometric properties~\cite{Halverson:2017ffz}. Understanding where the Standard Model fits into this String Landscape is difficult for a number of reasons, most notably because the landscape is so large. If the details of our vacuum are not entirely determined by the anthropic principle, then a cosmological mechanism must select vacua similar to ours from some subset of the broader landscape. \par

This string landscape problem can be considered as a set and graph problem in the multiverse picture of cosmology. This model considers an eternally inflating parent spacetime which nucleates vacua of other types (called bubbles) within it, and those vacua can further nucleate other vacua, all the way to the infinite future. If a vacuum selection mechanism exists, evidence will appear as a non-uniform bubble distribution at future infinity. Naturally, one would expect to incorporate information about transition rates, bubble decays and collisions, and additional topological data. At the present time, such information is unavailable. We make here a first attempt to study vacuum selection in an eternal inflation model of cosmology, combining for the first time network science, string theory, and cosmology in the same framework.

\section{Overview}
The following is an overview of the rest of this dissertation. Part~\ref{part:foundations} lays out important concepts in high performance computing, set theory, and graph theory. We begin in Chapter~\ref{chap:parallel} by discussing fundamental concepts in computer architecture and parallel programming. This clarifies the subsequent discussion about data structures and algorithm design, which usually reflects some constraints imposed by the processor function, memory layout and access, and peripheral device capabilities. We end the chapter by relating the hardware to specific parallelization and vectorization techniques in C/C++. Chapters~\ref{chap:sets} and~\ref{chap:graphs} then focus on compact data structures and efficient algorithms for sets and graphs. \par

Part~\ref{part:qg} brings the discussion to causal set quantum gravity, where causal sets can be modeled as partially ordered sets and as directed acyclic graphs. We discuss efficient methods to calculate the causal set action in Chapter~\ref{chap:causets} and then in Chapter~\ref{chap:chull} move toward a discussion about the convex hull of a causal set using computational geometry. \par

We then develop in Part~\ref{part:geodesics} new solutions to the geodesic differential equations in conformally flat Lorentzian spaces. These closed-form solutions and related numerical approximations are useful in a wide array of applications, but in this dissertation they are motivated by the information routing experiments conducted in the following part. \par

Part~\ref{part:apps} looks at interdisciplinary applications, including information routing in networks and vacuum selection in string theory. In Chapter~\ref{chap:navigation}, we study information routing in Lorentzian random geometric graphs using the results from Chapter~\ref{chap:geodesics} to accelerate numerical experiments. Then, Chapter~\ref{chap:multiverse} looks at how efficient set algorithms are used to construct a network using subset of the string landscape, which reveals a natural vacuum selection mechanism in an eternally inflating multiverse cosmology. \par

Finally, in Part~\ref{part:final}, we make concluding remarks in Chapter~\ref{chap:conclusion}, and then we report miscellaneous unpublished expressions for causal sets in Appendix~\ref{chap:useful}.

\section{Guide for Readers}
Since this dissertation covers a mixture of subjects, we provide here a guide for readers interested in specific material:\\
\textbf{Computer Architecture and Low-Level Optimizations:} Chapter~\ref{chap:parallel}, Sections~\ref{sec:bitset_struct}--\ref{sec:opt_poset_alg}, \ref{sec:construction},~\ref{sec:action}, and~\ref{sec:scaling}.\\
\textbf{GPU Programming:} Sections~\ref{sec:gpu_arch} and~\ref{sec:gpu_linking}.\\
\textbf{Causal Set Quantum Gravity:} Section~\ref{sec:posets},~\ref{sec:rggs}, Chapters~\ref{chap:causets}--\ref{chap:chull}, and Appendix~\ref{chap:useful}.\\
\textbf{General Relativity and Lorentzian Geometry:} Sections~\ref{sec:lorentzian_geometry},~\ref{sec:eh_action}, and~\ref{sec:emb_thm}, and Chapter~\ref{chap:geodesics}.\\
\textbf{Information Routing:} Chapter~\ref{chap:navigation}.\\
\textbf{F-Theory and Multiverse Cosmology:} Chapter~\ref{chap:multiverse}.

%\afterpage{\blankpage}

\part{High Performance Algorithms}
\label{part:foundations}
\thispagestyle{empty}
\afterpage{\blankpage}
\chapter{High Performance Computing}
\label{chap:parallel}
\thispagestyle{empty}

The rapid development of computer technology over the past half century has allowed computational science to flourish as a field in its own right. Processors today each have billions of transistors, individual computers can support terabytes of memory, and peripheral devices like graphics processing units (GPUs), coprocessors, and field-programmable gate arrays (FPGAs) are commonplace in high performance computing (HPC) systems across the world. While the impact of these developments has been felt across all segments of society, it is perhaps nowhere more obvious than in scientific computing, where researchers often  struggle to keep current with the latest devices and programming languages. \par

Today, computing and analytical skills are equally valuable. Often when we get stuck trying to solve a problem analytically, we may gain insight from large-scale simulations --- insight which then guides us in the right direction to solve the problem. In the past, it was commonplace to wait weeks or months for simulations to produce viable results, but today, if we are clever, we can redesign algorithms to get the same results in seconds or minutes. This ability to redesign algorithms is explicitly due to the availability of new hardware in HPC systems. \par

The consequence is far greater than simply an accelerated pace of scientific achievements: we can now study areas of science previously inaccessible. In particular, the rise of network science over the past two decades has transformed how we study complex systems such as the Internet, the brain, social networks, the power grid, and countless others. The availability of a nearly unlimited amount of empirical data has put network science on a pedestal at the center of applied science, and its theories grant us greater control and a better understanding of the world around us than ever before. As we strive to analyze larger and larger data in real-time, it is even more important that we write efficient algorithms and fully utilize our computational resources. \par

In this chapter, we review some of the more advanced concepts from computer science used in the rest of this dissertation, so that it becomes more clear why and how algorithms are designed in a specific manner. To begin, we review how the hardware works, including low-level details about the microarchitectures, instruction pipelines, and memory management. Then, we discuss how to leverage these low-level features using common examples. In doing so, we highlight common challenges one might face when optimizing algorithms, which we refer back to in later chapters.

\section{Processor Architectures}
\subsection{CPU Architecture}
\label{sec:cpu_arch}
The Central Processing Unit (CPU) in a computer has several important architectural features. Typically, when comparing two CPUs, we compare them by the number of \textit{physical cores}, or independent processing units, and the \textit{clock signal frequency}, i.e., the number of micro-operations (\muops) executed per second. Modern processors in HPC systems commonly have between four and 24 physical cores which run at between 2.0 and 3.8 GHz. Nevertheless, there are many other components inside a CPU which affect software performance. We review these briefly here. \par
\begin{figure}[!ht]
\centering
\includegraphics[width=\textwidth]{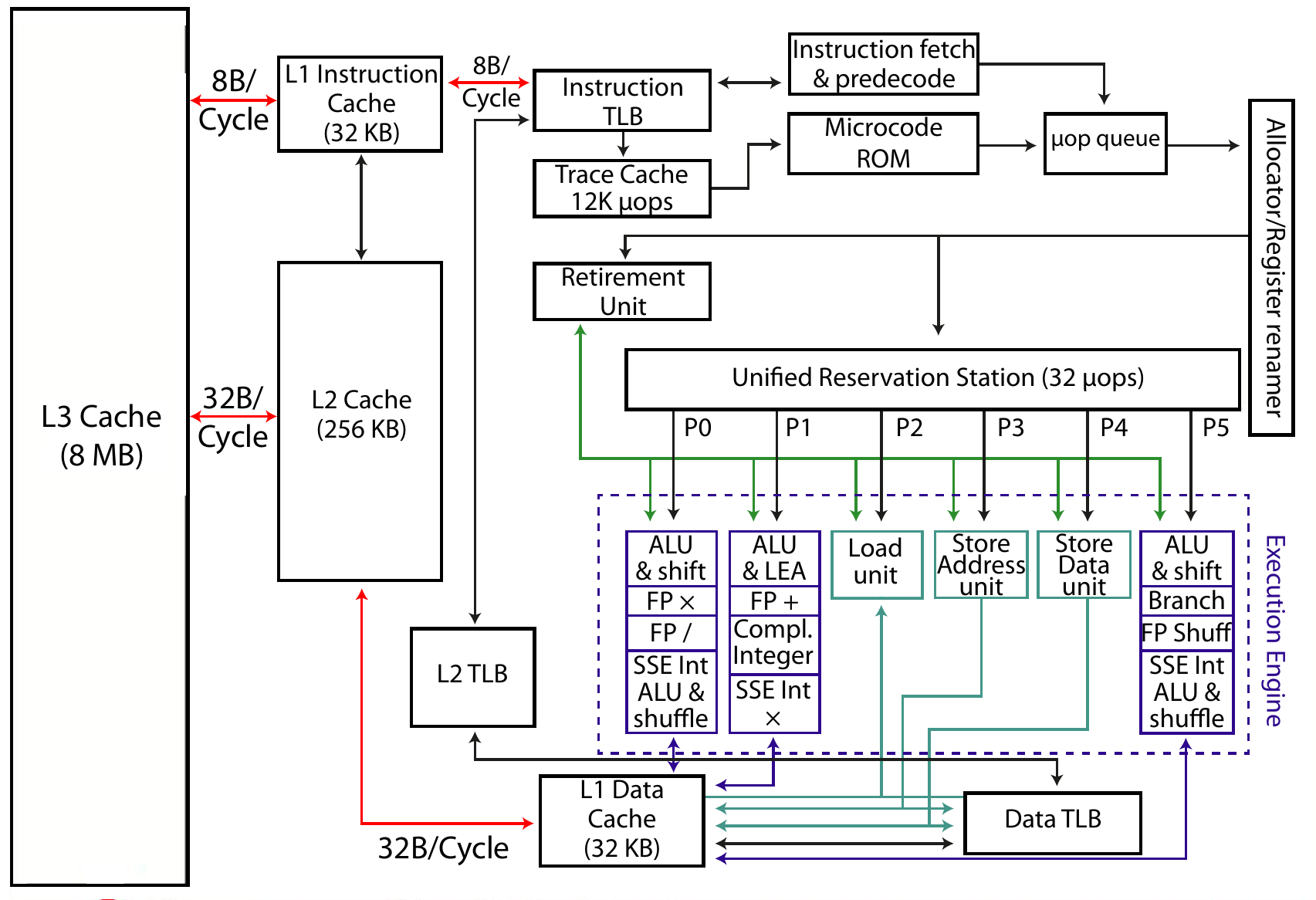}
\caption{\textbf{The block diagram of a single processor core.} Both instructions and data are loaded into their respective L1 caches via fetch operations, which pull from the RAM, L3, and L2 caches. Once decoded, operations enter the micro-operation queue, at which point register renaming and instruction reordering can occur depending on recommendations from the branch prediction table and out-of-order execution (OoOE) unit. Micro-operations are then executed in parallel by the dispatcher and scheduler. Once instructions and their operands are used in the execution unit, the retirement unit removes data from the reorder buffer when there are no further dependencies. The content of this diagram was taken from~\cite{nehalem2011block}.}
\label{fig:nehalem}
\end{figure}
\begin{figure}[!ht]
\centering
\includegraphics[width=\textwidth]{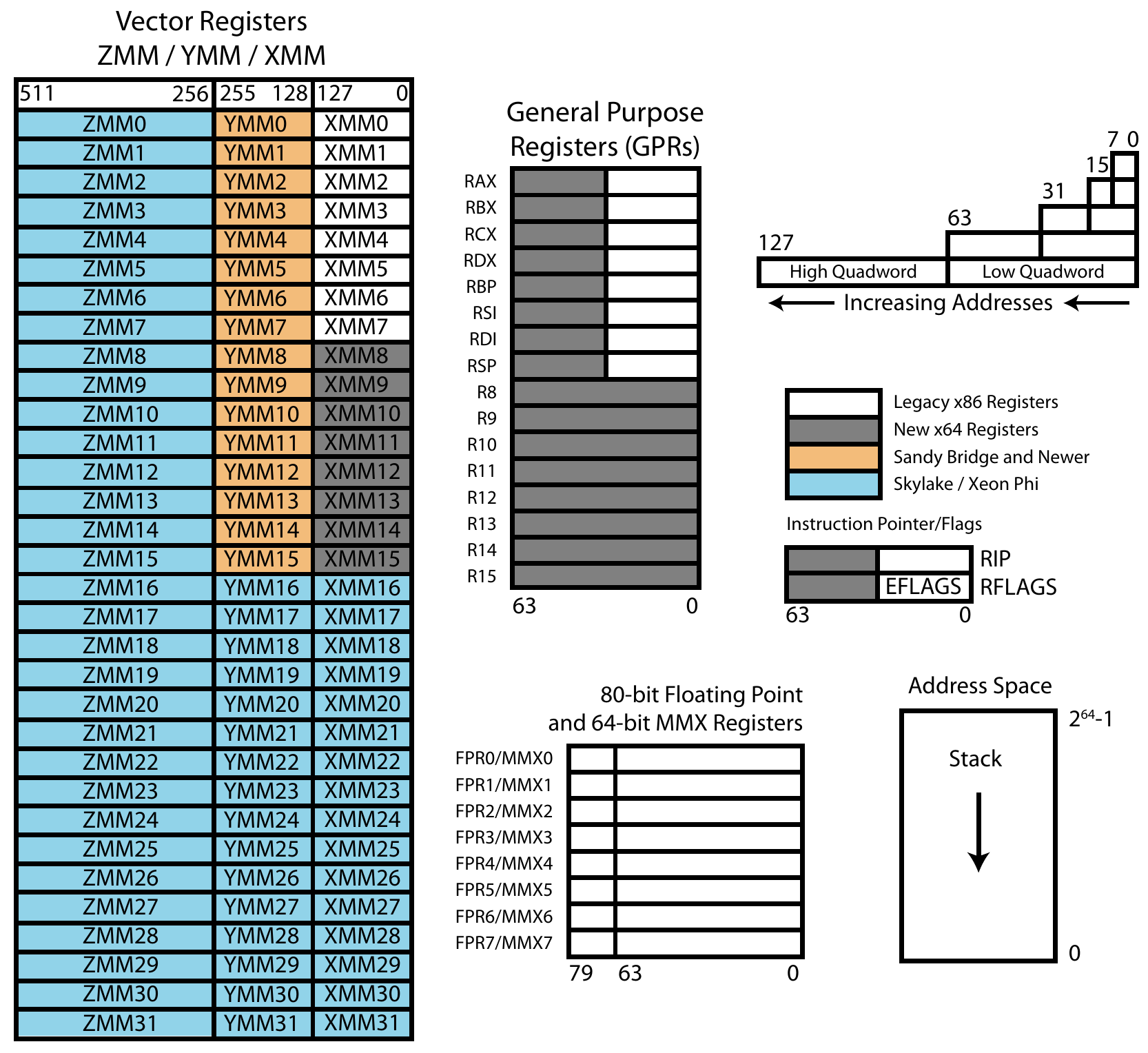}
\caption{\textbf{The Intel x64 processor registers.} The Intel x64 register set consists of 16 64-bit general purpose registers, 8 80-bit floating point and MMX registers, a 64-bit program counter register which points to the next instruction, a 64-bit instruction register which holds the current instruction, a 64-bit status register (RFLAGS) which holds results of executed instructions, 32 512-bit ZMM vector registers (which can also be addressed as 128-bit XMM or 256-bit YMM registers), as well as several other registers unavailable to the user. Adjusting how data enters and exits these registers can improve a program's runtime by orders of magnitude. Part of the information in diagram was taken from~\cite{intel2012introduction}.}
\label{fig:registers}
\end{figure}
Inside each physical core, we find many components, shown in Figure~\ref{fig:nehalem} for the Nehalem microarchitecture. A program's binary code begins in the RAM, and when executed by the operating system, a part of the control unit called the instruction unit fetches the binary instructions from the RAM into the L1 instruction cache (sometimes by way of the L3 and L2 caches). The translation lookaside buffer (TLB) and trace cache are special caches which help keep track of which memory cache to pull instructions and other data from to expedite this process. Once in the instruction cache, instructions are decoded into one or more \muops. At this point, modern processors place the decoded \muops~into a queue and reorganize them in the re-order buffer (ROB) via the out-of-order execution (OoOE) unit and branch-prediction table, described in more detail in Section~\ref{sec:x64}. The execution unit takes sequences of \muops~from the queue and executes them in parallel if possible. This is where mathematical and logical operations on operands from the L1 data cache occur. There are more integer (general purpose) registers than floating point registers, which is one of the reasons why floating point operations are slower. Once dependencies are removed, results are written to the L1 and/or L2 caches, and instructions are removed from the ROB by the retirement unit. When data is evicted from lower-level caches, any changes made by write operations are propagated upwards through the L3 cache to the RAM. \par
We now examine the lower-level details of the execution unit. A scheduler takes independent sets of instructions from the \muop~queue and the ROB and executes them in parallel if possible. At each step, the \muop~specified by the dispatcher is executed, often using one or more data operands residing in the registers. A schematic of the internal register set is shown in Figure~\ref{fig:registers}. Most data operands will enter either the 64-bit general purpose or 80-bit floating point registers. One should keep in mind that even when working with data less than 64 bits, such as 2-byte \texttt{short} or 4-byte \texttt{int} or \texttt{float} variables, the data consumes a full 64 bits, or 8 bytes. In such circumstances, one can sometimes achieve a speedup by packing data together to use the full register. For instance, when working on 32-bit data with any bitwise operations, such as the \texttt{AND}, one can achieve a 2$\times$ speedup by representing the underlying binary data in a 64-bit \texttt{long} instead of a 32-bit \texttt{int}. It is possible a compiler or instruction decoder would implement this speedup regardless of the programmer's implementation, but this is not guaranteed on all platforms, and so in the end it is better to explicitly design data structures which pack data efficiently. \par
Moreover, when working with large data arrays, one can ``vectorize'' instructions by using the vector registers, called the (128-bit) XMM, (256-bit) YMM, and (512-bit) ZMM registers. In a system with ZMM registers, the user would explicitly move data from several general purpose registers into two ZMM registers using \texttt{vmovdqa}, perform one \texttt{vpandd} operation, and then move the result back from a ZMM to several general purpose registers. Neglecting the data transfer operations, this appears to give an 8$\times$ speedup, but in practice the transfer time is on par, or even greater, than the bitwise operation itself. Hence, algorithms should be designed such that the number of $\mu$ops on data in vector registers should greatly exceed the number of data transfers.

\subsection{x64 Microarchitecture}
\label{sec:x64}
Aside from some of these basic components, it is worth reviewing some of the more advanced capabilities such as instruction latency and throughput, out-of-order execution, the cache access pattern, and prefetching. The exact behavior of these features is typically specific to a particular microarchitecture, and they can vary significantly among the different generations. They can be leveraged when writing code a particular way so that runtimes can sometimes decrease by orders of magnitude, as we find in Chapter~\ref{chap:causets}. A full review of Intel, AMD, and VIA microarchitectures can be found in~\cite{fog2017microarchitecture}. \par
\begin{algorithm}[!b]
\caption{Loop Unrolling}
\label{alg:unrolling}
\begin{algorithmic}[1]
\Input
\Statex $X$ \Comment Data vector
\Statex $N$ \Comment Length of $X$
\Statex $x$ \Comment Mask variable

\Procedure{no\_unrolling}{$X, N, x$}
\For {$i = 0;~i < N;~i\plusplus$}
\State $X[i]\xoreq x$ \Comment The operator $\veebar$ is the logical \texttt{XOR}
\EndFor
\EndProcedure

\Procedure{unrolling}{$X, N, x$}
\For {$i = 0;~i < N;~i\pluseq4$} \Comment $N$ should be divisible by $4$
\State $X[i]\xoreq x$
\State $X[i+1]\xoreq x$
\State $X[i+2]\xoreq x$
\State $X[i+3]\xoreq x$
\EndFor
\EndProcedure

\Output
\Statex $X$
\end{algorithmic}
\end{algorithm}
Instruction efficiency is measured by \textit{throughput} and \textit{latency}. The throughput of an instruction is the number of clock cycles it requires to execute, while the latency is the number of cycles before which output data is available to be used by another instruction. For example, in the Intel Skylake microarchitecture, the 64-bit \texttt{XOR} instruction has a latency of one cycle and a throughput of four cycles. As a result, if a user is performing many \texttt{XOR} operations on an array of data, the user can take advantage of \textit{instruction-level parallelism} by designing an algorithm which explicitly performs four independent \texttt{XOR} operations sequentially in an unrolled loop, and because the OoOE unit recognizes these instructions have no mutual dependencies, they will be executed concurrently. Hence, the overall runtime for $x$ instructions is $x+3$ clock cycles instead of $4x$ cycles. An example is of such a procedure is given in Algorithm~\ref{alg:unrolling}.\par

Instruction-level parallelism is possible in part due to the OoOE unit, which breaks sets of instructions into smaller interdependent groups called \textit{dependency chains}. If a program has many short dependency chains, it will run far faster than one with a few long ones. For instance, a \texttt{for} loop in which each action depends on the output of the previous action will be slow compared to one in which actions are mutually independent, since in the latter case the control unit can dispatch instructions across multiple ports in the execution unit, thereby increasing the effective throughput. Thus, a user can speed up code by manually reducing the length of dependency chains when possible. \par

A classic example is \texttt{if/else} branching. When a branch occurs in the instruction pipeline, the control unit must wait for the outcome of one of the instructions in order to decide which instruction to execute next. To accelerate this process, the CPU attempts to predict the outcome of the \texttt{if/else} statement using a \textit{branch predictor}. This unit executes the instruction predicted to come after the branch, and if it was incorrect, it modifies the branch prediction tables, which store the probabilities of outcomes, and returns the instructions to the beginning of the instruction pipeline, consequently causing a delay of 10-20 clock cycles~\cite{fog2017microarchitecture}. Since modern instruction pipelines have at least 14 stages, this can be a major interruption to other parts of the pipeline flow. \par

To avoid such a catastrophe, one can modify code by changing branch operators to mathematical ones when possible. For instance, when using a conditional accumulator, one can simply add zero when the condition is not met, as demonstrated in Algorithm~\ref{alg:branching}. In the first procedure, a variable is incremented by the vector's contents if a certain condition is met. Since that condition is mathematical, it can be re-written in the form shown in the second procedure to eliminate the branch. The key step is Operation~\ref{op:sign_shift}, where the right shift operator is used to extract the sign of $X[i]$ by looking at its 31$^{\mathrm{st}}$ bit. If non-negative, the original branching condition is met and a value $1$ is returned; otherwise $0$ is returned. The returned variable is multiplied by the vector contents, so that if the condition is not met then the accumulator gains zero. \par

\begin{algorithm}[t]
\caption{Branch Elimination}
\label{alg:branching}
\begin{algorithmic}[1]
\Input
\Statex $X$ \Comment Data vector
\Statex $N$ \Comment Length of $X$
\Statex $x$ \Comment Accumulator

\Procedure{branching}{$X, N, x$}
\For {$i = 0;~i < N;~i\plusplus$}
\If {$X[i] \geq 0$}
\State $x\pluseq X[i]$
\EndIf
\EndFor
\EndProcedure

\Procedure{no\_branching}{$X, N, x$}
\For {$i = 0;~i < N;~i\plusplus$}
\LineComment {$s$ holds the sign bit of $X[i]$}
\State $s\gets\, \sim\!(X[i] \gg 31)$ \Comment Note $\gg$ is the right shift operator \label{op:sign_shift}
\State $x\pluseq s\times X[i]$
\EndFor
\EndProcedure

\Output
\Statex $x$
\end{algorithmic}
\end{algorithm}

The way one reads memory in loops also affects efficiency. In these examples we saw data was read sequentially, but when it is read from random or irregular memory locations the performance will take a critical hit. When one requests a small variable from memory, the control unit pulls a full 64-byte \textit{cache line} through the memory hierarchy into the L1 data cache~\cite{fog2017microarchitecture}. When one subsequently reads data in adjacent memory slots, as in the \texttt{for} loops in Algorithms~\ref{alg:unrolling} and~\ref{alg:branching}, the TLB recognizes it can pull the data from the L1 cache instead of the main memory, which can be several orders of magnitude faster. This implies when working with a vector which exceeds the L1 cache size (e.g., 32 KB) it is essential that data access is sequential. When the control unit attempts to read data from the L1 cache which is not there, a \textit{cache miss} occurs and the next-highest level cache is checked. Algorithm~\ref{alg:fetch} demonstrates how this comes into play when manipulating a large matrix improperly. A 32 KB cache will fit at most 8000 32-bit integers, meaning the worst possible strategy would be to operate on data in memory locations separated by 8000 integers. This is demonstrated by accessing an $8000\times8000$ element matrix in column-major order when data is stored in row-major order (Operation~\ref{op:low_fetch}). This triggers a cache miss every single time. By instead accessing the data sequentially using row-major order (Operation~\ref{op:high_fetch}), every value in the L1 cache will be used. Typical L2 and L3 cache sizes are 512 KB and 8 MB, respectively, so all of matrix $X$ fits in this L3 cache, exactly half of it fits in this L2 cache, and just a single row fits in this L1 cache. Therefore, optimizing the memory access pattern can drastically reduce the number of cache misses when working with vector and matrix data.

\begin{algorithm}[!t]
\caption{Pipelined Cache Access}
\label{alg:fetch}
\begin{algorithmic}[1]
\Input
\Statex $X$ \Comment Data vector
\Statex $x$ \Comment Modifier variable
\Statex $N\gets8000$ \Comment Row size is 8000

\Procedure{column\_major\_access}{$A,N,x$}
\For {$i = 0;~i < N;~i\plusplus$}
\For {$j = 0;~j < N;~j\plusplus$}
\State $X[j*N+i]\pluseq x$ \Comment Low fetch utilization \label{op:low_fetch}
\EndFor
\EndFor
\EndProcedure

\Procedure{row\_major\_access}{$A,N,x$}
\For {$i = 0;~i < N;~i\plusplus$}
\For {$j = 0;~j < N;~j\plusplus$}
\State $X[i*N+j]\pluseq x$ \Comment High fetch utilization \label{op:high_fetch}
\EndFor
\EndFor
\EndProcedure

\Output
\Statex $X$
\end{algorithmic}
\end{algorithm}

Another way memory access is improved is by \textit{prefetching} instructions and data. When memory access is regular in a program, the control unit begins to fetch instructions and data it believes it will need from the higher-level caches, so they are ready in the L1 cache once other operations need them. This helps alleviate some of the delay associated with reading large amounts of data from the main memory. In general, either the compiler inserts prefetching instructions in the assembled code or the control unit issues its own prefetching instructions. It is somewhat rare that the user can manually insert prefetching statements which outperform the compiler or control unit. \par

\subsection{GPU Architecture}
\label{sec:gpu_arch}
\begin{figure}[!t]
\centering
\includegraphics[width=\textwidth]{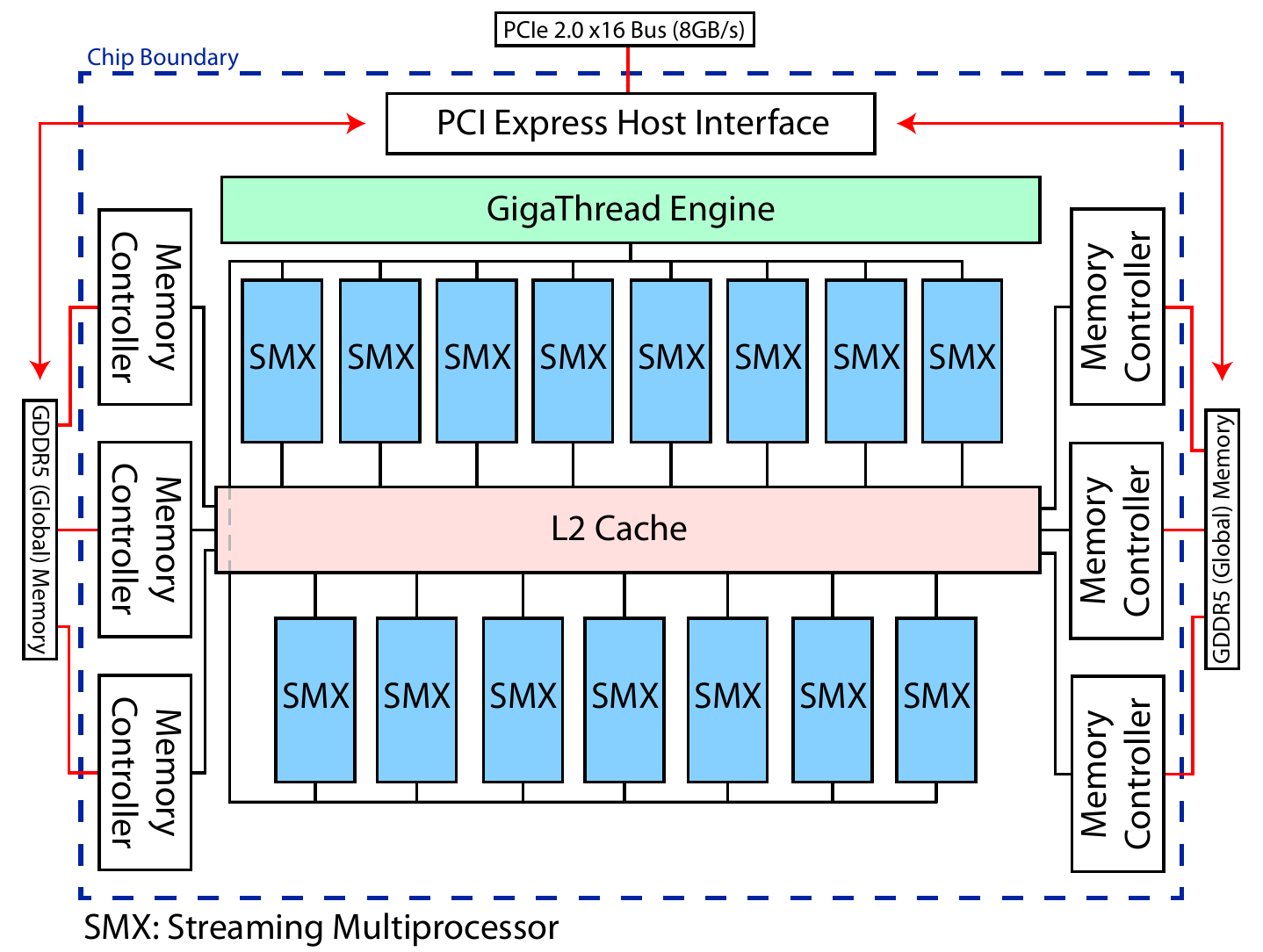}
\caption{\textbf{The block diagram of a GPU.} The NVIDIA Tesla K20X GPU has 15 streaming multiprocessors, each of which has its own L1 cache, scheduler and dispatcher, and 192 cores. The gigathread engine is the global scheduler which divides work across the multiprocessors. The information in this diagram was taken from~\cite{k20x2012block}.}
\label{fig:gpu_core}
\end{figure}

The architecture of a GPU is very different from that of a CPU. Most notably, there are far more cores and execution units, a much more diverse memory hierarchy, and a different instruction pipeline. Whereas a typical CPU has at most two dozen cores, a GPU can have several thousand cores, which places a much heavier emphasis on the role of the scheduler and dispatcher. The block diagram for a GPU is shown in some detail in Figure~\ref{fig:gpu_core}. Execution units are split into \textit{streaming multiprocessors} (SMPs or SMXs), the number of which varies among different GPUs. In the NVIDIA K20X processor, there are 15 SMPs, each of which has 192 cores. Though it initially may seem all GPU cores execute concurrently, in reality each SMP independently and asynchronously launches one or more thread blocks at a time, where thread blocks are the groups of threads guaranteed to execute concurrently. Threads in different thread blocks cannot communicate or synchronize with each other, and there is no guarantee about the order in which thread blocks will execute. \par
The user can write code for the GPU using the Compute Unified Device Architecture (CUDA) C/C++ library~\cite{cuda}. The independent processes described by the CUDA \textit{kernel function}, i.e., the programmer-defined function which runs on the GPU, are each executed by one of the cores. Within a kernel, one may access several types of memory caches on the GPU. The GDDR5 \textit{global memory} is typically several gigabytes, making it the device's largest cache. Data begins in the global memory when it is transferred from the RAM using a CUDA memory copy command. Local variables declared within a kernel are stored in memory \textit{registers}, which are small caches within each core. Within the kernel one most frequently reads and writes data directly between the global memory and the register cache, but the thread block paradigm makes the \textit{shared memory} useful as well. Shared memory is a reserved portion of the 64 KB L1 cache directly accessible by the programmer. It can be useful when many threads read the same value from global memory: instead, a thread block's master thread (thread 0) can move the data from the global memory to the L1 cache, after which the other threads in the block can read it from the spatially closer L1 cache. The thread block paradigm states that threads within the same block execute simultaneously, and that they may also synchronize with each other. The use of a synchronization statement allows one to be sure the rest of the threads do not attempt to read the shared memory until the master thread has written the data. This technique likewise is useful for writing from registers to the global memory by way of the L1 cache. An example of these procedures is given in Chapter~\ref{chap:graphs}. For a more complete description of all memory caches, refer to~\cite{cuda}.

\section{Parallel Programming}
Leveraging the architecture to optimize parallel programs is almost always challenging. Since modern HPC systems have high-end multicore processors and peripheral devices, it is important to try to design code in a way which can be parallelized, if possible at all. We will see later this can drastically change the size of physical simulations we can study in a fixed time period.

\subsection{Multithreading}
The first step in parallelizing an algorithm is identifying which parts may be performed independently, i.e., without any recursion or dependency chains. For instance, in a \texttt{for} loop in which all operations are independent of the results of all other operations, tasks within each iteration may be dispatched to different threads on the CPU using OpenMP, which is a C/C++ and Fortran library used to distribute parallel tasks~\cite{openmp}. There are other similar libraries, such as Cilk~\cite{cilk}, Thread Building Blocks~\cite{reinders2010intel}, and OpenACC~\cite{openacc}, which also parallelize code but are not studied here. \par

An example of parallel vector addition is shown in Algorithm~\ref{alg:vecadd}. The parallelization comes from Operation~\ref{op:omp}, added directly before the \texttt{for} loop, and it splits the $N$ additions evenly over all threads. If the vector size $N$ is too small, the process of forking and joining threads can take nearly an equal amount of time as the additions themselves, so often it is smart to add an \texttt{if} clause to the OpenMP code so parallelization only occurs above a certain size threshold. \par
\begin{algorithm}[!t]
\caption{Parallel Vector Addition}
\label{alg:vecadd}
\begin{algorithmic}[1]
\Input
\Statex $X$ \Comment First input vector
\Statex $Y$ \Comment Second input vector
\Statex $N$ \Comment Length of $X$ and $Y$

\Procedure{parallel\_for}{$X,Y,N$}
\State \texttt{\#pragma omp parallel for} \label{op:omp}
\For {$i=0;~i<N;~i\plusplus$}
\State $Z[i]=X[i]+Y[i]$
\EndFor
\EndProcedure

\Output
\Statex $Z$ \Comment Output vector

\end{algorithmic}
\end{algorithm}

\begin{algorithm}[!t]
\caption{Avoiding Write Conflicts}
\label{alg:wconf}
\begin{algorithmic}[1]
\Input
\Statex $X$ \Comment First input vector
\Statex $Y$ \Comment Second input vector
\Statex $N$ \Comment Length of $X$ and $Y$

\Procedure{critical\_write}{$X,Y,N$}
\State \texttt{\#pragma omp parallel for}
\For {$i=0;~i<N;~i\plusplus$}
\State \texttt{\#pragma omp critical} \Comment Useful for conditional or multi-line clauses
\If {$X[i]+Y[i]>z$}
\State $z\plusplus$
\EndIf
\EndFor
\EndProcedure

\Procedure{atomic\_write}{$X,Y,N$}
\State \texttt{\#pragma omp parallel for}
\For {$i=0;~i<N;~i\plusplus$}
\State \texttt{\#pragma omp atomic} \Comment Useful for single-line math
\State $z\pluseq X[i]+Y[i]$
\EndFor
\EndProcedure

\Procedure{reduction\_write}{$X,Y,N$}
\State \texttt{\#pragma omp parallel for reduction($+:z$)} \Comment Also good for math
\For {$i=0;~i<N;~i\plusplus$}
\State $z\pluseq X[i]+Y[i]$
\EndFor
\EndProcedure

\Procedure{local\_write}{$X,Y,N$}
\State \texttt{\#pragma omp parallel for}
\For {$i=0;~i<N;~i\plusplus$}
\State $Z[t]\pluseq X[i]+Y[i]$ \Comment Best: no write conflicts by construction
\EndFor
\For {$t=0;~t<T;~t\plusplus$}
\State $z\pluseq Z[t]$
\EndFor
\EndProcedure

\Output
\Statex $z$ \Comment Output scalar

\end{algorithmic}
\end{algorithm}
In instances when there are read/write dependencies, i.e., multiple threads writing to the same memory location or otherwise operating on the same piece of dynamically changing data, one must ensure writes occur sequentially by using a \textit{spinlock}. A spinlock is a mechanism within a scheduler which funnels parallel operations into a queue. OpenMP offers several methods to easily avoid \textit{write conflicts} by way of the \texttt{critical}, \texttt{atomic}, and \texttt{reduction} directives. The \texttt{critical} keyword is used to indicate the proceeding statement(s) should be executed by only one thread at a time. If misused, it can greatly slow down a parallel section by effectively making operations sequential, plus adding additional overhead for thread management. \par

When the statement is one of several numerical operations on integer or floating point data, one can instead use the \texttt{atomic} directive. This is implemented more efficiently in the hardware via the \texttt{lock} prefix in front of the (single) assembly instruction represented by the statement directly proceeding the directive. Since an atomic operation is fundamentally a read-modify-write sequence of operations, the \texttt{lock} prefix signals to the CPU the pipeline must be halted until the instruction completes, i.e., the cache cannot be read until the atomic operation finishes and the lock is released. \par

The \texttt{reduction} directive, which is placed in the OpenMP \texttt{parallel} clause, can be used in place of an atomic statement when modifying individual Plain-Old-Data (POD) variables, i.e., \texttt{ints} or \texttt{floats} but not arrays or classes. When the memory to which threads write is small, one can also make one copy for each of the $T$ threads, execute all write operations independently using thread-specific memory, and then perform a reduction operation, i.e., a sum, after the thread exits. This is always the best solution if enough memory is available. All four of these methods are demonstrated in Algorithm~\ref{alg:wconf}. It is not always clear which of these implementations will be most efficient, so in practice one should always benchmark. \par

\begin{algorithm}[!t]
\caption{Vectorized Addition}
\label{alg:avxadd}
\begin{algorithmic}[1]
\Input
\Statex $X$ \Comment First input vector
\Statex $Y$ \Comment Second input vector
\Statex $N$ \Comment Length of $X$ and $Y$

\Procedure{vectorized\_add}{$X,Y,N$}
\For {$i=0;~i<N;~i\pluseq 8$}
\State \texttt{ymm0} $\gets$ \texttt{\_mm256\_load\_si256}($\&X[i]$)
\State \texttt{ymm1} $\gets$ \texttt{\_mm256\_load\_si256}($\&Y[i]$)
\State \texttt{ymm0} $\gets$ \texttt{\_mm256\_add\_epi32}(\texttt{ymm0},\texttt{ymm1})
\State \texttt{\_mm256\_store\_si256}($\&Z[i]$, \texttt{ymm0})
\EndFor
\EndProcedure

\Output
\Statex $Z$ \Comment Output vector

\end{algorithmic}
\end{algorithm}

\subsection{Vectorization}
Another way to accelerate algorithms is to vectorize code using the XMM/YMM/ZMM registers described in Section~\ref{sec:cpu_arch}. Vectorization is possible whenever the same operation is performed on all elements of an array with at least $2^{10}$ elements. For instance, the vector addition of 32-bit unsigned integers is demonstrated in Algorithm~\ref{alg:avxadd}, supposing the CPU supports the vector instructions provided by the second Advanced Vector Extensions (AVX2) library~\cite{intel}. Vectorized operations usually consist of three stages: copy data to the vector registers, operate on one or more vector registers, and copy results from the vector registers to the general purpose ones. One should carefully consider the amount of time the memory copies take compared to the time consumed by the operations among the registers themselves, i.e., the second stage should take much longer than the first and third.

\afterpage{\blankpage}

\chapter{Set Algorithms}
\label{chap:sets}
\thispagestyle{empty}
%\fontfamily{mdugm}\selectfont

First formalized by Cantor~\cite{cantor1874ueber}, set theory is the branch of mathematics which describes the relations between items $x$ and collections of items $X=\{x_0,x_1,\ldots\}$. Since its inception, set theory has provided the mathematical foundations for topology, discrete geometry, and more recently quantum gravity. Hereafter, we study sequences of finite sets of geometric objects, i.e., points in a manifold, which we show converge in topology and conformal geometry to continuum spaces in the infinite-size limit. To demonstrate how such convergence occurs, we first examine the fundamental mathematical operations among sets and then construct data structures and algorithms suitable for efficient numerical experiments.

\section{Set Operations}
Sets can be formed from other sets using one or more of the four set operations, known as the \textit{intersection} $\cap$,
\begin{equation}
\label{eq:intersection}
X\cap Y=\{x : (x\in X)\,\wedge\,(x\in Y)\}\,,
\end{equation}
where $\wedge$ is the logical \texttt{AND} operator, the \textit{union} $\cup$, 
\begin{equation}
X\cup Y=\{x : (x\in X)\,\vee\,(x\in Y)\}\,,
\end{equation}
where $\vee$ is the logical \texttt{OR} operator, the \textit{disjoint union} $\sqcup$,
\begin{equation}
X\sqcup Y=\{x : ((x\in X)\,\wedge\,(x\notin Y))\,\vee\,((x\notin X)\,\wedge\,(x\in Y))\}\,,
\end{equation}
and the \textit{difference}, or \textit{relative complement} $\setminus$,
\begin{equation}
\label{eq:difference}
X\setminus Y=\{x : (x\in X)\,\wedge\,(x\notin Y)\}\,.
\end{equation}
Any more complicated set operations can be reduced to a combination of these, which in turn rely only on boolean operators. Though it is true the disjoint union could be decomposed to a union and difference operations, in the computer it is implemented as the single-cycle \texttt{XOR} instruction, so we still consider it to be fundamental. We are particularly interested in how these operations work on \textit{totally ordered} sets, i.e., sets $X$ endowed with a strict relational operator $\prec$. The order topology ($X$,$\prec$), which is the collection of all open subsets, has the properties of transitivity (if $x\prec y$ and $y\prec z$ then $x\prec z$) and irreflexivity ($x\nprec x$), and all distinct pairs of elements $\{(x,y) : (x\in X)\,\wedge\,(y\in X)\,\wedge\,x\neq y\}$ are comparable: $\forall\,\,(x,y)$ either $x\prec y$ or $y\prec x$. We index elements in a totally ordered set by a bijection to the set of non-negative integers, $X\mapsto\mathbb{N}_0$. Hereafter elements of set $X$ will be referred to by indices $i\in\mathbb{N}_0$. \par

\section{Partially Ordered Sets}
\label{sec:posets}
Along with totally ordered sets, we are also interested in the more general \textit{partially ordered} sets (posets), that is, the sets of objects within which only some pairs of elements are comparable. The topological base of finite posets which we study here is the Alexandroff topology~\cite{alexandroff1937diskrete}. Hence, an important class of subsets in a poset is the Alexandroff set $A_{ij}$, defined as the intersection of the set of elements proceeding some element $i$, denoted $\mathcal{J}^+(i)=\{j:i\prec j\}$, with the set of elements preceding a different element $j$, denoted $\mathcal{J}^-(j)=\{i:i\prec j\}$, so that $A_{ij}=\mathcal{J}^+(i)\cap \mathcal{J}^-(j)$. Notice $A_{ij}=\varnothing$ can mean two things: either $i\nprec j$ or $(i\prec j)\wedge(\nexists\, k : i\prec k \prec j)$. Since in numerical experiments we never calculate $A_{ij}$ for unrelated elements, we interpret $A_{ij}=\varnothing$ using the latter definition. We also assign a time-ordering to related elements, meaning if $i\prec j$ we say $i$ is to the past of $j$ and $j$ is to the future of $i$. \par

Partially ordered sets contain a wealth of information which can be characterized by subsets called chains and antichains. A {\it chain} is a subset in the poset $P$ which forms a clique, i.e., a total order. Conversely, an {\it antichain} is a subset of mutually unrelated elements. When a chain or antichain is inextendible with respect to the other elements in $P$ it is said to be {\it maximal}. Henceforward, all chains and antichains are understood to be maximal unless explicitly stated otherwise. We also refer later to the {\it maximum} chain and antichain in a poset, which are (one of) the largest maximal chain(s) and antichain(s), respectively. While there may exist more than one maximum chain and/or antichain, any methods described hereafter are independent of the one with which we choose to work. \par

The ends of chains form the set of {\it extremal elements}: the set of {\it minimal elements} which have no past relations, $\mathcal{P}=\{i \in P : \mathcal{J}^-(i)=\varnothing\}$, and the set of {\it maximal elements} which have no future relations, $\mathcal{F}=\{j\in P : \mathcal{J}^+(j)=\varnothing\}$. One can generate representations of posets consisting of chains and antichains using the following procedure. First, identify the maximum chain by measuring all possible chains with endpoints $(p,f)\in\mathcal{P}\times\mathcal{F}$, where a chain's two endpoints are those elements with either no past relations or no future relations. Then, exclude the {\it extremal pair} $(p,f)$ consisting of the endpoints of the maximum chain, and repeat the procedure to identify the pair which bounds the second-longest chain. This process continues until there are either no more minimal elements or no more maximal elements remaining. The set of chains which join each extremal pair is called the {\it chain representation}. This representation is an extension of Dilworth's theorem~\cite{dilworth1950decomposition}, which allows one to partition a partially ordered set into at most $W_P$ chains, where $W_P$ is the size of the largest antichain, i.e., the width of the poset. We add the additional constraints that the chains are maximal and their extremal elements do not overlap. \par

Whereas a chain is constructed by specifying two endpoints, an antichain can be generated by providing a single seed element. The most natural method to construct antichains uses the elements of the maximum chain as the seeds. By Mirsky's theorem~\cite{mirsky1971normal}, a partially ordered set of height $H_P$ may be partitioned into $H_P$ antichains, where $H_P$ is the size of the largest chain. By allowing these antichains to overlap except at the seed element, we ensure each will be maximal. This set of antichains form the {\it antichain representation} of the poset. While the chain and antichain representations are not always unique, the methods which use them remain valid, and sometimes even work better, for highly symmetric posets.

\section{Data Structures for Totally Ordered Sets}
\label{sec:bitset_struct}
In numerical experiments, we represent a totally ordered set most compactly using a binary representation, called a \textit{bitset} in computer science. A bitset can be implemented in several ways in C++, the programming language we exclusively consider hereafter. The naive approach is to use a {\tt std::vector<bool>} object. While this is a compact data structure, there is no guarantee memory is contiguously stored internally and, moreover, reading from and writing to individual locations is computationally expensive. Because the data is stored in binary, there is necessarily an internal conversion involving several bitwise and type-casting operations which make these seemingly simple operations take longer than they would for other data types. \par

The next best option is the {\tt std::bitset<>} object. This is a better option than the {\tt std::vector<bool>} because it has bitwise operators pre-defined for the object as a whole, i.e., to multiply two objects one need not use a {\tt for} loop; rather, operations like {\tt c = a \& b} are already implemented. Further, it has a bit-counting operation defined, making it easy to immediately count the number of bits set to `1' in the object. Still, there is no guarantee of contiguous memory storage and, worst of all, the size must be known at compile-time. These two limitations make this data structure impossible to use if we want to specify the size of the bitset at runtime. \par

Finally, the last option we'll examine is the {\tt boost::dynamic\_bitset<>} provided in the Boost C++ Libraries~\cite{boost}. While this is not a part of the ISO C++ Standard~\cite{cpp}, it is a well-maintained and trusted library. Boost is known for offering more efficient implementations of many common data structures and algorithms. The {\tt boost::dynamic\_bitset<>} can be dynamically sized, unlike the {\tt std::bitset<>}, the memory is stored contiguously, and it even has pre-defined bitwise and bit-counting operations. Still, it does not suit the needs of the problems we will study in Chapter~\ref{chap:causets} because it is not possible to access individual portions of the bitset: we are limited to work only with individual bits or the entire bitset. \par

Given these limitations, we present the {\tt FastBitset} class, which represents bitsets in the most efficient way for the non-local algorithms used later. Internally, the \texttt{FastBitset} is an array of 64-bit \texttt{unsigned integers} called blocks which contain matrix elements in the raw bits. We provide all four set operations (intersection, union, disjoint union, and difference) and several bit-counting operations, including variations which may be used on a proper subset of the entire object~\cite{cunningham2017generator}. The performance-critical algorithms are optimized using Intel x64 assembly with the Streaming SIMD Extensions (SSE) and Advanced Vector Extensions (AVX) instructions~\cite{intel}. \par

The posets we study can be represented by an upper-triangular matrix $\mathbf{P}$. A non-zero entry at row $i$ and column $j$ indicates the existence of the relation $i\prec j$ in the poset. For computational reasons described in Chapter~\ref{chap:graphs}, we use the symmetrized version of this matrix, i.e., $\mathbf{P}\gets\mathbf{P}+\mathbf{P}^T$. The matrix $\mathbf{P}$ is comprised of a {\tt std::vector} of {\tt FastBitset} objects, with each object corresponding to a row of the matrix.

\section{Optimized Bitset Algorithms}
\label{sec:opt_bitset_alg}
\subsection{Set Operations}
There are four set operations, but below we focus on the implementation of only one --- the set intersection --- while emphasizing that just one instruction changes in the implementations of the other three operations. The bitset intersection is simply a multiplication using the bitwise \texttt{AND} operator. The naive implementation uses a {\tt for} loop, but the optimized algorithm takes advantage of the 256-bit YMM registers located within each physical CPU core~\cite{intel}. In the following analysis, we consider processors with a Haswell or newer microarchitecture. For a review of x86 microarchitectures, see~\cite{weidendorfer2011intel,fog2017microarchitecture}. The larger width of the YMM registers means in a single CPU cycle we may perform a bitwise \texttt{AND} on four times the number of bits as in the naive implementation at the expense of moving data to and from these registers. The outline is described in Algorithm~\ref{alg:intersection}. It is important to note that for such an operation to be possible, the array of blocks must be 256-bit aligned. Any bits used as padding are always set to zero so they do not affect any results.

\begin{algorithm}[!t]
\caption{Vectorized Set Intersection}
\label{alg:intersection}
\begin{algorithmic}[1]
\Input
\Statex $X$ \Comment The first bitset
\Statex $Y$ \Comment The second bitset
\Statex $b$ \Comment The number of blocks

\Procedure{intersection}{$X, Y, b$}
\For { $i=0;~i < b;~i\pluseq4$ }
\State {\tt ymm0} $\gets X[i]$ \Comment Move data from cache to YMM registers
\State {\tt ymm1} $\gets Y[i]$
\State {\tt ymm0} $\gets$ ({\tt ymm0}) \& ({\tt ymm1}) \Comment Execute intersection \label{op:vpand}
\State $X[i] \gets$ {\tt ymm0} \Comment Move result from YMM register to cache
\EndFor
\EndProcedure

\Output
\Statex $X$ \Comment The first bitset now holds the result
\end{algorithmic}
\end{algorithm}

The implementation of the code shown inside Algorithm~\mbox{\ref{alg:intersection}'s} {\tt for} loop is written entirely in x64 assembly~\cite{cunningham2017generator}, with Operation~\ref{op:vpand} using the SIMD instruction {\tt vpand} provided by AVX2. Though it appears that only one in four bitset entries is used, the move operations imply four adjacent entries are moved, since these are 256-bit instructions. Therefore, for each set of 256 bits, we use two move operations from the L1 or L2 cache to the YMM registers, one bitwise {\tt AND} operation, and one final move operation of the result back to the general purpose registers. The bottleneck in this operation is not the bitwise operation, but rather the move instructions {\tt vmovdqu}, which limits throughput due to the bus bandwidth to these registers. As a result, it is not faster to use all of the YMM registers. It is possible the reason for this is that register renaming is already optimizing transfers by using more than just two YMM registers. Though certain prefetch instructions were tested we found no further speedup. For the other set operations,~Operation \ref{op:vpand} is replaced by SIMD instructions \texttt{vpor} in the union, \texttt{vpxor} in the disjoint union, and \texttt{vpxor} followed by \texttt{vpand} in the difference. \par

\begin{algorithm}[!t]
\caption{Vectorized Partial Intersection}
\label{alg:partial_intersection}
\begin{algorithmic}[1]
\Input
\Statex $X$ \Comment The first bit array
\Statex $Y$ \Comment The second bit array
\Statex $o$ \Comment Starting bit index
\Statex $n$ \Comment Length of subset
\Function {bitmask}{$z$}
\Return $(1\ll z)-1$
\EndFunction

\Procedure{partial\_intersection}{$X,Y,o,n$}
\LineComment Divide $o$ by 64 to get the block index
\State $x\gets o\gg 6$ \Comment $o\gg 6 \Leftrightarrow o/64$
\LineComment Indices within the blocks
\State $a\gets o \mathand 5$ \Comment $o\mathand 5\Leftrightarrow o\mathmod 64$
\State $b\gets (o+n) \mathand 5$
\If {range inside single block}
\parState {$X[x]\gets X[x] \mathand Y[x] \mathand$ \textsc{bitmask($a$)} $\mathand$ \textsc{bitmask($b$)}}
\State $u\gets 1$ \Comment Used one block
\Else
\LineComment{Intersection on full blocks}
\State $m\gets(n-1)\gg 6$ \Comment Number of full blocks
\State \textsc{intersection($X[x+1], Y[x+1], m$)}
\LineComment {Intersection on end blocks}
\State $X[x] \andeq Y[x] \mathand$ \textsc{bitmask($a$)}
\State $X[x+m] \andeq Y[x+m] \mathand$ \textsc{bitmask($b$)}
\State $u\gets m+2$ \Comment Used $m+2$ blocks
\EndIf

\LineComment {Set other blocks to zero}
\State $l\gets a$
\State $h\gets$ \textsc{$X$.get\_num\_blocks()}$-l-u$ \Comment This function is implemented in~\cite{cunningham2017generator}
\If {$l > 0$}
\State \textsc{memset($X,0,8l$)} \Comment \textsc{memset} is a standard C function
\EndIf
\If {$h > 0$}
\State \textsc{memset($X[l+u],0,8h$)}
\EndIf
\EndProcedure

\Output
\Statex $X$ \Comment The first bit array now holds the result
\end{algorithmic}
\end{algorithm}
One of the reasons the \texttt{FastBitset} data structure was developed was so we could perform these operations on a subset of two bitsets. We apply the same principle as in Algorithm~\ref{alg:intersection}, but with unwanted bits masked out, i.e., set to zero after the operation. For blocks which lie outside the range of interest, they are not even included in the {\tt for} loop. The new operation, denoted the {\it partial intersection}, is outlined in Algorithm~\ref{alg:partial_intersection}. \par

In the partial intersection algorithm, we consider two scenarios: in one the entire range of bits lies within a single block, and in the second it lies over some range of blocks, in which case the original intersection algorithm may be used on those full blocks. In either case, it is essential all bits outside the range of interest are set to zero, as indicated by the \textsc{memset} and \textsc{bitmask} function calls.

\subsection{The Bitcount}
We also want a fast way to calculate the \textit{bitcount} (or \textit{partial bitcount}), which returns the cardinality of a bitset (or a subset of the bitset), where here cardinality refers to the number of ones in the bitset. This is a well-studied operation which has many implementations and is strongly dependent on the hardware and compiler being used. The bitcount operation takes a binary string, usually in the form of an \texttt{unsigned int}, and returns the number of bits set to one. Because it is such a fundamental operation, some processors support a native assembly instruction called {\tt popcnt} which acts on a 32- or 64-bit  unsigned integer. Even on systems which support these instructions, the compiler is not always guaranteed to choose these instructions, and often it does not. For instance, the GNU function {\tt \_\_builtin\_popcount} actually uses a lookup table~\cite{gnu}, as does Boost's {\tt do\_count} method used in its {\tt dynamic\_bitset}~\cite{boost}. Both are fast, but they are not fully optimized, and for this reason we attempt to compose the fastest known implementation for the {\tt FastBitset}. When such an instruction is not supported, code should default to Boost's implementation. \par
\begin{algorithm}[!t]
\caption{Unrolled Bit Counting}
\label{alg:popcnt}
\begin{algorithmic}[1]
\Input
\Statex $X$ \Comment The bit array
\Statex $b$ \Comment The number of blocks

\Procedure{count\_bits}{$X,b$}
\LineComment {The counter variables}
\State $c[4]\gets \{0,0,0,0\}$
\For {{$i = 0;\, i < b;\, i \pluseq 4$}}
\State $X[i]\gets ${\tt popcntq($X[i]$)}
\State $c[0]\pluseq X[i]$
\State $X[i+1]\gets ${\tt popcntq($X[i+1]$)}
\State $c[1]\pluseq X[i+1]$
\State $X[i+2]\gets ${\tt popcntq($X[i+2]$)}
\State $c[2]\pluseq X[i+2]$
\State $X[i+3]\gets ${\tt popcntq($X[i+3]$)}
\State $c[3]\pluseq X[i+3]$
\EndFor
\EndProcedure

\Output
\Statex $c[0]+c[1]+c[2]+c[3]$ \Comment Number of ones in the bitset
\end{algorithmic}
\end{algorithm}
The fastest known implementation of the bitcount algorithm uses the native 64-bit CPU instruction {\tt popcntq}, where the trailing `q' indicates the instruction operates on a (64-bit) \textit{quadword} operand. While one could use a {\tt for} loop with a simple assembly call, this method would not take advantage of the modern pipeline architecture~\cite{fog2017microarchitecture} with just one call to one register. For this reason, one should unroll the loop (see Algorithm~\ref{alg:unrolling}) and perform the operation in pseudo-parallel fashion, i.e., in a way in which prefetching and prediction mechanisms will improve the instruction throughput by explicit suggestions to the out-of-order execution (OoOE) units in the CPU. We demonstrate how this works in Algorithm~\ref{alg:popcnt}. \par

This algorithm is as successful as it is because the instructions are not blocked nearly as much here as they would be if they were performed using a single register. This is because the {\tt popcnt} instruction has a latency of three cycles, but a throughput of just one cycle, meaning $x$ {\tt popcnt} instructions can be executed in $x+2$ cycles instead of $3x$ cycles when they are all independent operations~\cite{fog2017instruction}. As a result, the Intel instruction pipeline allows the four sets of operations to be performed nearly simultaneously (i.e., instruction-level parallelism) via the OoOE units. While it would be possible to extend this performance to use another four registers, this would then mean the bitset would need to be 512-bit aligned. \par

\section{Optimized Poset Algorithms}
\label{sec:opt_poset_alg}
\subsection{The Vector Product}
\label{sec:poset_vecprod}
Since the Alexandroff set is an important object, we want an efficient method to calculate the size of all Alexandroff sets within a given poset. In particular, this is motivated by numerical experiments described later in Chapter~\ref{chap:causets}. To do this, we need to successively calculate the set intersection followed by the bitcount, which together is just an inner product between two bitsets. To execute the vector product operation, we want to utilize the best features described above. If the {\tt popcnt} is performed directly after the intersection, a lot of time is wasted copying data to and from vector registers when the sum variable could be stored directly in a YMM register, for instance. Since the {\tt vmovdqu} operations are comparatively expensive, removing one out of three offers a great speedup. Furthermore, for large bitsets it is actually faster to use a vectorized implementation of the bitcount~\cite{mula2017faster}, shown in Algorithm~\ref{alg:vecprod}. Refer to ~\cite{mula2017faster} for an explanation of the low-level details of assembly operations.

\begin{algorithm}[!t]
\caption{Vectorized Inner Product}
\label{alg:vecprod}
\begin{algorithmic}[1]
\Input
\Statex $X$ \Comment The first bit array
\Statex $Y$ \Comment The second bit array
\Statex $b$ \Comment The number of blocks

\Procedure{inner\_product}{$X,Y,b$}
\State {\tt ymm2}$\gets${\tt table} \Comment Lookup table
\State {\tt ymm3}$\gets${\tt 0xf} \Comment Mask variable
\For {{$i=0;\, i<b;\, i\plusplus$}}
\State {\tt ymm0}$\gets X[i]$
\State {\tt ymm1}$\gets Y[i]$
\State {\tt ymm0}$\gets${\tt (ymm0) \& (ymm1)} \Comment Intersection
\State {\tt ymm4}$\gets${\tt (ymm0) \& (ymm3)} \Comment Lower Mask
\State {\tt ymm5}$\gets${\tt ((ymm0) $\gg 4$) \& (ymm3)} \Comment High Mask
\State {\tt ymm4}$\gets${\tt vpshufb(ymm2, ymm4)} \Comment Shuffle
\State {\tt ymm5}$\gets${\tt vpshufb(ymm3, ymm5)} \Comment Shuffle
\State {\tt ymm5}$\gets${\tt vpaddb(ymm4, ymm5)} \Comment Horiz. Add
\State {\tt ymm5}$\gets${\tt vpsadbw(ymm5, ymm7)} \Comment Horiz. Add
\State {\tt ymm6}$\gets${\tt ymm5}$+${\tt ymm6} \Comment Accumulator
\EndFor
\State $c\gets${\tt ymm6}
\EndProcedure

\Output
\Statex $c[0]+c[1]+c[2]+c[3]$ \Comment Vector product sum
\end{algorithmic}
\end{algorithm}

Algorithm~\ref{alg:vecprod} is among the best known SIMD algorithms for bit accumulation. At the very start, a lookup table and mask variable are each loaded into a YMM vector register. The lookup table is the first half of the Boost lookup table (see~\cite{mula2017faster}), stored as an {\tt unsigned char} array. These variables are essential for the instructions later to work properly, but their contents are not particularly interesting. Once the intersection is performed, two mask variables are created using the preset mask. The bits in these masks are then shuffled ({\tt vpshufb}) according to the contents of the lookup table in a way which allows the horizontal additions ({\tt vpaddb}, {\tt vpsadbw}) to store the sum of bits in each 64-bit range in the respective range. Finally, the accumulator saves these values in {\tt ymm6}. The instructions are once again paired in a way which allows the instruction throughput to be maximized via instruction-level parallelism, and the partial inner product uses a very similar setup to the partial intersection with respect to masking and {\tt memset} operations. If the bitset is too short, this algorithm will perform poorly due to the larger number of instructions, though it is easy to experimentally determine which to use on a particular system and then hard-code a threshold. \par

All of the algorithms mentioned so far may be easily modified for a system with (512-bit) ZMM registers, and we should expect the greatest speedup for the set operations. Using Intel Skylake X-series and newer processors, which support 512-bit SIMD instructions, we may replace something like {\tt vpand} with the 512-bit equivalent {\tt vpandd}. An optimal configuration today would use a Xeon E3 processor with a Kaby Lake microarchitecture, which can have up to a 3.9 GHz base clock speed, together with a Xeon Phi Knights Landing co-processor, where AVX-512 instructions may be used together with OpenMP to broadcast data over 72 physical (288 logical) cores.
\begin{algorithm}[!t]
\caption{Maximal Chain Length}
\label{alg:chain}
\begin{algorithmic}[1]
\Input
\Statex $A_{ij}$ \Comment Alexandroff set
\Statex $L$ \Comment Length array
\Statex $l$ \Comment Longest chain length
\Statex $i$ \Comment Minimal element index
\Statex $j$ \Comment Maximal element index
\Procedure{chain}{$A_{ij},L,l,i,j$}
\For {$k\in A_{ij}$} \Comment Recursively measure length from each $k$ to $j$ \label{op:k_in_A}
\State {$\kappa\gets0$}
\If {$L[k]=-1$} \Comment The distance $L_{kj}$ has not been calculated
\State {$A_{kj}\gets\mathcal{J}^+(k)\cap\mathcal{J}^-(j)$} \Comment Look at elements between $k$ and $j$
\If {$|A_{kj}|>0$} \Comment If the Alexandroff set is not empty
\State {$l^*\gets\Call{chain}{A_{kj}, L, l, k, j}$} \Comment Find the longest distance
\State {$L[k]\gets l^*$} \Comment And record the results
\State {$\kappa\gets l^*$}
\Else \Comment Otherwise, the distance is 1
\State {$L[k]\gets1$}
\State {$\kappa\gets1$}
\EndIf
\Else \Comment If it's already calculated, use the recorded value
\State {$\kappa\gets L[k]$}
\EndIf
\State {$l\gets$ \textsc{max($\kappa,l$)}} \Comment Record the largest length
\EndFor
\Return $l+1$
\EndProcedure
\Output
\Statex $l$ \Comment Length of the longest chain
\end{algorithmic}
\end{algorithm}
\subsection{Set Partitions}
\label{sec:set_partitions}
The method to identify the chain length, i.e., the longest path, between a pair of related elements $(i,j)$ is a recursive algorithm which moves from the future to the past elements in the Alexandroff set $A_{ij}\equiv\mathcal{J}^+(i)\cap\mathcal{J}^-(j)$, recording the largest distance from each element $k\in A_{ij}$ to the final element $j$ in an array $L$ during each iteration (Algorithm~\ref{alg:chain}). The distances in $L$ are initialized to $-1$ rather than $0$ to distinguish between paths which have already been traversed and those which have not. It is possible to perform these operations efficiently if the poset is stored in binary format and traversed using bitwise set operations. In a more complicated variation of Algorithm~\ref{alg:chain}, one may also extract the elements of the longest chain; see~\cite{cunningham2017generator} for details.\par

One of the more subtle parts of Algorithm~\ref{alg:chain} is Operation~\ref{op:k_in_A}, where we scan elements in the Alexandroff set $A_{ij}$. Every set, including an Alexandroff set, is stored as a \texttt{FastBitset}. These data structures are efficient to work with unless we are accessing individual elements, which means scanning for non-zero entries is inefficient if done using a \texttt{for} loop. One good alternative is to use the \texttt{bsf} (bit scan forward) instruction, which takes a 64-bit binary string, i.e., an \texttt{unsigned long}, and returns the index of the first non-zero entry. The \texttt{bsf} instruction has a latency of 3 cycles and a throughput of 1 cycle, making it ideal for writing an algorithm which uses instruction-level parallelism. In order to identify \textit{all} non-zero entries, one must reset each non-zero bit once identified until the entire bitset is empty. An example of this procedure is shown in Algorithm~\ref{alg:bsf}. \par
\begin{algorithm}[!t]
\caption{Non-Zero Elements in Alexandroff Set}
\label{alg:bsf}
\begin{algorithmic}[1]
\Input
\Statex $A_{ij}$ \Comment Alexandroff set
\Statex $b$ \Comment Number of blocks used to represent $A_{ij}$

\Procedure{scan\_bitset}{$A_{ij},b$}
\For {$k=0;\,k<b;\,k\plusplus$}
\While {$(B=A_{ij}$\texttt{.readBlock(k)}) $\neq0$}
\State $m\gets$ \texttt{bsfq($B$)}
\State $a=m+(k\ll6)$ \Comment Global non-zero index
\State Do something with $a\ldots$
\State $A_{ij}[a]\gets 0$
\EndWhile
\EndFor
\EndProcedure

\end{algorithmic}
\end{algorithm}
\begin{algorithm}[!b]
\caption{Maximal Antichain}
\label{alg:antichain}
\begin{algorithmic}[1]
\Input
\Statex $i$ \Comment Antichain seed
\Statex $\Xi$ \Comment Antichain candidates
\Procedure{antichain}{$i$, $\Xi$}
\State {$\mathcal{A}_\Xi\gets \{i\cup\Xi\}$} \Comment Initially consider $i$ and elements unrelated to $i$
\While {$|\Xi|>0$} \Comment Continue until no candidates remain
\State $\sigma\gets0$, $k\gets0$
\For {$j\in\Xi$} \Comment Consider each candidate
\State $c\gets|\mathcal{A}_\Xi-\mathcal{J}(j)|$ \Comment Find how many elements remain
\If {$c>\sigma$} \Comment Record the element which maximizes $c$
\State $\sigma\gets c$
\State $k\gets j$
\EndIf
\EndFor
\State $\mathcal{A}_\Xi\minuseq\mathcal{J}(k)$ \Comment Remove the neighbors $\mathcal{A}_\Xi$
\State $\Xi\minuseq\mathcal{J}(k)\cup k$ \Comment Remove neighbors plus the element from $\Xi$
\EndWhile
\State $\mathcal{A}\gets\mathcal{A}_\Xi$ \Comment When complete, $\mathcal{A}_\Xi$ will be the antichain
\Return $\mathcal{A}$
\EndProcedure
\Output
\Statex $\mathcal{A}$ \Comment A maximal antichain
\end{algorithmic}
\end{algorithm}
The antichain construction algorithm uses a slightly different procedure: it is a variation of the maximal independent set problem for transitively closed directed acyclic graphs~\cite{gavril1987algorithms}. Rather than implement the exact maximal antichain procedure, to save time in calculations we implement a \textit{greedy} variant~\cite{cormen1990introduction}, which uses local optimizations at each step. We first specify an initial seed element $i$, and then consider all other elements in the poset $P$ unrelated to $i$, $\Xi=P\setminus\mathcal{J}(i)$, $\mathcal{J}(i)\equiv\mathcal{J}^+(i)\cup\mathcal{J}^-(i)$, as potential candidates for the antichain $\mathcal{A}$, so that initially $\mathcal{A}_\Xi=\{i\cup\Xi\}$ and by the end of the procedure $|\Xi|\to0$ and $\mathcal{A}_\Xi\to\mathcal{A}$. In each step of the algorithm, for each $j\in\Xi$ we measure the number of elements $c$ which would remain in $\mathcal{A}_\Xi$ if the relations of $j$ were removed, i.e., $c=|\mathcal{A}_\Xi\setminus\mathcal{J}(j)|$. Keeping the element $j\,\,\in\mathcal{A}_\Xi$ which maximizes $c$ thus maximizes the size of the final antichain $\mathcal{A}$. This procedure continues until no candidates remain, $|\Xi|\to0$, at which point $\mathcal{A}_\Xi\to\mathcal{A}$ is a true maximal antichain. An implementation of this algorithm is provided in Algorithm~\ref{alg:antichain}. \par

Since Algorithm~\mbox{\ref{alg:antichain}} is a greedy algorithm, it uses a short-term optimization to avoid considering all possible antichains. While it will only very rarely produce the true maximum antichain with respect to the entire poset, it is still useful to measure width, since the true width is directly proportional to that given by this algorithm, as evidenced by results in Chapter~\ref{chap:chull}. Algorithm~\ref{alg:antichain} could easily be modified to a non-greedy version by taking all possible $j\in\Xi$ at each step, and then using a recursive method as in Algorithm~\ref{alg:chain}. These algorithms, as well as the others described in this dissertation, are implemented in C++ and Intel x64 Assembly with OpenMP and AVX optimization as part of the {\it Causal Set Generator} software package~\cite{cunningham2017generator}.
\afterpage{\blankpage}

\chapter{Graph Algorithms}
\label{chap:graphs}
\thispagestyle{empty}

Since Euler formulated the famous K\"onigsberg bridge problem in 1735~\cite{euler1736solutio}, graph theory has grown into one of the most prolific fields of mathematics, explaining the structure and evolution of many of the complex systems we encounter in our daily lives. A graph is defined as a set of objects $\{0,1,\ldots\}$ called elements together with relations among the objects $(i,j,\ldots)$ called relations, where the variables $i,j,k$ are used to index elements. We note that while the terminology for these components is different in set theory (elements and relations), graph theory (vertices and edges), and network science (nodes and links), in this dissertation we use the set theory vocabulary for consistency. The graphs we discuss here are simple directed acyclic graphs (DAGs), meaning elements cannot be related to themselves, relations are directed from one element to another, and there exist no cycles, so that the graphs can be topologically sorted, i.e., labeled elements can be ordered such that for every pairwise directed relation $(i,j)$ element $i<j$ in the ordering. Transitively closed DAGs, i.e., DAGs which possess relations $(i,k)$ when $(i,j)$ and $(j,k)$ are also present, are the natural graph representation of partial orders, since the relational operator $\prec$ translates to the directed relation $\rightarrow$ and irreflexivity is enforced by acyclicity. Therefore, the adjacency matrix for the DAG $G$ which represents a partial order $P$ is simply given by the upper triangular matrix $\mathbf{P}$ defined in Section~\ref{sec:bitset_struct}.

\section{Random Geometric Graphs}
\label{sec:rggs}
Random geometric graphs~\cite{dall2002random,penrose2003random,spodarev2013stochastic} formalize the notion of ``discretization'' of a continuous geometric space or manifold. Elements in these graphs are points, sprinkled randomly according to some sprinkling density, over the manifold, thus representing ``atoms'' of space, while links encode geometry --- two elements are related if they happen to lie close in the space. These graphs are also a central object in algebraic topology since their clique complexes~\cite{costa2015fundamental} are Rips complexes~\cite{hausmann1995vietoris,kahle2011random} whose topology is known to converge to the manifold topology under very mild assumptions~\cite{latschev2001vietoris}. \par

Given a compact region of any $d$-dimensional manifold $\mathbb{M}^d$, a geometric graph $G_{\mathbb{M}^d}(N,R_0)$ on it is a set of $N$ elements labeled $X=\{0,1,\ldots,N-1\}$ with coordinates $\mathbf{x}_N=\{x_0,x_1,\ldots,\ab x_{N-1}\}$, and undirected edges connecting pairs $(i,j)$ located at distance $d(x_i,x_j)<R_0$ in the manifold~\cite{penrose2003random}. Such a graph is called a random geometric graph (RGG) when the coordinates $\mathbf{x}$ are a realization of a Poisson or other symmetry preserving random point process, thereby defining an ensemble of RGGs. Directed Lorentzian RGGs, also known as causal sets~\cite{bombelli1987space}, converge to Lorentzian manifolds $\mathcal{L}$ in the thermodynamic limit $N\to\infty$, since the causal structure alone is enough to recover the topology and conformal geometry of a Lorentzian manifold~\cite{hawking1976new,malament1977class}. While the simplest base (open sets) of the manifold topology in the Riemannian case are open balls, this base in the Lorentzian case are Alexandroff sets, which are intersections of past and future light cones of points in the manifold~\cite{alexandroff1937diskrete,kronheimer1967structure}. Therefore, an undirected Lorentzian RGG is constructed by Poisson sprinkling points onto $\mathcal{L}$, and then linking those pairs which are timelike separated. To better understand the geometric structure of these graphs, we consider in the next section some of the finer details of Lorentzian geometry.

\subsection{Lorentzian Geometry}
\label{sec:lorentzian_geometry}
While Riemannian manifolds are manifolds with positive-definite metric tensors $g_{ij}$ defining geodesic distances $ds$ by $ds^2=\sum_{i,j=1}^dg_{ij}\,dx_i\,dx_j$, where $d$ is the manifold dimension, Lorentzian manifolds are manifolds whose metric tensors $g_\munu$, $\mu,\nu=\{0,1,\ldots,d\}$, have signature $(-++\ldots+)$, meaning that if diagonalized by a proper choice of the coordinate system, these tensors have one negative entry on the diagonal, while all other entries are positive. In general relativity, Lorentzian manifolds represent relativistic spacetimes, which are solutions of Einstein's equations. Typically, the dimension of a Lorentzian manifold is denoted by $d+1$, with the ``+1'' referring to the temporal (zeroth) dimension, while the other $d$ dimensions are spatial. The Lorentzian metric structure naturally defines spacetime's causal structure: timelike intervals with $\Delta s^2<0$ connect pairs of causally related events, i.e., timelike-separated points on a manifold. \par

Einstein's equations are a set of ten coupled non-linear partial differential equations:
\begin{equation}
\label{eq:einstein}
R_\munu - \frac12 Rg_\munu + \Lambda g_\munu = 8\pi T_\munu\,,
\end{equation}
where we use the natural units with the gravitational constant and speed of light set to unity. The Ricci curvature tensor $R_\munu$ and Ricci scalar $R$ measure the manifold curvature, the cosmological constant $\Lambda$ is proportional to the dark energy density in the spacetime, and the stress-energy tensor $T_\munu$ represents the matter content. Spacetimes which are homogeneous and isotropic are called Friedmann-Lema\^itre-Robertson-Walker (FLRW) spacetimes~\cite{griffiths2009exact}, which have a metric of the form $ds^2 = -dt^2 + a(t)^2d\Sigma^2$. The time-dependent function $a(t)$ in front of the spatial metric $d\Sigma$ is called the scale factor. This function characterizes the expansion of the volume form in a spatial hypersurface with respect to time; it alone tells whether there is a ``Big Bang'' at $t=0$, i.e., whether $a(0)=0$. The scale factor is derived explicitly as a solution to the $00$-component ($\mu=\nu=0$) of~\eqref{eq:einstein}, known as the first Friedmann equation:
\begin{equation}
\label{eq:friedmann}
\left(\frac{\dot a}{a}\right)^2 = \frac \Lambda 3 - \frac{\mathcal{K}}{a^2} + \frac{c}{a^{3g}}\,,
\end{equation}
where the variable $g$ parametrizes the type of matter in the spacetime and $c$ is a constant proportional to the matter density. The spatial curvature of the spacetime is captured by $\mathcal{K}$: $\mathcal{K}=\{+1,0,-1\}$ implies positive, zero, or negative spatial curvature, respectively. We typically study flat spacetimes, motivated by the observation that our universe is nearly spatially flat~\cite{komatsu2011seven}, or positively curved spacetimes, since spatial hypersurfaces can be constructed with no timelike boundaries and therefore simplify certain physical problems (see Chapter~\ref{chap:causets}). 

\section{Graph Construction}
\label{sec:construction}

Here we consider the numerical details of how to construct graphs. Once element coordinates are sprinkled into a particular geometric space, the graph structure is fixed. Yet identifying all pairwise relations is a computationally intensive process, so we also consider several linking algorithms, which address the problem of efficiently constructing a graph's adjacency matrix or edge list.

\subsection{Coordinate Generation}
\label{sec:coord_gen}
For a finite region of a particular Lorentzian manifold, coordinates are sampled via a Poisson point process with constant intensity $\nu$, using the normalized distributions given by the volume form of the metric.  For instance, for any $(d+1)$-dimensional FLRW spacetime with compact spatial hypersurfaces, i.e., positive spatial curvature with $\mathcal{K}=+1$, the volume form may be written
\begin{equation}
dV=a(t)^ddt\,d\Omega_d\,,
\end{equation}
where $d\Omega_d$ is the differential form for the $d$-dimensional sphere. From this expression, we find the normalized temporal coordinate distribution is $\rho(t)=a(t)^d/\int a(t^\prime)^d\,dt^\prime$, and spatial coordinates are sampled from the surface of the $d$-dimensional unit sphere. Because the $(d+1)$ coordinates of each of the $N$ elements sprinkled within a spacetime are all independent with respect to each other, these may easily be generated in parallel using OpenMP~\cite{openmp}. Refer to Section~\ref{sec:causet_sprinklings} for more detail on coordinate sampling in different spacetime regions.

\subsection[Data Structures]{Data Structures for Graphs}
A graph $G$ is described by a set of $N$ labeled elements along with a set of pairs $(i,j)$ which describe pairwise relations between elements, so the most straightforward representation uses an adjacency matrix $\mathbf{A}$ of size $N\times N$. When a graph is simple, i.e., there exist no self-loops or multiply-connected pairs, then this matrix contains only 1's and 0's, with each entry indicating the existence or non-existence of a relation between the pair of elements specified by a particular pair of row and column indices. If a graph is undirected, $\mathbf{A}$ is symmetric. If it is directed but topologically sorted, it will be upper triangular, and can therefore also be stored as a symmetric matrix if kept as $\mathbf{A}+\mathbf{A}^T$, as we discussed for a partial order $\mathbf{P}$ in Section~\ref{sec:bitset_struct}. Since the partial order $\mathbf{P}$ can be represented by a \texttt{std::vector<FastBitset>} object, a DAG can be as well.

\subsection{Pairwise Relations}
Once coordinates are assigned to the elements, the pairwise relations are found by identifying timelike-separated pairs of elements, and efficient storage requires the proper choice of the representative data structure. We represent naturally ordered partial orders as undirected graphs with topologically sorted elements, meaning elements are labeled such that an element with a larger index will never precede an element with a smaller index. In the context of a conformally flat embedding space, which is the only type we consider here, this simply means elements are sorted by their time coordinate before relations are identified. Yet this does not mean that the presented graph generation algorithms are impossible to adjust to generate RGGs in spacetimes that are not conformally flat. Indeed, in such spacetimes topological sorting can be used, as any partial order can be topologically sorted by the order-extension principle~\cite{brightwell1991counting}.

\subsubsection{Naive Linking Algorithm}
The naive implementation of the linking algorithm using the CPU uses a sparse representation in the compressed sparse row (CSR) format~\cite{sato1963techniques,tinney1967direct}. Because the elements are sorted, we require twice the memory to store sorted lists of both future-directed and past-directed relations, i.e., one list identifies relations to the future and the other those to the past. While identification of the relations is in fact only $O(N^2)$ in time, the data reformatting (list sorting) pushes it roughly to $O(N^{2.6})$, see Section~\ref{sec:convergence_times}.
\begin{algorithm}[!t]
\caption{Triangular Matrix Indexing}
\label{alg:nested_loop}
\begin{algorithmic}[1]
\Input
\Statex $N$ \Comment Matrix width

\Procedure{serial\_indexing}{$N$}
\For {$i=0;\,i<N;\,i\plusplus$}
\For {$j=i+1;\,j<N;\,j\plusplus$}
\LineComment Access element pair $(i,j)$
\EndFor
\EndFor
\EndProcedure

\Procedure{parallel\_indexing}{$N$}
\State $n\gets n+n\mathand 1$ \Comment Round $N$ up to the nearest even number
\State $p\gets n(n-1)/2$ \Comment Number of pairs
\State \texttt{\#pragma omp parallel for}
\For {$k=0;\,k<p;\,k\plusplus$}
\State $i\gets k/N$ \Comment Un-mapped indices
\State $j\gets k\mathmod N$
\State $m\gets i\geq j$ \Comment \texttt{bool} flag signals mapping
\State $i\pluseq m\times((((n\gg1)-i)\ll1)-1)$ \Comment Mapped indices
\State $j\pluseq m\times(((n\gg1)-j)\ll1)$
\If {$j=N$} \textbf{continue} \EndIf
\LineComment Access element pair $(i,j)$
\EndFor
\EndProcedure

\end{algorithmic}
\end{algorithm}
\subsubsection{Parallel Linking Algorithm}
The second implementation uses a dense graph representation and is parallelized using OpenMP. Using this dense representation for a sparse graph can waste a relatively large amount of memory compared to the information content; however, the nature of the problem described in Chapter~\ref{chap:causets} dictates a dense representation will permit much faster algorithms elsewhere. Moreover, the sparsity will depend greatly on the input parameters, so in many cases the binary adjacency matrix is the ideal representation. \par

Parallelizing the linking algorithm requires an indexing scheme for an upper-triangular matrix. It is almost always better to translate a double \texttt{for} loop into a single \texttt{for} loop before parallelization: if only the outer loop is parallelized there are still $O(N)$ jump statements (\texttt{jle} or \texttt{jnz}) which decrease instruction throughput, and if both are parallelized via nested OpenMP commands, more time is spent by the scheduler dispatching threads. To avoid these two issues, we map the upper triangular matrix to a rectangular matrix of size $(N/2)\times(N-1)$ using the formula shown in Algorithm~\ref{alg:nested_loop}. Note the bitshift operators ``$\gg$'' and ``$\ll$'' respectively half and double their operands.

\subsubsection{Naive Linking Algorithm Using the GPU}
\label{sec:gpu_linking}
While OpenMP offers a great speedup over the naive implementation, the procedure is several orders of magnitude faster when instead we use one or more Graphics Processing Units (GPUs) with the CUDA library~\cite{cuda}. Since they have many more cores than CPUs, GPUs are typically best at solving problems which require many thousands of independent low-memory tasks to be performed. There are many difficulties in designing appropriate algorithms to run on a GPU: one must consider size limitations of the global memory, which is the GPU equivalent of the RAM, and the GPU's L1 and L2 memory caches, as well as the most efficient memory access patterns. One particularly common optimization uses the shared memory, which is a reserved portion of up to 48 KB of the GPU's 64 KB L1 cache. This allows a single memory transfer from global memory to the L1 cache so that spatially local memory reads and writes by individual threads afterward are at least 10x faster. At the same time, an additional layer of synchronizations among threads in the same thread block (i.e., threads which execute concurrently) must be considered to avoid thread divergence~\cite{wong2010demystifying} and unnecessary \texttt{if/else} branching. It also puts constraints on data structures since it requires spatially local data or else the cache miss rate, i.e., the percent of time data is pulled from the RAM instead of the cache, will drastically increase. \par

The first GPU implementation offers a significant speedup by allowing each of the 2496 cores in the NVIDIA K80m (using a single GK210 processor) to perform a single comparison of two elements. The two elements are defined by the row and column indices of one of the upper triangular entries where the result will be stored, meaning one must first identify the map from a linear index $k$ to an $(i,j)$ pair, shown in detail in Algorithm~\ref{alg:idxmap}. Note that tuple variables \texttt{threadIdx} and \texttt{blockIdx} are defined by CUDA. Each thread operates on two index pairs: the first where $i<j<N/2$ or $N/2<i<j$ (Regions A/B) and the second where $i<N/2<j$ (Region C). The indices are broken up this way because the occupied portion of the lower-right quadrant of the upper-triangular adjacency matrix may be mapped to the unoccupied portion of the upper-left quadrant, thereby transforming a triangular matrix into a rectangular one in the same way as in Algorithm~\ref{alg:nested_loop}. \par
\begin{algorithm}[!t]
\caption{Triangular Matrix Indexing with CUDA}
\label{alg:idxmap}
\begin{algorithmic}[1]
\Input
\Statex $N$ \Comment Half matrix width

\Procedure{index\_cuda\_thread}{$N$}
\State $t\gets$\texttt{threadIdx.x} \Comment Thread index within thread block
\State $i\gets$\texttt{blockIdx.y} \Comment Original row index
\State $j\gets$\texttt{blockDim.x}$\times$\texttt{blockIdx.x}$+$\texttt{threadIdx.x}\Comment Original column index
\State $m\gets i\geq j$ \Comment If Region B, map to upper-left quadrant
\State $i_{ab}\gets i+m\times(((N-i)\ll 1)-1)$ \Comment Row index in Region A/B
\State $j_{ab}\gets j+m\times((N-j)\ll1)$ \Comment Column index in Region A/B
\State $i_c\gets i$\Comment Row index in Region C
\State $j_c\gets j+N$\Comment Column index in Region C
\EndProcedure

\Output
\Statex $t$ \Comment Local thread index
\Statex $(i_{ab},j_{ab})$ \Comment Mapped row/column pair (Region A/B) for a CUDA thread
\Statex $(i_c,j_c)$ \Comment Row/Column pair (Region C) for a CUDA thread

\end{algorithmic}
\end{algorithm}

We use 1-dimensional thread blocks, each with 128 threads, so that $t\in\{0,\,1,\,\ldots,\,127\}$, \texttt{blockDim.y} $=1$, and \texttt{threadIdx.y} $=0$ for all threads. These parameters are chosen for architectural reasons. Since the thread block is linear, one coordinate tuple is read by all threads in each thread block, meaning it is optimal to use the shared memory for reasons described in Section~\ref{sec:gpu_arch}. The master thread in each block (thread 0) reads data from the global memory into the L1 cache, after which there is a synchronization, and then all other threads (1-127) read from the cache. This procedure is explicitly shown in Algorithm~\ref{alg:shrmem}. \par
\begin{algorithm}[!t]
\caption{Reading from Shared Memory}
\label{alg:shrmem}
\begin{algorithmic}[1]
\Input
\Statex $\mathbf{x}$ \Comment Element coordinates
\Statex $t$ \Comment Local thread index
\Statex $(i_{ab},j_{ab})$ \Comment Row/column pair for Region A/B
\Statex $(i_c,j_c)$ \Comment Row/column pair for Region C
\State $m$ \Comment \texttt{bool} signaling mapping

\Procedure{cache\_read}{$\mathbf{x},t,i_{ab},j_{ab},i_c,j_c$}
\If {$t=0$}
\State $s^i_c\gets \mathbf{x}[i_c]$
\EndIf
\State \texttt{\_\_syncthreads()}
\State $r^i_c=s^i_c$
\State $r^j_c=\mathbf{x}[j_c]$
\State $r^i_{ab}=m\,?\,\mathbf{x}[i_{ab}]:r^i_c$
\State $r^j_{ab}=\mathbf{x}[j_{ab}]$
\EndProcedure

\Output
\Statex $(r^i_{ab},r^j_{ab})$ \Comment Coordinates for element pair in Region A/B
\Statex $(r^i_c,r^j_c)$ \Comment Coordinates for element pair in Region C

\end{algorithmic}
\end{algorithm}
While connections can be sparse, often there are multiple edges identified in a single thread block. This CUDA kernel serves to construct not only the adjacency matrix, but also the in- and out-degree vectors, which indicate the number of incoming (past) and outgoing (future) relations associated with each element. To avoid serializing writes to global memory with an atomic operation, it is best to perform a reduction operation in the L1 cache. This ensures the degree values are modified once per thread block (1 write) rather than once per thread (128 writes). The procedure which records degrees is shown in Algorithm~\ref{alg:wrdeg}. \par
\begin{algorithm}[!t]
\caption{Write Operations for Degrees}
\label{alg:wrdeg}
\begin{algorithmic}[1]
\Input
\Statex $e_{ab},e_c$ \Comment Indicators for relations
\Statex $t$ \Comment Local thread index
\Statex $m$ \Comment \texttt{bool} signaling mapping
\Statex $k_i,k_o$ \Comment In- and out-degree vectors in global memory
\Statex $n_a,n_b,n_c$ \Comment Arrays in L1 cache

\Procedure{write\_degrees}{$e_{ab},e_c,t,m,k_i,k_o,n_a,n_b,n_c$}
\State $n_a[t]\gets \,!m\times e_{ab}$
\State $n_b[t]\gets m\times e_{ab}$
\State $n_c[t]\gets e_c$
\State \texttt{\_\_syncthreads()}
\For {$s=1;\,s<128;\,s\ll=1$}
\If {$!(t\mathmod(s\ll1))$}
\State $n_a[t]\pluseq n_a[t+s]$
\State $n_b[t]\pluseq n_b[t+s]$
\State $n_c[t]\pluseq n_c[t+s]$
\EndIf
\State \texttt{\_\_syncthreads()}
\EndFor
\If {$e_{ab}=1$}
\State \texttt{atomicAdd($k_i[j_{ab}],1$)}
\EndIf
\If {$e_c=1$}
\State \texttt{atomicAdd($k_i[j_c],1$)}
\EndIf
\If {$t=0$}
\If {$n_a[0] > 0$}
\State \texttt{atomicAdd($k_o[i],n_a[0]$)}
\EndIf
\If {$n_b[0] > 0$}
\State \texttt{atomicAdd($k_o[i_{ab},n_b[0]$)}
\EndIf
\If {$n_c[0]>0$}
\State \texttt{atomicAdd($k_o[i_c],n_c[0]$)}
\EndIf
\EndIf
\EndProcedure

\Output
\Statex $k_i,k_o$ \Comment Updated degree vectors

\end{algorithmic}
\end{algorithm}
\begin{algorithm}[!t]
\caption{Write Operations for Relations}
\label{alg:wrrel}
\begin{algorithmic}[1]
\Input
\Statex $e_{ab},e_c$ \Comment Indicators for relations
\Statex $i_{ab},j_{ab},i_c,j_c$ \Comment Row/column indices
\Statex $\mathbf{E}$ \Comment Relation list
\Statex $g$ \Comment Global edge list index

\Procedure{write\_relations}{$\mathbf{E},e_{ab},e_c,i_{ab},j_{ab},i_c,j_c,g$}
\State $k\gets 0$
\If {$e_{ab}\,|\,e_c$}
\State \texttt{$k\gets$atomicAdd($g,e_{ab}+e_c$)}
\EndIf 
\If {$e_{ab}>0$}
\State $\mathbf{E}[k\plusplus]=(i_{ab}\ll32)\,|\,j_{ab}$
\EndIf
\If {$e_c>0$}
\State $\mathbf{E}[k]=(i_c\ll32)\,|\,j_c$
\EndIf
\EndProcedure

\Output
\Statex $\mathbf{E}$ \Comment Modified list holding new relations

\end{algorithmic}
\end{algorithm}
Finally, relations are written to the adjacency matrix in global memory using similar atomic operations to ensure there are no write conflicts. This procedure is shown in Algorithm~\ref{alg:wrrel}. The output is a sparse list $\mathbf{E}$ of 64-bit \texttt{unsigned ints}, so that the lower and upper 32 bits each contain a 32-bit \texttt{unsigned int} corresponding to a pair of related elements. After the list is fully generated, it is decoded on the GPU using a parallel bitonic sort~\cite{batcher1968sorting} to construct the past and future sparse lists. During this procedure, vectors containing degree data are also constructed by counting the number of writes to $\mathbf{E}$.

\subsubsection{Optimized GPU Linking Algorithm}
Despite the great increase in efficiency, this method fails if $N$ is too large for the list of relations to fit in global GPU memory or if $N$ is not a multiple of 256. The latter failure occurs because the thread block size, i.e., the number of threads guaranteed to execute concurrently, is set to 128 for architectural reasons~\footnote{On the NVIDIA K80m, which has a Compute Capability of 3.7, each thread block cannot have greater than 1024 threads, there can be at most 16 thread blocks per multiprocessor, and at the same time no greater than 2048 threads per multiprocessor.}, and the factor of two comes from the index mapping used internally which treats the adjacency matrix as four square submatrices of equal size. The second GPU implementation addresses these limitations by tiling the adjacency matrix, i.e., sending smaller submatrices to the GPU serially.  Further, when $N$ is not a round number these edge cases are handled by exiting threads with indices outside the proper bounds so that no improper memory accesses are performed. \par

This second implementation also greatly improves the speed by having each thread work on four pairs of elements instead of just one. Since each of the four pairs has the same first element by construction, the corresponding data for that element may be read into the shared memory, thereby reducing the number of accesses to global memory. Moreover, threads in the same thread block also use shared memory for the second element in each pair. Hence, since each thread block has 128 threads and each thread works on four pairs, there are only 132 reads (128+4) to global memory rather than 512 (128$\times$4), where each read consists of reading $(d+1)$ \texttt{floats} for a $(d+1)$-dimensional causal set. Finally, when the dense graph representation is used, the decoding step may be skipped, which offers a rather substantial speedup when the graph is dense. There are other optimizations to reduce the number of writes to global memory using similar techniques via the shared memory cache.

\subsubsection{Asynchronous GPU Linking Algorithm}
\label{sec:gpu_asynch}
A third version of the GPU linking algorithm also exists which uses asynchronous CUDA calls to multiple concurrent streams~\cite{cuda}. By further tiling the problem, simultaneously data can be passed to and from the GPU while another stream executes the kernel, i.e., the linking operations. This helps reduce the required bandwidth over the PCIe bus, which connects the GPU to the CPU and other devices, and can sometimes improve performance when the data transfer time is on par with the kernel execution time. We find in Section~\ref{sec:convergence_times} this does not provide as great a speedup as we expected, so this is one area for future improvement should this end up being a bottleneck in other applications.
%\afterpage{\blankpage}

\part[Applications to Causal Set Quantum Gravity]{Applications to Causal Set\\ Quantum Gravity}
\label{part:qg}
\thispagestyle{empty}
\afterpage{\blankpage}
\chapter{Causal Set Action Algorithms}
\label{chap:causets}
\thispagestyle{empty}

%\lettrine{\scalebox{2.1}{T}}{here} 
There exist a multitude of viable approaches to quantum gravity, among which causal set theory is perhaps the most minimalistic in terms of baseline assumptions. It is based on the hypothesis that spacetime at the Planck scale is composed of discrete ``spacetime atoms'' related by causality~\cite{bombelli1987space}. These ``atoms'', hereafter called elements, possess a partial order which encodes all information about the causal structure of spacetime, while the number of these elements is proportional to the spacetime volume---``Order + Number = Geometry''~\cite{sorkin2003notes}. One of the first successes of the theory was the prediction of the order of magnitude of the cosmological constant long before experimental evidence~\cite{sorkin1990spacetime}, while one of the most recent significant advances was the definition and study of a statistical partition function for the canonical causal set ensemble $\mathcal{C}$~\cite{surya2012evidence} based on the Benincasa-Dowker action~\cite{benincasa2010scalar}. This work, which examined the space of 2D orders $\mathcal{C}_{2D}\subseteq\mathcal{C}$ defined in~\cite{brightwell2008model}, provided a framework to study phase transitions and measure observables, with paths towards developing a dynamical theory of causal sets from which Einstein's equations could possibly emerge in the continuum limit. \par

\begin{figure}[!t]
\centering
\includegraphics[width=0.7\linewidth]{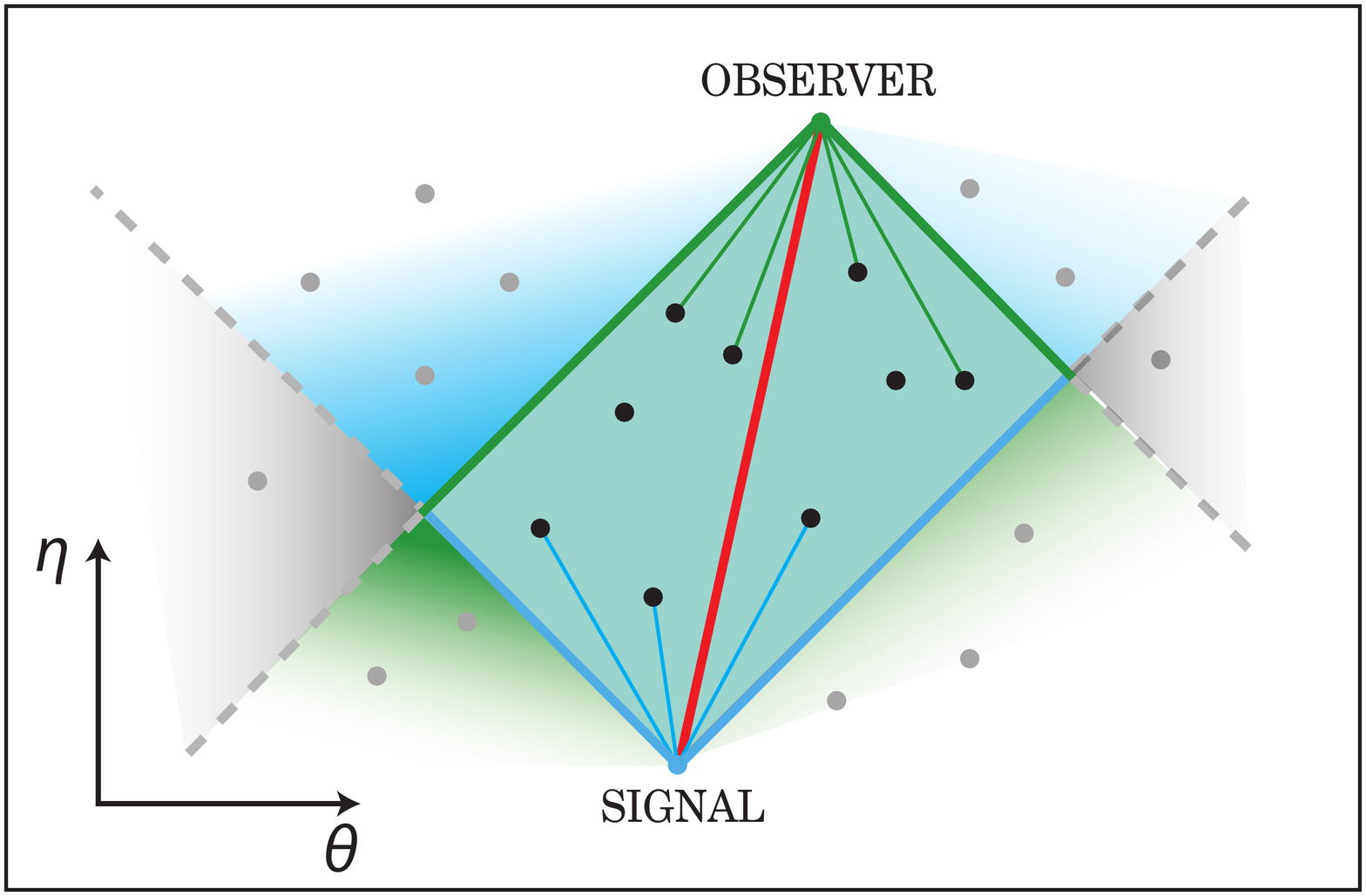}
\caption{{\bf The causal set as a random geometric graph.} \mbox{Elements} of the causal set are sprinkled uniformly at random with intensity $\nu$ into a particular region of spacetime, where $\eta$ and $\theta$ respectively refer to the temporal and spatial coordinates in $(1+1)$-dimensions. Light cones, drawn by 45-degree lines in these conformal coordinates, bound the causal future and past of each element. When light cones of a pair of elements (shown in blue and green) overlap, the elements are said to be causally related, or timelike separated, as indicated by the bold red line. The black elements both to the future of the signal and to the past of the observer form the pair's Alexandroff set shown by the teal color. Not all pairwise relations are drawn.}
\label{fig:alexandroff}
\end{figure}

Causal sets, or locally-finite posets, are the central object in the causal set approach to quantum gravity~\cite{bombelli1987space,wallden2010causal,surya2011directions}. These structures are modeled as DAGs, introduced in the previous chapter, with $N$ labeled elements $(0,1,\ldots,N-1)$ and directed pairwise relations $(i,j)$, where the direction $i\prec j$ is implied by ordering. If obtained by Poisson sprinkling onto a Lorentzian manifold, a causal set converges to the manifold in the continuum limit $N\to\infty$. These DAGs are a particular type of random geometric graph~\cite{penrose2003random}: elements are assigned coordinates in time and $d$-dimensional space via a Poisson point process with constant intensity $\nu$, and are linked pairwise if they are causally related, i.e., timelike-separated in the spacetime with respect to the underlying metric (Figure~\ref{fig:alexandroff}). As a side note, sprinkling onto a given Lorentzian manifold is definitely not the only way to generate random causal sets. The general definition of a causal set can be found in~\cite{bombelli1987space}, and random causal sets can also be obtained by sampling from the canonical ensemble $\mathcal{C}$~\cite{surya2012evidence}, or more generally, from the ensemble of random partial orders $P_{N,p}$~\cite{winkler1985random}. Due to the non-locality implied by the Lorentz invariant discretization and causal structure, causal sets have an information content which scales at least as $O(N^2)$ compared to that in competing theories of discrete spacetime which scales as $O(N)$~\cite{glaser2017finite,surya2017private,surya2017numerical}. As a result, by using the causal structure information contained in these DAG ensembles, one can recover the spacetime dimension~\cite{myrheim1978statistical,meyer1989dimension}, continuum geodesic distance~\cite{brightwell1991structure,rideout2009emergence}, spatial homology~\cite{major2006spatial,major2009stable}, differential structure~\cite{dowker2013causal,glaser2014closed,aslanbeigi2014generalized,belenchia2016continuum}, Ricci curvature~\cite{benincasa2010scalar}, and the Einstein-Hilbert action~\cite{benincasa2011random,benincasa2013action,buck2015boundary,glaser2017finite}, among other properties. \par

One of the most interesting open avenues of research is the development of the dynamical theory, which should explain both the growth of the causal set itself as well as the evolution of quantum fields and matter living on the causal set. One attempt is the Classical Sequential Growth model~\cite{rideout1999classical}, which provides a stochastic growth model for causal sets. Subsequent work has examined the dynamics of scalar fields propagating across causal sets~\cite{johnston2009feynman,sorkin2011scalar,afshordi2012distinguished,ahmed2017scalar}, including those with variable topology~\cite{buck2017sorkin}.  Other recent work uses a top-down approach with Monte Carlo dynamics to evaluate the gravitational partition function furnished by the Einstein-Hilbert action $S_{EH}$,
\begin{equation}
\label{eq:grav_partition_function}
Z_G=\int\mathcal{D}[g_{\mu\nu}]e^{iS_{EH}[g_{\mu\nu}]/\hbar}\,,
\end{equation}
using the causal set discretization of spacetime~\cite{surya2012evidence,henson2016onset,glaser2016hartle,glaser2017finite}. The causal set approach makes sense of the functional integral above by replacing it with a sum over a finite number of $N$-element causal sets $C$ belonging to some ensemble $\mathcal{C}(N)$, 
\begin{equation}
\label{eq:int_to_sum}
\int\mathcal{D}[g_{\mu\nu}]\rightarrow\sum\limits_{C\in\mathcal{C}}\,,
\end{equation}
where $\mathcal{C}$ is the collection of all causal sets, and some subset $\mathcal{C}_\mathbb{M}\subset\mathcal{C}$ will be manifold-like in the large-$N$ limit. For all $C\in\mathcal{C}_\mathbb{M}(N)$, we expect each converges when $N\to\infty$ to a Lorentzian manifold by the Hawking-Malament theorem~\cite{hawking1976new,malament1977class}, which states the causal structure alone is enough to recover a spacetime's conformal geometry and topology. At the same time, it is not known how the non-manifold-like Kleitman-Rothschild orders~\cite{kleitman1975asymptotic} are suppressed in this limit, since they entropically dominate the canonical ensemble. With these considerations, we expect all Lorentzian RGGs are manifold-like, though it is not yet known whether there exist manifold-like causal sets which cannot be obtained via a Poisson point process. Yet there is also active debate over what it means to analytically continue the action in~\eqref{eq:grav_partition_function}, since the causal set action is not an extensive property due to non-locality, and it cannot be Wick rotated~\cite{wick1954properties} because there is no time coordinate in the action, as we discuss in the following sections. To address these open questions, causal set researchers need a better understanding of both the canonical ensemble of causal sets $\mathcal{C}$ as well as the causal set action which replaces $S_{EH}$ in~\eqref{eq:grav_partition_function}. 

%\afterpage{\blankpage}

\section{The Einstein-Hilbert Action}
\label{sec:eh_action}
In many areas of physics, the action ($S$) plays the most fundamental role: using the principle of least action~\cite{maupertuis1744accord,gelfand1963calculus}, one can recover the dynamical laws of the theory as the Euler-Lagrange equations that represent the necessary condition for action extremization $\delta S = 0$. In general relativity, Einstein's field equations can be explicitly derived from the Einstein-Hilbert (EH) action,
\begin{equation}
\label{eq:eh_action}
S_{EH} = \frac{1}{2}\int\!R\left(x^\mu\right)\sqrt{-|g_{\mu\nu}|}\,dx^\mu\,,
\end{equation}
where $R$ is the Ricci scalar curvature, and then solved given a particular set of constraints~\cite{wald1984general}. However, this expression for the gravitational action is complete only for a compact manifold without boundary. It was originally realized by J.\ York~\cite{york1972role}, and later expanded upon by G.\ Gibbons and S.\ Hawking~\cite{gibbons1977action}, that integration by parts introduces new boundary terms associated with codimension-1 boundaries:
\begin{equation}
\label{eq:ghy_action}
S_{GHY}=\int_\Sigma\!K\left(x^i\right)\sqrt{|h_{ij}|}\,dx^i\,,
\end{equation}
where $K$ is the extrinsic curvature and $h_{ij}$ is the induced metric on the subspace $\Sigma$. These terms arise because the Ricci scalar contains terms linear in the second derivatives of the metric tensor, so there are also contributions from codimension-2 boundaries~\cite{hartle1981boundary,farhi1990possible,brill1992splitting,hayward1993gravitational,jubb2017boundary}. These contributions can dominate in particular spacetime regions, such as Minkowski spacetime where $S_{EH}=0$, meaning we must take them into account in numerical experiments. Finally, there can also exist a so-called non-dynamical action term which is introduced to renormalize~\eqref{eq:ghy_action} when the spatial size of a region becomes infinite. For a full treatment of such divergences, see~\cite{poisson2002advanced}. \par

If one hopes to develop a dynamical theory of quantum gravity, one would hope that either the discrete action in the quantum theory converges to~\eqref{eq:eh_action} in the large-$N$ limit, as we find with the Regge action for gravitation~\cite{regge1961general}, or an interacting theory leads to an effective action, as we see with the Wilson action in quantum chromodynamics~\cite{wilson1974confinement}. The numerical investigation of whether such a transition does indeed take place can be quite difficult: the quantum gravity scale is the Planck scale, so that if convergence is slow, it may be extremely challenging to observe it numerically. Furthermore, given the importance of boundary contributions, one must also ensure a discrete action either encapsulates all boundary terms in a single expression or there exists a separate expression for each boundary term, as recent work~\cite{buck2015boundary} has suggested. \par

In the next section, we study such a discrete action, which is one of an infinite family of solutions in the $N\to\infty$ limit~\cite{belenchia2016continuum}. Then, in Section~\ref{sec:action}, we examine efficient methods for calculating this action. These optimizations prove to be quite useful for numerical experiments, since they accelerate code by a factor of $1000$, as we find in Section~\ref{sec:simulations}.

\section{The Benincasa-Dowker Action}
\begin{figure}[!t]
\centering
\includegraphics[width=0.5\textwidth]{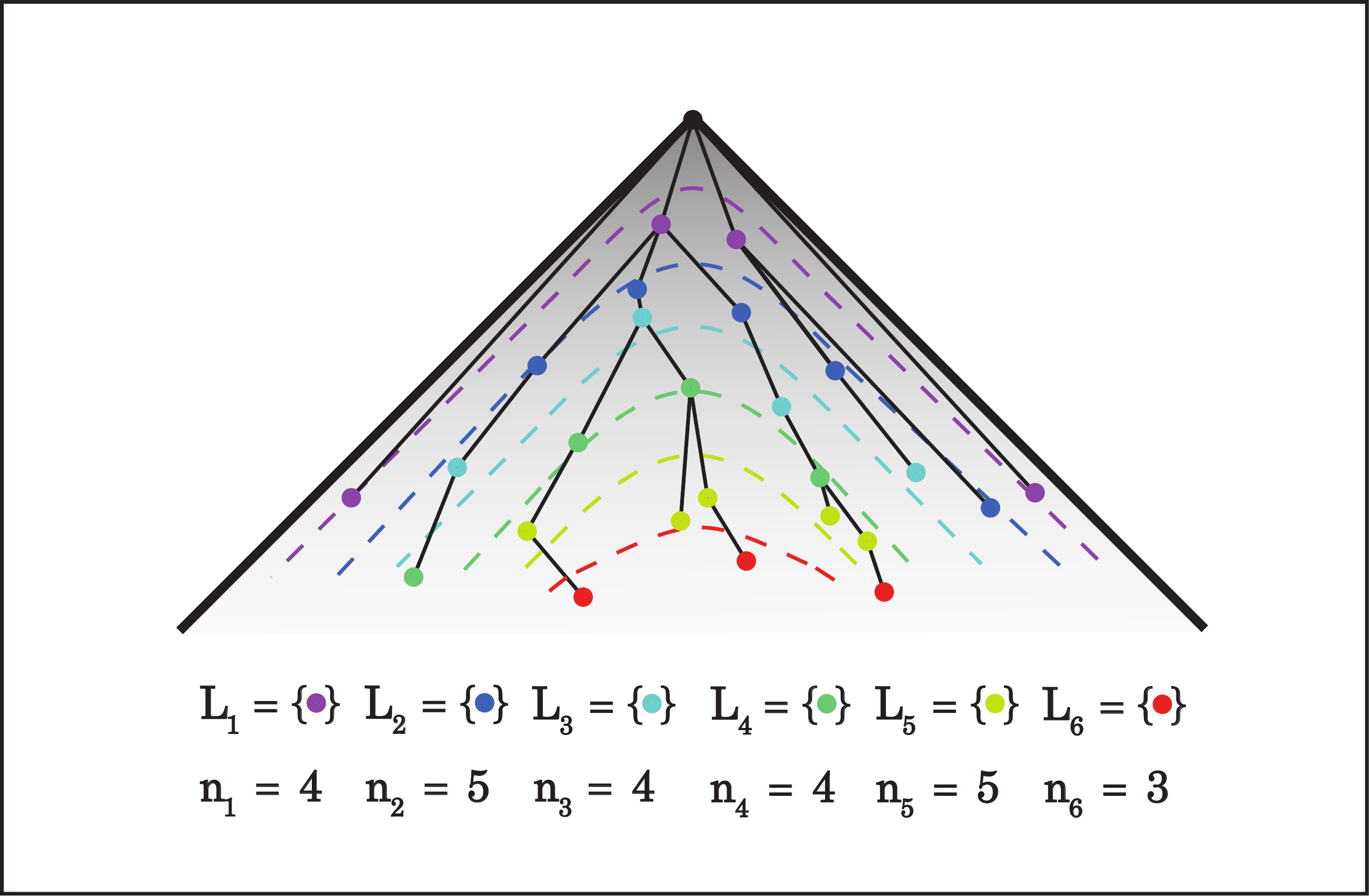}%
\includegraphics[width=0.5\textwidth]{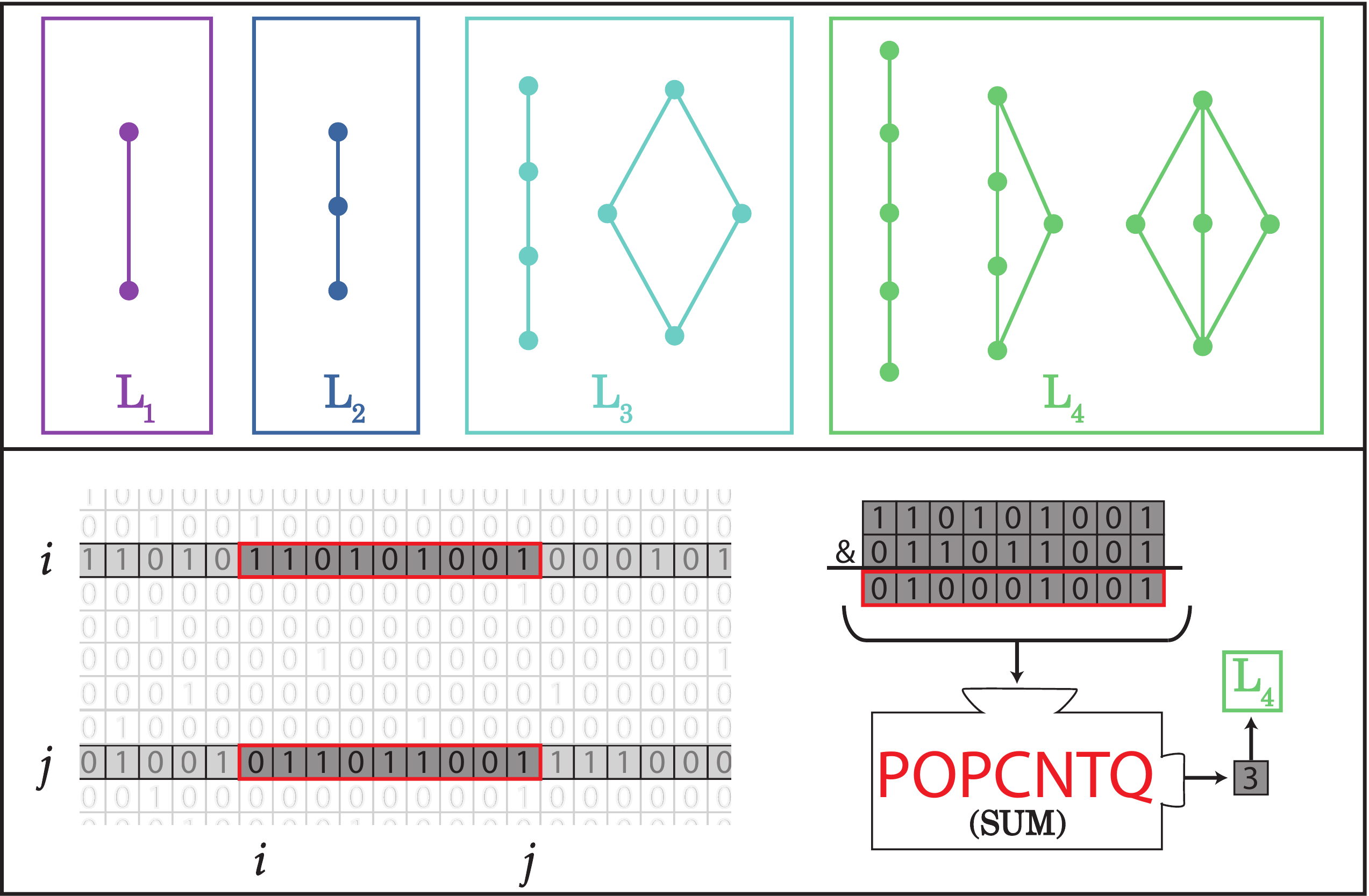}
\caption{{\bf Proper distance and the order intervals.} The left panel shows discrete hypersurfaces of constant proper time $\tau=\sqrt{x^2-t^2}$ (dashed) are approximated using the graph distance. If the black point is some element in a larger causal set, then the IOIs~\eqref{eq:IOIs} are found by counting the number of elements belonging to each hypersurface, i.e., $n_m = |L_m|$. In general the structure is not tree-like. The top of the right panel shows the subgraphs associated with each of the first four inclusive order intervals used in~(\ref{eq:s2d_local}-\ref{eq:s4d_local}), and the bottom part shows how they are detected using the causal (adjacency) matrix, assuming the graph has been topologically sorted, i.e., time-ordered. For each pair of timelike separated elements $(i,j)$, we take the inner product of rows $i$ and $j$ between columns $i$ and $j$ using the bitwise {\tt AND} in place of multiplication and the {\tt popcntq} instruction in place of a sum. The resulting value tells how many elements lie within the Alexandroff set $A_{ij}$. Details of the algorithm can be found in Section~\ref{sec:act_alg_avx}.}
\label{fig:intervals}
\end{figure}

The discrete causal set action, called the Benincasa-Dowker (BD) action, was discovered in the study of the discrete d'Alembertian ($B^{(d+1)}$), i.e., the discrete second-derivative approximating $\Box^{(d+1)}\equiv-\partial_t^2+\nabla^2$ on Lorentzian manifolds, defined in various dimensions~\cite{dowker2013causal} as
\begin{align}
B^{(1+1)}\phi(j)&=\frac{2}{\ell^2}\left[-\phi(j)+2\left(\sum\limits_{i\in L_1(j)}\phi(i)-2\sum\limits_{i\in L_2(j)}\phi(i)+\sum\limits_{i\in L_3(j)}\phi(i)\right)\right]\,,\label{eq:b2d_local}\\
B^{(2+1)}\phi(j)&=\frac{1}{\ell^2\Gamma(5/3)}\left(\frac{\pi}{3\sqrt{2}}\right)^{2/3}\left[-\phi(j)+\sum\limits_{i\in L_1(j)}\phi(i)-\frac{27}{8}\sum\limits_{i\in L_2(j)}\phi(i)+\frac{9}{4}\sum\limits_{i\in L_3(j)}\phi(i)\right]\,,\label{eq:b3d_local}\\
B^{(3+1)}\phi(j)&=\frac{4}{\ell^2\sqrt{6}}\left[-\phi(j)+\sum\limits_{i\in L_1(j)}\phi(i)-9\sum\limits_{i\in L_2(j)}\phi(i)+16\sum\limits_{i\in L_3(j)}\phi(i)-8\sum\limits_{i\in L_4(j)}\phi(i)\right]\,,\label{eq:b4d_local}
\end{align}
where $\phi(j)$ is a slowly-varying scalar field at element $j$ on the causal set, $\ell\equiv\nu^{-1/(d+1)}$ is the discreteness scale, and the $m^{th}$ order inclusive order interval (IOI) $L_m$ corresponds to the set of elements which precede $j$ with exactly $(m-1)$ elements $X$ within each open Alexandroff set,
\begin{equation}
\label{eq:IOIs}
L_m(j)=\{i : |A_{ij}|=m-1\}\,,
\end{equation}
shown in the upper-right panel of Figure~\ref{fig:intervals}. In~\cite{benincasa2010scalar} it was shown that in the continuum limit, (\ref{eq:b2d_local}-\ref{eq:b4d_local}) each converge in expectation to the continuum d'Alembertian plus another term proportional to the Ricci scalar curvature:
\begin{equation}
\label{eq:dalembertian}
\lim_{N\to\infty}\mathbb{E}\left[B\phi(i)\right] = \Box\phi(i)-\frac{1}{2}R(i)\phi(i)\,.
\end{equation}
\noindent From~(\ref{eq:b2d_local}-\ref{eq:b4d_local}) and~\eqref{eq:dalembertian} one can see when the field is constant everywhere, so that $\Box^{(d+1)}\phi=0$,~(\ref{eq:b2d_local}-\ref{eq:b4d_local}) converge to the Ricci curvature in the continuum limit, and therefore to $S_{EH}$ when summed over the entire causal set. \par

It was also shown in~\cite{benincasa2010scalar,benincasa2013action} that the corresponding expressions for the BD action are
\begin{align}
S_{BD}^{(1+1)}/\hbar&=2(N-2n_1+4n_2-2n_3)\,,\label{eq:s2d_local}\\
S_{BD}^{(2+1)}/\hbar&=\frac{1}{\Gamma(5/3)}\left(\frac{\pi}{3\sqrt{2}}\right)^{2/3}\frac{\ell}{l_p}\left(N-n_1+\frac{27}{8}n_2-\frac{9}{4}n_3\right)\,,\label{eq:s3d_local}\\
S_{BD}^{(3+1)}/\hbar&=\frac{4}{\sqrt{6}}\left(\frac{\ell}{l_p}\right)^2\left(N-n_1+9n_2-16n_3+8n_4\right)\,,\label{eq:s4d_local}
\end{align}
where $l_p$ is the Planck length and $n_m$ is the abundance of the $m^{th}$ order IOI, i.e., the cardinality of the set $L_m$ (Figure~\ref{fig:intervals}). Note we interchangeably use the terms \textit{IOI abundances}, \textit{interval abundances}, and \textit{cardinalities} to refer to the set $\{n_m\}$. While~(\ref{eq:s2d_local}-\ref{eq:s4d_local}) converge in expectation, any typical causal set tends to have a $S_{BD}$ far from the mean. This poses a serious problem for numerical experiments which already require large graphs, $N\gtrsim2^{16}$, to show convergence in curved high-dimensional spacetimes, and also suggests Monte Carlo experiments require relatively large mixing times. To partially alleviate this problem, it is not~(\ref{eq:s2d_local}-\ref{eq:s4d_local}) which one usually calculates, but rather another expression, called the \textit{smeared} or \textit{non-local} action ($S_\varepsilon$), which is obtained by averaging (or smearing) over subgraphs described by a mesoscale characterized by $\varepsilon\in(0,1)$. The new expressions which replace~(\ref{eq:s2d_local}-\ref{eq:s4d_local}) are
\begin{align}
S_\varepsilon^{(1+1)}/\hbar&=2\varepsilon\left[N-2\varepsilon\sum\limits_{m=1}^{N-1}n_mf_2(m-1,\varepsilon)\right]\,, \label{eq:s2d_smeared}\\
S_\varepsilon^{(2+1)}/\hbar&=\frac{1}{\Gamma(5/3)}\left(\frac{\pi\varepsilon}{3\sqrt{2}}\right)^{2/3}\frac{\ell}{l_p}\left[N-\varepsilon\sum\limits_{m=1}^{N-1}n_mf_3(m-1,\varepsilon)\right]\,,\\
S_\varepsilon^{(3+1)}/\hbar&=4\sqrt{\frac{\varepsilon}{6}}\left(\frac{\ell}{l_p}\right)^2\left[N-\varepsilon\sum\limits_{m=1}^{N-1}n_mf_4(m-1,\varepsilon)\right]\,,\label{eq:s4d_smeared}
\end{align}
where the smearing functions $f_{d+1}$ are given by
\begin{align}
f_2(m,\varepsilon)&=(1-\varepsilon)^m\left[1-\frac{2m\varepsilon}{1-\varepsilon}+\frac{m(m-1)\varepsilon^2}{2(1-\varepsilon)^2}\right]\,,\\
f_3(m,\varepsilon)&=(1-\varepsilon)^m\left[1-\frac{27m\varepsilon}{8(1-\varepsilon)}+\frac{9m(m-1)\varepsilon^2}{8(1-\varepsilon)^2}\right]\,,\\
f_4(m,\varepsilon)&=(1-\varepsilon)^m\left[1-\frac{9m\varepsilon}{1-\varepsilon}+\frac{8m(m-1)\varepsilon^2}{(1-\varepsilon)^2}-\frac{4m(m-1)(m-2)\varepsilon^3}{3(1-\varepsilon)^3}\right]\,.
\end{align}
The smeared action~(\ref{eq:s2d_smeared}-\ref{eq:s4d_smeared}) was shown to also converge to $S_{EH}$ in expectation, while fluctuations are greatly suppressed, so that numerical experiments with the same degree of convergence accuracy can be performed with orders of magnitude smaller graph sizes~\cite{belenchia2016continuum}. \par

\section{Action Algorithms}
\label{sec:action}
We now discuss several methods to calculate the action in numerical experiments. While one can easily implement the naive method, we find that the parallelization and vectorization techniques developed in this dissertation provide such a drastic speedup that they are worth explaining in detail. We also discuss methods to distribute these calculations among two or more computers, which is useful when $N$ becomes very large. 

\subsection{Naive Action Algorithm}
The optimizations described in the next sections which use OpenMP and AVX are orders of magnitude faster than the naive action algorithm, which we review here. The primary goal in the action algorithm is to identify the abundance $n_m$ of the subgraphs $L_m$ identified in Figure~\ref{fig:intervals}. When we use the smeared action rather than the local action, this series of subgraphs continues all the way up to those defined by the set of elements $L_{N-2}$, i.e., the largest possible subgraph is an open Alexandroff set containing $N-2$ elements. Therefore, the naive implementation of this algorithm is an $O(N^3)$ procedure which uses three nested {\tt for} loops to count the number of elements in the Alexandroff set of every pair of related elements. For each non-zero entry $(i,j)$ of the causal matrix, with $i<j$ due to time-ordering, we calculate the number of elements $k$ both the future of element $i$ and to the past of element $j$ and then add one to the array of interval abundances at index $k$. This algorithm is summarized in Algorithm~\ref{alg:naive_cardinalities}.

\begin{algorithm}[!t]
\caption{Naive Interval Abundance Measurement}
\label{alg:naive_cardinalities}
\begin{algorithmic}[1]
\Input
\Statex $A$ \Comment Adjacency matrix
\Statex $N$ \Comment Number of elements in causal set
\Statex $n$ \Comment Array for interval abundances

\Procedure{naive\_ioi\_measurement}{$A,N,n$}
\For {$i=0;\,i<N-1;\,i\plusplus$} \Comment We look at all Alexandroff sets $A_{ij}$
\For {$j=i+1;\,j<N;\,j\plusplus$}
\State $x\gets 0$
\If {$i\nprec j$} \textbf{continue} \EndIf
\For {$k=i+1;\,k<j;\,k\plusplus$}
\If {$i\prec k$ \textbf{and} $k\prec j$} \Comment Check if $k$ is in the Alexandroff set $A_{ij}$
\State $x\plusplus$
\EndIf
\EndFor
\State $n[x+1]\plusplus$
\EndFor
\EndFor
\EndProcedure

\Output
\Statex $n$ \Comment The interval abundances
\end{algorithmic}
\end{algorithm}

\subsection{Parallel Action Algorithm}
The most obvious optimization of Algorithm~\ref{alg:naive_cardinalities} uses OpenMP to parallelize the two outer loops of the naive action algorithm, since the properties of each Alexandroff set in the causal set are mutually independent. Therefore, we combine the two outer loops into a single loop of size $N(N-1)/2$ which is parallelized with OpenMP, and then keep the final inner loop serialized. When we do this, we must make sure we avoid write conflicts to the interval abundance array: if two or more threads try to modify the same spot in the array, some attempts may fail. To avoid this, we generate $T$ copies of this array so that each of the $T$ threads can write to its own array. After the action algorithm has finished, we perform a reduction on the $T$ arrays to add all results to the first array in the master thread. This algorithm still scales like $O(N^3)$ since the outer loop is still $O(N^2)$ in size. \par

\subsection{Vectorized Action Algorithm}
\label{sec:act_alg_avx}
The partial vector product algorithms described in Section~\ref{sec:poset_vecprod} naturally provide a highly efficient modification to the naive action algorithm. The partial intersection returns a binary string where indices with 1's indicate elements both to the future of element $i$ and to the past of element $j$, and then a bitcount returns the total number of elements within this interval. This algorithm can be further optimized by using OpenMP followed by a reduction (which prevents write conflicts) to accumulate the cardinalities. In turn, each physical core vectorizes instructions via AVX, and then each CPU parallelizes instructions by distributing tasks in the outer loop to each core. While it is typical to use the number of logical cores during OpenMP parallelization, we instead use the number of physical cores (typically half the logical cores, or a quarter in a Xeon Phi co-processor) because it is not always efficient to use hyperthreading alongside AVX. A summary of this procedure is given in Algorithm~\ref{alg:cardinalities}. \par

\begin{algorithm}[!t]
\caption{Optimized Interval Abundance Measurement}
\label{alg:cardinalities}
\begin{algorithmic}[1]
\Input
\Statex $A$ \Comment Adjacency matrix
\Statex $N$ \Comment Number of elements in causal set
\Statex $n$ \Comment Array for interval abundances
\Statex $p$ \Comment Number of element pairs

\Procedure{optimized\_ioi\_measurement}{$A,N,n,p$}
\State \texttt{\#pragma omp parallel for}
\For {$k=0;\, k < p;\, k\plusplus$}
\State $t\gets$ thread ID
\LineComment {Convert the pair index to two element indices}
\State $\{i,j\}\gets$ \textsc{convert\_index($k$)} \Comment Use mapping from Alg.~\ref{alg:nested_loop}
\If {$i\nprec j$}
\State \textbf{continue}
\EndIf
\LineComment {Cardinality for pair $(i,j)$}
\State $m\gets A[i].$\textsc{partial\_vecprod(}$A[j],i,j-i+1${\tt )}
\State $n[(t\times N)+m+1]\plusplus$
\EndFor
\LineComment Reduction sums results from each thread
\For {$k=1;\,k<T;\,k\plusplus$}
\For {$m=0;\,m<N;\,m\plusplus$}
\State $n[m]\pluseq n[(k\times N)+m]$
\EndFor
\EndFor
\EndProcedure

\Output
\Statex $n$ \Comment The interval abundances
\end{algorithmic}
\end{algorithm}

\subsection{MPI Optimization: Static Design}
\label{sec:mpi_static}
Another method of algorithm optimization is to distribute tasks among multiple computers using one of the variants of the Message Passing Interface (MPI) protocol~\cite{mpi}. When the causal set is small, so that the entire adjacency matrix fits in memory on each computer, we can simply split the {\tt for} loop in Algorithm~\ref{alg:cardinalities} evenly among all the cores on all computers using a hybrid OpenMP and Platform MPI approach. But when the graph is extremely large, e.g., $N\gtrsim 2^{21}$, we cannot necessarily fit the entire adjacency matrix in memory. To address this limitation, we use MPI to split Algorithm~\ref{alg:cardinalities} among $2^x$ computers, where $x\in\mathbb{N}$. Each computer generates some fraction of the element coordinates, and after sharing them among all other computers, generates its portion of the adjacency matrix, hereafter referred to as the \textit{adjacency submatrix}. In general, these steps are fast compared to the action calculation. \par

The MPI version of the action algorithm is performed in several steps. It begins by performing every pairwise operation possible on each adjacency submatrix, without any memory swaps among computers. Afterward, each adjacency submatrix is labeled by two numbers: the first refers to the first half of rows of the adjacency submatrix on that computer while the second corresponds to the second half, so that there are $2^{x+1}$ groups of rows labeled $\{0,\ldots,2^{x+1}-1\}$. There is never an odd number of rows, since the matrix is 256-bit aligned. We then wish to perform the minimal number of swaps of these row groups necessary to operate on every pair of rows of the original matrix. Within each row group all pairwise operations have already been performed, so moving forward only operations among rows of different groups are performed. \par

\begin{table}[t]
\centering
\begin{tabularx}{\linewidth}{*{8}{|>{\centering\arraybackslash}X}|}
\hline
\multicolumn{2}{|c|}{Rank 0} &%
\multicolumn{2}{c|}{Rank 1} &%
\multicolumn{2}{c|}{Rank 2} &%
\multicolumn{2}{c|}{Rank 3} \\
\hline
0 & 1 & 2 & 3 & 4 & 5 & 6 & 7 \\ \hline
0 & 3 & 2 & 5 & 4 & 7 & 6 & 1 \\ \hline
0 & 5 & 2 & 7 & 4 & 1 & 6 & 3 \\ \hline
0 & 7 & 2 & 1 & 4 & 3 & 6 & 5 \\ \hline
0 & 2 & 1 & 3 & 4 & 6 & 5 & 7 \\ \hline
0 & 4 & 1 & 5 & 2 & 6 & 3 & 7 \\ \hline
0 & 6 & 1 & 7 & 4 & 2 & 5 & 3 \\
\hline
\end{tabularx}
\captionof{figure}{{\bf Permutations of MPI buffers using four computers.} Each of four computers, identified by its rank, holds a quarter of the adjacency matrix. Two buffers on each computer each hold an eighth of the entire matrix, labeled $\{0,\ldots,7\}$, so that all pairwise row operations may be performed using the minimal number of inter-rank transfers. Each of the seven rows is a non-trivial permutation of the eight buffers, indicating only six rounds of MPI data transfers are necessary to calculate the action when the algorithm is split over four computers.}
\label{tbl:perms}
\end{table}

We label all possible permutations except those which provide trivial swaps, i.e., moves which would swap the submatrix rows in memory buffers within a single computer, or moves which swap buffers in only some (rather than all) computers. The non-trivial configurations are shown for four computers in Figure~\ref{tbl:perms}. By organizing the data in this way, we can ensure no computer will be idle after each data transfer. We use a cycle sort to determine the order of permutations so that we perform the minimal number of total buffer swaps. We simulate this using a simple array of integers populated by a given permutation, after which the actual operation takes place. By starting at the current permutation and sorting to each unvisited permutation, we record how many steps each would take. Often it is the case that several will use the same number of steps, in which case we move from the current permutation to any of the others which use the fewest number of swaps. Once all pairwise partial vector products have completed on all computers for a particular permutation, that permutation is removed from the global list of unused permutations shared across all computers. Thus, using these techniques it becomes straightforward to distribute Algorithm~\ref{alg:cardinalities} among two or more computers.

\subsection{MPI Optimization: Load Balancing}
\label{sec:mpi_balanced}
The MPI algorithm described in the previous section grows increasingly inefficient when the pairwise partial inner product operations are not load-balanced across all computers. In Algorithm~\ref{alg:cardinalities}, there is a {\tt continue} statement which can dramatically reduce the runtime when the subgraph studied by one computer is less dense than that on another computer. When the entire adjacency matrix fits on all computers, this is easily addressed by identifying a random graph automorphism by performing an $O(N)$ Fisher-Yates shuffle~\cite{fisher1948statistical} of labels. This allows each computer to choose unique random pairs, though it introduces a small amount of overhead. \par
\begin{figure}[!hpt]
\centering
\includegraphics[width=0.8\textwidth]{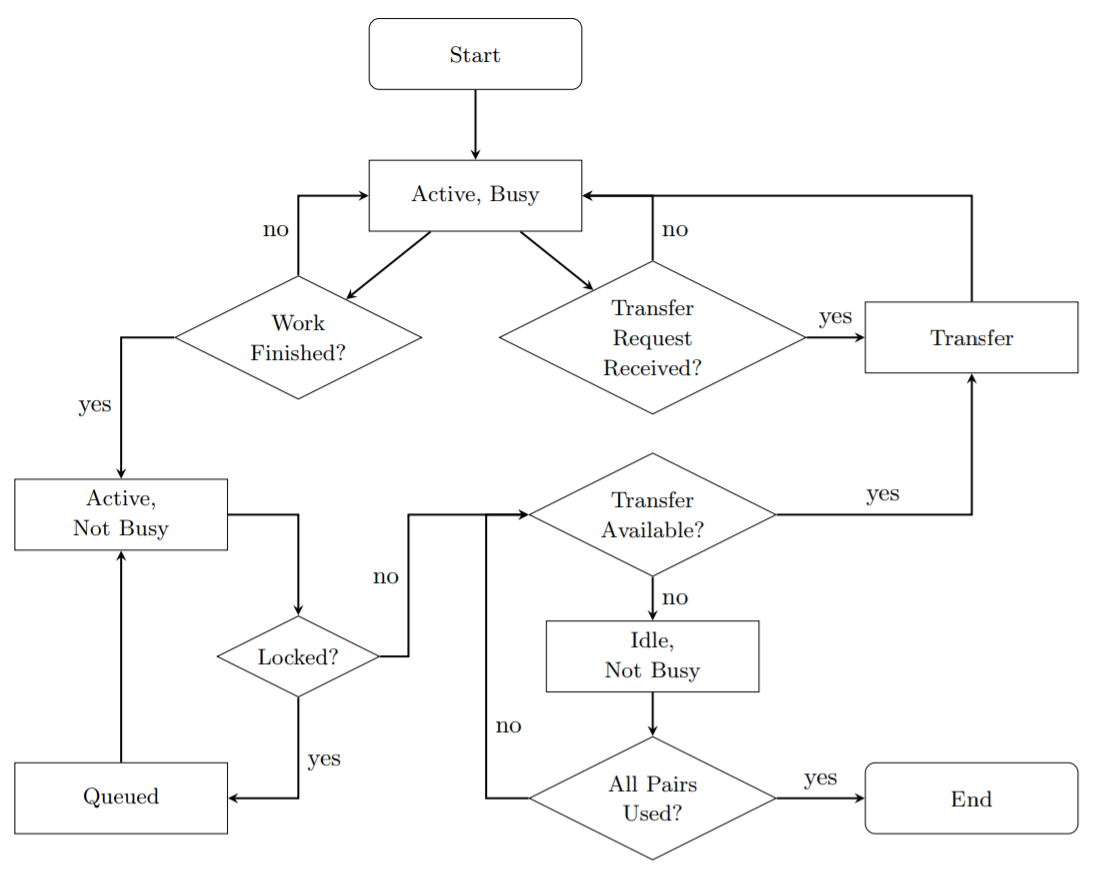}
\caption{{\bf Load-balanced action algorithm using MPI.} When the adjacency matrix is split among multiple computers, we want to make sure no computers end up idle for long periods of time. Yet to move from an Idle to Busy state at least one other computer must have finished its action calculations. Initially, all computers are Active and Busy, indicating they are not waiting for another task to finish and are currently executing the action algorithm. If two other computers have requested an exchange, an Active, Busy computer allows them to use part of its memory for temporary storage (Transfer). Once a computer finishes its portion of work on the action calculation, it enters the Active, Not Busy state, at which point it will add its pair of buffer indices to the global list of available buffers. An MPI spinlock, developed specifically for this algorithm, is implemented to ensure only one computer can manage a transfer. If another pair of computers is exchanging data, the Active, Not Busy computer enters a Queued state, where it remains until other transfers have completed. Otherwise, it attempts a memory transfer if possible by checking the list of available buffers. If no other buffers are available, or if any available transfers would lead to redundant calculations, the computer enters the Idle, Not Busy state, where it waits for another computer to initiate a transfer. Once all buffer pairs have been used, the algorithm ends.}
\label{fig:mpi_flowchart}
\end{figure}

On the other hand, if the adjacency matrix must be split among multiple computers, load balancing is much more difficult. If we suppose that in a four-computer setup the {\tt for} loops on two computers finish long before those on the other two, it would make sense for the idle computers to perform possible memory exchanges and resume work rather than remain idle. The dynamic design in Figure~\ref{fig:mpi_flowchart} addresses this flaw by permitting transfers to be performed independently until all operations are finished. \par

The primary difficulty with such a design is that for this problem, MPI calls require all computers to listen and respond, even if they do not participate in a particular data transfer. The reason for this is that the temporary storage used for an exchange is spread across all computers to minimize overhead and balance memory requirements. Therefore, each computer launches two POSIX~\cite{posix} threads: a master thread listens and responds to MPI calls, and also monitors whether the computer is active or idle with respect to action calculations, while a slave thread performs all tasks related to those calculations. A shared flag variable indicates the active/idle status on each computer. \par

As opposed to the static MPI action algorithm (Section~\ref{sec:mpi_static}), where whole permutations are fundamental, buffer pairs are fundamental in the load-balanced implementation. This means there is a list of unused pairs as well as a list of pairs available for trading, i.e., those pairs on idle computers. When two computers are both idle, they check to see if a buffer swap would give either an unused pair, and if so they perform a swap. After a swap to an unused pair, the computer moves back from an idle to an active status. 

\section{Simulations and Scaling Evaluations}
\label{sec:simulations}
\subsection{Spacetime Region Considered}
In benchmarking experiments, we choose to study a $(1+1)$-dimensional compact region of de Sitter spacetime. The de Sitter manifold is one of the three maximally symmetric solutions to Einstein's equations, and it is well-studied because its spherical foliation has compact spatial slices (i.e., no timelike boundaries), constant curvature everywhere, and most importantly, a non-zero value for $S_{EH}$. We study a region bounded by some constant conformal time $\eta_0$ so that the majority of elements, which lie near the minimal and maximal spatial hypersurfaces, are connected to each other in a bipartite-like graph.  While normally one would need to consider the Gibbons-Hawking-York boundary terms which contribute to the total gravitational action, it is known that spacelike boundaries do not contribute to the BD action~\cite{buck2015boundary}. \par

The $(1+1)$-dimensional de Sitter spacetime using the spherical foliation is defined by the metric
\begin{equation}
\label{eq:ds_metric}
ds^2 = \sec^2\eta(-d\eta^2 + d\theta^2)\,,
\end{equation}
and volume element $dV = \sec^2\eta\,d\eta\,d\theta$. Elements are sampled using the probability distributions $\rho(\eta|\eta_0) = \sec^2\eta/\tan\eta_0$ and $\rho(\theta) = 1/2\pi$, so that $\eta\in[-\eta_0,\eta_0]$ and $\theta\in[0,2\pi)$. Finally, the form of~\eqref{eq:ds_metric} indicates elements are timelike-separated when $d\theta^2 < d\eta^2$, i.e., $\pi - |\pi - |\theta_1 - \theta_2|| < |\eta_1 - \eta_2|$ for two particular elements with coordinates $(\eta_1,\theta_1)$ and $(\eta_2,\theta_2)$. This condition is used in the CUDA kernel which constructs the causal matrix in the asynchronous GPU linking algorithm, which was introduced in Section~\ref{sec:gpu_asynch}. \par

\begin{figure}[!t]
\centering
\includegraphics[width=\textwidth]{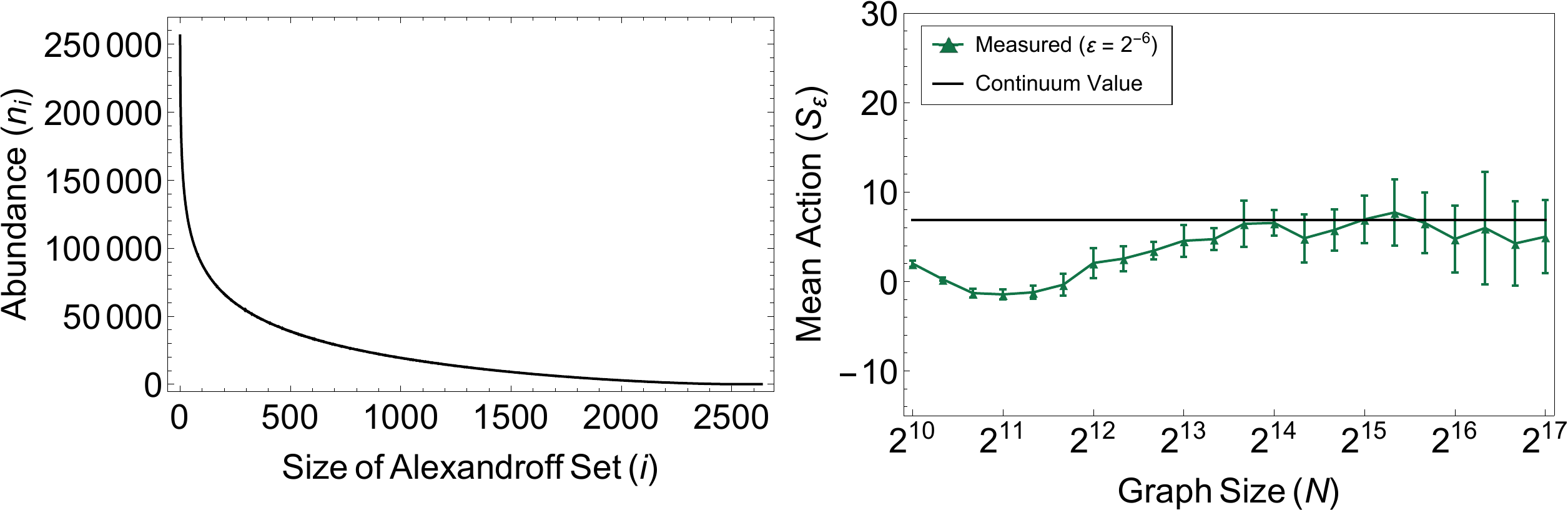}
\caption{{\bf The action in $(1+1)$-dimensional de Sitter spacetime.} The left panel shows the interval abundance distribution for a $(1+1)$-dimensional de Sitter slab with $N=2^{15}$ and $\eta_0=0.5$. The right panel shows the smeared BD action (green) tends toward the EH action (black) as the graph size increases. We take a symmetric temporal cutoff $\eta_0=\pm0.5$ and a small smearing parameter $\varepsilon=2^{-6}\ll 1$ so the onset of convergence appears as early as possible. Remarkably, the terms in the series~\eqref{eq:s2d_smeared} are several orders of magnitude larger than the continuum result $S\approx 6.865$, yet the standard deviation about the mean is quite small in comparison, shown by the error bars in the second panel. The error increases with the graph size because the smearing parameter $\varepsilon$ is fixed while the discreteness scale $\ell=\sqrt{V/N}$ decreases. All data shown is averaged over ten graphs.}
\label{fig:action}
\end{figure}

We expect the precision of the results to improve with the graph size, so we study the convergence over the range $N\in[2^{10},2^{17}]$ in these experiments. Larger graph sizes are typically used to study higher-dimensional spacetimes and, therefore, are not considered here. We choose a cutoff $\eta_0=0.5$ in particular because for $\eta_0$ too small we begin to see a flat Minkowski manifold, whereas for $\eta_0$ too large, a larger $N$ is needed for convergence, since the discreteness scale $\ell=\sqrt{V/N}$ is larger.

\subsection{Convergence and Running Times}
\label{sec:convergence_times}
Initial experiments conducted to validate the BD action show the interval abundance distribution takes the form as that for manifold-like causal sets (versus in Kleitman-Rothschild partial orders)~\cite{glaser2013towards}, and the mean begins to converge to $S_{EH}$ around $N\gtrsim 2^{14}$, Figure~\ref{fig:action}. The standard deviation $\sigma_S$ increases like $\sqrt{N}$ because we have chosen to keep the smearing parameter $\varepsilon$ fixed as $N$ increases, as is the more common practice, but if we had instead chosen to let $\varepsilon\to\varepsilon/N$, then $\sigma_S\to0$ as $N\to\infty$~\cite{benincasa2013action}. The Ricci curvature for the constant-curvature de Sitter manifold is $R=d(d+1)$ so that $S_{EH}$ is simply
\begin{equation}
S_{EH}=\frac{d(d+1)}{2}V(\eta_0) = 4\pi\tan\eta_0\,.
\end{equation}
\begin{figure}[!t]
\centering
\includegraphics[width=\linewidth]{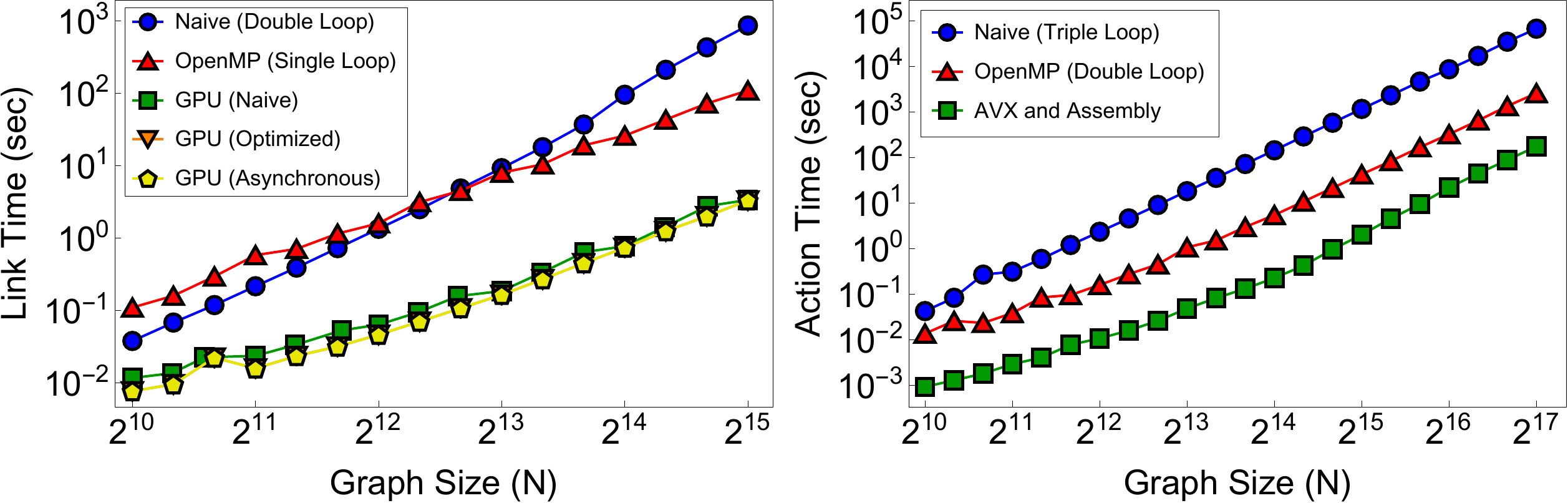}
\caption{{\bf Performance of the linking and action algorithms.} We benchmark the $O(N^2)$ node linking algorithm (left) and the $O(N^3)$ action algorithm (right) over a wide range of graph sizes. The left panel shows moving from a sparse (blue) to a dense (red) representation improves the scaling of the linking algorithm, though it can still take several minutes to generate causal sets of modest size. When the NVIDIA K80m GPU is used, we find a dramatic speedup compared to the original implementation, which allows us to generate much larger causal sets in the same amount of time. We find the three variations of the GPU algorithm (green, orange, yellow) provide nearly identical run times. The right panel shows the benefits of using both OpenMP and AVX instructions to parallelize. The optimal OpenMP scheduling scheme varies according to the problem size, though in general a static schedule is best, since it has the least overhead.}
\label{fig:link_action_times}
\end{figure}
\begin{figure}[!t]
\centering
\includegraphics[width=\linewidth]{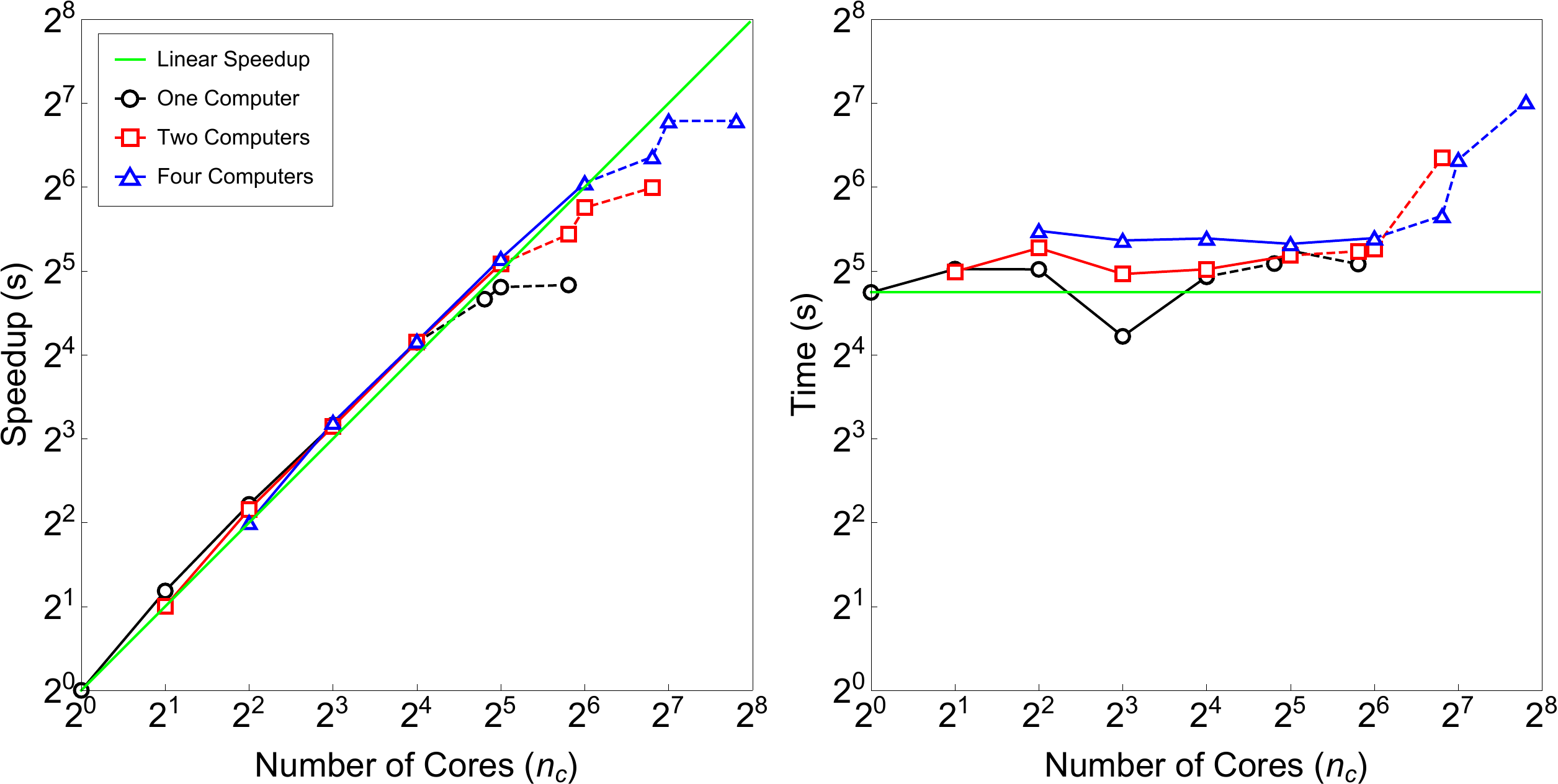}
\caption{{\bf Strong and weak scaling of the action algorithm.} The action algorithm exhibits nearly perfect strong and weak scaling, shown by the straight green lines in each panel. The {\tt for} loop in Algorithm~\ref{alg:cardinalities} is parallelized using OpenMP, while the partial inner product is vectorized using AVX. When multiple computers are used, pairs identified by the loop are evenly distributed among all computers. We find the best speedups when the total number of cores used is a power of two and hyperthreading is disabled (solid lines). When we use all 28 physical cores, or we use 32 or 56 logical cores in our dual Xeon E5-2680v4 CPUs, we find a modest increase in speedup (dashed lines). In the right panel, the runtime should remain constant while the number of processors is increased as long as the amount of work per processor remains fixed. The constant increase in runtime when more computers are added is likely due to a high MPI communication latency over a 10Gb TCP/IP network.}
\label{fig:benchmarking}
\end{figure}
The generation and study of causal sets is extremely efficient when the GPU is used for element linking and AVX is used on top of OpenMP to find the action (Figure~\ref{fig:link_action_times}). The GPU and AVX optimizations offer nearly a $1000\times$ speedup compared to the naive linking and action algorithms, which in turn allows us to study larger causal sets in the same amount of time. The decreased performance of the naive implementation of the linking algorithm, shown in the first panel of Figure~\ref{fig:link_action_times}, reflects the extra overhead required to generate the sparse lists for both future and past relations. There is a minimal speedup from using asynchronous CUDA calls because the memory transfer time is already much smaller than the kernel execution time. \par

\subsection{Scaling: Amdahl's and Gustafson's Laws}
\label{sec:scaling}
We analyze how Algorithm~\ref{alg:cardinalities} performs as a function of the number of CPU cores to show both strong and weak scaling properties (Figure~\ref{fig:benchmarking}). Amdahl's Law, which measures strong scaling, describes speedup as a function of the number of cores at a fixed problem size~\cite{amdahl1967validity}. Since no real problem may be infinitely subdivided, and some finite portion of any algorithm is serial, such as cache transfers, we expect at some finite number of cores the speedup will no longer substantially increase when more cores are added. In particular, strong scaling is important for Monte Carlo experiments, where the action must be calculated many thousands of times for smaller causal sets. We find, remarkably, a superlinear speedup when the number of cores is a power of two and hyperthreading is disabled, shown by the solid lines. The dashed lines in Figure~\ref{fig:benchmarking} indicate the use of 28, 32, and 56 logical cores on dual 14-core processors. \par

We also measure the weak scaling, described by Gustafson's Law~\cite{gustafson1988reevaluating}, which tells how runtime varies when the number of computations, $O(N^3)$, per processor is constant (Figure~\ref{fig:benchmarking}(right)). This is widely considered to be a more accurate measure of scaling, since we usually limit our experiments by the runtime and not by the problem size. Weak scaling is most relevant for convergence tests, where the action of extremely large causal sets must be studied in a reasonable amount of time. Our results show nearly perfect weak scaling, again deviating when the number of cores is not a power of two or hyperthreading is enabled. We get slightly higher runtimes overall when more computers are used for two reasons: the computers are connected via a 10Gb TCP/IP cable rather than Infiniband and the load imbalance becomes more apparent as more computers are used.  Since the curves have a nearly constant upward shift, we believe the likely explanation is the high MPI latency. For each data point in these experiments, we ``warm up'' the code by running the algorithm three times, and then record the smallest of the next five runtimes. All experiments were conducted using dual Intel Xeon E5-2680v4 processors running at 2.4 GHz on a Redhat 6.3 operating system with 512 GB RAM, and code was compiled with nvcc 8.0.61 and linked with g++/mpiCC 4.8.1 with Level 3 optimizations enabled.

\section{Summary}
By using low-level optimization techniques which take advantage of modern CPU and GPU architectures (Chapters~\ref{chap:parallel}--\ref{chap:graphs}), we have shown it is possible to reduce runtimes for causal set action experiments by a factor of 1000. We used OpenMP to generate the element coordinates in parallel in $O(N)$ time and used the GPU to link elements much faster than with OpenMP. By tiling the adjacency matrix and balancing the amount of work each CUDA thread performs with the physical cache sizes and memory accesses, we allowed the GPU to generate causal sets of size $N\gtrsim 2^{20}$ in just a few hours. Using the compact data structures developed in Section~\ref{sec:bitset_struct} and the optimized bitset algorithms from Sections~\ref{sec:opt_bitset_alg} and~\ref{sec:opt_poset_alg}, we constructed Algorithm~\ref{alg:cardinalities} to efficiently measure the interval abundances needed for the action calculation. The MPI algorithms described in Sections~\ref{sec:mpi_static} and~\ref{sec:mpi_balanced} provide a rigorous protocol for asynchronous information exchange in the most efficient way when the adjacency matrix is too large to fit on a single computer. Finally, we demonstrated superlinear scaling of the action algorithm with the number of CPU cores, indicating that the code is well-suited to run in its current form on large computer clusters.

%\afterpage{\blankpage}

\twolinechapter
\chapter[Inference of Causal Set Boundaries]{\texorpdfstring{Inference of Causal\\[-0.8cm] Set Boundaries}{Inference of Causal Set Boundaries}}
\chaptermark{Inference of Boundaries}
\mainchapter
\label{chap:chull}
\thispagestyle{empty}

\begin{figure}[!t]
\centering
\includegraphics[width=0.4\linewidth]{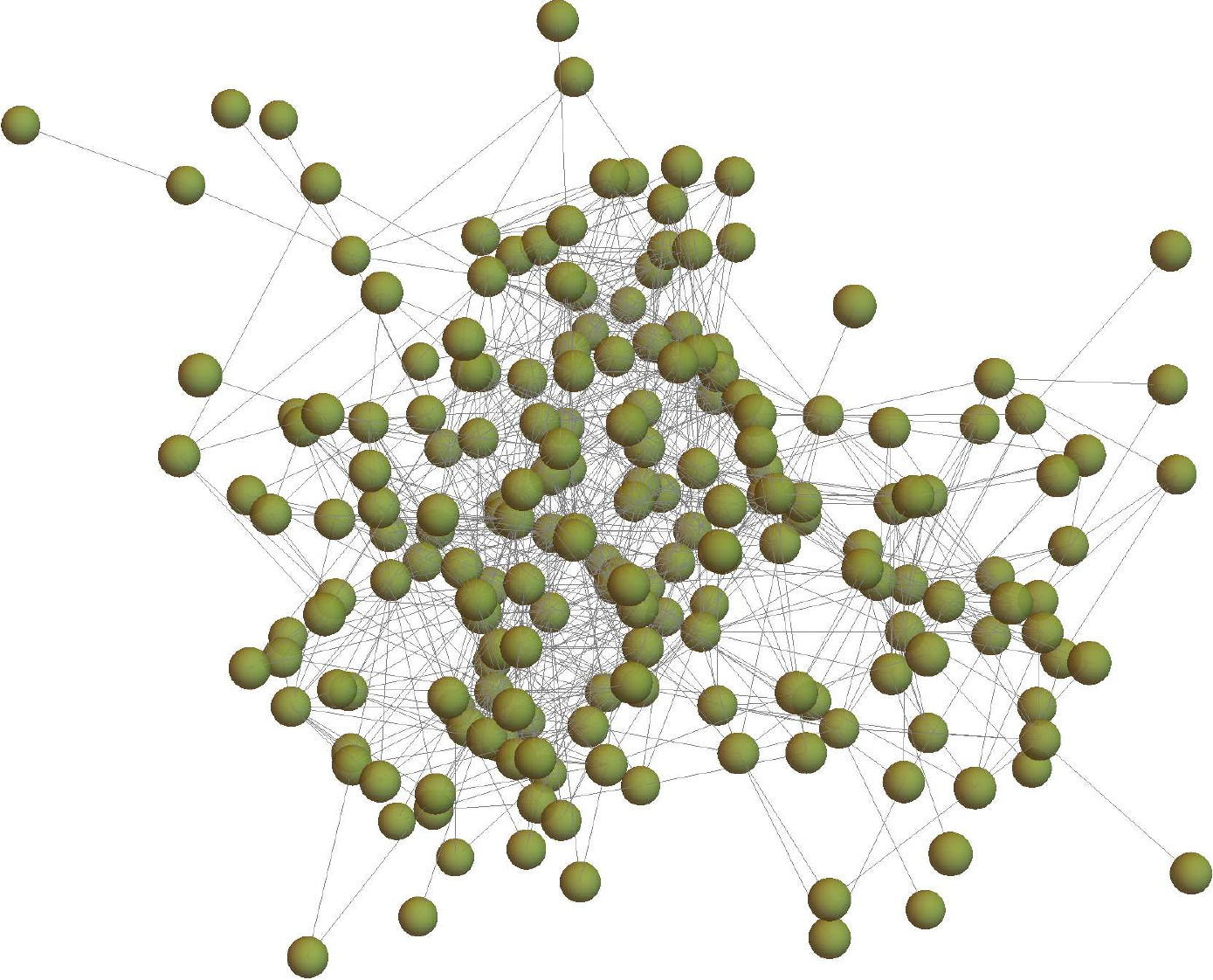}\hspace{2cm}%
\includegraphics[width=0.35\linewidth]{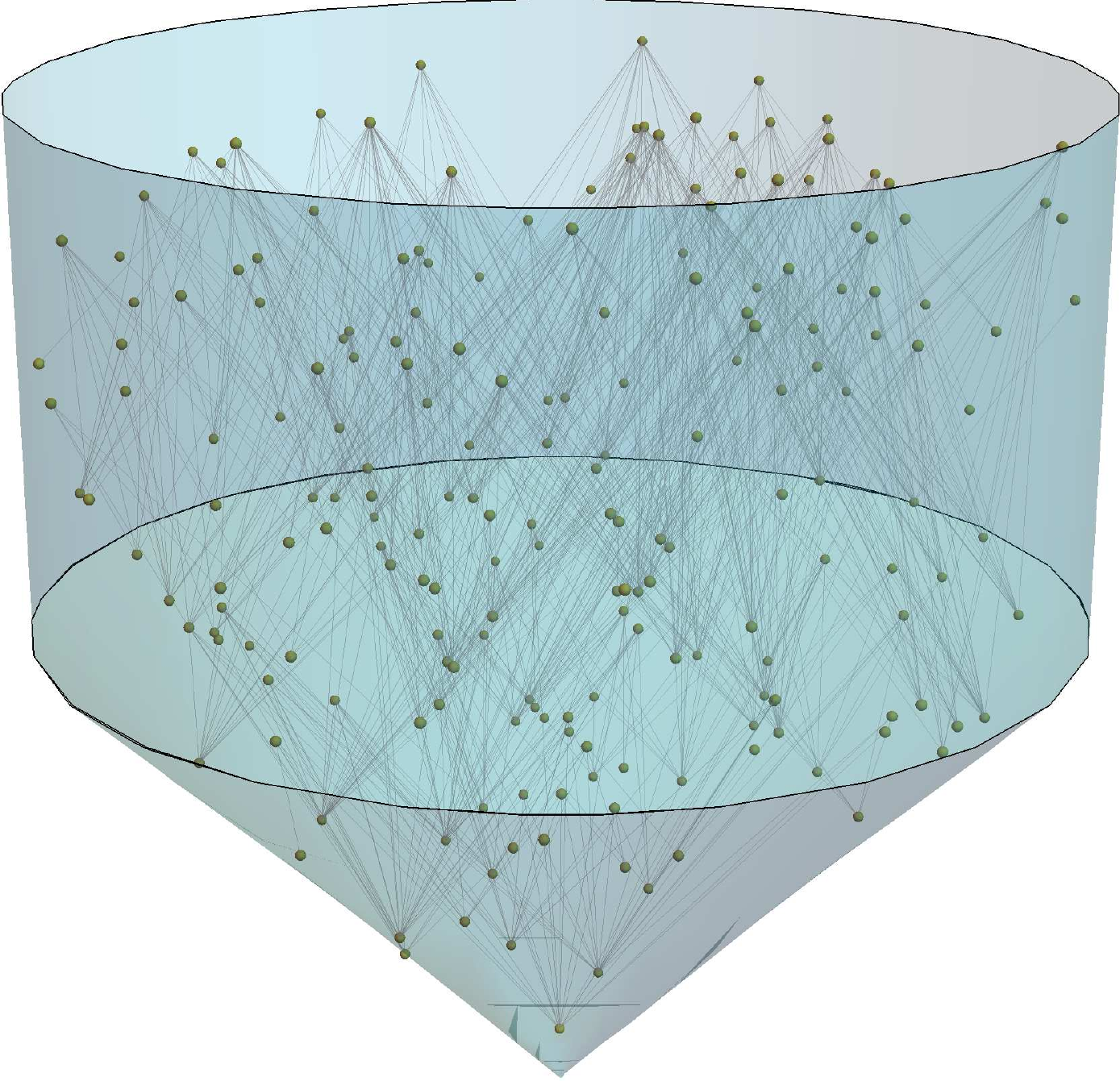}
\caption{{\bf The faithful embedding of a causal set.} An unlabeled causal set (left) with $N=200$ spacetime elements, indicated by the green points, and $5373$ causal relations, indicated by the gray lines, is faithfully embedded into the blue region (right). The causal set is bounded below by a null boundary and above by a constant-time hypersurface, with a timelike boundary of constant radius separating the two. This particular region demonstrates how in practice we can encounter causal sets with a non-trivial combination of boundaries. A general embedding algorithm for a given causal set is unknown, but it may be possible to extract information from the causal set structure about the types of hypersurfaces which form the bounding region.}
\label{fig:embedding}
\end{figure}

%\lettrine{\scalebox{2.1}{T}}{he}
The causal set program~\cite{bombelli1987space} is centered around the {\it Hauptvermutung}~\cite{bombelli1989origin,brightwell1991structure,sorkin2003notes}, which claims that two different uniform embeddings of the same locally-finite partial order, called a causal set, into a Lorentzian manifold are nearly isometric in the Gromov-Hausdorff sense~\cite{edwards1975structure,gromov1981structures,gromov1981groups}. However, this conjecture can be understood in a much simpler way: a causal set contains all geometric and topological information about a spacetime above the discreteness scale $\ell$, up to a conformal rescaling. Since the Hauptvermutung describes an embedding problem (Figure~\ref{fig:embedding}), one would hope to eventually discover an embedding method, either in the form of an analytic expression or an algorithm, to test the conjecture under certain mild assumptions (see~\mbox{\cite{clough2016embedding}} for recent progress). \par
While this is a difficult problem, one can take a first step by building a set of tools to measure extrinsic properties of causal sets with respect to an embedding space. One potential avenue has opened in the study of the Benincasa-Dowker (BD) action~\cite{benincasa2010scalar}, i.e., the discrete analogue to the Einstein-Hilbert action for general relativity. Though the BD action was developed during the study of intrinsic properties --- the d'Alembertian and the Ricci curvature --- it was soon noticed~\cite{benincasa2011random} it captures one of the Gibbons-Hawking-York (GHY) boundary terms~\cite{york1972role,gibbons1977action} which measures the contribution to the classical action from boundaries in the embedding space. In the case of the {\it causal interval}, defined as the past light cone of one element intersected with the future light cone of another element, the BD action measures both the bulk term as well as the volume of the \mbox{codimension-2} surface defined by the intersection of the two light cones~\cite{benincasa2011random,benincasa2013action}. While it was known the BD action cannot measure the spacelike boundary terms, this observation gave hope that perhaps other boundary terms were also hidden within the expression~\cite{benincasa2011there}, which would be a good indication it held information about extrinsic geometry. Yet more recent numerical experiments have shown no other codimension-2 boundary terms are measured, and the BD action even diverges upon encountering timelike boundaries rather than recovering extrinsic geometric information. \par

The recent discovery of the discrete boundary term for spacelike boundaries~\cite{buck2015boundary} was a significant step forward, because it indicates the possibility of the existence of boundary terms for each type of codimension-1 and codimension-2 boundary just as in continuum physics. Yet even if there were to exist an expression akin to the spacelike boundary term for each type of boundary, it would remain unclear when such terms should be included when one calculates the full discrete action for some manifold-like causal set. This problem is compounded by the fact that in the continuum the action of a region whose boundaries approach null surfaces becomes infinite, yet the limit is finite, so there is an ambiguity over which limit any discrete structure should choose in the continuum limit. This paradox will not be studied here, but should be kept in mind. \par

In this chapter, we study several methods which allow one to infer the boundary geometry of finite causal sets. We first review the classical Lorentzian embedding theorems in order to understand which results one should hope to recover once the Hauptvermutung is proven. However, since we cannot yet solve the entire embedding problem, we consider what information we can extract from finite causal sets. We list the necessary assumptions in Section~\ref{sec:finite_causets}. Then, we attempt to classify boundary geometry between null and non-null classes by first characterizing the geometry of a causal interval (Section~\ref{sec:distributions}), and then examining what causal sets look like as their boundaries approach null ones. The main result is an algorithm to measure timelike boundaries, given by Algorithms~\ref{alg:candidates} and~\ref{alg:timelike_measurement} in Section~\ref{sec:volume}. Finally, we conclude with several illustrative examples in Section~\ref{sec:examples5}.

\section{Lorentzian Embedding Theorems}
\label{sec:emb_thm}
A Lorentzian manifold $\mathcal{L}$ is a manifold which admits a metric tensor $g_{\mu\nu}$ with just one negative eigenvalue. If for every point $x\in\mathcal{L}$ there exists a local coordinate system in which $g_{\mu\nu}$ is proportional to the Minkowski metric, then $\mathcal{L}$ is also conformally flat, a property we assume of spacetimes hereafter. Manifolds with this general form are called Friedmann-Lema\^itre-Robertson-Walker (FLRW) manifolds, which correspond to isotropic spacetimes with homogeneous matter content. To understand the classical side of the Hauptvermutung, we review here several embedding theorems for analytic manifolds.

 The extrinsic geometry of analytic manifolds in the context of embedding spaces can be traced back to Schl\"afli's 1873 conjecture~\cite{schlafli1873nota} about the embeddability of Riemannian manifolds, i.e., those with strictly positive-definite metrics, into Euclidean spaces whose metrics are simply the unit matrix (in the proper choice of coordinates). Several decades later, the work of Janet~\cite{janet1926possibilite}, Cartan~\cite{cartan1927possibilite}, and Burstin~\cite{burstin1931beitrag} formalized these ideas in the following theorem:
\begin{adjustwidth}{1cm}{}
\textbf{Theorem 1:} Any analytic $n$-dimensional Riemannian manifold may be analytically and isometrically embedded into an $m$-dimensional Euclidean space $\mathcal{E}^m$, where $m=n(n+1)/2$.
\end{adjustwidth}
Therefore, a $4$-dimensional Riemannian manifold could require a $10$-dimensional embedding space. This result was extended by A.\ Friedman~\cite{friedman1961local,friedman1965isometric} to include $n$-dimensional pseudo-Riemannian manifolds $\mathcal{R}^n_{p,q}$ whose metric tensors have $p$ positive and $q$ negative eigenvalues:
\begin{adjustwidth}{1cm}{}
\textbf{Theorem 2:} Any pseudo-Riemannian manifold $\mathcal{R}^n_{p,q}$ with analytic metric can be analytically and isometrically embedded in $\mathcal{E}^m_{r,s}$ where $m=n(n+1)/2$ and $r,s$ are any prescribed integers satisfying $r\geq p\,,s\geq q$.
\end{adjustwidth}
where $\mathcal{E}^m_{r,s}$ denotes the $m$-dimensional pseudo-Euclidean manifolds whose metric tensors have $r$ positive and $s$ negative eigenvalues. However, if the embedding space is Ricci flat, then the number of extra dimensions needed for the embedding is reduced to just one. The Campbell-Magaard theorem~\cite{campbell1926course,magaard1963einbettung} states
\begin{adjustwidth}{1cm}{}
\textbf{Theorem 3:} Any analytic $n$-dimensional Riemannian space can be locally embedded in a $(n+1)$-dimensional Ricci-flat space.
\end{adjustwidth}
By a similar method used to demonstrate Theorem 2, this result is generalized in~\cite{romero1996embedding,lidsey1997applications,dahia2002embedding} to pseudo-Riemannian spaces of one higher spatial or temporal dimension:
\begin{adjustwidth}{1cm}{}
\textbf{Theorem 4:} Any analytic pseudo-Riemannian space $\mathcal{R}^n_{s,t}$ can be locally embedded in a Ricci-flat pseudo-Riemannian space $\mathcal{R}^{n+1}(\tilde{s},\tilde{t})$, where either $\tilde{s}=s$ and $\tilde{t}=t+1$ or $\tilde{s}=s+1$ and $\tilde{t}=t$.
\end{adjustwidth}
In general relativity, we employ this theorem to gain an extra spatial dimension: $\mathcal{R}^5_{4,1}=\mathcal{M}^5$. The metric of the embedding space can be written in terms of the $4$-dimensional metric $g_{\mu\nu}$ along with a new fifth dimension represented by the coordinate $\psi$ and some function $\Phi(x^\mu,\psi)$:
\begin{equation}
\label{eq:emb0}
ds^2 = g_{\mu\nu}\,dx^\mu\,dx^\nu + \varepsilon\phi^2\,d\psi^2\,,
\end{equation}
where $\varepsilon\equiv n_\mu n^\mu=\pm 1$ indicates the orientation of the normal $n^\mu$ of the $4$-dimensional surface within the embedding space. The higher-dimensional metric coefficients $\eta_{\mu\nu}$ may be found if there exist a set of functions $\Omega_{\mu\nu}$ related to the extrinsic curvature which satisfy
\begin{align}
\Omega_{\mu\nu}&=\Omega_{\nu\mu}\,,\label{eq:emb1}\\
\Omega^\mu_{\,\,\nu;\mu}&=\Omega_{\,,\nu}\,,\label{eq:emb2}\\
\Omega_{\mu\nu}\Omega^{\mu\nu}-\Omega^2&=-\varepsilon R\,,\label{eq:emb3}
\end{align}
on a hypersurface $\psi=\psi_0$, where $\Omega^\mu_\nu\equiv g^{\mu\lambda}\Omega_{\lambda\nu}\,,\Omega\equiv g^{\mu\nu}\Omega_{\mu\nu}\,,$ and $R\equiv g^{\mu\nu}R_{\mu\nu}$. At the same time, there are the dynamical constraints
\begin{align}
\frac{\partial g_{\mu\nu}}{\partial\psi}&=-2\Phi\Omega_{\mu\nu}\,,\label{eq:emb4}\\
\frac{\partial \Omega^\mu_\nu}{\partial\psi}&=\Phi\left(\Omega\Omega^\mu_\nu - \varepsilon R^\mu_\nu\right)+\varepsilon g^{\mu\lambda}\Phi_{\,;\lambda\nu}\,.\label{eq:emb5}
\end{align}
It can be shown~(\ref{eq:emb1}-\ref{eq:emb5}) provide a local embedding into a vacuum $(n+1)$-dimensional Lorentzian manifold $\mathcal{L}^{n+1}$ with the metric~\eqref{eq:emb0}, supposing~(\ref{eq:emb4},\ref{eq:emb5}) are integrable~\cite{romero1996embedding,lidsey1997applications}. Thus, we should hope to identify a discrete analogue for such a set of functions once the Hauptvermutung is better understood.

\section{Embedding Finite Causal Sets}
\label{sec:finite_causets}
In the rest of this chapter, we examine a simpler problem than the Hauptvermutung: we discuss methods which allow one to infer the extrinsic geometry of boundaries in a particular causal set using computational methods. We achieve this by partitioning a causal set in two ways --- the timelike (chain) and spacelike (antichain) representations defined in Section~\ref{sec:set_partitions} --- and then using two new algorithms to identify and measure different boundaries. \par

The \textit{spacetime representation}, which is the union of the timelike and spacelike representations, admits a natural scheme for ordering chains and antichains. Antichains are labeled according to the graph distance of the seed element from the minimal element in the maximum chain, i.e., they are time-ordered with respect to the seed element. The chain ordering is performed by ranking the elements which intersect with the maximal antichain by an inferred spatial distance from the {\it representation-induced origin} of the causal set, defined as the element at the intersection of the maximum chain and maximum antichain. The algorithm to determine this inferred spatial distance is discussed in more detail in Section~\ref{sec:volume}. \par

The causal sets we study here are realized as random geometric graphs in $(1+1)$-dimensional Minkowski spacetime, and are generated using the methods described in Chapter~\ref{chap:graphs}. For the following results to hold, we assume the size of the maximum chain is large, $H_P\gtrsim2^{6}$, the size of the maximum antichain is large, $W_P\gtrsim2^{6}$, and the causal set is relatively large, $N\gtrsim2^{10}$. Furthermore, when the inverse extrinsic curvature $K^{-1}$ of a non-null boundary is on the order of the discreteness scale $\ell$ of the causal set, the boundary is indistinguishable from a null boundary (Section~\ref{sec:distributions}), so in the following measurement of the boundary volume (Section~\ref{sec:volume}) we assume any non-null boundary is smooth, continuous, has an inverse extrinsic curvature much larger than the discreteness scale, $\ell K\ll1$, and is otherwise well-behaved.

\section{Characteristics of the Causal Interval}
\label{sec:distributions}
\subsection{Chain and Antichain Profiles}
\begin{figure*}[!t]
\centering
\vspace{-1cm}%
\includegraphics[width=\linewidth]{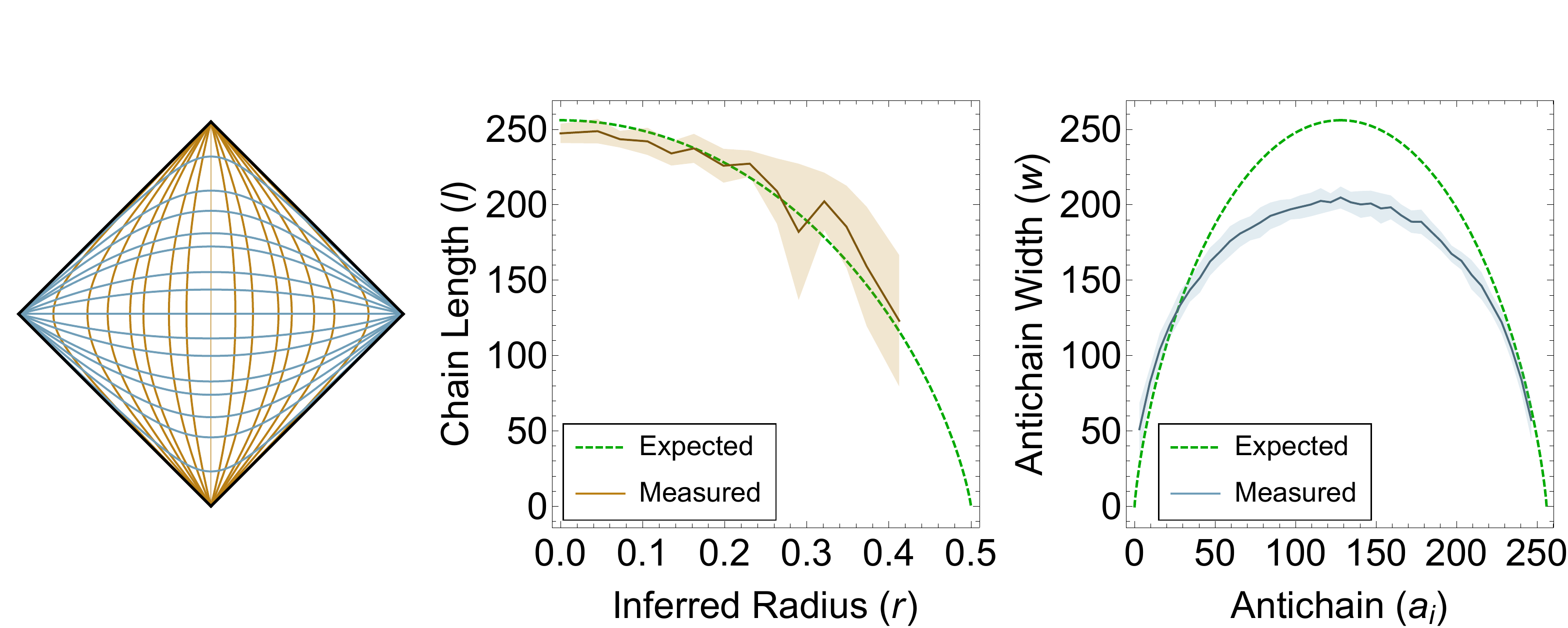}
\caption{{\bf The spacetime representation for the causal interval.} The representation of the causal interval in the $(1+1)$-dimensional Minkowski spacetime is shown in the left panel, where the orange curves correspond to the expected paths of chains and the blue curves correspond to the expected paths of antichains. The orange and blue straight lines crossing the center, denoted the representation-induced origin, respectively represent the maximum chain and antichain. In the center panel, the empirical chain lengths (orange) fall nearly perfectly across the expected values (green), given by~(\ref{eq:geo_to_chain}). The radial coordinates are inferred by averaging over values sampled from the marginal distribution~(\ref{eq:marginal_x}).  Fluctuations increase with radial distance due to finite-size effects. The right panel shows the antichain widths, i.e., cardinalities, for the same causal sets, ranked by a time coordinate inferred from the intersection of each antichain with the maximum chain. All data is averaged over ten graphs with unit height and size $N=2^{14}$, and the shaded regions indicate the standard deviation of the mean.}
\label{fig:null_profiles}
\end{figure*}

The first challenge in characterizing a timelike or spacelike boundary is distinguishing it from a null boundary. We can characterize the ordered sets of chain and antichain sizes, hereafter called {\it profiles}, for an interval of height $l_0$ in $(d+1)$-dimensional Minkowski spacetime using the spacetime representation. The continuum limit of the chain representation can be modeled by the family of hyperbolic curves which pass through the bottom of the interval, $T_p$, the top of the interval $T_f$, and some point $(0,r)$, shown by the orange curves in Figure~\ref{fig:null_profiles}(left). The geodesic length of a chain passing through the waist ($t=0$) at radius $r$ is
\begin{equation}
l(r)=\frac{1}{2}\sqrt{4\zeta(r)^2-l_0^2}\ln\left(\frac{4\zeta(r)}{2\zeta(r)-l_0}-1\right)\,,
\end{equation}
where $\zeta(r)\equiv r/2+l_0^2/(8r)$. \par

The continuum length $l(r)$ is directly proportional to the discrete graph distance $L(r)$~\cite{brightwell1991structure}. For instance, in $(1+1)$-dimensional Minkowski spacetime, 
\begin{equation}
\label{eq:geo_to_chain}
\mathbb{E}\left[L(r)\right]=\sqrt{2}\,l(r)/\ell\,.
\end{equation}
Since the spatial distribution of maximal elements $\mathcal{F}\subseteq C$ is not uniform, we approximate the spatial distribution $\rho(x)$ for $i\in\mathcal{F}$ by considering a Poisson point process inside the region between the future null boundary and the hyperbolic surface at proper time $\ell$ to the past of the boundary. This gives a marginal distribution
\begin{equation}
\label{eq:marginal_x}
\rho(x) = \left(1/2-x-\zeta(x)+\sqrt{x^2+\zeta(x)^2-L^2/4}\right)/\tilde{V}\,,
\end{equation}
where $\tilde V$ is the volume of the region described. Hence, when comparing measured chain lengths to the theoretical profile for a known region, one can sample $x$ from this distribution both to rank chains in a profile and to infer spatial separations of chains in an embedding space. \par

By symmetry, the same arguments can be used to calculate the width of an antichain centered about the origin. The continuum width $w(t)$ of an antichain passing through $r=0$ at time $t$ is equal to the length of a chain passing through $t=0$ at spatial distance $|t|$, multiplied by half the volume of the $(d-1)$-sphere $S_{d-1}$,
\begin{equation}
w(t)=l(|t|)S_{d-1}/2\,,
\end{equation}
where $S_d=(d+1)\pi^{(d+1)/2}/\Gamma((d+1)/2+1)$. The antichain width is translated to the discrete setting in the same way as the chain length:
\begin{equation}
\label{eq:expected_antichain}
\mathbb{E}\left[W(t)\right] = l(|t|)S_{d-2}/(\sqrt{2}\ell)\,,
\end{equation}
where $W(t)$ is the discrete antichain width. Using these expressions, the chain and antichain profiles are shown in Figure~\ref{fig:null_profiles}(center, right). Recall that while the antichain width measured by Algorithm~\ref{alg:antichain} does not exactly match that given by~\eqref{eq:expected_antichain}, the functional form is the same, making it a good enough measure of width for the purposes of the following experiments. It is believed the ratio of the peaks is a constant dependent only on dimension, as mentioned at the end of Chapter~\ref{chap:sets}, which will be left as an open problem for future study. \par

\begin{figure*}[!t]
\centering
\includegraphics[width=\linewidth]{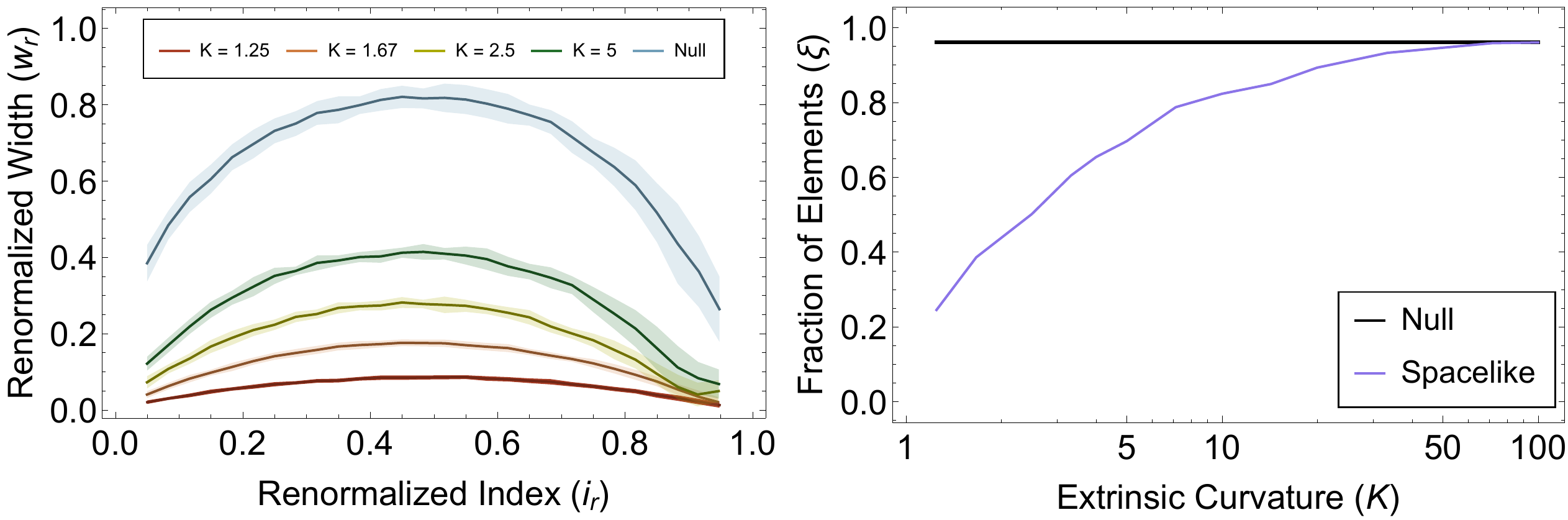}
\caption{{\bf Timelike and spacelike boundaries.} The left panel compares the renormalized antichain widths, $w_r=W/H_P$, for causal sets in several regions bounded by constant-curvature timelike hypersurfaces (red, orange, yellow, green) to those of causal sets with a null boundary (blue). The values are renormalized to lie in the range $[0,1]$ to account for changes in volume of different regions. The right panel shows the fraction of elements $\xi\equiv |A_{pf}|/N$ which lie in the Alexandroff set $A_{pf}$ defined by the extremal pair $(p,f)$ of the maximum chain for causal sets in regions bounded by spacelike hypersurfaces with variable extrinsic curvature $K$ (purple). In both cases, the boundaries are indistinguishable from null ones when $\ell K\to1$. All data is averaged over ten causal sets of size $N=2^{14}$, and the shaded regions indicate the standard deviation of the mean.}
\label{fig:profiles}
\end{figure*}

\subsection{Comparison of Timelike and Null Boundaries}
Using the two profiles in Figure~\ref{fig:null_profiles} for reference, one can compare causal sets from a region with timelike boundaries to those from one with a null side. In general, the chain profile is used to detect the top and bottom corners of the interval, and the antichain profile to detect the side corners. Therefore, to study timelike boundaries we focus on the antichain profile in particular. By studying a family of causal sets bounded by constant-curvature timelike surfaces one can show their antichain profiles converge toward the profile for the null boundary as $\ell K\to1$ (Figure~\ref{fig:profiles}(left)). The {\it renormalized index} $i_r\equiv a_i/H_P$ is simply the antichain index $a_i$ rescaled by the length of the maximum chain $H_P$, that is, the antichain ordering introduced in Section~\ref{sec:set_partitions} allows antichain labels $a_i$ to range from $0$ to $1$. The renormalized width $w_r=W_i/H_P$ likewise is the rescaled size of each antichain. Consequently, it becomes straightforward to distinguish timelike from null boundaries.

\subsection{Comparison of Spacelike and Null Boundaries}
One method to characterize spacelike boundaries is to examine the chain profile, but not very many chains are selected compared to the number of antichains, since chains cannot share extremal elements, and the fluctuation in lengths tends to increase for chains with $r\sim r_{max}$. Another way to characterize the boundary is to consider the size of the Alexandroff set $A_{pf}$ of the extremal pair $(p,f)$ of the maximum chain. For causal sets embedded in a causal interval, $|A_{pf}|$ converges to the size of the entire causal set as $N\to\infty$, whereas in a region with spacelike boundaries it does not. The right panel of Figure~\ref{fig:profiles} shows the fractional cardinality $\xi\equiv|A_{pf}|/N$ for causal sets in regions with constant-curvature spacelike boundaries as well as for those in the causal interval. Thus, we can sufficiently distinguish spacelike from null boundaries as well.

\section{The Boundary Volume}
\label{sec:volume}
Once we have distinguished the types of boundaries of an embedded causal set, we can confidently measure their volumes and other properties. In the following analysis, we assert $\ell K\ll1$, as mentioned in Section~\ref{sec:finite_causets}, to avoid further discussion about ambiguities. We begin by reviewing the analytic expression for the volume of spacelike boundaries, and then discuss algorithms for measuring the volume of timelike boundaries. In $(1+1)$-dimensional Minkowski spacetime, \mbox{codimension-2} corners enter as $0$-dimensional points, so the following discussion only covers numerical methods for their identification.
\subsection{Review of Spacelike Boundaries}
The volume of spacelike boundaries in causal sets was first reported in~\cite{buck2015boundary}. Given the number of minimal elements $P_0=|\mathcal{P}|$ and maximal elements $F_0=|\mathcal{F}|$ one may write the volumes of the past and future boundaries, $\Sigma^-$ and $\Sigma^+$ respectively, as
\begin{eqnarray}
V_{\Sigma^-} &=& \left(\frac{\ell}{l_p}\right)^d\frac{b_d}{\Gamma(\frac{1}{d+1})}F_0\,, \\
V_{\Sigma^+} &=& \left(\frac{\ell}{l_p}\right)^d\frac{b_d}{\Gamma(\frac{1}{d+1})}P_0\,,
\end{eqnarray}
where
\begin{equation}
b_d = (d+1)\left(\frac{S_{d-1}}{d(d+1)}\right)^{1/(d+1)}\,.
\end{equation}
In practice, the continuum volume is compared to $l_p^dV_{\Sigma^\pm}$. The convergence of these expressions is studied in Section~\ref{sec:examples5}.

\subsection{Timelike Boundaries}
While it is easy to identify and measure spacelike boundaries in a causal set, it is challenging to do the same for timelike boundaries. The following procedure solves this problem by first detecting these elements and then building chains which cover the boundary.

\begin{figure*}[!t]
\centering
\includegraphics[width=\linewidth]{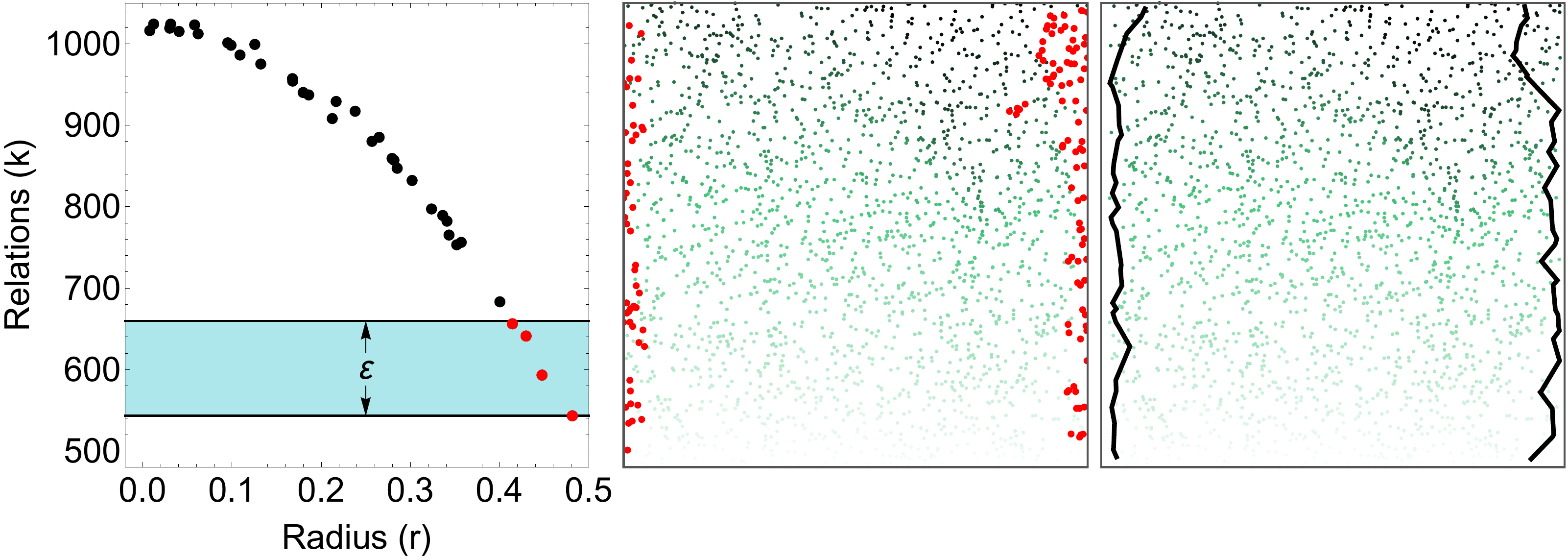}
\caption{{\bf Measurement of timelike boundary volume.} The causal set is partitioned into antichains, each of whose elements lie at a constant graph distance to the minimal elements $\mathcal{P}$. In each antichain, elements near the timelike boundary have the fewest number of relations (left). Those with a number of relations in the range $k\in[k_{min},k_{min}+\epsilon)$ are selected as candidates (red). The center panel shows the resulting set of candidates $\mathcal{T}$ on top of the antichain partitions, where each partition's elements are the same shade of green. Maximal chains $\mathcal{B}\in\mathfrak{B}$, which are proxies for timelike geodesics, are then constructed by maximizing the number of elements $\mathcal{T}$ in each chain, shown by the bold black lines (right). The origin is always taken to be at the center of the region to suggest a natural extension to higher dimensions.}
\label{fig:timelike}
\end{figure*}

\subsubsection{Boundary Element Detection}
Unlike the set of extremal elements which cover a spacelike boundary, the subset of elements $\mathcal{T}$ in the causal set $C$ which cover a timelike boundary is not trivial to quickly identify. We distinguish these elements from the {\it internal elements} located deep in the bulk first by observing that they have far fewer relations in expectation. In a faithful embedding into curved spacetime, the number of relations of elements along the timelike boundary also varies in the temporal direction. Therefore, we only compare elements on a spatial hypersurface, or antichain. This is not one of the maximal antichains described previously, but rather one of the set partitions generated when the causal set is partitioned into antichains. \par

The antichains are constructed by assigning to each element the maximum graph distance from that element to any one of the minimal elements, i.e., for element $n$ the distance is $t_n=$ \textsc{max}(\textsc{chain}($p,n$)) for all minimal elements $p\in\mathcal{P} : p\prec n$, where \textsc{chain}($p,n$) indicates the length of the longest chain between elements $p$ and $n$. Hence, each antichain is defined by the set of elements with equal $t_n$. The correlation between the number of relations and spatial distance from the origin is shown for the causal set embedded into a square in Figure~\ref{fig:timelike}(left). The elements with degree $k\in[k_{min},k_{min}+\epsilon)$, where the degree is the number of relations, are selected from each antichain as potential candidates to cover the timelike boundary, shown in the center panel of Figure~\ref{fig:timelike}. The depth $\epsilon$ adjusts the algorithm to select elements within a variable spatial distance from the boundary. The algorithm which selects a causal subset $\mathcal{T}\subset C$ is shown in Algorithm~\ref{alg:candidates}.\par

\begin{algorithm}[!t]
\caption{Timelike Boundary Candidates}
\label{alg:candidates}
\begin{algorithmic}[1]
\Input
\Statex $C$ \Comment A causal set
\Statex $\mathcal{P}$ \Comment Minimal elements
\Statex $k$ \Comment Number of relations per element
\Statex $\epsilon$ \Comment Boundary depth

\Procedure{chain}{$i$, $j$} \Comment This is a helper function for the procedure below
\State $A_{ij}\gets\mathcal{J}^+(i)\cap\mathcal{J}^-(j)$
\State $L\gets\{\}$ \Comment Empty array
\Return $\Call{chain}{A_{ij},L,0,i,j}$ \Comment The longest chain between $i$ and $j$ in $C$
\EndProcedure

\Procedure{candidates}{$C$, $\mathcal{P}$, $k$, $\epsilon$}
\State $\mathcal{T}\gets\{\}\,,t_n\gets -1\,\forall\,n$
\For {$p\in\mathcal{P}$ \textbf{and} $n\notin\mathcal{P}$} \Comment $p$ is a minimal element; $n$ is not
\If {$p\nprec n$}
\State \textbf{continue}
\EndIf
\State $t_n\gets$ \textsc{max}($t_n, \Call{chain}{p,n}$) \Comment Record the longest distance from $t_n$ to the $p$'s
\EndFor
\State $\kappa\gets\{\infty,\ldots,\infty\}$
\For {$i\in\{0,\ldots,\textsc{max}(t)-1\}$} \Comment In each of the antichain partitions$\ldots$
\For {$n\in C$} \Comment Record the fewest relations
\If {$t_n = i$}
\State $\kappa[i]\gets$ \textsc{min}($\kappa[i],k[n]$)
\EndIf
\EndFor
\For {$n\in\mathcal{C}$} \Comment Record elements with few relations
\If {$t_n = i$ {\bf and} $k[n] < \kappa[i]+\epsilon$} \Comment i.e., within the minimum plus $\epsilon$
\State $\mathcal{T}$.append($n$)
\EndIf
\EndFor
\EndFor
\EndProcedure
\Output
\Statex $\mathcal{T}$ \Comment The candidate elements
\end{algorithmic}
\end{algorithm}

\subsubsection{Timelike Boundary Measurement}
The second part of the procedure uses the candidate elements $\mathcal{T}$ to build a collection of chains $\mathfrak{B}$ which cover the timelike boundary. The method is similar to the one described in Section~\ref{sec:posets} which formed the set of extremal pairs. Using the maximal and minimal elements within the subset $\mathcal{T}$, maximal chains are formed using only the candidates in $\mathcal{T}$. For each adjacent pair of elements in a chain, i.e., $\{(i,j)\in \mathcal{B}(\mathcal{T}) : A_{ij}=\varnothing\}$ and $\mathcal{B}(X)$ is a chain along the boundary of $C$ consisting only of elements $X\subset C$, a maximal chain is constructed between $i$ and $j$ using the elements $\{m\in C\setminus\mathcal{T}\}$. This guides the chain along the boundary, enabling us to measure the boundary by incorporating elements in the full causal set rather than just the candidate elements. The longest chain $\mathcal{B}_{max}$ is taken to be a good cover of the boundary in a particular region, and then the elements which form that chain are removed from $\mathcal{T}$. \par

\begin{algorithm}[!t]
\caption{Timelike Boundary Measurement}
\label{alg:timelike_measurement}
\begin{algorithmic}[1]
\Input
\Statex $C$ \Comment A causal set
\Statex $\mathcal{T}$ \Comment Boundary candidates
\Statex $\delta$ \Comment Chain length threshold
\Procedure{ax\_set}{$X$, $i$, $j$} \Comment This is a helper function for the procedure below
\Return $\mathcal{J}^+_X(i)\cap\mathcal{J}^-_X(j)$ \Comment The Alexandroff set using elements $i$, $j$ in set $X$
\EndProcedure

\Procedure{timelike\_volume}{$C$,$\mathcal{T}$,$\delta$}
\State $\mathcal{P}\gets \{t\in\mathcal{T} : \mathcal{J}^-_\mathcal{T}(t)=\varnothing\}$ \Comment The minimal boundary elements \label{op:first}
\State $\mathcal{F}\gets \{t\in\mathcal{T} : \mathcal{J}^+_\mathcal{T}(t)=\varnothing\}$ \Comment The maximal boundary elements
\State $L\gets\{\},\,\mathcal{B}_{max}\gets\{\},\,l_{max}\gets0$
\For {$p\in\mathcal{P}$ {\bf and }$f\in\mathcal{F}$} \Comment For all pairs of minimal/maximal elements
\State $A_{pf}\gets$ \Call{ax\_set}{$\mathcal{T},p,f$} \Comment The Alexandroff set using only $\mathcal{T}$
\State $\{l_{pf},\mathcal{B}\}\gets\Call{chain}{A_{pf},L,0,p,f}$ \Comment The longest chain $\mathcal{B}$ and its length
\For {$(m, n)\in\mathcal{B} :$ \Call{ax\_set}{$\mathcal{T},m,n$} $= \varnothing$} \Comment For each link in $\mathcal{B}_X$
\State $A_{mn}\gets$ \Call{ax\_set}{$C,m,n$} \Comment Find the longest chain using $C$
\State $l_{pf}\pluseq\Call{chain}{X_{mn},L,0,m,n}$
\EndFor
\If {$l_{pf} > l_{max}$} \Comment Record the longest chains and lengths
\State $l_{max}\gets l_{pf}$
\State $\mathcal{B}_{max}\gets\mathcal{B}$
\EndIf
\EndFor
\If {$l_{max}>\delta$} \Comment If the chain is long enough, it is a good cover
\State $\tau\pluseq l_{max}$
\State $\mathcal{T}\mathrel{\setminus}=\mathcal{B}_{max}$
\State \Goto{op:first} \Comment Continue until no good covers remain
\EndIf
\EndProcedure

\Output
\Statex $\tau$ \Comment The boundary volume
\end{algorithmic}
\end{algorithm}

In $(1+1)$-dimensions, only the two longest chains are taken to cover the timelike boundaries, $|\mathfrak{B}|=2$, but in higher dimensions the procedure is repeated while $|\mathcal{B}_{max}|>\delta$ for some $\delta$. If the procedure continues until $\mathcal{T}=\varnothing$, one can see a sharp drop in the {\it chain weight}, defined as the number of elements in $\mathcal{T}$ occurring in $\mathcal{B}\in\mathfrak{B}$, and this transition can be used to pick $\delta$. Those chains with size smaller than $\delta$ typically cover regions already covered by longer chains. The algorithm describing this procedure is given in Algorithm~\ref{alg:timelike_measurement} and the result is shown in Figure~\ref{fig:timelike}(right). Once the total number of elements $\tau=\sum_i|\mathcal{B}_i|$ in the set of chains has been measured, the continuum length may be recovered via~(\ref{eq:geo_to_chain}).

\subsubsection{Convergence}
We claim in the $N\to\infty$ limit the chains $\mathcal{B}\in\mathfrak{B}$ perfectly cover the timelike boundaries. This can only occur if $\epsilon$ is controlled in a way that the number of elements in $\mathcal{T}$ grows like the \mbox{codimension-1} volume of the timelike boundary, in units of $\ell$, rather than the number of elements $N$. Hence, if $\epsilon$ is chosen such that the number of candidate elements per antichain grows like $N^{d-1}$, and $\epsilon$ is as small as possible such that the causal subset $C_\mathcal{T}$ defined by the elements of $\mathcal{T}$ is percolated, i.e., $C_\mathcal{T}$ has two connected components in $(1+1)$ dimensions or one connected component in higher dimensions, then the elements of $\mathcal{T}$ will always remain close to the timelike boundary. Since it is known maximal chains converge to timelike geodesics as $N\to\infty$~\cite{brightwell1991structure}, then the measured boundary volume will converge to the continuum volume when $\epsilon$ is bounded using this prescription.

\subsection{Corners}
\label{sec:corners}
The codimension-2 boundaries are known as corners, and they arise due to the intersections of codimension-1 boundaries. Detecting corners induced by spacelike-timelike boundary intersections is easy, since the corner elements $\Upsilon$ are simply the extremal elements of the chains covering the timelike boundaries, i.e., $\Upsilon=(\mathcal{P}\cap\mathcal{B})\cup(\mathcal{F}\cap\mathcal{B})$. When a corner has an obtuse angle, it is difficult to infer its presence from the chain and antichain profiles alone. It is helpful to use another profile as well, called the {\it Alexandroff profile}, to characterize the hypersurface. One measures the size of the Alexandroff set $A_{pf}$ using each chain's extremal pair $(p,f)$, and watches how its size changes with the inferred radial distance, which was described in Section~\ref{sec:distributions}. To detect these corners, the graph density must be very large to get an accurate measurement of the derivative of the Alexandroff and chain profiles for small renormalized index. If they never tend to zero, we can remain confident a corner actually exists. \par

\begin{figure}[!t]
\centering
\includegraphics[width=\linewidth]{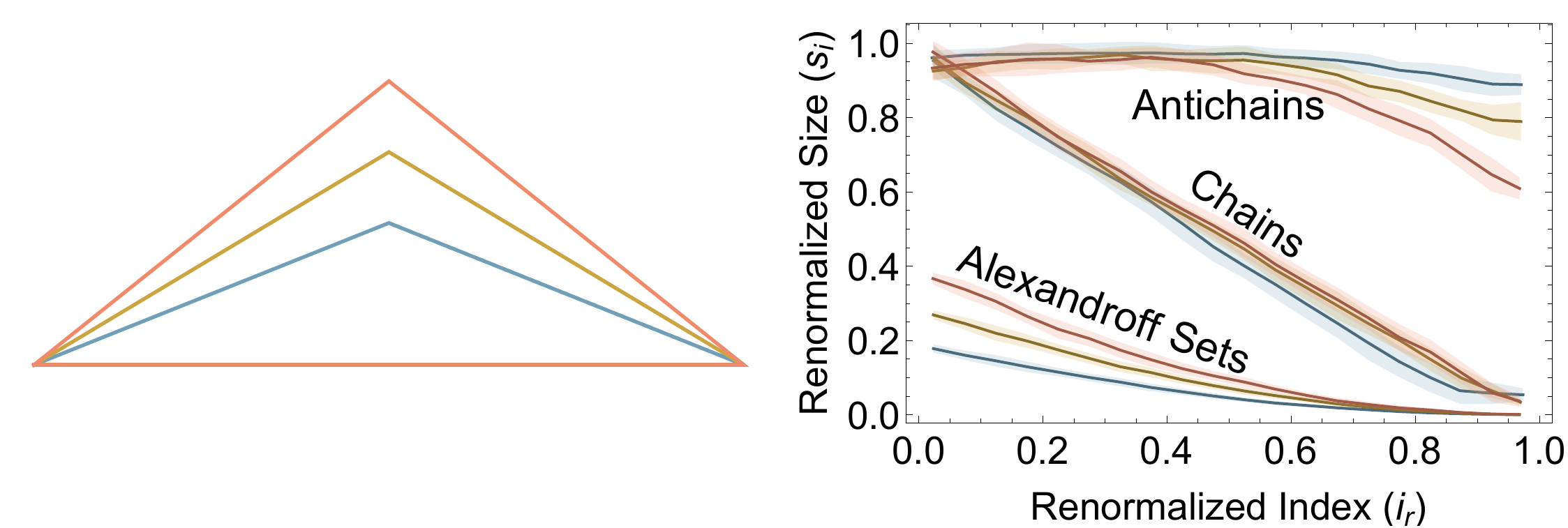}
\caption{{\bf Detection of codimension-2 corners.} Causal sets embedded into different triangular regions (left) have three codimension-2 corners. The corners formed by two spacelike hypersurfaces intersecting at acute angles are characterized by a large difference in the chain and antichain sizes at large renormalized index (right). In particular, the antichain size never decreases toward zero unless one of the hypersurfaces approaches the null limit. The renormalized size is equivalent to the renormalized width $w_r$ for antichains, the renormalized length $l_r$ for chains, and fractional Alexandroff set size $\xi$ for Alexandroff sets. Data is averaged over ten causal sets of size $N=2^{13}$, and the shaded regions indicate the standard deviation of the mean.}
\label{fig:corners}
\end{figure}
When the corner's angle is acute, it is somewhat easier to determine its presence. Figure~\ref{fig:corners} demonstrates what the chain, antichain, and Alexandroff profiles look like for an isosceles triangle defined by the points $\{(0,-1),(0,1),(t_0,0)\}$. The renormalized size $s_i$ for chains and antichains refers to the renormalized length and width, respectively, whereas for the Alexandroff profile $s_i=|A_{pf}|/N$. The acute angle is characterized by the large difference in the two profiles: chains whose lengths go to zero at large inferred radius combined with antichains which are always large indicate there is no timelike or null boundary. While it may appear the slope of the chain profile in Figure~\ref{fig:corners} could measure the angle, preliminary experiments indicate neither the slope of the chain profile nor that of the Alexandroff profile are reliable metrics.

\section{Examples}
\label{sec:examples5}
To demonstrate the approaches described heretofore, we consider several examples in various regions of $(1+1)$-dimensional Minkowski spacetime. In each case, we look at how the chain, antichain, and Alexandroff profiles can be used together to identify the shape of a bounding region in a flat embedding space and estimate the boundary volume. It is important to emphasize that the following arguments are useful as a first step towards characterizing the boundary, and in practice it is best to compare results to the profiles of causal sets with known boundaries which are generated from sprinklings. \par

Each example highlights a certain difficulty or ambiguity which one might encounter in practice. When we measure timelike boundaries, we take the smallest $\epsilon$ such that at least two elements are selected from each antichain partition, and we take the largest two chains in $\mathfrak{B}$. All data shown is averaged over ten graphs of size $N=2^{11}$ unless otherwise indicated. \par

\subsection{The Square and the Cylinder}
\begin{figure}[!p]
\centering
\includegraphics[width=\linewidth]{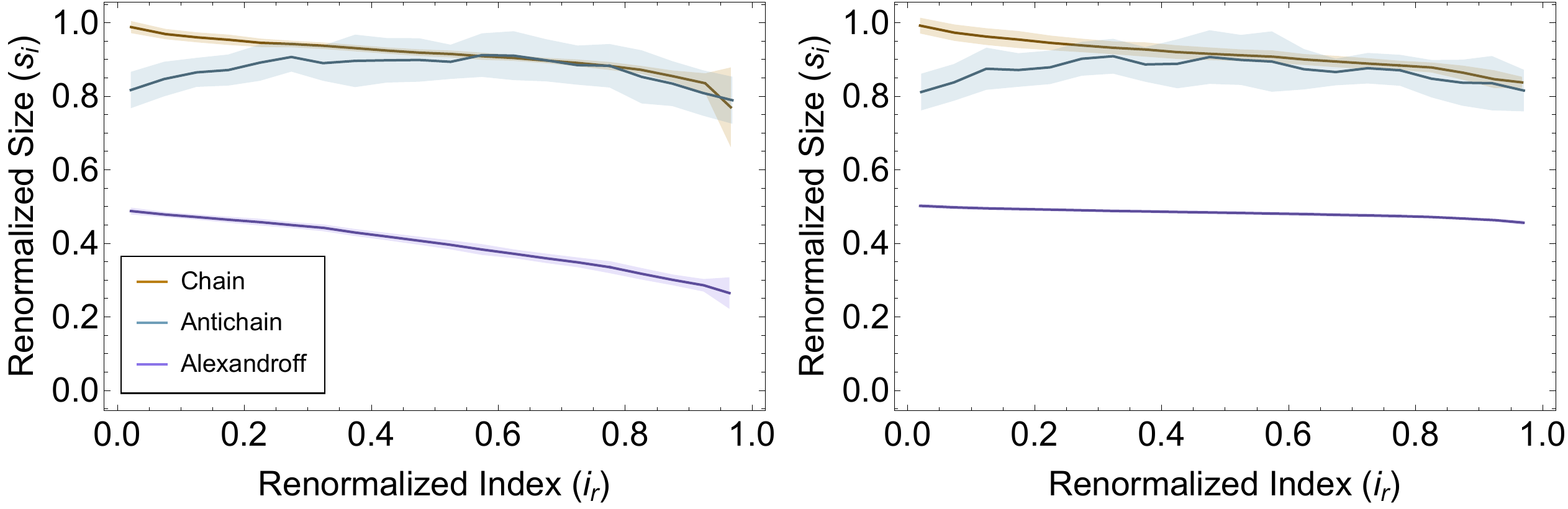}
\caption{{\bf The square versus the cylinder.}}
\label{fig:sq_v_cyl}
\end{figure}

The first example demonstrates how one might differentiate between a causal set in a square region with flat timelike boundaries and one in a region with no spatial boundaries, i.e., the surface of a 2-cylinder. The chain, antichain, and Alexandroff profiles are shown for the causal sets in the square in Figure~\ref{fig:sq_v_cyl}(left). The chain and antichain profiles remain nearly constant, indicating the boundary shape is likely flat and symmetric. The chain profile always decreases slightly even when the spacelike boundaries are flat and constant, since the chain distribution can never be uniform when there are Poisson fluctuations near a boundary. Further, the renormalized size of the Alexandroff profile decreases from about a half to a quarter, which is a characteristic of the square region. All three of the square's profiles are distinct from those of the null boundary (Figures~\ref{fig:null_profiles},~\ref{fig:profiles}), leaving no ambiguity over the existence of at least a spacelike boundary. Compared to those of the square, the profiles for the cylinder (Figure~\ref{fig:sq_v_cyl}(right)) are nearly the same, except for the Alexandroff profile. When the chain length between the past and future spacelike hypersurfaces is spatially independent, likewise there should be no dependence of the Alexandroff profile on spatial position. \par

\subsection{The Deformed Square}
\begin{figure}[!p]
\centering
\includegraphics[width=\linewidth]{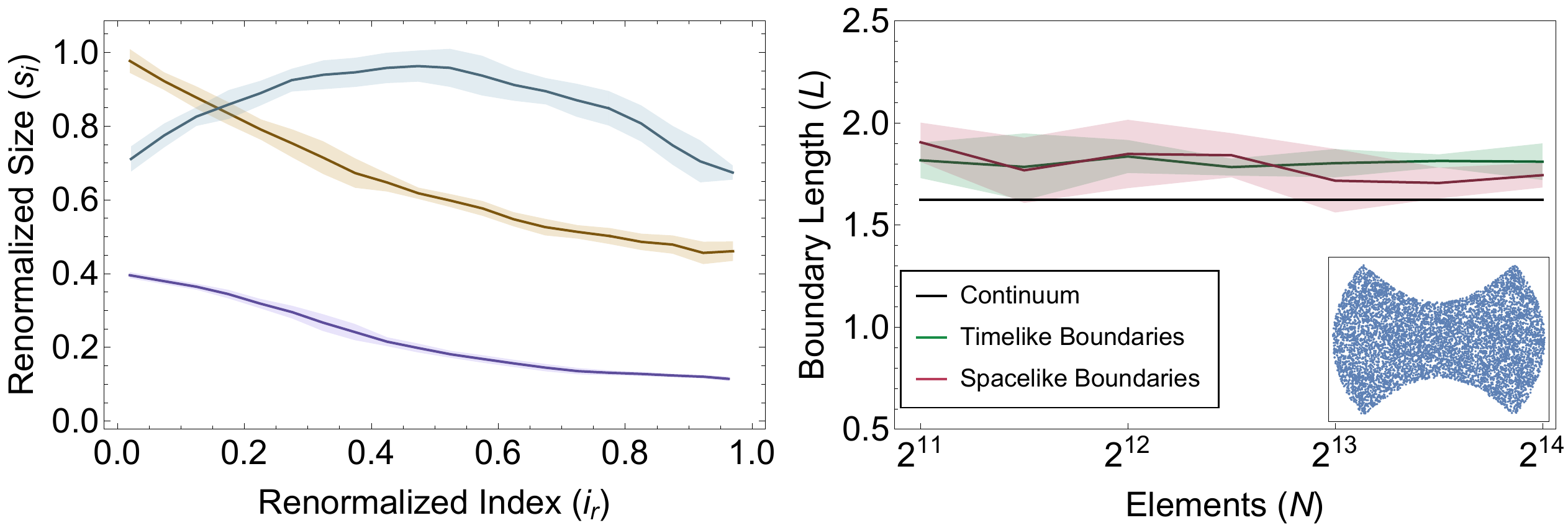}
\caption{{\bf The deformed square.}}
\label{fig:def_sq}
\end{figure}

The second example demonstrates what happens when there is a mixture of convex and concave boundaries, shown by the deformed square in the inset of Figure~\ref{fig:def_sq}(right). The left panel of the figure shows the three profiles for this region. The antichain profile indicates the timelike boundary is convex but non-null, since the renormalized size always remains far above zero, i.e., there are no small antichains. The Alexandroff profile differs from the previous example in values but not in behavior, indicating the presence of timelike boundaries and curved spacelike boundaries. \par

The most notable difficulty here is that the longest chain no longer runs through the center, but rather through two of the four corners. When a spacelike boundary is concave, the chains in the chain profile are ordered differently, so the method which detects elements near a timelike boundary has some trouble, especially when the extrinsic curvature is large. The monotonically decreasing chain profile could lead one to believe the spacelike boundary is actually convex. Despite this apparent ambiguity, the sign of the spacelike boundary term of the action on expectation gives the sign of the boundary curvature~\cite{buck2015boundary}. \par

\begin{figure}[!p]
\centering
\includegraphics[width=\linewidth]{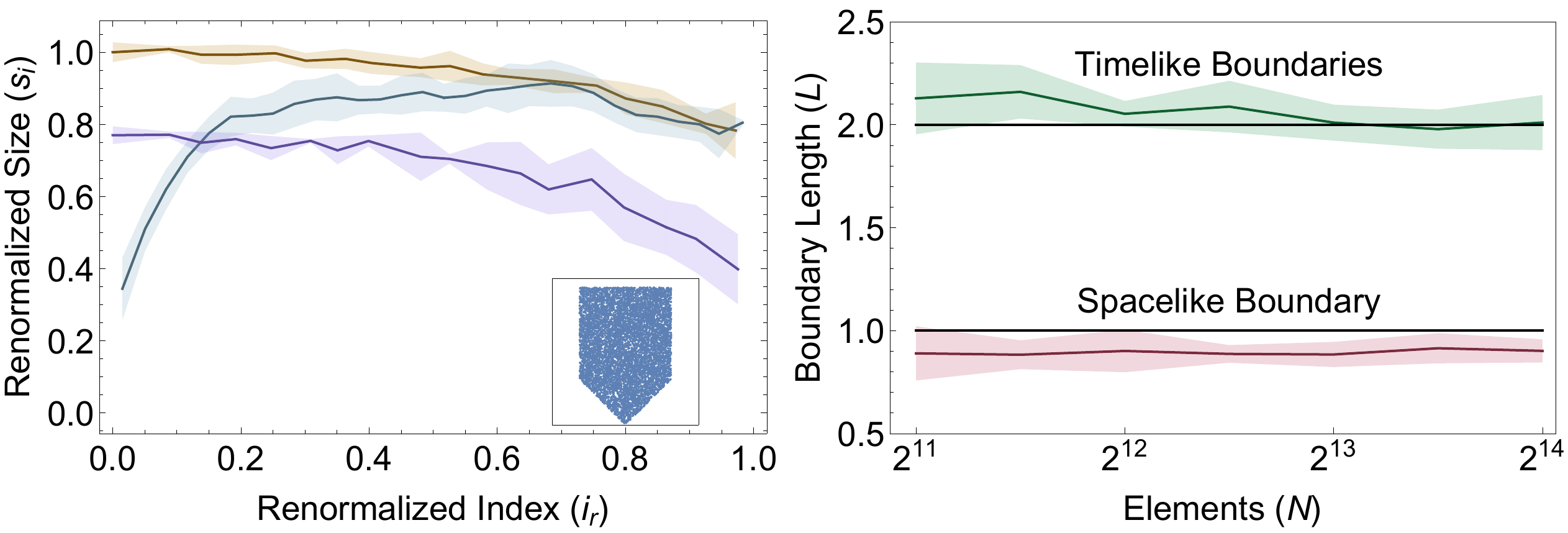}
\caption{{\bf The isosceles right pentagon.}}
\label{fig:pentagon}
\end{figure}
The right panel of Figure~\ref{fig:def_sq} shows the results of the timelike and spacelike boundary volume estimation using the methods described in Section~\ref{sec:volume}. All four sides of this region have the same length in the continuum. While the results are in close agreement for the range of causal set sizes shown, they do not yet converge precisely to the continuum limit. It is expected at larger $N$ this convergence occurs, but it is not yet clear what order of magnitude is required. Surprisingly, the value measured for the timelike boundary volume is very close to that for the spacelike boundary volume, despite the fact that the former is an algorithm and the latter an analytic result. One might think it would be worse, since the longest chain is no longer at the spatial origin and some of the original assumptions do not hold. For instance, the correlation between radius and number of relations (Figure~\ref{fig:timelike}(left)) is positive rather than negative in this region, so in the first few antichain partitions, candidates are selected close to $r=0$. For this particular choice of width and height, the boundary measurement algorithm still succeeds, but this flaw implies that when the extrinsic curvature of a concave boundary is too large, the algorithm builds a chain directly through the center. Since one may easily test the sign of the boundary curvature, future work will focus on modifications for these cases.

\subsection{The Pentagon}
In the concluding example, we return to the region shown in Figure~\ref{fig:embedding}. For simplicity we take the $(1+1)$-dimensional version, i.e., an isosceles right pentagon with three equal sides, shown by the inset in the left panel of Figure~\ref{fig:pentagon}. The left panel once again shows the three profiles for the region. The chain profile is nearly uniform, indicating the spacelike boundaries are either flat or null and are radially symmetric. The Alexandroff profile is not extremely helpful in this case; it only suggests the existence of timelike boundaries, as it did in the other two examples. The key feature which clarifies the extrinsic geometry of this region is the antichain profile: the curve grows quickly in the first third of the region, and then remains roughly constant in the upper two-thirds. The fact that there are small antichains, along with the shape of the growth, indicates there is a null boundary (see Figure~\ref{fig:null_profiles}). The uniformity in the upper two-thirds strongly suggests that portion of the boundary is flat and timelike. Together with the chain profile, these results also suggest the future spacelike boundary is flat, and this can be confirmed by studying the spacelike boundary term of the action. \par

The right panel of Figure~\ref{fig:pentagon} shows the measurements of the spacelike and timelike boundary volumes. Not surprisingly, the timelike boundary measurement algorithm performs well for flat boundaries, even in the presence of a null boundary. The spacelike boundary volume measurement appears to be consistently below the continuum value, indicating these causal sets are not yet large enough to show convergence.

\section{Summary}
By constructing and examining the chain, antichain, and Alexandroff profiles, we have learned how to identify the different types of boundaries of a causal set. We developed a spacetime representation to build maximal chains and antichains as a way to qualitatively describe the causal set. After looking at the profiles for the null boundary in Section~\ref{sec:distributions}, we could distinguish a null surface from both a spacelike and timelike boundary, provided it has a small enough extrinsic curvature. The timelike boundary element detection algorithm presented in Algorithm~\ref{alg:candidates} led to a method for the measurement of such a boundary via the guided chain construction in Algorithm~\ref{alg:timelike_measurement}. Finally, we studied the properties of causal sets embedded into three spacetime regions in Section~\ref{sec:examples5}. While results here focused on $(1+1)$-dimensional Minkowski spacetime, the techniques can easily be generalized to study higher-dimensional conformally flat spacetimes in future work.
\afterpage{\blankpage}

%\part{Navigation and the Multiverse}
%\label{part:apps}
\part{Geodesics in Conformally Flat Manifolds}
\label{part:geodesics}
\thispagestyle{empty}
\afterpage{\blankpage}
\twolinechapter
\chapter[Geodesics in Conformally Flat Manifolds]{\texorpdfstring{Geodesics in Conformally\\[-0.8cm] Flat Manifolds}{Geodesics in Conformally Flat Manifolds}}
\mainchapter
\label{chap:geodesics}
\thispagestyle{empty}

Cosmic microwave background experiments such as COBE~\cite{smoot1992structure}, WMAP~\cite{hinshaw2013nine}, and Planck~\cite{ade2016planck} provide evidence for both early time cosmic inflation~\cite{guth1981inflationary,linde1982new} and late time acceleration~\cite{perlmutter1998discovery,riess1998observational}, with interesting dynamics in between explaining many features of the universe, many of which are remarkably accurately predicted by the $\Lambda$CDM model~\cite{peebles1982large,turner1984flatness,blumenthal1984formation,davis1985evolution}. These and other experiments in recent decades have demonstrated that, to a high degree of precision, at large scales the visible universe is spatially homogeneous, isotropic, and flat, i.e., its spacetime is described by the Friedmann-Lema\^itre-Robertson-Walker (FLRW) metric. FLRW spacetimes are therefore of particular interest in modern cosmology. \par

Here we develop a method for the exact calculation of the geodesic distance between any given pair of events in any flat FLRW spacetime. Geodesics and geodesic distances naturally arise in a wide variety of investigations not only in cosmology, but also in astrophysics and quantum gravity, with topics ranging from the horizon and dark energy problems, to gravitational lensing, to evaluating the observational signatures of cosmic bubble collisions and modified gravity theories, to the AdS/CFT correspondence~\cite{ellis1999deviation,gron2007space,albareti2012focusing,demianski2003approximate,hirata2005superhorizon,bikwa2012photon,doplicher2013quantum,melia2013proper,bhattacharya2017cosmic,pyne1996lens,park2008rigorous,sereno2009role,mukohyama2009dark,mukohyama2009caustic,traschen1986large,futamase1989light,cooperstock1998influence,caldwell2001shortcuts,dappiaggi2008stable,kaloper2010mcvitties,melia2013ct,bahrami2017saturating,koyama2001strongly,dong2012frw,dong2012aspects,minton2008new,wainwright2014simulating,hagala2016cosmological}. Closed-form solutions of the geodesic equations are also quite useful in validating a particular FLRW model by investigating how curvature, quintessence, local shear terms, etc., affect observational data.  These solutions are of perhaps the greatest and most direct utility in large-scale N-body simulations, e.g., studying the large scale structure formation, which can benefit greatly from using such solutions by avoiding the costly numerical integration of the geodesic differential equations~\cite{adamek2016general,koksbang2015methods,bibiano2017pairwise}. In this dissertation, these closed-form solutions are motivated by the greedy information routing problems presented in the following chapter. \par

For a general spacetime, solving the geodesic equations exactly for given initial-value or boundary-value constraints is intractable, although it may be possible in some cases. For example, in $(3+1)$-dimensional de Sitter space, which represents a spacetime with only dark energy and is a maximally symmetric solution to Einstein's equations, it turns out to be rather simple to study geodesics by embedding the manifold into flat $(4+1)$-dimensional Minkowski space $\mathcal{M}^5$. This construction was originally realized by de Sitter himself~\cite{desitter1917einstein}, and was later studied by Schr\"odinger~\cite{schrodinger1956expanding}. In Section~\ref{sec:desitter} we review how geodesics may be found using the unique geometric properties of this manifold. \par

However, it is not so easy to explicitly calculate geodesic distances in other FLRW spacetimes except under certain assumptions. One approach would be to follow de Sitter's philosophy by finding an embedding into a higher-dimensional manifold. Such an embedding always exists due to the Campbell-Magaard theorem, which states that any analytic $n$-dimensional Riemannian manifold may be locally embedded into an $(n+1)$-dimensional Ricci-flat space~\cite{campbell1926course,magaard1963einbettung}, combined with a theorem due to A.~Friedman extending the result to pseudo-Riemannian manifolds~\cite{friedman1961local,friedman1965isometric}. In fact, the embedding map is given explicitly by J.~Rosen in~\cite{rosen1965embedding}. However, it has since been shown that the metric in the embedding space is block diagonal with respect to the embedded surface, i.e., when the geodesic is constrained to the $(3+1)$-dimensional subspace we regain the original $(3+1)$-dimensional geodesic differential equations and we learn nothing new~\cite{romero1996embedding}. \par

Instead, in Section~\ref{sec:exact_solutions} we solve directly the geodesic differential equations for a general FLRW spacetime in terms of the scale factor and a set of initial-value or boundary-value constraints. The final geodesic distance can be written as an integral which is a function of the boundary conditions and one extra constant $\mu$, defined by a transcendental integral equation. This constant proves to be useful in a number of ways: it tells us if a manifold is geodesically connected provided only the scale factor. The solution of the integral equation defining this constant exists only if a geodesic exists for a given set of boundary conditions, and it helps one to find the geodesic distance, if such a geodesic exists. \par

We then give some examples in Section~\ref{sec:examples8} to show that for many scale factors of interest, we can find a closed-form solution. In cases where no closed-form solution exists, we can still transform the problem into one which is suitable for fast numerical integration. Numerical approximations used in the following chapter are also explained in detail in Section~\ref{sec:num_approx}.

\section[Review of Friedmann-Lema\^itre-Robertson-Walker Spacetimes and de Sitter Embeddings]{Review of FLRW Spacetimes and de Sitter Embeddings}
\sectionmark{Review of FLRW Spacetimes}
\label{sec:frw_spacetimes}
Friedmann-Lema\^itre-Robertson-Walker (FLRW) spacetimes are spatially homogeneous and isotropic $(3+1)$-dimensional Lorentzian manifolds which are solutions to Einstein's equations~\cite{griffiths2009exact}. These manifolds have a metric $g_\munu$ with $\mu$, $\nu \in \{0,1,2,3\}$ that, when diagonalized in a given coordinate system, gives an invariant interval $ds^2=g_\munu\,dx^\mu\,dx^\nu$ of the form
\begin{equation}
\label{eq:invariant_interval}
ds^2=-dt^2+a(t)^2\,d\Sigma^2\,,
\end{equation}
where $a(t)$ is the scale factor, which describes how space expands with time $t$, and $d\Sigma$ is the spatial %part of the
metric given by $d\Sigma^2=dr^2+r^2(d\theta^2+\sin^2\theta\,d\phi^2)$ for flat space in spherical coordinates $(r,\theta,\phi)$ that we use hereafter. The scale factor is found by solving Friedmann's equation, the differential equation given by the $\mu=\nu=0$ component of Einstein's equations:
\begin{equation}
\left(\frac{\dot a}{a}\right)^2 = \frac \Lambda 3 + \frac{c}{a^{3g}}\,,
\end{equation}
where $\Lambda$ is the cosmological constant, $g$ parametrizes the type of matter within the spacetime,
$c$ is a constant proportional to the matter density, and we have assumed spatial flatness in our choice of $d\Sigma$. The scale factors for manifolds which represent spacetimes with dark energy ($\Lambda$), dust ($D$), radiation ($R$), a stiff fluid ($S$),%
\footnote{Stiff fluids are exotic forms of matter which have a speed of sound equal to the speed of light. They have been studied in a variety of models of the early universe, including kination fields, self-interacting (warm) dark matter, and Ho\^rava-Lifshitz cosmologies~\cite{dutta2010big}.}
or some combination (e.g., $\Lambda D$ for dark energy and dust matter) are given by~\cite{griffiths2009exact}
\begin{subequations}
\label{eq:scale_factors}
\begin{align}
a_\Lambda(t) &= \lambda e^{t/\lambda} \label{eq:scale_factor_lambda}\,, \\
a_D(t) &= \alpha\left(\frac{3t}{2\lambda}\right)^{2/3} \label{eq:scale_factor_D}\,, \\
a_R(t) &= \alpha^{3/4}\left(\frac{2t}{\lambda}\right)^{1/2} \label{eq:scale_factor_R}\,, \\
a_S(t) &= \alpha^{1/2}\left(\frac{3t}{\lambda}\right)^{1/3} \label{eq:scale_factor_S}\,, \\
a_{\Lambda D}(t) &= \alpha\sinh^{2/3}\left(\frac{3t}{2\lambda}\right) \label{eq:scale_factor_lambda_D}\,, \\
a_{\Lambda R}(t) &= \alpha^{3/4}\sinh^{1/2}\left(\frac{2t}{\lambda}\right) \label{eq:scale_factor_lambda_R}\,, \\
a_{\Lambda S}(t) &= \alpha^{1/2}\sinh^{1/3}\left(\frac{3t}{\lambda}\right) \label{eq:scale_factor_lambda_S}\,,
\end{align}
\end{subequations}
where $\lambda$ and $\alpha\equiv(c\lambda^2)^{1/3}$ are the temporal and spatial scale-setting parameters.
In manifolds with dark energy, i.e., $\Lambda>0$, $\lambda\equiv\sqrt{3/\Lambda}$.

\subsection{de Sitter Spacetime}
\label{sec:desitter}
The de Sitter spacetime is one of the first and best studied spacetimes: de Sitter himself recognized that the $(3+1)$-dimensional manifold d$\mathcal{S}^4$ can be visualized as a single-sheet hyperboloid embedded in $\mathcal{M}^5$, defined by
\begin{equation}
-z_0^2+z_1^2+z_2^2+z_3^2+z_4^2=\lambda^2\,,
\end{equation}
where $\lambda$ is the pseudo-radius of the hyperboloid~\cite{desitter1917einstein}. The injection $\chi\,:\,\mathrm{d}\mathcal{S}^4\hookrightarrow\mathcal{M}^5\,,\,\chi(x)\mapsto z$ is
\begin{align}
\begin{aligned}
\frac{\lambda^2+s^2}{2\eta}&\mapsto z_0\,,\\
\frac{\lambda^2-s^2}{2\eta}&\mapsto z_1\,,\\
\frac{\lambda}{\eta}r\cos\theta&\mapsto z_2\,,\\
\frac{\lambda}{\eta}r\sin\theta\cos\phi&\mapsto z_3\,, \\
\frac{\lambda}{\eta}r\sin\theta\sin\phi&\mapsto z_4\,,
\end{aligned}
\end{align}
where $s^2\equiv r^2-\eta^2$, and the conformal time $\eta$ is defined as
\begin{equation}
\label{eq:conformal_time}
\eta(t)=\int^t\!\frac{dt'}{a(t')}\,.
\end{equation}
This embedding is a particular instance of the fact that any analytic $n$-dimensional pseudo-Riemannian manifold may be isometrically embedded into (at most) a $(n(n+1)/2)$-dimensional pseudo-Euclidean manifold (i.e., a flat metric with arbitrary non-Riemannian signature)~\cite{friedman1961local,friedman1965isometric}. The minimal $(n+1)$-dimensional embedding is most easily obtained using group theory by recognizing that the Lorentz group \textit{SO}(1,3) is a stable subgroup of the de Sitter group \textit{dS}(1,4) while the pseudo-orthogonal group \textit{SO}(1,4) acts as its group of motions, i.e., \textit{dS}(1,4)~=~\textit{SO}(1,4)/\textit{SO}(1,3), thereby indicating the minimal embedding is into the $\mathcal{M}^5$ space~\cite{aldrovandi1995introduction}. \par

\begin{figure*}[!t]
\vspace*{-5.5mm}
\includegraphics[width=\textwidth]{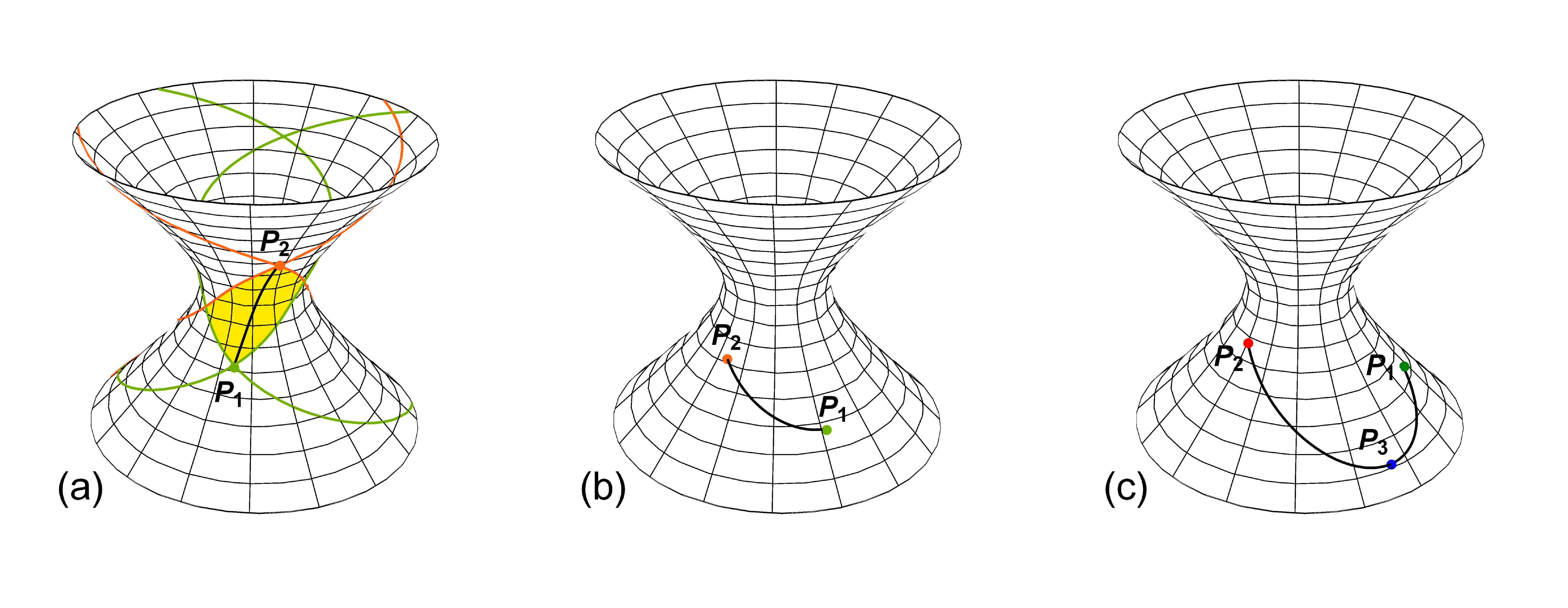}
\centering
\vspace*{-10mm}
\caption{\textbf{Geodesics on the 1+1 de Sitter manifold.} There are three classes of non-null geodesics on the de Sitter manifold. In (a), we see a future-directed timelike geodesic emanating from $P_1$ and terminating at $P_2$. These geodesics map out physical trajectories of subluminal objects within spacetime because the two points lie within each other's light cones, shown by the green and red lines. The Alexandroff set of points causally following $P_1$ and preceding $P_2$ is shown in yellow.
A spacelike geodesic joining two points with no causal overlap, shown in (b), ``bends away from the origin,'' meaning that in the plane defined by the origin of $\mathcal{M}^3$ and points $P_1,P_2$, this geodesic is farther from the origin than the Euclidean geodesic between the same points. If a spacelike geodesic extends far enough, there will exist an extremum, identified as $P_3$ in (c). As a result, it is simplest to use the spatial distance $\omega$ to parametrize these geodesics, though time can be used as well so long as those geodesics with turning points are broken into two parts at the point $P_3$.
}
\label{fig8-1}
\end{figure*}

If a spacelike geodesic extends far enough, there will exist an extremum, identified as $P_3$ in Figure~\ref{fig8-1}(c). As a result, it is simplest to use the spatial distance $\omega$ to parametrize these geodesics, though time can be used as well so long as those geodesics with turning points are broken into two parts at the point $P_3$.
Furthermore, it can be shown that geodesics on a de Sitter manifold follow the lines defined by the intersection of the hyperboloid with a hyperplane in $\mathcal{M}^5$ containing the origin and both endpoints of the geodesic~\cite{asmus2009duality}. An illustration of both timelike and spacelike geodesics constructed this way in d$\mathcal{S}^2$ embedded in $\mathcal{M}^3$ can be found in Fig.~\ref{fig8-1}. This construction implies that the geodesic distance $d(x,y)$ in d$\mathcal{S}^4$ between two points $x$ and $y$ can be found using their inner product $\langle x,y\rangle=-x_0y_0+x_1y_1+x_2y_2+x_3y_3+x_4y_4$ in $\mathcal{M}^5$ via the following expression:
\begin{equation}
\label{eq:ds_5d}
d(x,y) = \begin{cases}\lambda\arccosh\frac{\langle x,y\rangle}{\lambda^2}&\text{if } x - y \text{ is timelike,} \\
0&\text{if } x - y \text{ is lightlike,} \\
\infty&\text{if } \langle x,y\rangle\leq -\lambda^2\text{ and } x\neq -y\,, \\
\lambda\arccos\frac{\langle x,y\rangle}{\lambda^2}&\text{otherwise.} \end{cases}
\end{equation}
\\
While there are many ways to find geodesic distances on a de Sitter manifold, this is perhaps the simplest one.

\section{The Geodesic Equations in Four Dimensions}
\label{sec:exact_solutions}
While de Sitter symmetries cannot be exploited in a general FLRW spacetime, it is still possible to solve the geodesic equations. A geodesic is defined in general by the variational equation
\begin{equation}
\delta\int ds = 0\,,
\end{equation}
which, if parametrized by parameter $\sigma$ ranging between two points $\sigma_1$ and $\sigma_2\,,$ becomes
\begin{equation}
\label{eq:geodesic_variation}
\delta\int_{\sigma_1}^{\sigma_2}\!\sqrt{g_{\mu\nu}\frac{\partial x^\mu}{\partial\sigma}\frac{\partial x^\nu}{\partial\sigma}}\,d\sigma = 0\,.
\end{equation}
The corresponding Euler-Lagrange equations obtained via the variational principle yield the well-known geodesic differential equations:
\begin{equation}
\label{eq:geodesic_diff_eq}
\nabla_X\frac{\partial x^\mu}{\partial\sigma} = \frac{\partial^2x^\mu}{\partial\sigma^2} + \Gamma_{\rho\tau}^\mu\frac{\partial x^\rho}{\partial\sigma}\frac{\partial x^\tau}{\partial\sigma} = \gamma\left(\sigma\right)\frac{\partial x^\mu}{\partial\sigma}\,,
\end{equation}
for the geodesic path $x^\mu(\sigma)$ with some as yet unknown function $\gamma(\sigma)$, where $\Gamma_{\rho\tau}^\mu$ are the Christoffel symbols defined by
\begin{equation}
\Gamma_{\rho\tau}^\mu = \frac{1}{2}g^{\mu\nu}\left(\frac{\partial g_{\nu\rho}}{\partial x^\tau} + \frac{\partial g_{\nu\tau}}{\partial x^\rho} - \frac{\partial g_{\rho\tau}}{\partial x^\nu}\right)\,,
\end{equation}
and $\nabla_X$ indicates the covariant derivative with respect to the tangent vector field $X$~\cite{oneill1983semi}. If the parameter $\sigma$ is affine, then $\gamma(\sigma)=0$. To solve a particular problem with constraints, we must use both~\eqref{eq:geodesic_variation} and~\eqref{eq:geodesic_diff_eq}.

\subsection{The Differential Form of the Geodesic Equations}
If only the non-zero Christoffel symbols are kept, then \eqref{eq:geodesic_diff_eq} can be broken into two differential equations written in terms of the scale factor:
\begin{subequations}
\label{eq:geodesic2}
\begin{align}
\frac{\partial^2t}{\partial\sigma^2} + a\frac{da}{dt}h_{ij}\frac{\partial x^i}{\partial\sigma}\frac{\partial x^j}{\partial\sigma} &= \gamma\frac{\partial t}{\partial\sigma}\label{eq:geodesic2a}\,, \\
\frac{\partial^2x^i}{\partial\sigma^2}+\frac{2}{a}\frac{da}{dt}\frac{\partial t}{\partial\sigma}\frac{\partial x^i}{\partial\sigma} + \Gamma_{jk}^i\frac{\partial x^j}{\partial\sigma}\frac{\partial x^k}{\partial\sigma} &= \gamma\frac{\partial x^i}{\partial\sigma}\,, \label{eq:geodesic2b}
\end{align}
\end{subequations}
where $h_{ij}$ is the first fundamental form, i.e., the induced metric on a constant-time hypersurface, and the Latin indices are restricted to $\{1,2,3\}$. \par

To solve these, consider the spatial (Euclidean) distance $\omega$ between two points $\sigma_1$ and $\sigma_2$:
\begin{align}
\omega &= \int_{\sigma_1}^{\sigma_2}\!\sqrt{h_{ij}\frac{\partial x^i}{\partial\sigma}\frac{\partial x^j}{\partial\sigma}}\,d\sigma\,, \nonumber \\
\left(\frac{\partial\omega}{\partial\sigma}\right)^2 &= h_{ij}\frac{\partial x^i}{\partial\sigma}\frac{\partial x^j}{\partial\sigma}\,. \label{eq:omega_def}
\end{align}
This relation implies the spatial coordinates obey a geodesic equation with respect to the induced metric $h_{ij}$. Now,~\eqref{eq:geodesic2a} may be written in terms of $\omega$ using~\eqref{eq:omega_def}. The transformation needed for~\eqref{eq:geodesic2b} is found by multiplying by $h_{ij}(\partial x^j/\partial\sigma)$ and substituting the derivative of~\eqref{eq:omega_def} with respect to $\omega$:
\begin{align}
\begin{aligned}
\frac{\partial\omega}{\partial\sigma}\frac{\partial^2\omega}{\partial\sigma^2} &- \frac{1}{2}\frac{\partial h_{ij}}{\partial x^k}\frac{\partial x^i}{\partial\sigma}\frac{\partial x^j}{\partial\sigma}\frac{\partial x^k}{\partial\sigma} + \frac{2}{a}\frac{da}{dt}\frac{\partial t}{\partial\sigma}\left(\frac{\partial\omega}{\partial\sigma}\right)^2 \\
&+ h_{ij}\Gamma_{kl}^i\frac{\partial x^j}{\partial\sigma}\frac{\partial x^k}{\partial\sigma}\frac{\partial x^l}{\partial\sigma} = \gamma\left(\frac{\partial\omega}{\partial\sigma}\right)^2\,.
\end{aligned}
\end{align}
The second and fourth terms cancel by symmetry and, supposing $(\partial\omega/\partial\sigma)\neq 0$, the pair of equations~\eqref{eq:geodesic2} may be written as
\begin{subequations}
\label{eq:geodesic3}
\begin{align}
\frac{\partial^2 t}{\partial\sigma^2}+a\frac{da}{dt}\left(\frac{\partial\omega}{\partial\sigma}\right)^2 &= \gamma\frac{\partial t}{\partial\sigma}\,, \label{eq:geodesic3a} \\
\frac{\partial^2 \omega}{\partial\sigma^2} + \frac{2}{a}\frac{da}{dt}\frac{\partial t}{\partial\sigma}\frac{\partial\omega}{\partial\sigma} &= \gamma\frac{\partial\omega}{\partial\sigma}\,. \label{eq:geodesic3b}
\end{align}
\end{subequations}
We now proceed by parametrizing the geodesic by the Euclidean spatial distance, i.e., $\sigma\equiv\omega$. This yields $\partial\omega/\partial\sigma=1$, $\partial^2\omega/\partial\sigma^2=0$, and then~\eqref{eq:geodesic3b} gives $\gamma=(2/a)(da/dt)(\partial t/\partial\omega)$. Using these new relations,~\eqref{eq:geodesic3a} can be written as
\begin{equation}
\label{eq:geodesic3.5}
\frac{\partial^2 t}{\partial\omega^2} + \frac{da}{dt}\left(a - \frac{2}{a}\left(\frac{\partial t}{\partial\omega}\right)^2\right) = 0\,.
\end{equation}
While neither the spatial distance $\omega$ nor time $t$ are affine parameters along all Lorentzian geodesics, the results will not be affected, since the differential equations no longer refer to $\gamma$. We can see that if $\partial t/\partial\omega=0$ then the second derivative of $t$ is always negative for $t>0$ and positive for $t<0$, since $da/dt>0$ for expanding spacetimes:
\begin{equation}
\frac{\partial^2 t}{\partial\omega^2}=-a\frac{da}{dt}\,.
\end{equation}
If there exists a critical point exactly at $t=0$, it is a saddle point. From these facts, we conclude that any extremum found along a geodesic on a Friedmann-Lema\^itre-Robertson-Walker manifold is a local maximum in $t>0$ and a local minimum in $t<0$ with respect to $\omega$.\footnote{This statement is true under the assumption that the scale factor is a well-behaved monotonic function, as it is for most physical solutions. If this condition does not hold, the following analysis must be reinspected.} An example of such a curve with an extremum is shown in Fig.~\ref{fig8-1}(c). \par

The second-order equation~\eqref{eq:geodesic3.5} may be simplified by multiplying by $2a^{-4}(\partial t/\partial\omega)$ and integrating by parts to get a non-linear first-order differential equation and a constant of integration $\mu$:
\begin{align}
0 &= \frac{\partial}{\partial\omega}\left[a^{-4}\left(\left(\frac{\partial t}{\partial\omega}\right)^2-a^2\right)\right]\,, \\
\frac{\partial\omega}{\partial t} &=\pm\left(a^2\left(t\right) + \mu a^4\left(t\right)\right)^{-1/2} \equiv G(t;\mu)
\,, \label{eq:geodesic4}
\end{align}
the right hand side of which is hereafter referred to as the geodesic kernel $G(t;\mu)$. We may neglect the sign by noting that the spatial distance $\omega$ should always be an increasing function of $t$, so that any integration of the geodesic kernel should be always be performed from past to future times. It will prove necessary to know the value of $\mu$ to find the final value of the geodesic length between two events.
\par

\subsection{The Integral Form of the Geodesic Equations}
To find the geodesic distance between a given pair of points/events, we need to use~\eqref{eq:geodesic4} in conjunction with the integral form of the geodesic equation, given in~\eqref{eq:geodesic_variation}. We begin by defining the integrand in~\eqref{eq:geodesic_variation} as the distance kernel $D(\sigma)$:
\begin{align}
D\left(\sigma\right) &\equiv \frac{ds}{d\sigma} = \sqrt{g_\munu\frac{\partial x^\mu}{\partial\sigma}\frac{\partial x^\nu}{\partial\sigma}}\,, \\
\intertext{so that the geodesic distance is}
d(\sigma_1,\sigma_2)&=\int_{\sigma_1}^{\sigma_2}\!D\left(\sigma\right)\,d\sigma\,. \label{eq:distance_general}
\end{align}
The invariant interval~\eqref{eq:invariant_interval} tells us that $D^2$ is negative for timelike-separated pairs and positive for spacelike-separated ones, assuming $\sigma$ is monotonically increasing along the geodesic. Therefore, we always take the absolute value of $D^2$ so that the distance kernel is real-valued, while keeping in mind which type of geodesic we are discussing. \par

Depending on the particular scale factor and boundary values, we might sometimes parametrize the system using the spatial distance and other times using time. If we parametrize the geodesic with the spatial distance we find
\begin{align}
D\left(\omega\right) &= \sqrt{-\left(\frac{\partial t}{\partial\omega}\right)^2+a^2\left(t\left(\omega\right)\right)}\,, \label{eq:distances_general_spatial} \\
\intertext{and if we instead use time we get}
D\left(t\right) &= \sqrt{-1+a^2\left(t\right)\left(\frac{\partial\omega}{\partial t}\right)^2}\,, \label{eq:distances_general}
\end{align}
where the function $t(\omega)$ in the former equation is the inverted solution $\omega(t)$ to the differential equation~\eqref{eq:geodesic4}. Since the distance kernel is a function of the geodesic kernel, we will need to know the value $\mu$ associated with a particular set of constraints. \par

If we insert~\eqref{eq:geodesic4} into~\eqref{eq:distances_general}, we can can see what values the constant $\mu$ can take:
\begin{equation}
D\left(t;\mu\right) = \sqrt{\frac{-\mu a^2\left(t\right)}{1+\mu a^2\left(t\right)}}\,.
\end{equation}
If $D^2<0$ for timelike intervals, then $\mu>0$. If $\mu=0$, we obtain a lightlike geodesic, since the distance kernel becomes zero. Hence, spacelike intervals correspond to $-a^{-2}(t)<\mu<0$. We do not consider $\mu<-a^{-2}(t)$ because this corresponds to an imaginary $\partial\omega/\partial t$, which we consider non-physical.

\par

\subsection{Geodesic Constraints and Critical Points}
We would like to find geodesics for both initial-value and boundary-value problems. If we have Cauchy boundary conditions, i.e., the initial position and velocity vector are known, then finding $\mu$ is simple: since the left hand side of~\eqref{eq:geodesic4} is just the speed $v_0\equiv|v^i(t_0)|$, where $v^i$ is the velocity vector defined by our initial conditions, we have
\begin{equation}
\label{eq:cauchy}
\mu=a_0^{-2}\left(v_0^{-2}a_0^{-2}-1\right)\,,
\end{equation}
where $a_0\equiv a(t_0)$. This allows for simple solutions to cases with Cauchy
boundary conditions.

However, if we have Dirichlet boundary conditions, i.e., the initial and final positions are known, then we must integrate~\eqref{eq:geodesic4} instead. The bounds of such an integral need to be carefully considered: if we have a spacelike geodesic which starts and ends at the same time, for instance, then it is not obvious how to integrate the geodesic kernel. In fact, we face an issue with the boundaries whenever we have geodesics with turning points. This feature occurs whenever $\partial t/\partial\omega = 0$, i.e.,
\begin{equation}
\label{eq:max_time_constraint}
a(t_c) = \pm\sqrt{-\mu^{-1}}\,.
\end{equation}
Since in this case $\mu < 0$, we see that turning points only occur for spacelike
geodesics. Furthermore, since all of the scale factors given by~\eqref{eq:scale_factors} are monotonic, this situation occurs in such spacetime only at a single point along a geodesic, if at all, identified as $P_3$ in Fig.~\ref{fig8-1}(c). Specifically, if $t_1,t_2,t_3$ respectively correspond to the times at points $P_1,P_2,P_3$, then the integral of the geodesic kernel is found by integrating from $t_3$ to $t_1$ as well as from $t_3$ to $t_2$, since time is not monotonic along the geodesic. If no such turning point $P_3$ exists along the geodesic, a single integral from $t_1$ to $t_2$ may be performed. The integral of the distance kernel should be performed in the same way for the same reasons. \par

To determine if a turning point exists along a spacelike geodesic, we begin by noting that there is a corresponding critical spatial distance $\omega_c$ which corresponds to the critical time defined in~\eqref{eq:max_time_constraint}. If we suppose $t_2>t_1>0$, then the geodesic kernel is maximized when $\mu=\mu_c\equiv-a^{-2}(t_2)$, i.e., when $\mu$ attains its minimum value. This is the minimum value of $\mu$ along the geodesic, since
$a(t)$ is monotonically increasing. The critical spatial distance is defined by this
$\mu_c$ and is given by
\begin{equation}
\label{eq:critical_separation}
\omega_c = \int_{t_1}^{t_2}\!\frac{1}{a(t)}\left(1-\left(\frac{a(t)}{a(t_2)}\right)^2\right)^{-1/2}\,dt\,.
\end{equation}
Since $\mu_c$ maximizes the geodesic kernel, it is impossible for a spacelike-separated pair to be spatially farther apart without their geodesic having a turning point. We then conclude that if $\omega<\omega_c$ for a particular pair of spacelike-separated points, then the geodesic is of the form shown in Fig.~\ref{fig8-1}(b), and if $\omega>\omega_c$ it is of the form shown in Fig.~\ref{fig8-1}(c). In other words, if the geodesic is of the latter type, then the solution to~\eqref{eq:geodesic4} is
\begin{align}
\omega&=\int_{t_1}^{t_c}\!G\left(t;\mu\right)\,dt + \int_{t_2}^{t_c}\!G\left(t;\mu\right)\,dt\,,\label{eq:omega_coupled} \\
\intertext{while the solution to~\eqref{eq:distance_general} using~\eqref{eq:distances_general} is}
d(t_1,t_2;\mu)&=\int_{t_1}^{t_c}\!D\left(t;\mu\right)\,dt+\int_{t_2}^{t_c}\!D\left(t;\mu\right)\,dt\,,
\end{align}
again supposing $t_2>t_1>0$. The bounds on the integral are chosen this way due to
the change of sign in the geodesic kernel on opposite sides of the critical point.
If $0>t_2>t_1$ then the bounds on the integrals are reversed so that $\omega,d>0$. \par

\subsection{Geodesic Connectedness}
\label{sec:geo_conectedness}
\begin{figure*}[!htb]
\includegraphics[width=\textwidth]{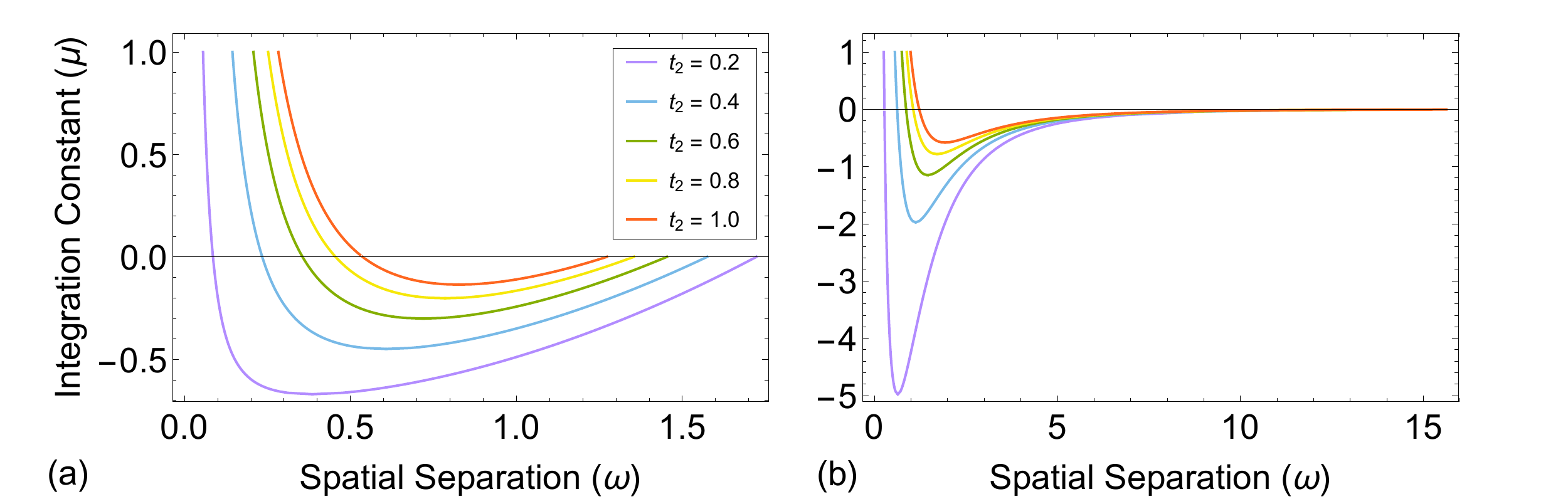}
\centering
\caption{\textbf{Evidence of geodesic horizons in FLRW manifolds.} Certain FLRW manifolds are not spacelike-geodesically-connected, such as the de Sitter manifold. In (a) we see the relation between the integration constant $\mu$, first defined in~\eqref{eq:geodesic4}, and the spatial separation between two points on the de Sitter manifold. The initial point is located at $t_1=0.1$ and the curves show the behavior for several choices of the final time $t_2$. For small $\omega$, the pair of points is timelike-separated and $\mu$ is positive. As $\omega$ tends to zero, $\mu$ tends to infinity, indicating the manifold is timelike-geodesically-complete. As $\omega$ increases and the geodesic becomes spacelike, it will ultimately have a turning point at $\omega_c$, located at the minimum of each curve and defined by~\eqref{eq:critical_separation}. Ultimately, for manifolds which are spacelike-geodesically-incomplete the curve terminates at some maximum spatial separation $\omega_m$ defined by~\eqref{eq:maximum_separation}. In (b), showing the Einstein-de Sitter manifold case, the curves extend to infinity on the right because the manifold is geodesically complete.}
\label{fig8-2}
\end{figure*}

Certain FLRW manifolds are not spacelike-geodesically-connected, meaning not all pairs of spacelike-separated points are connected by a geodesic. For a given pair of times $t_1,t_2$ there exists a maximum spatial separation $\omega_m$ past which the two points cannot be connected by a geodesic. To determine this maximum spatial distance $\omega_m$ for a particular pair of points, we use~\eqref{eq:omega_coupled}, this time taking the limit $\mu\to 0^-$. This limit describes a spacelike geodesic which is asymptotically becoming lightlike. If the critical time $t_c$ remains finite in this limit, the manifold is geodesically connected and $\omega_m=\infty$, whereas if it becomes infinite then $\omega_m$ remains finite, shown in detail in Fig.~\ref{fig8-2}. The equation~\eqref{eq:omega_coupled} in the limit $\mu\to 0^-$ is
\begin{equation}
\label{eq:maximum_separation}
\omega_m = \int_{t_1}^{t_c}\!\frac{dt}{a(t)} + \int_{t_2}^{t_c}\!\frac{dt}{a(t)}\,.
\end{equation}
Comparing~\eqref{eq:maximum_separation} to~\eqref{eq:conformal_time} we notice that $\omega_m$ is simply a combination of conformal times using the boundary points $t_1$ and $t_2$: $\omega_m\propto\eta_c\equiv\eta(t_c)$, and so if $\eta_c$ is finite, then $\omega_m$ will be finite as well. Therefore, we conclude that a Friedmann-Lema\^itre-Robertson-Walker manifold is geodesically complete if
\begin{align}
\lim_{\mu\to 0^-}\left|\eta_c\right|&=\infty\,, \label{eq:geo_compl}\\
\intertext{where $\eta_c$ is obtained by inverting}
a(t(\eta_c))&=\pm\sqrt{-\mu^{-1}}\,,
\end{align}
using the appropriate $a(t)$ and $t(\eta)$ for the given manifold. \par

As an example, consider the de Sitter manifold:
\begin{equation}
\frac{\lambda}{\eta_c} = \pm\sqrt{-\mu^{-1}}\,,
\end{equation}
so that the limit maximum conformal time in terms of $\mu$ is
\begin{equation}
\lim_{\mu\to 0^-}\left|\eta_c\right| = \lim_{\mu\to 0^-}\lambda\sqrt{-\mu} = 0\,.
\end{equation}
Therefore, in the flat foliation, there exist pairs of points on the de Sitter manifold which cannot be connected by a geodesic. On the other hand, if we consider the Einstein-de Sitter manifold, which represents a spacetime with dust matter, the scale factor is proportional to $\eta^2$:
\begin{equation}
\lim_{\mu\to 0^-}\left|\eta_c\right| \propto \lim_{\mu\to 0^-}\left(-\mu^{-1}\right)^{1/4} = \infty\,,
\end{equation}
so that every pair of points may be connected by a geodesic.

\section{Examples}
\label{sec:examples8}
Here we apply the results above to calculate geodesics in the FLRW manifolds defined by each of the scale factors in~\eqref{eq:scale_factors}, using two type of constraints: the Dirichlet and Cauchy boundary conditions. The former conditions specify two events or points in a given spacetime that can be either timelike or spacelike separated, as in Fig.~\ref{fig8-1}. The latter conditions specify just one point and a vector of initial velocity. If the initial speed is below the speed of light, then the resulting geodesic is timelike, and corresponds to a possible world line of a massive particle. If the initial speed is above the speed of light, i.e., the initial tangent vector is spacelike, then the resulting geodesic is spacelike, and corresponds to a geodesic of a hypothetical superluminal particle. Even though tachyons may not exist, spacelike geodesics are well defined mathematically.
The last example that we consider illustrates how to apply these techniques to find numerical values for geodesic distances in our physical universe.
\subsection{Dark Energy}
Suppose we wish to find the geodesic distance using the Dirichlet boundary conditions $\{t_1,\ab t_2,\ab\omega\}$. The geodesic kernel in a flat de Sitter spacetime is
\begin{equation}
G_\Lambda\left(t;\mu\right) = \lambda^{-1}\left(e^{2t/\lambda}+\mu e^{4t/\lambda}\right)^{-1/2}\,,
\end{equation}
where $\mu$ has absorbed a factor of $\lambda^2$ and we use $\eta\in[-1,0)$ so that $t\geq0$. We can easily transform the kernel into a polynomial equation by using the conformal time:
\begin{equation}
G_\Lambda\left(\eta;\mu\right) = \left(1+\frac{\mu}{\eta^2}\right)^{-1/2}\,.
\end{equation}
If the minimal value of $\mu$ is inserted into this kernel, the turning point $\omega_c$ can be found exactly:
\begin{align}
\mu_c &= -\eta_2^2\,, \\
G_\Lambda\left(\eta;\mu_c\right) &= \left(1-\left(\frac{\eta_2}{\eta}\right)^2\right)^{-1/2}\,, \\
\omega_c\left(\eta_1,\eta_2;\mu_c\right) &= \int_{\eta_1}^{\eta_2}\!G_\Lambda\left(\eta;\mu_c\right)\,d\eta\,, \nonumber \\
  &= \sqrt{\eta_1^2-\eta_2^2}\,.
\end{align}
The geodesic kernel may now be integrated both above and below the turning point:
\begin{equation}
\label{eq:ds_mu}
\omega = \begin{cases}
\sqrt{\eta_1^2+\mu}-\sqrt{\eta_2^2+\mu}&\text{if } \omega<\omega_c\,, \\
\sqrt{\eta_1^2+\mu}+\sqrt{\eta_2^2+\mu}&\text{if } \omega>\omega_c\,.
\end{cases}
\end{equation}
The variable $\mu$ is then found by inverting one of these equations. Finally, substitution of the scale factor and numerical value $\mu$ into~\eqref{eq:distances_general} gives the geodesic distance for a pair of coordinates defined by $\{\eta_1,\ab \eta_2,\ab\omega\}$:
\begin{subequations}
\label{eq:ds_4d}
\begin{align}
d_\Lambda\left(t_1,t_2;\mu\right) &= \sinh^{-1}\left(\frac{\sqrt\mu}{\eta_1}\right)-\sinh^{-1}\left(\frac{\sqrt\mu}{\eta_2}\right)\,,\\
\intertext{for timelike-separated pairs, and}
d_\Lambda\left(t_1,t_2;\mu\right) &=
\begin{cases}
\sinh^{-1}\left(\frac{\sqrt{-\mu}}{\eta_1}\right)-\sinh^{-1}\left(\frac{\sqrt{-\mu}}{\eta_2}\right)&\text{if } \omega<\omega_c\,, \\
\sinh^{-1}\left(\frac{\sqrt{-\mu}}{\eta_1}\right)+\sinh^{-1}\left(\frac{\sqrt{-\mu}}{\eta_2}\right)+\pi&\text{if } \omega>\omega_c\,,
\end{cases}
\end{align}
\end{subequations}
for spacelike-separated pairs. \par

\subsubsection{Equivalence of de Sitter Solutions}
We now show this solution is equivalent to the solution found using the embedding in Sec.~\ref{sec:desitter}. Let us refer to~\eqref{eq:ds_5d} as $d_1$ and~\eqref{eq:ds_4d} as $d_2$. The conformal time in the de Sitter spacetime is $\eta(t) = -e^{-t/\lambda}$, with $\eta\in[-1,0)$ so that the cosmological time $t$ remains positive. Since the geodesic distance depends on the spatial distance, but not the individual spatial coordinates, we can assume without loss of generality that the initial point is located at the origin, $r=\theta=\phi=0$, and the second point is located at some distance $\omega$ from the origin, $r=\omega,\,\theta=\phi=0$. Further, to simplify the proof, suppose the initial point is at time $t=0$ ($\eta=-1$) and the second point at some $t=t_0>0$ ($\eta=\eta_0\in(-1,0)$). We are allowed to make these assumptions due to the spatial symmetries associated with the dS(1,3) group and the existence of a global timelike Killing vector in the flat foliation of the de Sitter manifold~\cite{ref:podolsky1993lorentz}. In addition, suppose the geodesic is timelike so that $\omega\in[0,\eta_0+1)\subseteq[0,1)$. This same method may be applied to spacelike geodesics. \par

Using these values, the embedding coordinates in $\mathcal{M}^5$ are
\begin{align}
x&=((1-\lambda^2)/2,\,-(1+\lambda^2)/2,\,0,\,0,\,0)\,,\\
y&=((\lambda^2+\omega^2-\eta_0^2)/2\eta_0,\,(\lambda^2-\omega^2+\eta_0^2)/2\eta_0,\,\lambda\omega/\eta_0,\,0,\,0)\,.
\end{align}
These equations give a geodesic distance
\begin{equation}
d_1 = \lambda\arccosh\left(\frac{\omega^2-\eta_0^2-1}{2\eta_0}\right)\,.
\end{equation}
On the other hand, we can use the solution provided by~\eqref{eq:ds_4d} using the value of $\mu$ in~\eqref{eq:ds_mu}:
\begin{equation}
\label{eq:app_mu}
\mu=\frac{\left(\omega+\eta_0+1\right)\left(\omega+\eta_0-1\right)\left(\omega-\eta_0+1\right)\left(\omega-\eta_0-1\right)}{4\lambda^2\omega^2}\,,
\end{equation}
in the geodesic distance expression
\begin{equation}
d_2 = \lambda\left(\arcsinh\left(\frac{\lambda\sqrt{\mu}}{-\eta_0}\right) - \arcsinh\left(\lambda\sqrt{\mu}\right)\right)\,.
\end{equation}
If we apply $\cosh(d/\lambda)$ to each of these expressions, and use the identities $\cosh(x-y) = \cosh x\cosh y - \sinh x\sinh y$ and $\cosh\arcsinh x = \sqrt{x^2+1}$, we may equate them to get
\begin{equation}
\frac{\omega^2-\eta_0^2-1}{2\eta_0} = \sqrt{\left(\lambda^2\mu+1\right)\left(\frac{\lambda^2\mu}{\eta_0^2}+1\right)}+\frac{\lambda^2\mu}{\eta_0}\,.
\end{equation}
Using~\eqref{eq:app_mu} and some algebra, the right hand side may be simplified to give the result on the left hand side, thereby proving they are equal.

\subsection{Dust}
\label{sec:dust}
In this example, let us suppose we have Cauchy boundary conditions and we want an expression for the geodesic distance in terms of spatial distance traveled $\omega$. First, knowing the values $(t_0,r_0,\theta_0,\phi_0)$ and $|v_0|$, we can find the parameter $\mu$ via~\eqref{eq:cauchy}. Because the manifold has a
singularity at $t=0$, we assert $t_0\neq0$ to avoid a nonsensical value for $\mu$. We proceed by parametrizing the geodesic equation by the spatial distance, following~\eqref{eq:distances_general_spatial}, so that the distance kernel for this spacetime is
\begin{equation}
D_D\left(\omega;\mu\right) = \alpha^2\left|\mu\right|^{1/2}\left(\frac{3t\left(\omega\right)}{2\lambda}\right)^{4/3}\,.
\end{equation}
We use the geodesic kernel to find $t(\omega)$ directly, by solving~\eqref{eq:geodesic4} for $\omega(t)$ and inverting the solution. In the spacetime with dust matter and no cosmological constant the geodesic kernel is
\begin{equation}
G_D\left(t;\mu\right) = \left(\alpha^2\left(\frac{3t}{2\lambda}\right)^{4/3}+\mu\alpha^4\left(\frac{3t}{2\lambda}\right)^{8/3}\right)^{-1/2}\,,
\end{equation}
\\
\noindent which, using the transformations $x\equiv(3t/2\lambda)^{1/3}$ and $\mu\to\alpha^2\mu$, becomes
\begin{equation}
G_D\left(x;\mu\right) = \frac{2\lambda}{\alpha}\left(1+\mu x^4\right)^{-1/2}\,.
\end{equation}
The value of $\omega$ where the turning point occurs is then
\begin{equation}
\omega_c(x_0;\mu) = \frac{2\lambda}{\alpha}\left(\frac{\sqrt{\pi}\,\Gamma(5/4)}{\Gamma(3/4)}\left(-\mu\right)^{-1/4}
- x_0\,{}_2F_1\left(\frac{1}{4},\frac{1}{2};\frac{5}{4};-\mu x_0^4\right)\right)\,,
\end{equation}
where $x_0\equiv x(t_0)$ and ${}_2F_1(a,b;c;z)$ is the Gauss hypergeometric function. \par

The final expression $\omega(t)$ still depends on the existence of a critical point along the geodesic. To demonstrate how piecewise solutions are found, hereafter we suppose we are studying a superluminal inertial object moving fast and long enough to take a geodesic with a turning point. The spatial distance $\omega(x;x_0)$, with $x>x_0$ and $\mu<0$, which we know because the geodesic is spacelike, is
\begin{subequations}
\begin{align}
\omega^{(1)}\left(x;x_0,x_c\right) &= \frac{2\lambda}{\alpha}x_c\left(F\left(\arcsin\left(\frac{x}{x_c}\right)\bigg|-1\right)-F\left(\arcsin\left(\frac{x_0}{x_c}\right)\bigg|-1\right)\right)\,,\\
\intertext{before the critical point, and}
\omega^{(2)}\left(x;x_0,x_c\right) &= \frac{2\lambda}{\alpha}x_c\bigg(2K\left(-1\right) - F\left(\arcsin\left(\frac{x_0}{x_c}\right)\bigg|-1\right) - F\left(\arcsin\left(\frac{x}{x_c}\right)\bigg|-1\right)\bigg)\,,
\end{align}
\end{subequations}
afterward, where $x_c=(-\mu)^{-1/4}$, and $K(m)$ and $F(\phi|m)$ respectively are the complete and incomplete elliptic integrals of the first kind with parameter $m$. These expressions $\omega(x;x_0,x_c)$ are slightly different for $\mu>0$. Despite the apparent complexity of the above expressions, they are in fact easy to invert via the Jacobi elliptic functions. The distance for a geodesic with a turning point is
\begin{equation}
d\left(\omega;\mu\right) = \int_0^{\omega_c}\!D\left(\omega^{(1)};\mu\right)\,d\omega + \int_{\omega_c}^{\omega}\!D\left(\omega^{(2)};\mu\right)\,d\omega\,,
\end{equation}
giving the final result
\begin{equation}
\begin{split}
d_D\left(\omega;\mu\right) = \frac{\alpha^2\left|\mu\right|^{1/2}x_c^4}{3\beta_1}\Bigg(&\beta_1\left(2\omega_c-\omega\right) + \frac{x_0}{x_c}\sqrt{1-\left(\frac{x_0}{x_c}\right)^4} + \fn\left(\beta_3-\beta_1\omega_c|-1\right)\\
& - \fn\left(\beta_1\omega_c+\beta_2|-1\right) - \fn\left(\beta_1\omega-\beta_3|-1\right)\Bigg)\,,
\end{split}
\end{equation}
where we have used the auxiliary variables
\begin{align}
\beta_1 &\equiv \frac{\alpha}{2\lambda x_c}\,, \\
\beta_2 &\equiv F\left(\arcsin\left(\frac{x_0}{x_c}\right)\bigg|-1\right)\,, \\
\beta_3 &\equiv 2K\left(-1\right) - \beta_2\,, \\
\fn\left(\phi|m\right) &\equiv \sn\left(\phi|m\right)\cn\left(\phi|m\right)\dn\left(\phi|m\right)\,,
\end{align}
and the three functions in the last definition are the Jacobi elliptic functions with parameter $m$.

\subsection{Radiation}
Here we suppose we have Cauchy boundary conditions, but the particle will take a timelike geodesic, i.e., $\mu>0$. Using the transformations $x\equiv\sqrt{2t/\lambda}$ and $\mu\to\alpha^{3/2}\mu$, we can write the geodesic kernel as
\begin{equation}
G_R\left(x;\mu\right) = \frac{\lambda}{\alpha^{3/4}}\left(1+\mu x^2\right)^{-1/2}\,.
\end{equation}
If this kernel is integrated over $x$ to find the spatial distance $\omega(x)$, the result can be inverted to give
\begin{equation}
x\left(\omega;\mu,x_0\right) = \mu^{-1/2}\sinh\left(\beta_1\omega+\beta_2\right)\,,
\end{equation}
where $\beta_1\equiv\alpha^{3/4}\mu^{1/2}/\lambda$ and $\beta_2\equiv\arcsinh(\mu^{1/2}x_0)$. Since the geodesic distance is more easily found when we parametrize with the spatial distance $\omega$, we can write the distance kernel as
\begin{equation}
D_R\left(\omega;\mu\right) = \alpha^{3/2}\mu^{1/2}x^2\left(\omega;\mu,x_0\right)\,,
\end{equation}
and the geodesic distance as
\begin{align}
d_R\left(\omega;\mu,x_0\right) &=\int_0^\omega\!D_R\left(\omega^\prime\right)\,d\omega^\prime\,, \\
&=\frac{\alpha^{3/2}}{4\mu^{1/2}\beta_1}\left(\sinh\left(2\left(\beta_1\omega+\beta_2\right)\right) - \sinh\left(2\beta_2\right) - 2\beta_1\omega\right)\,.
\end{align}
Typically, timelike geodesics are parametrized by time: since there exists a closed-form solution for $\omega(x(t))$ this expression can be substituted here, though it would needlessly add extra calculations. Therefore, in practice it is computationally simpler to use a spatial parametrization.

\subsection{Stiff Fluid}
Suppose we have a spacetime containing a homogeneous stiff fluid, and we wish to find a timelike geodesic using Dirichlet boundary conditions. Using the transformation $x\equiv(3t/\lambda)^{1/3}$, we can write the geodesic kernel as
\begin{equation}
G_S\left(x;\mu\right) = \frac{\lambda}{\alpha^{1/2}}\frac{x}{\left(1+\mu x^2\right)^{1/2}}\,,
\end{equation}
where $\mu$ has absorbed a factor of $\alpha$. This kernel can easily be integrated to find
\begin{equation}
\omega\left(x_0,x_1;\mu\right) = \frac{\lambda}{\alpha^{1/2}\mu}\left(\sqrt{1+\mu x_1^2} - \sqrt{1+\mu x_0^2}\right)\,.
\end{equation}
\\
\noindent The constant $\mu$ may be found provided the initial conditions $\{x_0,x_1,\omega\}$ as:
\begin{equation}
\mu = \frac{x_0^2+x_1^2}{\omega^2} - 2\sqrt{\frac{x_0^2x_1^2 + \omega^2}{\omega^4}}\,.
\end{equation}
Finally, if the geodesic is parametrized by $x(t)$ we arrive at
\begin{equation}
d_S(x_0,x_1;\mu) = \frac{\lambda}{3\mu}\bigg(2\sqrt{x_1^2 + \mu^{-1}} - 2\sqrt{x_0^2 + \mu^{-1}} - x_1^3\sqrt{\mu\left(x_1^{-2} + \mu\right)} + x_0^3\sqrt{\mu\left(x_0^{-2} + \mu\right)}\bigg)\,.
\end{equation}

\subsection{Dark Energy and Dust}
\label{sec:mixed}
None of the spacetimes with a mixture of dark energy and some form of matter have closed-form solutions for geodesics, because the scale factors are various powers of the hyperbolic sine function, so it becomes cumbersome to work with the geodesic and distance kernels. However, by using the right transformations, it is still possible to make the problem well-suited for fast numerical integration. In this example, we use the mixed dust and dark energy spacetime, following the same procedure as before; for other spacetimes with mixed contents the same method applies. This time, the geodesic kernel is
\begin{equation}
G_{\Lambda D}\left(t;\mu\right) = \left(\sinh^{4/3}\left(\frac{3t}{2\lambda}\right)+\mu\sinh^{8/3}\left(\frac{3t}{2\lambda}\right)\right)^{-1/2}\,.
\end{equation}
Once again, the kernel can be written as a polynomial expression, this time using the square root of the scale factor as the transformation:
\begin{align}
x\left(t\right) &\equiv \sinh^{1/3}\left(\frac{3t}{2\lambda}\right)\,, \nonumber \\
G_{\Lambda D}\left(x;\mu\right) &= 2\left(\left(1+x^6\right)\left(1+\mu x^4\right)\right)^{-1/2}\,.\label{eq:G_lambdaD}
\end{align}
There is no known closed-form solution to the integral of $G_{\Lambda D}$. The distance kernel is best represented as a function of $t$ to simplify numerical evaluations:
\begin{equation}
D_{\Lambda D}\left(t;\mu\right) = \sqrt{\frac{-\mu\sinh^{2/3}\left(3t/\lambda\right)}{1+\mu\sinh^{2/3}\left(3t/\lambda\right)}}\,.
\end{equation}
There is no known closed-form solution to this kernel's integral either, but it can be quickly computed numerically, since the hyperbolic term needs to be evaluated only once for each value of $t$. In general, the numeric evaluations of such integrals can be quite fast if the kernels take a polynomial form, and a Gauss-Kronrod quadrature can be used for numeric evaluation of these integrals. Since the solution of this particular system will be used frequently in the next chapter, we will come back to numerical approximations in Section~\ref{sec:num_approx}.

\subsection{Dark Energy, Dust, and Radiation}
Typically in cosmology one studies one particular era, whether the early
inflationary phase, the radiation-dominated phase, the matter-dominated phase
after recombination, or ultimately today's period of accelerated
expansion. Perhaps the most important spacetime which we have not looked at
yet is the FLRW spacetime which most closely models our own physical universe,
in its entirety. In this section we will show how to most efficiently find geodesics in our (FLRW $\Lambda$DR) universe. \par

Because the scale factor $a(t)$ is a smooth, monotonic, differentiable, and bijective function of time, it, instead of time $t$ or spatial distance $\omega$, can parametrize geodesics, so long as we remember to break up expressions when there exists a turning point in long spacelike geodesics. In what follows we will restrict the analysis to timelike geodesics for simplicity. To find spacelike geodesics, refer to the steps performed in Sec.~\ref{sec:dust}. Using the scale-factor parametrization, the geodesic and distance kernels are
\begin{align}
&G_{\Lambda DR}\left(a;\mu\right) = \lambda\left[\left(1+\mu a^2\right)\left(\frac{\Omega_R}{\Omega_\Lambda}+\frac{\Omega_D}{\Omega_\Lambda}a+a^4\right)\right]^{-1/2}\,, \label{eq:full_gkernel} \\
&D_{\Lambda DR}\left(a;\mu\right) = \lambda\left[\left(\frac{-\mu a^4}{1+\mu a^2}\right)\left(\frac{\Omega_R}{\Omega_\Lambda}+\frac{\Omega_D}{\Omega_\Lambda}a+a^4\right)^{-1}\right]^{-1/2}\,, \label{eq:full_dkernel}
\end{align}
where $\Omega_\Lambda$, $\Omega_D$, and $\Omega_R$ respectively are the fractions of dark energy, dust, and radiation energy densities. As we saw in Sec.~\ref{sec:mixed}, integrands such as these produce no closed-form solutions, but they are easily evaluated numerically due to their polynomial form. \par

We now provide a simple example of computing an exact geodesic distance between a pair of events in our physical universe using these results. Suppose we are to measure the timelike geodesic distance between an event in the early universe, where $t_1=10^{11}$s, and another event near today, $t_2=4.3\times10^{17}$s. Let the spatial distance of this geodesic be $\omega=4.1\times10^{13}$km, roughly the distance to Alpha Centauri. Taking relevant experimental values from recent measurements~\cite{calabrese2017cosmological}, we find the Hubble constant is $H_0=100h$ km/s/Mpc, where $h=0.705$, and the density parameters are $\Omega_\Lambda=0.723$, $\Omega_D=0.277$, and $\Omega_R=9.29\times10^{-5}$. The leading constant $\lambda$ in the above equations can be expressed as $\lambda=H_0^{-1}\Omega_\Lambda^{-1/2}$, thereby completing the set of all the relevant physical parameters used in~\eqref{eq:full_gkernel} and~\eqref{eq:full_dkernel}.
We then integrate the geodesic kernel~\eqref{eq:full_gkernel}, inserting the speed of light $c$ where needed, to numerically solve for the integration constant $\mu$, which we find to be $\mu=2.53\times10^{23}$. Inserting this value into the distance kernel~\eqref{eq:full_dkernel} and evaluating numerically gives a final geodesic distance of $d=2.22\times10^{23}$km.

\section{Numerical Approximations}
\label{sec:num_approx}
In the numerical experiments in the following chapter, we will need some approximations to avoid using the bisection method on the integral of~\eqref{eq:G_lambdaD}:
\begin{equation}
\omega\left(x;\mu\right) = 2\int\!\left(\left(1+x^6\right)\left(1+\mu x^4\right)\right)^{-1/2}\,dx\,.
\end{equation}
The first term $(1+x^6)^{-1/2}$ may be expanded using a series for three regions of $x$. We cannot expand the second term $(1+\mu x^4)^{-1/2}$ in a series because $\mu$ can take very large or very small values. Therefore, we will find three approximate solutions of this integral for the following three regions. For $x\ll 1$ (Region I) we use a binomial expansion:
%%%%%
\begin{equation}
\label{eq:geodesic_binomial1}
\left(1+x^6\right)^{-1/2} = \sqrt{\pi}\sum_{k=0}^\infty \frac{x^{6k}}{k!\Gamma\left(\frac{1}{2}-k\right)}\,.
\end{equation}
%%%%%
\noindent Similarly, for $x\gg 1$ (Region III), we use a different\footnote{$\left(1+x^6\right)^{-1/2} = x^{-3}\left(1+x^{-6}\right)^{-1/2}$} binomial expansion:
%%%%%
\begin{equation}
\label{eq:geodesic_binomial2}
\left(1+x^6\right)^{-1/2} = \frac{\sqrt{\pi}}{x^3}\sum_{k=0}^\infty \frac{1}{k!\Gamma\left(\frac{1}{2}-k\right)x^{6k}}\,.
\end{equation}
%%%%%
\noindent Finally, in the regime where $x\approx 1$ (Region II) we use a Taylor expansion, including enough terms so that the overlap among the three approximations produces a sufficiently low error:
%%%%%
\begin{equation}
\label{eq:geodesic_taylor}
\left(1+x^6\right)^{-1/2}\approx \frac{1}{\sqrt{2}} - \frac{3}{2\sqrt{2}}\left(x-1\right) - \frac{3}{8\sqrt{2}}\left(x-1\right)^2 + \frac{55}{16\sqrt{2}}\left(x-1\right)^3 + \mathcal{O}\left(x^4\right)\,.
\end{equation}
%%%%%

\subsection{Region I}
\noindent Region I has the simplest solution:
%%%%%
\begin{equation}
\tilde{\omega}_I = 2\sqrt{\pi}\sum_{k=0}^\infty \frac{x^{6k+1}}{k!\Gamma\left(\frac{1}{2}-k\right)\left(6k+1\right)} {}_2F_1\left(\frac{1}{2},\frac{6k+1}{4};\frac{6k+5}{4};-\mu x^4\right)\,,
\end{equation}
%%%%%
where $\tilde\omega\equiv(\alpha/\lambda)\omega$.

\subsection{Region II}
\noindent Region II requires a bit more work. We split the solution into two cases depending on the sign of $\mu$:
%%%%%
\begin{equation}
\label{eq:omegaII}
\begin{split}
\tilde{\omega}_{II}^+ = &\frac{55\sqrt{1+\mu x^4} + 153\sqrt{\mu}\sinh^{-1}\left(\sqrt{\mu}x^2\right)}{16\sqrt{2}\mu} + \frac{21}{16\mu^{1/4}}\left(1+i\right)F\left(\phi^+,i\right) + \\
  & \frac{171}{16\mu^{3/4}}\left(1-i\right)\left[F\left(\phi^+,i\right) - E\left(\phi^+,i\right)\right]\,, \\
\tilde{\omega}_{II}^- = &\frac{55\sqrt{1+\mu x^4} - 153\sqrt{-\mu}\sin^{-1}\left(\sqrt{-\mu}x^2\right)}{16\sqrt{2}\mu} - \frac{21}{8\sqrt{2}\left(-\mu\right)^{1/4}}F\left(\phi^-,i\right) + \\
  & \frac{171}{8\sqrt{2}\left(-\mu\right)^{3/4}}\left[F\left(\phi^-,i\right) - E\left(\phi^-,i\right)\right]\,,
\end{split}
\end{equation}
%%%%%
\noindent where $F(\phi,k)$ and $E(\phi,k)$ are elliptic integrals of the first and second kind, respectively, and the kernels of these functions are defined as
%%%%%
\begin{equation}
\begin{split}
\phi^+ &\equiv i\sinh^{-1}\left(\left(-\mu\right)^{1/4}x\right)\,, \\
\phi^- &\equiv \sin^{-1}\left(\left(-\mu\right)^{1/4}x\right)\,.
\end{split}
\end{equation}
%%%%%%%%%%
Elliptic integrals are in general not especially difficult to approximate numerically~\cite{luke1968approximations,luke1970further}, but when $\mu > 0$ the kernel function is complex-valued, and so it is not immediately apparent how the final result is a real number. However, it is possible to split the final two terms in $\tilde{\omega}_{II}^+$ into real and imaginary components, at which point it is easy to show the imaginary components cancel. \par
%%%%%%%%%%
The incomplete elliptical integral of the third kind is
\begin{equation}
F\left(\phi,k,\nu\right) = \int_0^\phi\!\left(1-\nu^2\sin^2\alpha\right)^{-1}\left(1-k^2\sin^2\alpha\right)^{-1/2}\,d\alpha\,,
\end{equation}
%%%%%
\noindent for $\nu\neq 0$.
When $\nu = 0$ this becomes the incomplete elliptic integral of the first kind. \par
%%%%%%%%%%
We now use a Pad\'e approximation of the square root term~\cite{luke1968approximations}, indicating the $n^{\mathrm{th}}$ order approximation and its corresponding error in the following way:
%%%%%
\begin{equation}
F\left(\phi,k,\nu\right) = F_n\left(\phi,k,\nu\right) + \epsilon_n\left(\phi,k,\nu\right)\,.
\end{equation}
%%%%%
\noindent The approximation is now
%%%%%
\begin{equation}
\begin{split}
F_n\left(\phi,k,\nu\right) = \frac{1}{2n+1}&\left[A\left(\phi,\nu\right)\left\{1-2\nu^2\sum_{m=1}^n\frac{1}{k^2\sin^2\theta_m-\nu^2}\right\} + \right. \\
  & \left.2k^2\sum_{m=1}^n\frac{\sin^2\theta_m\tan^{-1}\left(\sigma_m\tan\phi\right)}{\sigma_m\left(k^2\sin^2\theta_m-\nu^2\right)}\right]\,,
\end{split}
\end{equation}
%%%%%
\noindent where the following definitions have been used:
%%%%%
\begin{align}
A\left(\phi,\nu\right) &= F\left(\phi,0,\nu\right)\,, \\
\theta_m &= \frac{m\pi}{2n+1}\,, \\
\sigma_m &= \sqrt{1-k^2\sin^2\theta_m}\,.
\end{align}
%%%%%
\noindent For the expression used in~\eqref{eq:omegaII}, we have
%%%%%
\begin{equation}
\label{eq:Fn}
\begin{split}
F_n\left(\phi,i,0\right) &= \frac{1}{2n+1}\left[\phi+2\sum_{m=1}^n\frac{\tan^{-1}\left(\sigma_m\tan\phi\right)}{\sigma_m}\right]\,, \\
\sigma_m &= \sqrt{1+\sin^2\theta_m}\,.
\end{split}
\end{equation}
%%%%%%%%%%
When $\mu>0$, we split this into real and imaginary components. Taking the given definition of $\phi^+$:
%%%%%
\begin{equation}
\phi^+ = \phi_R^+ + i\phi_I^+ = i\sinh^{-1}\left(i^{1/2}\mu^{1/4}x\right)\,,
\end{equation}
%%%%%
\noindent where the real and imaginary components are determined by the transcendental equations\footnote{Note: $\phi_R^+\in(-\pi/4,0]$ and $\phi_I^+\in[0,\infty).$}
%%%%%
\begin{align}
\frac{\mu^{1/2}x^2}{2} &= \frac{\sin^2\phi_R^+}{1-\tan^2\phi_R^+}\,, \\
\frac{\mu^{1/2}x^2}{2} &= \frac{\sinh^2\phi_I^+}{1+\tanh^2\phi_I^+}\,.
\end{align}
%%%%%
\noindent Then, since $\upsilon\equiv\tan\phi^+$, we have~\footnote{Note: $\upsilon_R\in(-\infty,0]$ and $\upsilon_I\in[0,1)$.}
%%%%%
\begin{align}
\upsilon_R &= \frac{\sin\left(2\phi_R^+\right)}{\cos\left(2\phi_R^+\right)+\cosh\left(2\phi_I^+\right)}\,, \\
\upsilon_I &= \frac{\sinh\left(2\phi_I^+\right)}{\cos\left(2\phi_R^+\right)+\cosh\left(2\phi_I^+\right)}\,.
\end{align}
%%%%%%%%%%
Finally, the expression in the summation of~\eqref{eq:Fn} may be split apart by writing
%%%%%
\begin{equation}
\xi\left(\tau_m\right) = \xi_R\left(\tau_m\right) + i\xi_I\left(\tau_m\right) = \tan^{-1}\left(\tau_m\upsilon\right)\,,
\end{equation}
%%%%%
\noindent where we have defined\footnote{Note: $\tau_m = \{\sigma_m,\rho_m\}$}
%%%%%
\begin{align}
\cos\left(2\xi_R\right)\pm\sqrt{1+\left(\frac{\upsilon_I}{\upsilon_R}\right)^2\sin^2\left(2\xi_R\right)} &= \frac{\sin\left(2\xi_R\right)}{\tau_m\upsilon_R}\,, \\
\cosh\left(2\xi_I\right)\pm\sqrt{1-\left(\frac{\upsilon_R}{\upsilon_I}\right)^2\sinh^2\left(2\xi_I\right)} &= \frac{\sinh\left(2\xi_I\right)}{\tau_m\upsilon_I}\,.
\end{align}
%%%%%
We arrive at the expression
%%%%%
\begin{equation}
\label{eq:ReFn}
\mathrm{Re}\left[\left(1\pm i\right)F_n\left(\phi^+,i\right)\right] = \frac{1}{2n+1}\left[\left(\phi_R^+\mp\phi_I^+\right) + 2\sum_{m=1}^n\frac{\xi_R\left(\sigma_m\right)\mp\xi_I\left(\sigma_m\right)}{\sigma_m}\right]\,.
\end{equation}
%%%%%%%%%%
Likewise, we do the same procedure for the incomplete integral of the second kind, given by
%%%%%
\begin{equation}
E\left(\phi,k\right) = \int_0^\phi\!\sqrt{1-k^2\sin^2\alpha}\,d\alpha\,.
\end{equation}
%%%%%
\noindent Again we take the $n^{\mathrm{th}}$ order approximation
%%%%%
\begin{equation}
E\left(\phi,k\right) = E_n\left(\phi,k\right) + \epsilon_n\left(\phi,k\right)\,,
\end{equation}
%%%%%
\noindent where the approximation is given by
%%%%%
\begin{align}
E_n\left(\phi,k\right) &= \left(2n+1\right)\phi - \frac{2}{2n+1}\sum_{m=1}^n\frac{\tan^2\theta_m\tan^{-1}\left(\rho_m\tan\phi\right)}{\rho_m}\,, \\
\rho_m &= \sqrt{1-k^2\cos^2\theta_m}\,.
\end{align}
%%%%%
\noindent In~\eqref{eq:omegaII}, we use
%%%%%
\begin{equation}
\rho_m = \sqrt{1+\cos^2\theta_m}\,.
\end{equation}
%%%%%
\noindent This expression may be split into real and imaginary components using the same techniques, resulting in the final expression
%%%%%
\begin{equation}
\label{eq:ReEn}
\mathrm{Re}\left[\left(1\pm i\right)E_n\left(\phi^+,i\right)\right] = \left(2n+1\right)\left(\phi_R^+\mp\phi_I^+\right) - \frac{2}{2n+1}\sum_{m=1}^n\frac{\tan^2\theta_m}{\rho_m}\left[\xi_R\left(\rho_m\right)\mp\xi_I\left(\rho_m\right)\right]\,.
\end{equation}
%%%%%
Therefore, the solution for Region II is given by~\eqref{eq:omegaII}, where we substitute~\mbox{(\ref{eq:ReFn},~\ref{eq:ReEn})} when $\mu>0$ to ensure all mathematics is real-valued.

\subsection{Region III}
%%%%%%%%%%
We wish to solve the integral
%%%%%
\begin{equation}
I_k = \int\!x^{-3\left(2k+1\right)}\left(1+\mu x^4\right)^{-1/2}\,dx\,.
\end{equation}
%%%%%
Unfortunately there is no general solution, but we can find one solution for even $k$ and another for odd $k$. To differentiate these two solutions, we index using the variable $l$ for even values of $k$, $l = \{0,2,4,\ldots\}$ and $m$ for odd values of $k$, $m = \{1,3,5,\ldots\}$.
For $I_l$,
%%%%%
\begin{equation}
\label{eq:Il}
I_l = -\frac{x^{-2\left(3l+1\right)}}{2\left(3l+1\right)} {}_2F_1\left(\frac{1}{2},-\frac{3l+1}{2};\frac{-3l+1}{2};-\mu x^4\right)\,.
\end{equation}
%%%%%
\noindent If we use this expression for all values of $k$, we would have a hypergeometric function with negative $b$ and $c$ values, with $c=b+1$. If it were the case that $c<b$ we could use a transformation to remove the singularity due to the Gamma function hidden in~\eqref{eq:Il}, but in this particular case any hypergeometric solution will evaluate to $\tilde\infty$ despite the fact that this is not the case for any given odd $k$ inserted into the original expression. Therefore, a more creative approach is required. \par
%%%%%%%%%%
We define the new variables $z\equiv\sqrt{1+\mu x^4}$ and $n\equiv\frac{m+1}{2}\in\mathbb{N}$. The expression $I_m$ is now
%%%%%
\begin{equation}
\label{eq:In}
I_n = \frac{1}{2}\mu^{3n-1}\int\!\left(z^2-1\right)^{-3n}\,dz\,.
\end{equation}
%%%%%
\noindent This expression can be solved using the method of partial fractions. This is trivial for any explicit value of $n$ but in general it is more complicated. We will ultimately obtain a solution of the following form, where the coefficients $A_i$ and $B_i$ are independent of the boundary conditions:
%%%%%
\begin{equation}
\label{eq:In_sol}
I_n = \frac{1}{2}\mu^{3n-1}\left[A_1\ln\left|z+1\right|+B_1\ln\left|z-1\right|+\sum_{i=2}^{3n}\left(1-i\right)^{-1}\left[\frac{A_i}{\left(z+1\right)^{i-1}}+\frac{B_i}{\left(z-1\right)^{i-1}}\right]\right]
\end{equation}
%%%%%
The partial fraction expansion of the integrand in~\eqref{eq:In} is
%%%%%
\begin{equation}
\begin{split}
\left(z^2-1\right)^{-3n} &= \sum_{i=1}^{3n}\left[\frac{A_i}{\left(z+1\right)^i}+\frac{B_i}{\left(z-1\right)^i}\right]\,, \\
1 &= \sum_{i=1}^{3n}\left[A_i\alpha_i\left(z\right) + B_i\beta_i\left(z\right)\right]\,.
\end{split}
\end{equation}
%%%%%
\noindent The goal is to solve a system of equations which is formed by matching powers of $z$, ultimately producing numerical values for $A_i$ and $B_i$.
The expression $\alpha_i(z)$ is given by
%%%%%
\begin{align}
\alpha_i &= \left(z+1\right)^{3n-i}\left(z-1\right)^{3n}\,, \\
  &= \left[\sum_{j=0}^{3n-i}\binom{3n-i}{j}z^j\right]\left[\sum_{j=0}^{3n}\left(-1\right)^{3n-j}\binom{3n}{j}z^j\right]\,, \\
  &= \sum_{j=0}^{6n-i}\gamma_{ij}z^j\,.
\end{align}
%%%%%
\noindent Here we have used the binomial expansion along with the Cauchy product of finite series.
The coefficients $\gamma_{ij}$ may be found using (a modified form of) Vandermonde's identity:
%%%%%
\begin{align}
\gamma_{ij} &= \sum_{r=0}^j \left(-1\right)^{r+3n-j}\binom{3n-i}{r}\binom{3n}{j-r}\,, \\
  &= \begin{cases} \gamma_{ij}^{\left(1\right)}&\mbox{if }\quad j\in\{0,\ldots,3n-1\}\,, \\
     \gamma_{ij}^{\left(2\right)}&\mbox{if }\quad j\in\{3n,\ldots,6n-i\}\,, \end{cases}
\end{align}
%%%%%
\noindent where
%%%%%
\begin{align}
\gamma_{ij}^{\left(1\right)} &\equiv \left(-1\right)^{3n-j}\binom{3n}{j} {}_2F_1\left(i-3n,-j;3n-j+1;-1\right)\,, \\
\gamma_{ij}^{\left(2\right)} &\equiv \left(-1\right)^{6n-j}\binom{3n}{j-3n} {}_2F_1\left(i-3n,3n-j;6n-j+1;-1\right)\,.
\end{align}
%%%%%%%%%%
Similarly, for $\beta_i\left(z\right)$ we find
%%%%%
\begin{align}
\beta_i &= \left(z+1\right)^{3n}\left(z-1\right)^{3n-i}\,, \\
  &= \left[\sum_{j=0}^{3n}\binom{3n}{j}z^j\right]\left[\sum_{j=0}^{3n-i}\left(-1\right)^{3n-i-j}\binom{3n-i}{j}z^j\right]\,, \\
  &= \sum_{j=0}^{6n-i}\delta_{ij}z^j\,,
\end{align}
%%%%%
\noindent with the coefficients
%%%%%
\begin{align}
\delta_{ij} &= \sum_{r=0}^j \left(-1\right)^{r+3n-i-j}\binom{3n}{r}\binom{3n-i}{j-r}\,, \\
  &= \begin{cases} \delta_{ij}^{\left(1\right)}&\mbox{if }\quad j\in\{0,\ldots,3n-1\}\,, \\
     \delta_{ij}^{\left(2\right)}&\mbox{if }\quad j\in\{3n,\ldots,6n-i\}\,, \end{cases}
\end{align}
%%%%%
\noindent where
%%%%%
\begin{align}
\delta_{ij}^{\left(1\right)} &\equiv \left(-1\right)^{3n-i-j}\binom{3n-i}{j} {}_2F_1\left(-j,-3n;3n-i-j;-1\right)\,, \\
\delta_{ij}^{\left(2\right)} &\equiv \left(-1\right)^{6n-i-j}\binom{3n-i}{j-3n} {}_2F_1\left(3n-j,-3n;6n-i-j+1;-1\right)\,.
\end{align}
%%%%%%%%%%
Finally, we can now construct the matrix
%%%%%
\begin{equation}
\mathbf{C} = \left(\begin{array}{cc}
\gamma_{ij}^{\left(1\right)} & \delta_{ij}^{\left(1\right)} \\
\gamma_{ij}^{\left(2\right)} & \delta_{ij}^{\left(2\right)} \end{array}\right)\,.
\end{equation}
%%%%%
\noindent If we define $\Psi = (A_i, B_i)^T$ then the system can be written
%%%%%
\begin{equation}
\mathbf{C}\Psi=\left(1,0,\ldots,0\right)^T\,,
\end{equation}
%%%%%
\noindent and the coefficients $A_i$ and $B_i$ are found by solving for $\Psi$:
%%%%%
\begin{equation}
\Psi = \mathbf{C}^{-1}\left(1,0,\ldots,0\right)^T\,.
\end{equation}
%%%%%%%%%%
We now have a complete solution for both $I_l$,~\eqref{eq:Il}, and $I_m$~\eqref{eq:In_sol}, where again $n=(m+1)/2$, leading to the final expression for Region III:
%%%%%
\begin{equation}
\tilde{\omega}_{III} = 2\sqrt{\pi}\left[\sum_{l=0}^\infty\frac{I_l}{l!\Gamma\left(\frac{1}{2}-l\right)} + \sum_{m=1}^\infty\frac{I_m}{m!\Gamma\left(\frac{1}{2}-m\right)}\right]\,.
\end{equation}
%%%%%%%%%%
At first glance this solution for Region III appears very cumbersome and impractical for numerical implementation.
However, the coefficients $A_i$ and $B_i$ can be solved beforehand and stored in a lookup table, so numerical experiments are efficient so long as one can efficiently calculate the Gauss hypergeometric function.
%%%%%%%%%%

\subsection{Full Solution}
When solving for a value $\mu$ given $\omega_{12}$, $\tau_1$, and $\tau_2$, a root-finding algorithm must be used to invert these expressions.
In most cases at most ten terms in each series are required for convergence with an error of $\mathcal{O}(10^{-10})$, thus demonstrating this is an efficient approach. Furthermore, much of the work can be done beforehand by creating lookup tables. The spatial distance is given by
%%%%%
\begin{equation}
\tilde{\omega}_{12} = \tilde{\omega}_X\left(t_2;\mu\right) - \tilde{\omega}_Y\left(t_1;\mu\right)\,,
\end{equation}
%%%%%
\noindent for timelike ($\mu > 0$) intervals and
%%%%%
\begin{equation}
\tilde{\omega}_{12} = 2\tilde{\omega}_Z\left(t_m\left(\mu\right);\mu\right) - \tilde{\omega}_Y\left(t_1;\mu\right) - \tilde{\omega}_X\left(t_2;\mu\right)\,,
\end{equation}
%%%%%
\noindent for spacelike ($\mu < 0$) intervals, where the $X$, $Y$, and $Z$ indicate we use the region indicated by the parameter $t$, i.e., they will indicate regions I, II, or III. Most importantly, when $X\neq Y$, we need to combine the approximations and extract the discontinuity at the (arbitrary) boundary by subtracting the difference between the two functions at this point. For instance, suppose we are searching a timelike interval for $\mu$ where $x(t_1) = 0.1$ and $x(t_2) = 1.0$ and Region I is defined as $x\in[0,0.9)$ and Region II as $x\in[0.9,1.1)$. We would find $\tilde{\omega}_{12}$ with the following expression:
%%%%%
\begin{equation}
\tilde{\omega}_{12} = \tilde{\omega}_{II}^+\left(x=1.0,\mu\right) - \tilde{\omega}_I\left(x=0.1,\mu\right) - \left[\tilde{\omega}_{II}^+\left(x=0.9,\mu\right) - \tilde{\omega}_I\left(x=0.9,\mu\right)\right]\,.
\end{equation}
%%%%%
\noindent A similar method is used at the upper boundary. If the two boundary points lie in Regions I and III, respectively, then both discontinuities must be extracted. \par

\section{Summary}
By integrating the geodesic differential equations~\eqref{eq:geodesic_diff_eq} we have shown for spacetimes with dark energy, dust, radiation, or a stiff fluid, that it is possible to find a closed-form solution for the geodesic distance provided either initial-value or boundary-value constraints. Furthermore, by studying the form of the first-order differential equation~\eqref{eq:geodesic4} we found that extrema along spacelike geodesic curves will always point away from the origin. This insight provides a better understanding of how to integrate the geodesic and distance kernels~(\ref{eq:geodesic4},~\ref{eq:distances_general_spatial},~\ref{eq:distances_general}) for different types of boundary conditions. Moreover, our other important result in Sec.~\ref{sec:geo_conectedness} demonstrates how, using~\eqref{eq:conformal_time},~\eqref{eq:max_time_constraint} and~\eqref{eq:geo_compl}, we are able to tell, using only the scale factor, whether or not all points on a flat FLRW manifold can be connected by a geodesic. This observation is particularly useful in numeric experiments and investigations that can study only a finite portion of a spatially flat manifold. Finally, in Section~\ref{sec:examples8} and~\ref{sec:num_approx} we provided several examples of how these results might be applied to some of the most well-studied FLRW manifolds, including the manifold describing our universe. While not all spacetimes have closed-form solutions for geodesics, it is still possible to reframe the problem in a way which may be solved efficiently using numerical methods in existing software libraries.

%\afterpage{\blankpage}

\part{Applications to Network Science and Cosmology}
\label{part:apps}
\thispagestyle{empty}
\afterpage{\blankpage}
\twolinechapter
\chapter[Navigation in Random Geometric Graphs]{\texorpdfstring{Navigation in Random\\[-0.8cm] Geometric Graphs}{Navigation in Random Geometric Graphs}}
\chaptermark{Navigation in RGGs}
\mainchapter
\label{chap:navigation}

\thispagestyle{empty}
In network science and applied mathematics, random geometric graphs have attracted increasing attention over recent years~\cite{serrano2012uncovering,kleineberg2016hidden,allard2017geometric,bianconi2015interdisciplinary,ostilli2015statistical,bianconi2015complex,wu2015emergent,bianconi2016network,bianconi2017emergent,zhang2014opinion,newman2015generalized,henderson2011geometric,roberts2016contribution,javarone2013perception,xie2015modeling,xie2015random,jin2017coupling,clough2016what,clough2016embedding,asta2015geometric,gugelmann2012random,fountoulakis2015geometrization,bode2015largest,candellero2016bootstrap,candellero2016clustering,abdullah2015typical,fountoulakis2016law,bringmann2015geometric,bringmann2016average,bringmann2016greedy,bradonjic2010efficient,bubeck2014testing,dhara2016solvable,friedrich2015cliques,friedrich2015diameter,blasius2016hyperbolic,penrose2013connectivity}, since it was shown that if the space defining these graphs is not Euclidean but negatively curved, i.e., hyperbolic, then these graphs share many common structural and dynamical properties of many real networks, including scale-free degree distributions, strong clustering, community structure, and network growth dynamics~\cite{krioukov2010hyperbolic,papadopoulos2012popularity,zuev2015emergence}.%
Yet more interesting is how these graphs explain the optimality of many network functions related to finding paths in the network without global knowledge of the network structure~\cite{boguna2009navigability,boguna2009navigating}. Random hyperbolic graphs appear to be optimal, that is, maximally efficient, with respect to the greedy path finding strategy that uses only spatial geometry to navigate through a complex network structure by moving at each step from a current element to its neighbor closest to the destination in the space~\cite{krioukov2010hyperbolic,bringmann2016greedy}. The efficiency of this process is called network navigability~\cite{kleinberg2000navigation}. High navigability of random hyperbolic graphs has led to practically viable applications, including the design of efficient routing in the future Internet~\cite{boguna2010sustaining,lehman2016experimental}, and have demonstrated that the spatiostructural organization of the human brain is nearly as needed for optimal information routing between different parts of the brain~\cite{gulyas2015navigable}. Yet if random hyperbolic graphs are truly geometric, meaning that if the sprinkling density is indeed constant with respect to the hyperbolic volume form, then the exponent $\gamma$ of the probability distribution $P(k)\sim k^{-\gamma}$ of element degrees $k$ in the resulting graphs is exactly $\gamma=3$~\cite{krioukov2010hyperbolic}. In contrast, in random geometric graphs in de Sitter spacetime, which is asymptotically the spacetime of our accelerating universe, or indeed in the spacetime representing the exact large-scale Lorentzian geometry of our universe, this exponent asymptotically approaches $\gamma=2$~\cite{krioukov2012network}, as in many real networks~\cite{boccaletti2006complex}. Yet it remains unclear if these random Lorentzian graphs are as navigable as random hyperbolic graphs. \par

Here we study the navigability of undirected random geometric graphs in three FLRW Lorentzian manifolds. We review the geometry of these spaces in Section~\ref{sec:flrw_geom}, and then discuss graph construction in Section~\ref{sec:flrw_rggs}. One manifold is de Sitter spacetime, corresponding to a universe filled with dark energy only, and no matter. Another manifold is the other extreme, a universe filled only with dust matter, and no dark energy. The third manifold is a universe like ours, containing both matter and dark energy. This last manifold interpolates between the other two. At early times and small graph sizes, it is matter-dominated and ``looks'' like the dust-only spacetime. At later times and large graph sizes, it is dark-energy-dominated and ``looks'' increasingly more like de Sitter spacetime. \par

We find in Section~\ref{sec:navigability} that random geometric graphs are navigable only in manifolds with dark energy. Specifically, if there is no dark energy, that is, in the dust-only spacetime, there is a finite fraction of paths for which geometric path finding fails, and this fraction is constant---it does not depend on the cutoff time, i.e., the present cosmological time in the universe, if the average degree in the graph is kept constant. In contrast, in spacetimes with dark energy, i.e., de Sitter spacetime and the spacetime of our universe, the fraction of unsuccessful paths quickly approaches zero as the cutoff time increases. \par

We then discuss these results in depth in Section~\ref{sec:nav_discussion} and the methodology used in experiments in Section~\ref{sec:methodology}. For network science this finding implies that in terms of navigability, random geometric graphs in Lorentzian spacetimes with dark energy are as good as random hyperbolic graphs. For physics, this finding establishes a connection between the presence of dark energy and navigability of the discretized causal structure of spacetime.

\section[Geometry of Friedmann-Lema\^itre-Robertson-Walker Spacetimes]{Geometry of FLRW Spacetimes}
\sectionmark{Geometry of FLRW Spacetimes}
\label{sec:flrw_geom}
The geometric structure of random geometric graphs in FLRW spacetimes is directly related to the spacetimes' matter content, which we review in this section. For more background on Lorentzian geometry, we refer back to Section~\ref{sec:lorentzian_geometry}. \par

The total energy density in our universe is known to come from four sources: the matter (dark and baryonic) density $\rho_M$, the dark energy density $\rho_\Lambda$, the radiation energy density $\rho_R$, and the curvature $\mathcal{K}$. The densities may be rescaled by a critical density: $\Omega\equiv\rho/\rho_c$, where $\rho_c\equiv3H_0^2/8\pi$; $H_0\equiv\dot{a}_0/a_0$ is the Hubble constant and $a_0\equiv a(t_0)$, i.e., the scale factor at the present time. Similarly, the curvature density parameter may be written as $\Omega_\mathcal{K}\equiv-\mathcal{K}/(a_0H_0)^2$ so that we obtain the state equation $\Omega_M + \Omega_\Lambda + \Omega_R + \Omega_\mathcal{K} = 1$. This allows us to rewrite Friedmann's equation~\eqref{eq:friedmann} in the integral form~\cite{weinberg2008cosmology}
\begin{equation}
\label{eq:friedmann2}
H_0t = \int_0^{a/a_0}\!\frac{dx}{x\sqrt{\Omega_\Lambda + \Omega_\mathcal{K}x^{-2} + \Omega_Mx^{-3} + \Omega_Rx^{-4}}}\,.
\end{equation}

In the flat universe, the curvature energy density contribution is zero: $\Omega_\mathcal{K}=0$. Furthermore, except for a short period in the early universe, the radiation energy density is also negligible compared to the other terms: $\Omega_R\approx 0$. Therefore, we study manifolds defined only by $\Omega_\Lambda$ and $\Omega_M$: the de Sitter (dark energy only) manifold ($\Lambda > 0, g=c=0$), the Einstein-de Sitter (dust only) manifold ($\Lambda = 0, g=1, c>0$), and the mixed dark energy and dust manifold ($\Lambda,c > 0, g=1$). Hereafter, these three manifolds are respectively referred to as the energy (\emph{E}), dust (\emph{D}), and mixed (\emph{M}) manifolds. Defining rescaled time $\tau=t/\lambda$, the scale factors in these spacetimes are solutions to~\eqref{eq:friedmann2}, respectively using non-zero $\Omega_\Lambda$, $\Omega_M$, or both:
\begin{equation}
\label{eq:scale_factors7}
a_E(\tau) = \lambda e^{\tau}\,, \quad
a_D(\tau) = \alpha\left(\frac{3}{2}\tau\right)^{2/3}\,, \quad
a_M(\tau) = \alpha\sinh^{2/3}\left(\frac{3}{2}\tau\right)\,.
\end{equation}
The parameters $\lambda$ and $\alpha$ respectively define the temporal and spatial scales. In a de Sitter manifold, there is no distinction between temporal and spatial scales, so that there is no $\alpha$, because the generators of the Lorentz group $SO(1,3)$ form a proper subset of those of the de Sitter group $SO(1,4)$, thereby removing a degree of freedom in the model. In manifolds which represent spacetimes with dust matter, this symmetry is broken, and relative rescalings between $\lambda$ and $\alpha$ are equivalent to an isotropic rescaling of space with respect to time. \par

The spatial scale of a mixed manifold, such as the one approximating our real universe, arises naturally from~\eqref{eq:friedmann2} when dimensionless variables are used; it is defined as $\alpha\equiv a_M(t_0)(\Omega_M/\Omega_\Lambda)^{1/3}$, related to the relative amount of dark energy~\cite{krioukov2012network}. The scale factor $a_M(\tau)$ asymptotically matches $a_D(\tau)$ at earlier times (a hot, matter-dominated universe) and $a_E(\tau)$ at later times (a cold, dark energy-dominated universe), so that the mixed manifold can be characterized by the dark energy density parameter $\Omega_\Lambda$. This way, the dark energy density is a measure of time via $\tau = (2/3)\arctanh\sqrt{\Omega_\Lambda}$. Using the present-day value of $\Omega_{\Lambda,0}\approx0.737$ in our universe gives the current rescaled cosmological time $\tau_0=t_0/\lambda\approx 0.473$, so that $\lambda$ sets the spacetime's timescale~\cite{astier2006supernova}. \par

In the FLRW spacetimes defined by~\eqref{eq:scale_factors7}, the scale factor and the metric tensor are used to find the volume form of the manifold:
\begin{equation}
\label{eq:diff_volume}
dV = \sqrt{-|g_\munu|}\sin\theta\,dt\,dr\,d\theta\,d\phi = a(t)^3r^2\sin\theta\,dt\,dr\,d\theta\,d\phi\,,
\end{equation}
where $r$ is the dimensionless radial coordinate and $\theta$ and $\phi$ are the polar and azimuthal angular coordinates. To study a particular spacetime in simulations below, it is necessary to consider its compact region, bounded by a temporal cutoff $t\in[0,t_0]$ and radial cutoff $r\in[0,r_0]$. Using rescaled temporal and spatial cutoffs $\tau_0=t_0/\lambda$ and $\rho_0=\tilde\alpha r_0$, where $\tilde\alpha=\alpha/\lambda$, except de Sitter spacetime where $\rho_0=r_0$, the volume of such a region in each spacetime is easily obtained via the integration of~\eqref{eq:diff_volume} within the corresponding bounds:
\begin{equation}
\label{eq:volumes}
\begin{split}
V_E\left(\tau_0,\rho_0\right) &= \frac{4\pi}{9}\lambda^4 \rho_0^3\left(e^{3\tau_0}-1\right)\,,\\
V_D\left(\tau_0,\rho_0\right) &= \pi\lambda^4\rho_0^3\tau_0^3\,, \\
V_M\left(\tau_0,\rho_0\right) &= \frac{2\pi}{9}\lambda^4\rho_0^3\left(\sinh\left(3\tau_0\right) - 3\tau_0\right)\,.
\end{split}
\end{equation}
We will also use conformal time $\eta$, defined as $\eta(t) = \int^tdt^\prime/a(t^\prime)$, which is
\begin{equation}
\begin{split}
\eta_E\left(\tau\right) &= -e^{-\tau}\,, \\
\eta_D\left(\tau\right) &= \frac{1}{\tilde\alpha}\left(12\tau\right)^{1/3}\,, \\
\eta_M\left(\tau\right) &= \frac{2}{\tilde\alpha}\sinh^{1/3}\left(\frac{3}{2}\tau\right){}_2F_1\left(\frac{1}{6},\frac{1}{2};\frac{7}{6};-\sinh^2\left(\frac{3}{2}\tau\right)\right)\,,
\end{split}
\end{equation}
where ${}_2F_1$ is the Gauss hypergeometric function. This transformation is particularly useful for distinguishing between timelike and spacelike intervals, since in these coordinates, the scale factor may be factored out: $ds^2 = a^2(t(\eta))(-d\eta^2+d\Sigma^2)$, so that timelike and spacelike intervals with $\Delta s^2<0$ and $\Delta s^2>0$ correspond to intervals with $\Delta\eta^2>\Delta\Sigma^2$ and $\Delta\eta^2<\Delta\Sigma^2$, respectively.

%\section[Constructing RGGs]{Constructing Random Geometric Graphs in\\[-0.3cm]\hspace*{1.2cm} Lorentzian Manifolds}
\section{Constructing Random Geometric Graphs in Lorentzian Manifolds}
\begin{figure}[!t]
\includegraphics[width=0.5\textwidth]{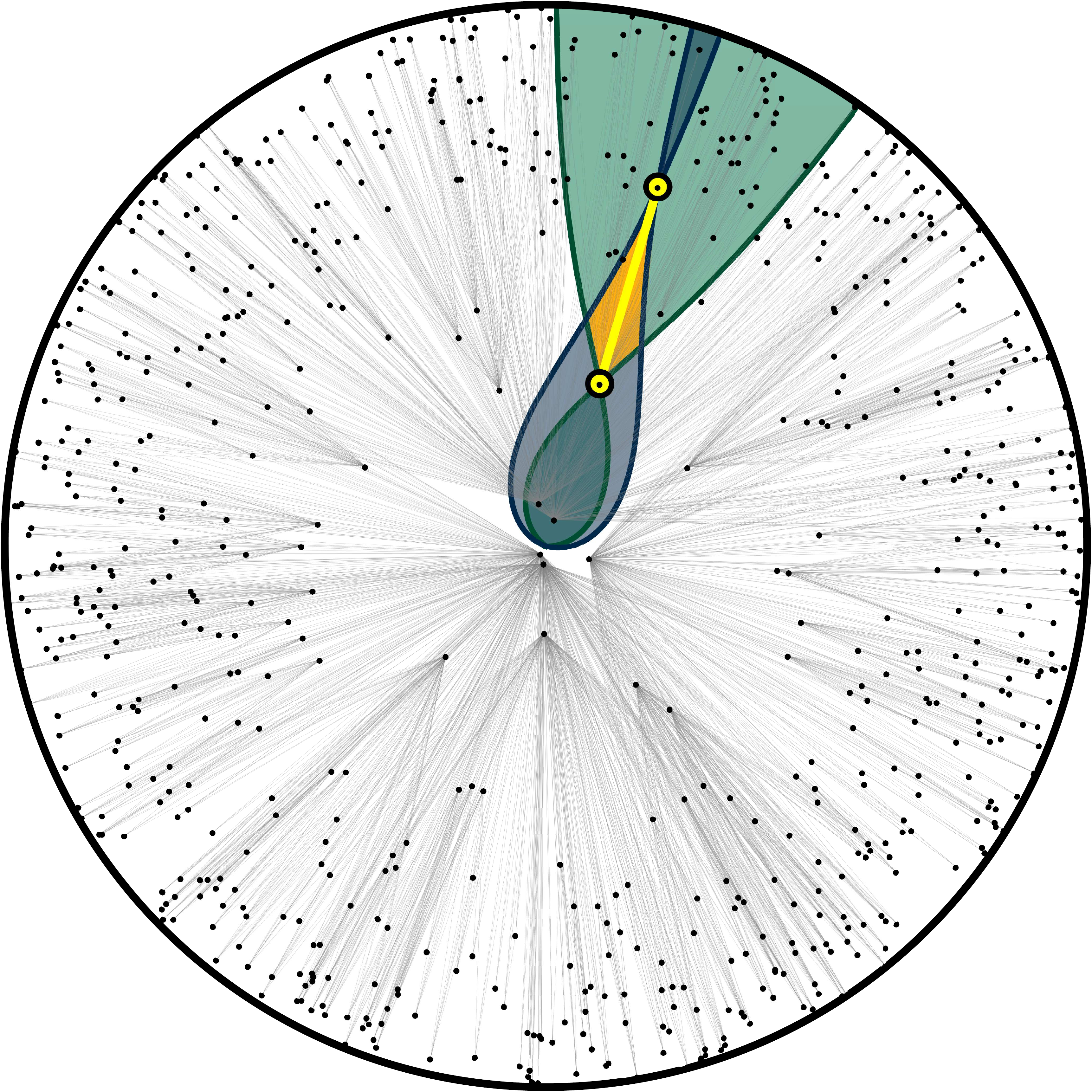}
\centering
\caption{\textbf{Random geometric graph in (1+1)-dimensional de Sitter spacetime.} The graph is realized by Poisson sprinkling $700$ elements onto a $(1+1)$-dimensional de Sitter manifold, with compact spatial foliation by circles, which are hypersurfaces of constant time. The temporal cutoff is $\tau_0=5.94$, which is the radius of the disk shown. In the figure, the graph has been mapped from the de Sitter manifold to a disk of this radius by equating the time coordinates of all points in de Sitter spacetime with the radial coordinates in the shown disk. A pair of elements, shown in yellow, is chosen and their light cones are shown in gray and green. The yellow elements are related to all other elements that happen to lie in their corresponding light cones. In particular, the yellow elements are related to each other since they lie within each other's light cones. The overlap between the past and future light cones of the higher-$t$ and lower-$t$ yellow elements respectively, shown in orange, is their Alexandroff set. The full set of gray relations is obtained by iterating over all element pairs.}
\label{fig9-0}
\end{figure}

\sectionmark{Constructing RGGs}
\label{sec:flrw_rggs}
We construct RGGs in Lorentzian manifolds by sampling three spatial coordinates and one temporal coordinate for $N$ elements in a particular region using a Poisson point process: $N$ is a random variable sampled from the Poisson distribution with mean $\bar{N}$, giving a sprinkling density $\nu\equiv N/V$. Given volumes~\eqref{eq:volumes}, and using the rescaled sprinkling density $q=\nu\lambda^4$, the numbers of elements in the three spacetimes are given by
\begin{equation}
\label{eq:nodes}
\begin{split}
N_E\left(\tau_0,\rho_0\right) &= \frac{4\pi}{9} q \rho_0^3\left(e^{3\tau_0}-1\right)\,, \\
N_D\left(\tau_0,\rho_0\right) &= \pi q \rho_0^3\tau_0^3\,, \\
N_M\left(\tau_0,\rho_0\right) &= \frac{2\pi}{9} q \rho_0^3\left(\sinh\left(3\tau_0\right) - 3\tau_0\right)\,,
\end{split}
\end{equation}
where all the parameters $q,\rho_0,\tau_0$ are dimensionless. A pair of elements $(i,j)$ is timelike related and, therefore, linked in the resulting graph if the following inequality is true:
\begin{equation}
\label{eq:spatial_form}
\Delta\Sigma_{ij}^2 = r_i^2 + r_j^2 - 2r_ir_j\left(\cos\theta_i\cos\theta_j + \sin\theta_i\sin\theta_j\cos\left(\phi_i - \phi_j\right)\right) < \left(\eta_i - \eta_j\right)^2\,,
\end{equation}
where the law of cosines has been used for the spatial distance $\Delta\Sigma_{ij}$ between the two elements in three dimensions. Figure~\ref{fig9-0} visualizes a random geometric graph in $(1+1)$-dimensional de Sitter spacetime, where $\Delta\Sigma_{ij}^2=(\theta_i-\theta_j)^2$ instead of~\eqref{eq:spatial_form}. \par
In simulations in the next section, we will also need to generate graphs with a given average degree. To find the expected average degree in RGGs in our Lorentzian regions, we observe that the volume of the past and future light cones emanating from any given element, and bounding regions timelike-related to the element, is directly proportional, with the proportionality coefficient $1/\nu$, to the expected number of sprinkled elements in them, and consequently, to the expected past and future degrees of the element. Integrating the expressions for these volumes, weighted by the element density in the space, over the entire region provides a theoretical expression for the expected degree as a function of the rescaled sprinkling density $q=\nu\lambda^4$ and the rescaled temporal cutoff $\tau_0=t_0/\lambda$, we get:
\begin{align}
\label{eq:degrees}
\begin{aligned}
\bar{k}_E(\tau_0) &= \frac{4\pi q}{9}\frac{\left(e^{-\tau_0}-1\right)\left(13-e^{-\tau_0}\left(14-13e^{-\tau_0}\right)\right)+6\tau_0\left(e^{-3\tau_0}+1\right)}{1-e^{-3\tau_0}}\,, \\
\bar{k}_D(\tau_0) &= \frac{18\pi q}{385}\tau_0^4\,, \\
\bar{k}_M(\tau_0) &= \frac{8\pi q}{\sinh\left(3\tau_0\right) - 3\tau_0}\int_0^{\tau_0}\!d\tau^\prime\int_0^{\tau_0}\!d\tau^\dprime\,\sinh^2\left(\frac{3\tau^\prime}{2}\right)\sinh^2\left(\frac{3\tau^\dprime}{2}\right)|\tilde{\eta}_M\left(\tau^\prime\right) - \tilde{\eta}_M\left(\tau^\dprime\right)|^3\,,
\end{aligned}
\end{align}
where rescaled conformal time $\tilde{\eta}\equiv\tilde\alpha\eta$ is used for convenience. These expressions do not depend on spatial cutoff $\rho_0$ because they are approximations for spatially large regions with $\rho_0\gg\tau_0$, so that boundary effects, i.e., the contributions to the average degree from elements with $\rho$s close to $\rho_0$, are negligible. \par

\begin{figure}[!t]
\includegraphics[width=\textwidth]{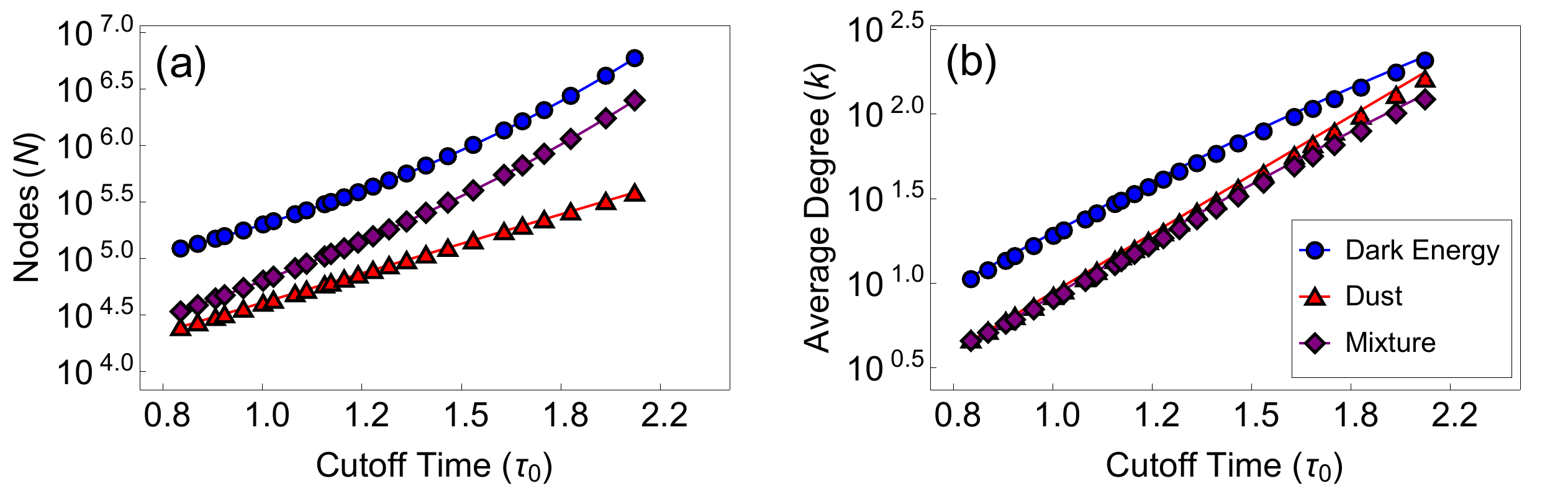}
\centering
\caption{\textbf{Graph size and average degree as functions of the cutoff time.} The figure shows the graph size $N$ and average degree $\bar{k}$ in simulations versus theoretical predictions, the solid curves, given by~(\ref{eq:nodes},\ref{eq:degrees}), for the constant rescaled sprinkling density $q=60$ and spatial cutoff $\rho_0=6$.}
\label{fig9-1}
\end{figure}
It is evident from the exposition above including (\ref{eq:nodes},~\ref{eq:degrees}) that only three out of the original five parameters defining the RGG ensemble with $N$ elements and average degree $\bar{k}$---sprinkling density $\nu\equiv N/V$, temporal scale $\lambda$, spatial scale $\alpha$, and temporal and spatial cutoffs $t_0$ and $r_0$---are independent because $N$ depends only on three dimensionless parameters, $q$, $\rho_0$, and $\tau_0$, while $\bar{k}$ depends only on two, $q$ and $\tau_0$. This is because the sprinkling density $\nu$ sets the discreteness scale, which can be rescaled by $\lambda$: two graph ensembles with different $\nu$s and $\lambda$s are the same if their rescaled sprinkling density $q=\nu\lambda^4$ is the same. Similarly, two graph ensembles with different $\lambda$s and $t_0$s are the same if their $\tau_0$s are the same, and two graph ensembles, and even spacetime regions, with different $\alpha$s and $r_0$s are the same if their $\rho_0$s are the same. Therefore the parameters $q,\rho_0,\tau_0$ form one natural choice of independent parameters, which is the one we use in simulations below. Yet, any three independent functions of these parameters is an equivalent choice. In particular, $N$ and $\bar{k}$ are two such independent functions, so that $N,\bar{k},\tau_0$ is another choice of parameters that we also use in simulations. We note that one parameter in these two sets of three parameters is not entirely independent, because the spatial cutoff $\rho_0$ must be such that $\rho_0\gg\tau_0$, so that the spatial boundary effects are negligible, and approximations~\eqref{eq:degrees} are valid, see Figure~\ref{fig9-1} and Section~\ref{sec:methodology}. \par

\begin{figure}[!t]
\includegraphics[width=\textwidth]{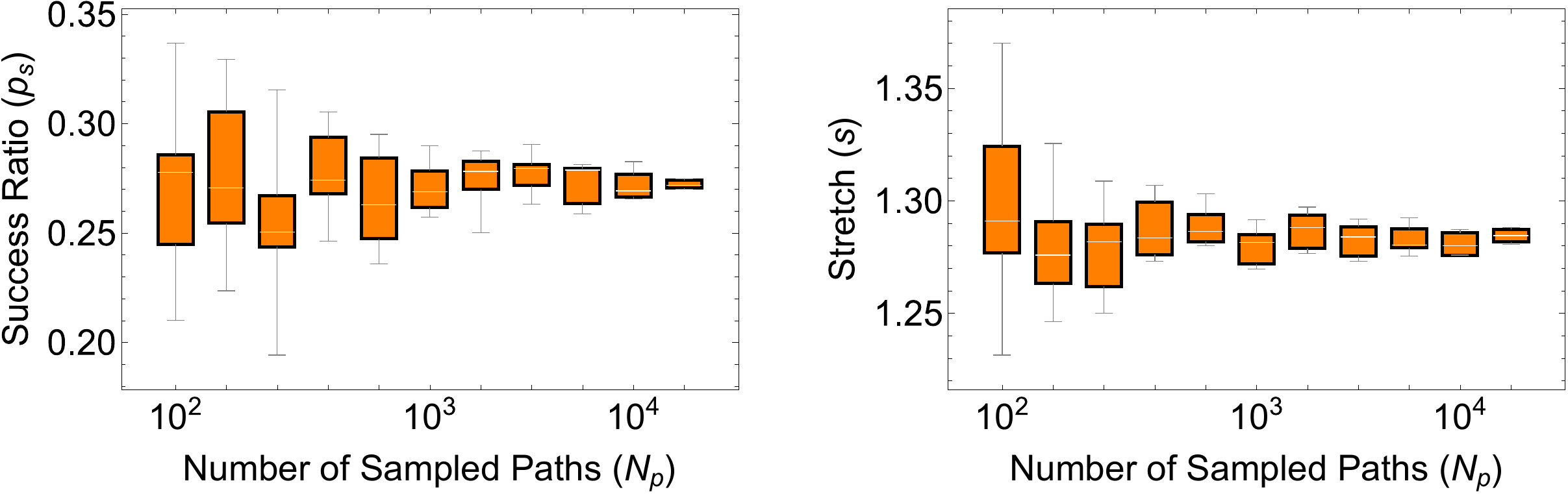}
\centering
\caption{\textbf{Convergence of success ratio and stretch.} The box plots summarize the distributions of the success ratio (a) and stretch (b) as functions of the number $N_p$ of random source-destination element pairs sampled in $10$ random geometric graphs ($N_p$ pair samples in each graph) in the Einstein-de Sitter (dust) manifold with $\tau_0=4.64$, $\bar{k}=10$, and $N=2^{20}$. The orange boxes range from the first to third quartiles, while the bars are minima and maxima. The distributions stabilize at $N_p\ll N$.}
\label{fig9-2}
\end{figure}
\section[Navigability]{Navigability of Random Geometric Graphs in Lorentzian Manifolds}
\label{sec:navigability}
The navigability of a geometric graph is the efficiency of greedy geometric path finding on it. This path finding strategy uses only local nearest-neighbor information to find a path in the graph between a given source element and a given destination element. Starting with the source element, the next element on the path is determined as the element's neighbor closest to the destination element according to geodesic distances in the manifold. When the closest neighbor has already been visited, the greedy path enters a loop. It does not reach the destination and is thus unsuccessful. This situation occurs when the two elements forming the loop, also called a local minimum, do not have any third element that would be closer to the destination than the two elements. The success ratio $p_s$ is defined as the fraction of greedy paths which successfully reach their destination, across a given set of source-destination element pairs in the graph. Here we select $N$ such pairs uniformly at random, where $N$ is the graph size. Increasing the number of pairs above $N$ does not noticeably affect the results, as can be seen from Figure~\ref{fig9-2}.
Another navigability metric is the stretch. The stretch of a successful greedy path is the ratio of the length of the path, measured as the number of hops, to the length of the shortest path between the same source and destination in the graph. The average stretch is the average of this quantity across successful paths between a given set of source-destination element pairs. \par

\begin{figure}[!t]
\includegraphics[width=\textwidth]{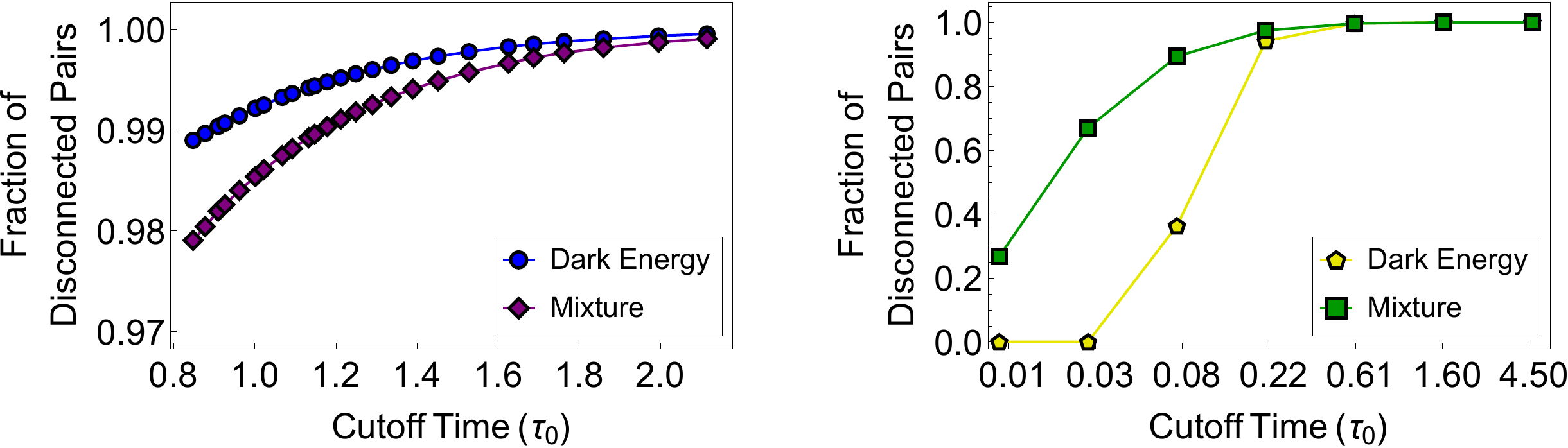}
\centering
\caption{\textbf{Fraction of geodesically disconnected element pairs.} Panels~(a,b) correspond to the graphs in the de Sitter (dark energy) and mixed manifolds with $q=60, \rho_0=6$ and $N=2^{20}, \bar{k}=10$, respectively. The graphs in the Einstein-de Sitter (dust) manifold have trivially no geodesically disconnected element pairs since the manifold is geodesically connected.}
\label{fig9-3}
\end{figure}
The geodesic distance between a pair of elements on the underlying manifold is found by integrating the geodesic differential equations~\ref{eq:geodesic_diff_eq}. The general solution takes the form
\begin{equation}
\begin{split}
d_{ij} &= \int_{t_i}^{t_j}\!\sqrt{\left|\frac{-\mu a^2\left(t\right)}{1+\mu a^2\left(t\right)}\right|}\,dt\,, \\
\Delta\Sigma_{ij} &= \int_{t_i}^{t_j}\!\left(a^2\left(t\right)+\mu a^4\left(t\right)\right)^{-1/2}\,dt\,,
\end{split}
\end{equation}
where the parameter $\mu$ is found by solving the second transcendental equation provided $\Delta\Sigma_{ij}$ and $(t_i,t_j)$. The full procedure is described in detail in Chapter~\ref{chap:geodesics}, with numerical approximations used for the mixed manifold described in Section~\ref{sec:num_approx}. \par

As opposed to Riemannian manifolds, Lorentzian manifolds can be geodesically incomplete, i.e., there can exist pairs of spacelike separated points between which a geodesic does not exist~\cite{oneill1983semi}. For such geodesically disconnected source-destination pairs, geodesic distances and consequently geodesic routing are \emph{undefined}, so that we exclude such pairs from our calculations. The fractions of geodesically disconnected element pairs in random graphs in the experiments below are reported in Figure~\ref{fig9-3}. \par

\begin{figure}[!t]
\includegraphics[width=\textwidth]{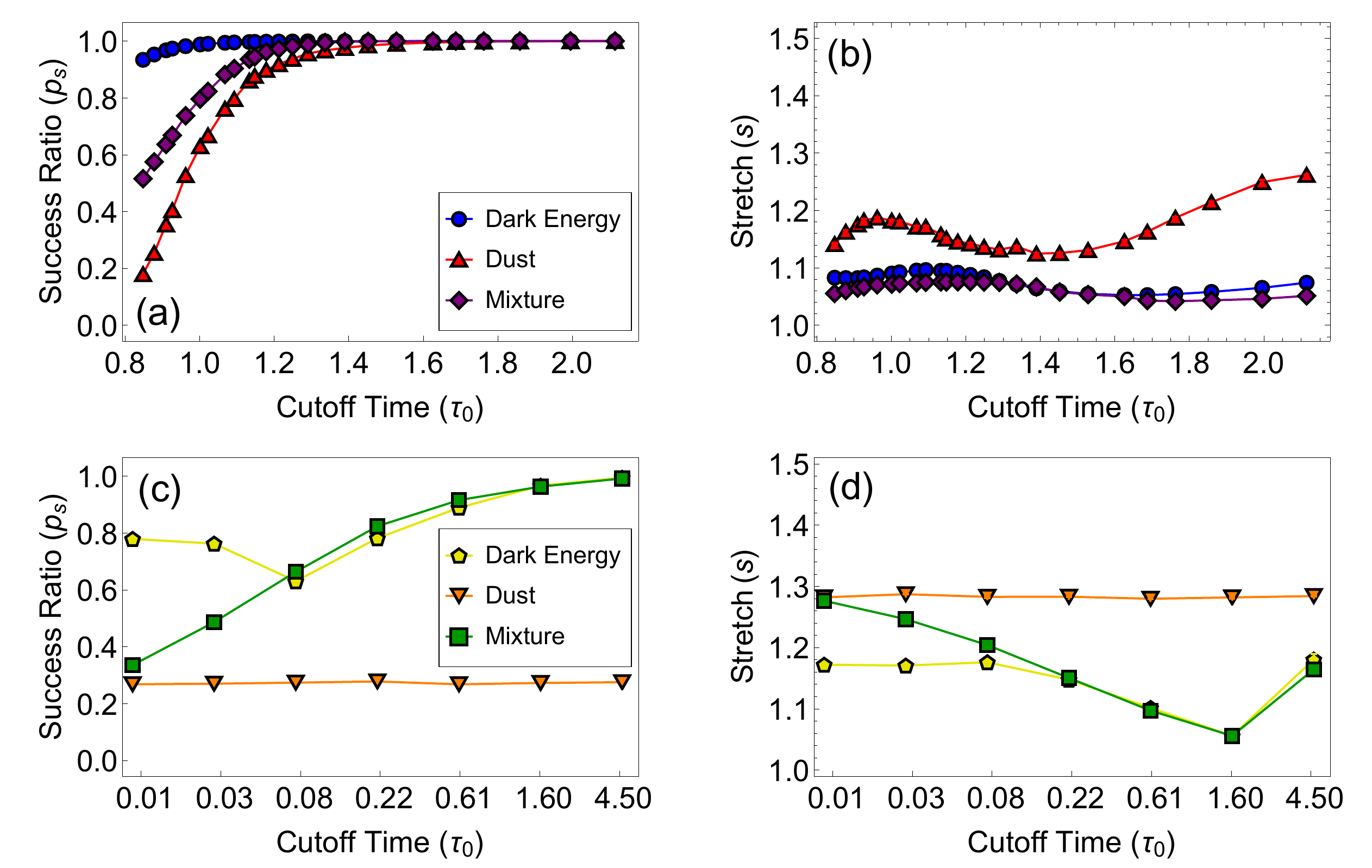}
\centering
\caption{\textbf{Navigability of random geometric graphs in the three manifolds.} In (a,b), corresponding to graphs in panels (a,b) in Fig.~\ref{fig9-1} where the sprinkling density and spatial cutoff are held constant at $q=60$ and $\rho_0=6$, the success ratio increases toward $100\%$ as the temporal cutoff increases, while the average stretch remains low and close to $1$, especially for spacetimes with dark energy. In (c,d), the graph size and average degree are kept constant $N=2^{20}$ and $\bar{k}=10$ as described in Section~\ref{sec:methodology}. The success ratio and stretch in this case depend only on the manifold geometry. The average stretch is still low, especially for the manifolds with dark energy. However, the success ratio increases to $100\%$ only for spacetimes with dark energy, while for the dust manifold it is a constant below $100\%$, which does not depend on the cutoff time.}
\label{fig9-4}
\end{figure}
%\clearpage
%
Figures~\ref{fig9-4}(a,b) show that if the dimensionless sprinkling density $q$ is held constant as the temporal cutoff increases, the success ratio $p_s$ increases to $100\%$ in all three manifolds, while the average stretch remains low and close to its minimum value~$1$, especially in the manifolds with dark energy. However, the average degree grows quickly with the temporal cutoff in this case,~\eqref{eq:degrees} and Figure~\ref{fig9-1}, and the success ratio and stretch depend on both the manifold geometry and the average degree. Indeed, all other things equal, e.g., the same patch of the same manifold with the same spatial and temporal cutoff, the higher the average degree, the higher the navigability, i.e., the higher the success ratio and the lower the stretch, because the larger the number of neighbors that each element has, the higher the chances that the element has a neighbor that does not lead to a loop, and the higher the chances that the next-hop neighbor is closer to the geodesic to the destination in the manifold, thus minimizing the stretch. \par

To disentangle the dependency of navigability on manifold geometry from its dependency on the graph properties, the average degree, and the graph size, we select for different temporal cutoffs, different sprinkling densities and spatial cutoffs such that the average degree and graph size stay constant as the temporal cutoff increases, see Section~\ref{sec:methodology}. In this case, the navigability metrics depend only on the geometry of the manifold. \par

The results in Figure~\ref{fig9-4}(c,d) show that in this case, while the average stretch remains low, especially in the manifolds with dark energy, the success ratio depends strongly on the presence of dark energy in the spacetime. In spacetimes with dark energy, the success ratio still quickly reaches $100\%$, while in the dust-only spacetime, it is a constant below $100\%$, i.e., does not increase with time. \par

\begin{figure}[!t]
\includegraphics[width=0.5\textwidth]{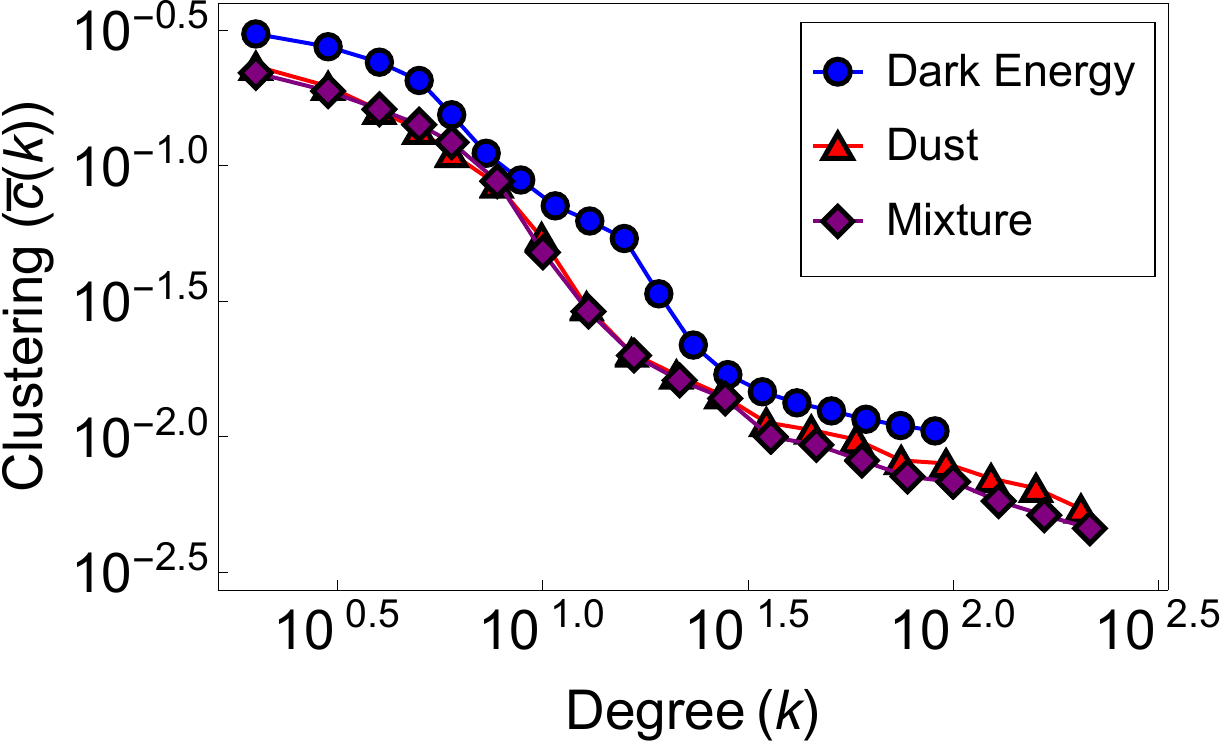}
\centering
\caption{\textbf{Clustering in Lorentzian RGGs.} The figure shows the average clustering $\bar{c}(k)$ of elements of degree $k$ in random geometric graphs with $q=60,\rho_0=6,\tau_0=0.84$ in the three studied manifolds. The mean clustering excluding elements with $k=\{0,1\}$ in the de Sitter, Einstein-de Sitter, and mixed manifolds are $\bar{c}_E=0.145$, $\bar{c}_D=0.164$, and $\bar{c}_M=0.166$, respectively.}
\label{fig9-5}
\end{figure}
We thus conclude that unless dark energy is present, random graphs in Lorentzian geometries are not navigable as their success ratio is a constant below $100\%$, independent of the temporal cutoff. Only in spacetimes with dark energy and asymptotically de Sitter geometry does the success ratio quickly reaches its maximum value of $100\%$, so that such spacetimes, including the spacetime of our universe, are fully navigable with respect to all geodesically connected pairs of elements. This result deserves a discussion.

\section{Discussion}
\label{sec:nav_discussion}
\begin{figure}[!pt]
\includegraphics[width=\textwidth]{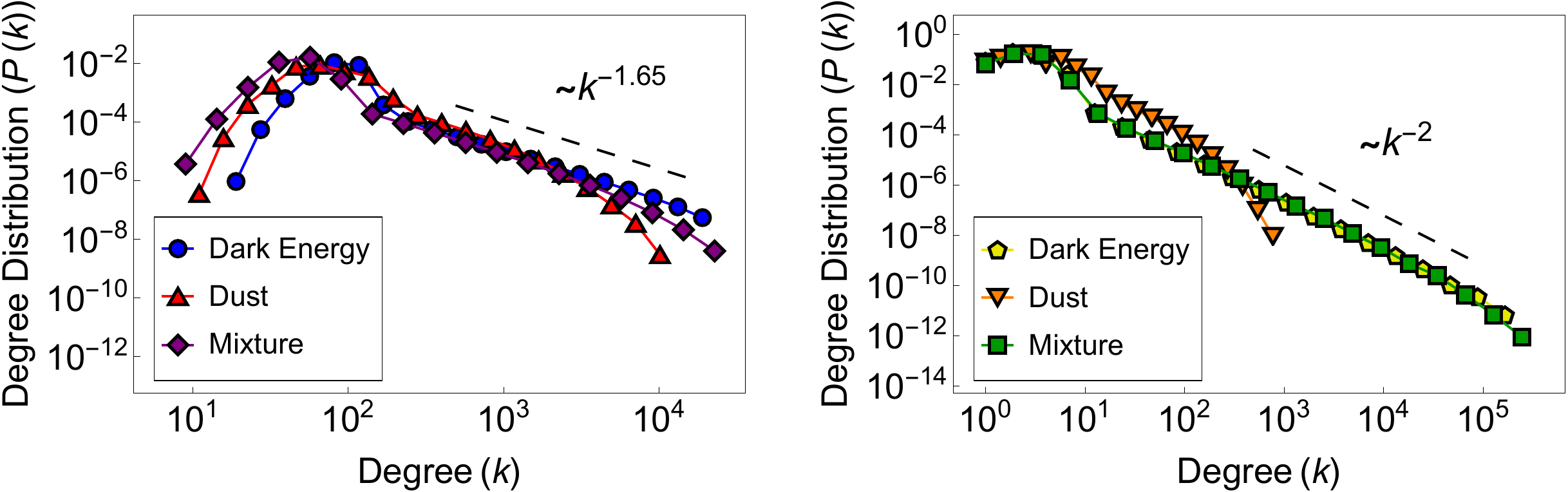}
\centering
\caption{\textbf{Degree distribution in Lorentzian RGGs.} Panels (a) and (b) show the degree distribution in the random geometric graphs in the three considered manifolds in the constant-$q$ and constant-$N,\bar{k}$ experiments, respectively, at the largest considered cut-off times $\tau_0$. Specifically, in panel (a) $q=60$, $\bar{k}=130$, $N=2518528$, $\tau_0=2.11$, and $\rho_0=6$, while in panel (b) $q=0.564$, $\bar{k}=10$, $N=2^{20}$, $\tau_0=4.64$, and $\rho_0=1.68$.}
\label{fig9-6}
\end{figure}

The higher the navigability of random hyperbolic graphs and real networks, the lower the power-law degree distribution exponent $\gamma$, and the stronger the clustering~\cite{boguna2009navigability,krioukov2010hyperbolic}. Clustering in Lorentzian random geometric graphs considered here is not so strong (Figure~\ref{fig9-5}) primarily because of their higher dimensionality~\cite{dall2002random,dhara2016solvable} (3+1 versus 1+1) and small cut-off times, but the tails of the degree distributions (Figure~\ref{fig9-6}) of the graphs in the manifolds with dark energy follow power laws in full agreement with the earlier results~\cite{krioukov2012network}, hence showing random geometric graphs in asymptotically de Sitter spacetimes have double power-law degree distributions with $\gamma=3/4$ at low degrees $k<q$ and $\gamma\to2$ at high degrees $k>q$. We note, however, that those results were derived only for the two limits $\tau_0 \ll 1$ and $\tau_0 \gg 1$.\par

\begin{figure}[!t]
\includegraphics[width=\textwidth]{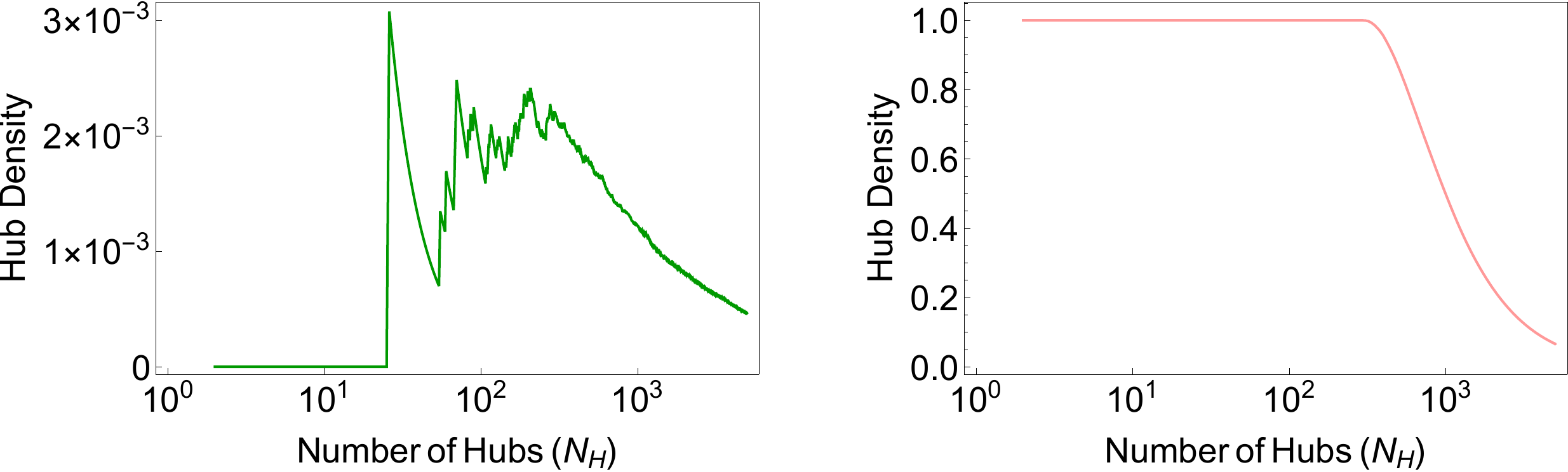}
\centering
\caption{\textbf{Hub density in Lorentzian and hyperbolic random graphs.} The hub density is defined as the number of links among the $N_H$ elements with largest degrees, divided by the maximum possible number $N_H\choose2$ of such links. Panels~(a,b) compare the hub density in two random graphs of the same size $N=2^{20}$ and average degree $\bar{k}=10$. Panel~(a) shows the data for the mixed-content (M) Lorentzian manifold graph with $\rho_0=1.68$ and $\tau_0=4.64$, while panel~(b) shows the same data for the hyperbolic graph generated using \url{http://named-data.github.io/Hyperbolic-Graph-Generator/} with parameters $N=2^{20}$, $\bar{k}=10$, $\gamma=2$, and $T=0$ (the resulting radial cutoff is $\rho_0=32.36$). There are exactly zero links between 25 largest-degree elements in the Lorentzian graph, while the subgraph induced by the first 103 highest-degree elements in the hyperbolic graph is the complete graph.}
\label{fig9-7}
\end{figure}
More interestingly, as evident from Figure~\ref{fig9-0}, hubs, i.e., the highest-degree elements, in random geometric graphs in Lorentzian manifolds are \emph{not} densely interconnected (Figure~\ref{fig9-7}) compared to random hyperbolic graphs and real networks which exhibit strong rich club effects~\cite{zhou2004rich,boccaletti2006complex}. This hub disconnectedness is a characteristic feature of any Lorentzian random geometric graphs, because elements with similar degree have similar time coordinates, and thus tend to be not connected, since they do not lie within each other's light cones with high probability. This observation may be puzzling, as it brings up the question of how Lorentzian graphs can be navigable at all, since one might intuitively think that geometric routing paths must go through the network core~\cite{boguna2009navigability}, and if the hubs in this core are not all densely interconnected, then routing should fail with high probability. \par

\begin{figure}[!t]
\includegraphics[width=0.5\textwidth]{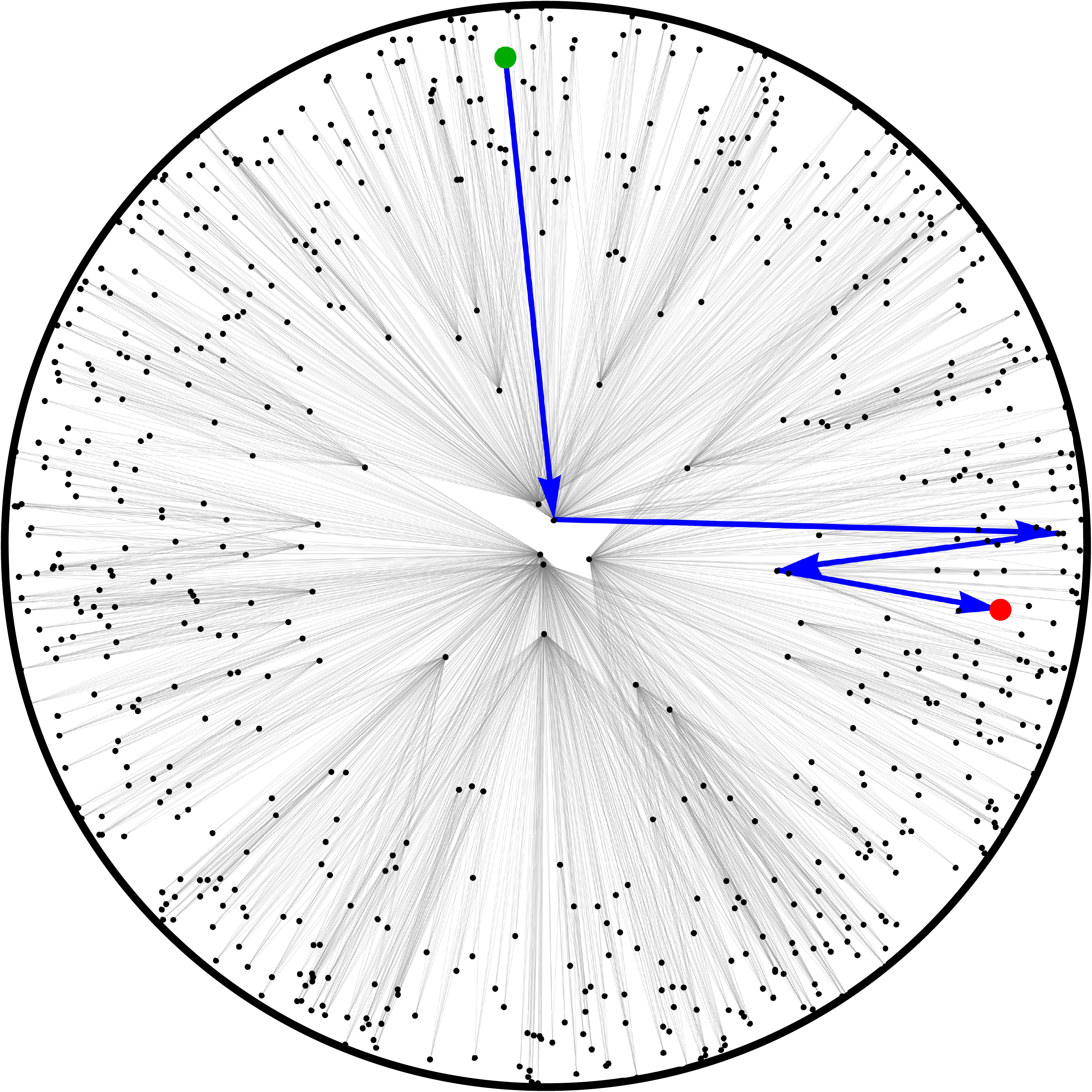}
\centering
\caption{\textbf{A typical navigation path in a Lorentzian RGG.} The figure shows the greedy geometric routing navigation path from the spacelike-separated green source and red destination in the same graph as in Figure~\ref{fig9-1}. The greedy path, which is also the shortest (stretch-$1$) path in the graph, alternates between hubs and peripheral elements. Any timelike-separated pairs of elements are directly linked, resulting in trivial one-hop stretch-$1$ paths.}
\label{fig9-8}
\end{figure}
This intuition turns out to be wrong, and the resolution of this puzzle lies in that the structure of geometric routing paths in Lorentzian graphs is completely different from that in Riemannian graphs~\cite{boguna2009navigability,krioukov2010hyperbolic}. Specifically, the Lorentzian path structure exhibits a peculiar periphery-core zigzagging pattern, illustrated in Figure~\ref{fig9-8}. This pattern, in which subsequent hops tend to lie close to light cone boundaries, is caused by the completely different nature of Lorentzian geometry and the structure of geodesics in it, versus the Riemannian case, making the graphs navigable even though their cores are sparse. \par

As a final remark, this navigation pattern also shows that the navigability of \emph{directed} causal sets based on random geometric graphs in Lorentzian manifolds is not so interesting. If links are directed in the past$\to$future time direction, then geometric routing respecting link direction and starting from a given source element succeeds only for destination elements lying in the future light cone of the source. All such destinations are directly connected to the source. Navigation fails for any other source-destination pairs, including all spacelike-separated pairs of elements, because paths between them necessarily involve hops in the future $\to$ past direction.

\section{Methodology}
\label{sec:methodology}
\subsection{Parameter Range Selection}
The three parameters of the studied graph ensembles are the rescaled sprinkling density $q=\nu\lambda^4$, (rescaled) spatial cutoff $\rho_0=(\alpha/\lambda)r_0$ ($\rho_0=r_0$ in de Sitter spacetime), and rescaled cutoff time $\tau_0=t_0/\lambda$, which taken together determine the graph size $N$ and average degree $\bar{k}$ via (\ref{eq:nodes},\ref{eq:degrees}). In simulations, especially in navigability experiments, we have the following constraints: 1)~the graphs cannot be too large so that they fit into memory, $N\lesssim2^{21}$; 2)~the average degree cannot be too low so that the graphs are above the percolation threshold, $\bar{k}\gtrsim5$; 3)~the spatial cutoff must be sufficiently larger than the temporal cutoff, so that the spatial boundary effects are negligible and we can rely on~\eqref{eq:degrees}; 4)~we want to explore the most interesting region of $\tau_0\sim1$, corresponding to the rescaled dark energy density $\Omega_\Lambda$ changing over essentially an entire range of its values between $0$ and $1$. \par

In experiments with constant $q=60$, Figures~\ref{fig9-1} and~\ref{fig9-4}(a,b), we select constant $\rho_0=6$ such that the average degree observed in simulations is within the error bound of $5\%$ from~\eqref{eq:degrees} for the largest considered value of $\tau_0>1$. This largest value of $\tau_0$ and the value of $q=60$ are determined in turn by the rest of the constraints above---decreasing $q$ would decrease the graph sizes, but would also decrease the average degree. The largest considered value of $\tau_0$ correspond to the largest graph sizes that fit into the memory, while the lowest value of $\tau_0$ is determined by the average degree value just above the percolation threshold. \par

\begin{figure*}[!t]
\includegraphics[width=\textwidth]{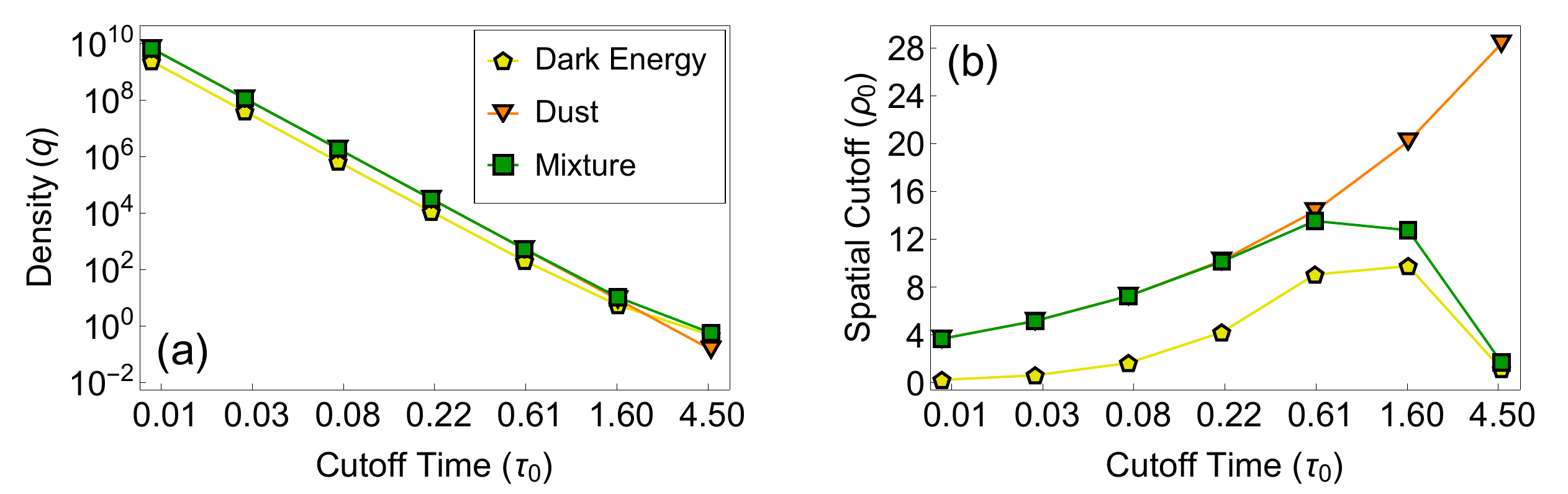}
\centering
\caption{\textbf{Rescaled sprinkling density and spatial cutoff as functions of the temporal cutoff in Fig.~\ref{fig9-4}(c,d).}}
\label{fig9-9}
\end{figure*}

In experiments with constant $\bar{k}=10$ and $N=2^{20}$, Figure~\ref{fig9-4}(c,d), $q$ and $\rho_0$ as functions of $\tau_0$ are varied as solutions of the systems of equations~(\ref{eq:nodes},\ref{eq:degrees}), Figure~\ref{fig9-9}. For all the considered values of the temporal cutoff $\tau_0$, the spatial cutoff $\rho_0$ is sufficiently larger than $\tau_0$, so that the average degree is within the $5\%$ error bound from its theoretical fixed value $\bar{k}=10$, except for the largest value of $\tau_0=4.64$, where the average degrees in the de Sitter and mixed manifold cases are $8.13$ and $8.48$, respectively. \par

The non-monotonic dependency of the success ratio $p_s$ on the cutoff time $\tau_0$ in the dark energy manifold in Figure~\ref{fig9-4}(c) is likely due to an interplay between increasing $\tau_0$, tending to increase $p_s$, and decreasing $q$, Figure~\ref{fig9-9}(a), tending to decrease $p_s$, in the absence of a spacetime singularity at $\tau_0$. The exact reason why this interplay is not important in the other two spacetimes that have this singularity is unclear. The non-monotonic behavior of stretch in Figure~\ref{fig9-4}(b,d) is not surprising, since stretch is computed for successful paths only, whose percentages vary as shown in Figure~\ref{fig9-4}(a,c). In particular, we have verified that the stretch increase in spacetimes with dark energy for the largest value of $\tau_0$ in Figure~\ref{fig9-4}(d) is not due a below-the-borderline value of $\rho_0$: we have densely sampled the region of $\tau_0\in[1.6,4.5]$ (not shown), and found that the intermediate stretch values for these two manifolds lie on smooth curves connecting the two shown data points, while for most of these intermediate values of $\tau_0$, the value of $\rho_0$ is above the $5\%$ $\bar{k}$-accuracy borderline discussed above. \par

\subsection{Greedy Routing Algorithm}
The greedy routing algorithm used in simulations is a parallel graph guided-exploration process. Since this process is non-local, due to the existence of transitive relations in a DAG, and unbalanced, since path lengths and element degrees are variable, this is a very challenging algorithm to optimize. As a result, we consider load balancing techniques as we did in Chapter~\ref{chap:causets}. \par

The greedy routing algorithm used in the abovementioned simulations is shown in Algorithm~\ref{alg:routing}. There are several places where this algorithm may be optimized. First, Operation~\ref{op:trav_for} can be implemented using the \texttt{bsf} operation described in Algorithm~\ref{alg:bsf}. Though that procedure was originally defined for iterating over elements in an Alexandroff set, it works just as well for any set. Here we use row $m$ of the adjacency matrix. \par

\begin{algorithm}[!t]
\caption{Greedy Routing in Lorentzian Spaces}
\label{alg:routing}
\begin{algorithmic}[1]
\Input
\Statex $\mathbf{A}$ \Comment Adjacency matrix
\Statex $\mathbf{x}$ \Comment Element coordinates
\Statex $N$ \Comment Number of graph elements
\Statex $N_p$ \Comment Number of pairs to traverse

\Procedure{greedy\_routing}{$\mathbf{A},\mathbf{x},N,N_p$}
\State $p\gets N(N-1)/2$
\State $z,N_z\gets0$
\For {$k=0;\,k<N_p;\,k\plusplus$} \label{op:gr_for}
\State $m\gets \mathfrak{u}p$ \Comment $\mathfrak{u}\in[0,1)$ is a uniform random variable
\State $(i,j)\gets$ \textsc{map\_index($m$)} \Comment Index mapping returns source/destination pair
\State $\mathbf{u}\gets\{0,\ldots,0\}$
\State $\sigma\gets$ \textsc{traverse($i,j,\mathbf{u}$)}
\If {$\sigma>-1$}
\State $N_z\plusplus$
\EndIf
\If {$\sigma>0$}
\State $z\plusplus$
\EndIf
\EndFor
\State $p_s\gets z/N_z$
\EndProcedure

\Procedure{traverse}{$i,j,\mathbf{u}$}
\If {$d(i,j)=\infty$} \Comment $d(i,j)$ is the geodesic distance between elements $i$ and $j$
\Return $-1$
\EndIf
\State $k\gets i$
\While {$k\neq j$}
\State $\mathbf{u}[k]\gets1$
\State $M\gets\infty\,,m^*\gets -1$
\For {$m\in\mathcal{J}(k)$} \label{op:trav_for}
\If {$m\prec j$}
\Return $1$
\EndIf

\If {$d(m,j)<M$}
\State $M\gets d(m,j)\,,m^*\gets m$
\EndIf
\EndFor

\If {$M<\infty$ \textbf{and} $\mathbf{u}[m^*]= 0$}
\State $k\gets m^*$
\Else
\Return $0$
\EndIf
\EndWhile
\EndProcedure

\Output
\Statex $p_s$ \Comment Success ratio

\end{algorithmic}
\end{algorithm}

This algorithm is not suited for vectorization, but it can be parallelized with some care. The most obvious place to start is the parallelization of Operation~\ref{op:gr_for}, i.e., each thread attempts to route an information packet across its own source-destination pair. This is an unbalanced operation, meaning it will take longer on some threads than others, so we use a \textit{dynamic} OpenMP scheduling protocol. This indicates to the scheduler that threads should receive new work as soon as they are finished, rather than dividing the work evenly before execution as dictated by the default \textit{static} protocol. We can be sure the dynamic protocol is appropriate, since at each iteration we calculate many geodesic distances, each of which requires a substantial number of mathematical operations (Chapter~\ref{chap:geodesics}), meaning the extra overhead for dynamic scheduling is greatly overshadowed by the work done by each thread. \par

Another possible optimization strategy is to parallelize the inner loop (Operation~\ref{op:trav_for}). This prevents the use of the \texttt{bsf} instruction, making it ideal only when $|\mathcal{J}(k)|\gg 1$. Parallelizing both loops indicates we must enable \textit{nested parallelism} with a call to the OpenMP library. When done properly, one in four of the total $T$ threads work with a source-destination pair, and then each of those $T/4$ threads launch four threads to calculate in parallel geodesic distances $d(m,j)$ between neighbors $m$ and destination $j$. The counter variables $z$ and $N_z$ used to calculate the success ratio are then modified using a \texttt{reduction} clause to avoid write conflicts, see Algorithm~\ref{alg:wconf}. This type of optimization often fails to increase performance due to the great increase in overhead associated with forking and joining nested threads, but due to the extreme load imbalance in this problem, it works out so long as the average degree is large, $\bar{k}\gg T$.

\subsection{Statistics and Simulations}
All the data shown in Figures~\ref{fig9-1} and~\ref{fig9-4} is averaged over ten random graphs if $N<2^{20}$, over five graphs if $N=2^{20}$, or over three graphs if $N>2^{20}$. All the error bars in these figures are smaller than the symbol sizes. To generate graphs efficiently, we use OpenMP to generate element coordinates in parallel (Section~\ref{sec:coord_gen}). Nodes are then linked using an NVIDIA K20m GPU via the CUDA library, since this step is the slowest when $N$ is large (Section~\ref{sec:gpu_linking}). While the linking algorithm is still $O(N^2)$, GPU parallelization offers a speedup of several orders of magnitude (Figure~\ref{fig:link_action_times}(right)). The full details of efficient graph construction are described in Section~\ref{sec:construction}.
%\afterpage{\blankpage}

\twolinechapter
\chapter[Vacuum Selection in String Theory and Cosmology]{\texorpdfstring{Vacuum Selection in\\[-0.8cm] String Theory and Cosmology}{Vacuum Selection in String Theory and Cosmology}}
\chaptermark{Vacuum Selection}
\mainchapter
%\chapter{Vacuum Selection in Cosmology}
\label{chap:multiverse}
\thispagestyle{empty}

\section{Introduction}
String theory is an ultraviolet complete theory of quantum gravity
that is a strong candidate for a unified theory of particle physics
and cosmology.
However, string theory requires the existence of extra dimensions. Their
geometric structure and discrete objects such as fluxes
give rise to
a vast landscape of metastable four-dimensional vacua. Originally estimated
lower bounds of $10^{500}$ possible flux vacua on fixed geometries~\cite{Douglas:2003um, Ashok:2003gk} have grown to $10^{272,000}$~\cite{Taylor:2015xtz}. Furthermore, the number of geometries themselves has
grown significantly; there is a now an exact lower bound of $4/3\times 2.96 \times 10^{755}$ on the number of geometries~\cite{Halverson:2017ffz}, which grows to $10^{3000}$ using estimates in~\cite{Taylor:2017yqr}.
The magnitude of these numbers, together with associated computational complexity~\cite{Denef:2006ad, Cvetic:2010ky,Denef:2017cxt}, makes it difficult to study the string landscape, though machine learning or other data science techniques may lead to breakthroughs~\cite{Abel:2014xta,He:2017aed,Ruehle:2017mzq,Carifio:2017bov}.

It is in this vast landscape that the physics of our Standard Model vacuum is expected to be found; therefore understanding the
landscape is of central importance for applications of string theory
in both particle physics and cosmology. 
If the details of our vacuum are not entirely determined by the
anthropic principle~\cite{Susskind:2003kw,Schellekens:2013bpa}, then a cosmological mechanism must select vacua similar to ours.
One possibility is that cosmology selects vacua from a relatively
flat distribution, but a final understanding of string 
theory will show that vacua similar to ours are typical~\cite{Grassi:2014zxa}.
Another possibility is that cosmological dynamics
prefers certain vacua over others, which is necessary if vacua similar to ours are strongly atypical in
the landscape~\cite{Kofman:2004yc,Bousso:2007er}.
A model of such vacuum selection is our main result.

More broadly, we introduce network science as a new tool for studying
the string theory landscape.  We represent coarse structures in it as a graph or network---a collection of nodes and edges. In one natural network, we let nodes be metastable vacua, with edges between all nodes weighted by tunneling rates. Since calculating all such tunneling rates is 
computationally infeasible at the current time, we instead study two networks that are concrete coarse-grained approximations to the full weighted network. 
In both, nodes are associated with smooth six-manifolds that are string geometries, and an edge exists between two nodes when they are related by a specific topological transition known as a blowup, in which the number of scalar fields in the low-energy 4D theory, known as K\"ahler moduli,  changes by one.  These networks are global topological structures that exist in the landscape independent of any physical interpretation.

After defining and constructing these networks,
we study vacuum selection in Coleman and de Luccia's cosmological model of bubble nucleation~\cite{PhysRevD.21.3305} in the context of
eternal inflation~\cite{Vilenkin:1983xq,1402-4896-1987-T15-024,Linde:1993xx, Linde:1993nz}.
In this cosmology, nucleation events occur successively in local patches,
yielding a multiverse with many different bubbles occupying numerous vacua. The distribution of occupation numbers provides a notion of vacuum selection determined by the transition rates between vacua, as well as model-dependent features, such as bubble collisions and collapses.

It remains an open question as to whether bubble nucleation rates derived from string theory will lead to a trivial or non-trivial distribution of vacua. We provide strong evidence that the distribution is highly non-trivial, i.e., some vacua are selected over others. In our context, the dependence of bubble cosmology on the transition physics follows from the structure of the network. Specifically, we apply a standard model of bubble nucleation to both of our networks of geometries and demonstrate that a network structure naturally provides a mechanism for vacuum (or, in this case, geometry) selection. The mechanism is most effective when transitions with small topology changes dominate over transitions with large topology changes. This model provides a concrete dynamical mechanism for vacuum selection, and it is an exciting prospect for a future understanding of how and why our vacuum might be selected in string theory.

\section{A Cosmological Model of Bubble Nucleation}
Cosmological bubble nucleation is well-studied. We consider a canonical model of bubble cosmology introduced in~\cite{1475-7516-2006-01-017}.
Consider the fraction of comoving volume $f_j$ occupied by a particular vacuum $j$ as a function of time, given the vacuum transition probabilities from vacuum $j$ to vacuum $i$, denoted $\Gamma_{ij}$. The dynamics of $f_j$ can be written as
\begin{equation}\label{eq:vdyn}
\frac{d\mathbf{f}}{dt} = \mathbf{M} \mathbf{f}\,,
\end{equation}
where $M_{ij} = \kappa_{ij} - \delta_{ij} \sum_r \kappa_{ri}$, with $\kappa_{ij} = \frac{4\pi}{3}\Gamma_{ij}H_j^{-4}$, and $H_j$ is the Hubble constant of vacuum $j$. The asymptotic solution to Eq.~\ref{eq:vdyn} takes the form
\begin{equation}
\mathbf{f}(t) = \mathbf{f^{(0)}} + \mathbf{s} e^{-q t} + \dots\, ,
\end{equation} 
where $-q$ is the (negative) spectral gap of $\mathbf{M}$, or the smallest-magnitude non-zero eigenvalue of $\mathbf{M}$, and $\mathbf{s}$ is the corresponding eigenvector, which we denote the \textit{dominant eigenvector}. By relating the volume fractions to the number of bubbles $N_j$ in vacuum $j$ as $t\rightarrow \infty$, one finds
\begin{equation}
N_j = \frac{3}{4\pi} \frac{1}{3-q} \epsilon^{-(3-q)}\sum\limits_\alpha H_\alpha^q \kappa_{j\alpha}s_\alpha\, ,
\end{equation}
where $\epsilon$ is a cutoff that bounds the minimum bubble size, and the index $\alpha$ hereafter ranges over non-terminal vacua, that is, vacua which can nucleate additional bubbles. As $\epsilon \rightarrow 0$, the number of bubbles $N_j$ in vacuum $j$ goes to infinity, so the authors of~\cite{1475-7516-2006-01-017} normalize the vector $N_j$ by dividing by the total number of vacua, in order to define a probability $p_j$. We therefore have
\begin{equation}\label{eq:sumprob}
p_j \propto \sum\limits_\alpha H_\alpha^q \kappa_{j\alpha}s_\alpha \, .
\end{equation}
To compute the probability distribution $p_j$, we need to compute the spectral gap of $\mathbf{M}$ and corresponding eigenvector $s_\alpha$, and subsequently compute the sum in Eq.~\ref{eq:sumprob}.  It is important to note that the consistency of this model requires that the set of non-terminal vacua cannot be split into disconnected groups, and that there exists at least one terminal vacuum with a non-zero transition amplitude to it. With this in mind, the matrix $\mathbf{M}$ can be written as
\begin{equation}
\label{eq:M_matrix}
\mathbf{M} =  \left(
\begin{array}{cc}
\mathbf{R} &  0 \\ 
\mathbf{S} & 0 \\
\end{array} \right)\, ,
\end{equation}
where $\mathbf{R}$ is the (non-terminal)-(non-terminal) block, and $\mathbf{S}$ is the (non-terminal)-(terminal) block. In this case $-q$ is the spectral gap of $\mathbf{R}$, and $s_\alpha$ the corresponding eigenvector. Hence, an analysis of $\mathbf{R}$ is sufficient to determine the probability distribution $\mathbf{p}$.

\section{Networks of String Geometries}
In this section, we study two networks of string geometries: one in the setting of F-theory,
and the other in weakly coupled type IIb compactifications.
In both, a node is a smooth six-manifold that provides the extra spatial dimensions in a four-dimensional compactification. Edges represent simple
topological transitions between geometries,  such as blowups. The set of edges is represented by the adjacency matrix $\mathbf{A}$ of the network, which has entry 1 if two geometries are directly connected by a topological transition and 0 otherwise. 

Though the exact size of the landscape is unknown, these networks are in a context larger than previously studied. Both are large ensembles of topologically connected geometries, and each geometry may support many flux vacua. Critically, F-theory also includes non-trivial string coupling corrections and gives rise to additional effects that may be more representative of the landscape as a whole than weakly coupled compactifications.

\subsection{The Tree Network}
The first network we construct has nodes that are $\frac{4}{3}\times 2.96\times 10^{755}$ bases for elliptically fibered Calabi-Yau fourfolds considered in~\cite{Halverson:2017ffz}, the vast majority of which contain strong coupling regions~\cite{Halverson:2017vde,Halverson:2016vwx}. Each geometry is generated by a series of topological transitions known as blowups from a six-manifold that is a weak Fano toric variety. A sequence of blowups in a local patch is represented diagrammatically as tree-like structure over a polytope, and we therefore refer to such a sequence of blowups as a ``tree''. We emphasize that this is descriptive, and does not 
mean tree in the sense of graph theory.
The key fact that makes studying a network with $\frac{4}{3}\times 2.96\times 10^{755}$ nodes possible is that the full network is a Cartesian product of smaller, more tractable networks. A Cartesian product $G\Box H$ of
two graphs $G$ and $H$ is a graph such that the vertices of $G \Box H$
are the Cartesian product of the vertices of $G$ and $H$, and any
two vertices $(u,u'), (v,v') \in G\Box H$ are adjacent if and only 
if $u=v$ and $u'$ is adjacent to $v'$ in $H$, or the converse. The ensemble is overwhelmingly composed of trees built over two reflexive polytopes, $\Delta^\circ_1$ and $\Delta^\circ_2$, each of which has 108 edges and 72 faces when triangulated. The network $G_T$ of tree geometries can then be written as $G_T = G_E^{\Box 108} \Box G_F^{\Box 72}$, where $G_E$ is the network of edge trees built over a single edge, which has $82$ nodes and $1386$ edges, and $G_F$ is the network of face trees built over a single face, which has $41,873,645$ nodes and $100,136,062$ edges.

\subsection{The Hypersurface Network}
In a similar vein, one can consider compactifications on
six-manifolds that are Calabi-Yau threefolds ($CY_3$s). We consider
 $CY_3$ hypersurfaces that are associated with a triangulation of a 4D reflexive polytope $\Delta^\circ$ as in~\cite{1993alg.geom.10003B}.
  Here we consider topological transitions from one $CY_3$ $X_a$ to another $X_b$ that can be encoded in the corresponding polytopes $\Delta^\circ_a$ and $\Delta^\circ_b$ in a simple manner: the nodes corresponding to $X_a$ and $X_b$ are connected by an edge in the hypersurface network if and only if $\Delta_a^\circ$ and $\Delta_b^\circ$ are related by the deletion of one or more vertices, without passing through an intermediate $\Delta_c^\circ$, followed by a $GL(4,\bZ)$ rotation. These correspond to blowups in the $CY_3$.
There are 473,800,776 reflexive polyhedra in 4 dimensions~\cite{Kreuzer:2000xy, 2000math......1106K}, and constructing the full network is currently out of reach. We therefore limit ourselves to the 11,626,070 polytopes with $\leq10$ vertices. This network has 43,545,632 edges.
As there are many ways to move from a Calabi-Yau threefold to an $\mathcal{N}=1$ string compactification, including the heterotic and type II string theories, our results are applicable in many settings.

\subsection{Construction Algorithms}
Constructing these graphs is an involved process, first requiring an efficient representation of triangulated polytopes, or configurations, and then an efficient method to determine whether two configurations are related. The naive method of storing an adjacency matrix is infeasible due to the problem size: the tree network's adjacency matrix $\mathbf{A}$ would require nearly $210$ TB RAM, and any subsequent analysis would require even more. Therefore, we employ a sparse solution, which ultimately uses only $1.5$ GB. \par

In the tree network, the maximum height label is just six, which restricts the number of possible cones, i.e., configuration parameters, to $N_C=349$ after all possible face and edge blowups. Since two configurations are related if their cone sets differ slightly, it makes sense to represent each configuration by a $349$-bit \texttt{FastBitset} object, where each bit indicates the presence or absence of a particular cone. Using this representation, it is then possible to reduce the relational operator to one using only the set operations described in Chapter~\ref{chap:sets}. \par
\begin{algorithm}[!t]
\caption{Tree Network Construction}
\label{alg:tree_construction}
\begin{algorithmic}[1]
\Input
\Statex $X_i$ \Comment First configuration
\Statex $X_j$ \Comment Second configuration
\Statex $\mathcal{Q}$ \Comment Set of cones $Q$

\Procedure{face\_blowup}{$X_i,X_j,\mathcal{Q}$}
\State $X\gets X_i\veebar X_j$ \Comment All common cones removed
\If {\textsc{count\_bits}($X,349$) $\neq4$}
\Return {$(-1,-1)$}
\EndIf

\State $x_0\gets$ \texttt{bsf($X\cap X_i$)} \Comment Original cone which will blow up

\For {$k=1;\,k\leq3;\,k\plusplus$}
\State $x_k\gets$ \texttt{bsf($X\cap X_j$)} \Comment Three new cones
\State $X_j[x_k]\gets 0$
\EndFor

\State $Y\gets\varnothing,Q_0\gets\mathcal{Q}[x_0]$ \Comment Expected new vertex from blowup
\While {$Q_0\neq \varnothing$} \Comment Extract old vertices
\State \texttt{$y^*\gets$ bsf($Q_0$)}
\State $Q_0[y^*]\gets0$
\State $Y\,\cup\!= y^*$ \Comment New vertex is sum of old ones
\EndWhile

\State $Q_k\gets\mathcal{Q}[x_k]$ \textbf{for} $k\in\{1,2,3\}$
\If {\textsc{count\_bits($Q_k\setminus Y$)} $\booleq 2$ \textbf{for} $k\in\{1,2,3\}$}
\Return $(i,j)$
\EndIf

\EndProcedure

\Procedure{edge\_blowup}{$X_1,X_2,Q$}
\State $X\gets X_1\veebar X_2$ \Comment All common cones removed
\If {\textsc{count\_bits}($X,349$) $\neq6$}
\Return {$(-1,-1)$}
\EndIf

\For {$k=0;\,k\leq 1;\,k\plusplus$}
\State $x_k\gets$ \texttt{bsf($X\cap X_i$)} \Comment Original two cones
\State $X_i[x_k]\gets 0$
\EndFor

\For {$k=2;\,k\leq 5;\,k\plusplus$}
\State $x_k\gets$ \texttt{bsf($X\cap X_j$)} \Comment Four new cones
\State $X[x_k]\gets 0$
\EndFor

\State $Y\gets\varnothing,Q_0\gets\mathcal{Q}[x_0],Q_1\gets\mathcal{Q}[x_1]$ \Comment Expected new vertices from blowup
\While {$Q_0\cap Q_1\neq \varnothing$} \Comment Extract old vertices
\State \texttt{$y^*\gets$ bsf($Q_0\cap Q_1$)}
\State $Q_0[y^*]\gets0$
\State $Y\,\cup\!= y^*$ \Comment New vertex is sum of old ones
\EndWhile

\State $Q_k\gets\mathcal{Q}[x_k]$ \textbf{for} $k\in\{2,3,4,5\}$
\If {\textsc{count\_bits($Q_k\setminus Y$)} $\booleq 2$ \textbf{for} $k\in\{2,3,4,5\}$}
\Return $(i,j)$
\EndIf

\EndProcedure

\Output 
\Statex $(e_i,e_j)$ \Comment Edge list entry, if not $(-1,-1)$

\end{algorithmic}
\end{algorithm}
The procedure which determines whether two configurations are related in the tree network is described in Algorithm~\ref{alg:tree_construction}. For each pair of configurations $X_i,X_j$, one checks for face and edge blowups using similar methods. The first step is always to construct the disjoint union of the configurations, $X=X_i\veebar X_j$, which holds only the cones not in common. It is then easy to recognize a face blowup replaces one cone with three while and edge blowup replaces two cones with four, so that the number of non-zero entries in $X$ should be four or six, respectively. In the case of a face blowup, one can extract the old cone $Q_0$, calculate the expected new vertex $Y$ by summing the entries of the three vertices (non-zero entries) of $Q_0$, and then study whether this new vertex belongs to all three new cones $Q_1$, $Q_2$, and $Q_3$. For an edge blowup, the common vertices of the two old cones $Q_0$ and $Q_1$ sum to form the new vertex $Y$, which should then belong to all four new cones $Q_2$, $Q_3$, $Q_4$, and $Q_5$. If $X_i$ and $X_j$ satisfy either of these conditions, we add the entry $(i,j)$ to the edge list. Since these operations are already vectorized (Sections~\ref{sec:opt_bitset_alg},~\ref{sec:opt_poset_alg}), the obvious optimization is parallelization via OpenMP. Since the workload is not balanced across threads, due to the early \textsc{return} statements, one should use a \texttt{dynamic} scheduling scheme.

\section{Cosmological Selection of Geometries}
We now consider a simple model of cosmology on each of our networks. As stated above, the model of~\cite{1475-7516-2006-01-017} requires the presence of both terminal and non-terminal vacua. In general, it is expected that each geometry supports a large number of vacua; we consider a simplified model in which each geometry supports two vacua: one terminal and one non-terminal. In addition, in order to isolate the effect of the graph structure on the cosmological dynamics we set $H_\alpha = 1$ for all $\alpha$. 

We now argue that topologically connected vacua are more likely to transition to one another than to vacua realized in geometries separated by multiple topological transitions. A complete argument requires a generalization of Coleman-de Luccia result beyond a single effective field theory. However, it is quite natural to assume that such a generalization still depends on a  generalized notion of distance in field space. Recall that these Calabi-Yau geometries lie in a connected supersymmetric moduli space. If the leading-order instantons between the vacua interpolate along this moduli space, as opposed to over hills with non-zero energy cost, then the graph structure indeed naturally characterizes field space distance. While the graph information is currently too coarse-grained to determine these distances numerically, it is clear that distance increases upon traversing the graph, i.e., a transition along many edges requires traversing a greater distance in field space than one along fewer edges.

Such a transition model could also be justified if the dominant transition mechanism is a (de Sitter) thermal fluctuation. For example, consider a stabilized string compactification with branes:
if the temperature of the branes is higher than the Kaluza-Klein scale associated
with the topological transition, then the branes could potentially fluctuate thermally to a configuration 
on a different geometry. In either case, the dominant transitions would be between adjacent nodes in the network.

 In our simple model of cosmology, therefore, there are two transition effects:  leading effects described by the matrix $\mathbf{\Gamma}^{l}$, and
subleading effects by the matrix $\mathbf{\Gamma}^{sl}$. The actual values for these matrices are determined by the microphysics of vacua, such as their cosmological constants, which at this point are incalculable in a large ensemble. Without further information, we consider an agnostic model, where transitions can happen in either direction along any edge of the graph, governed by some overall constant $\beta_1$ that determines the leading transition rates. At the level of pure geometry, the tunneling rates from the non-terminal to non-terminal vacua and from the non-terminal to the terminal vacua are the same, and hence for both we take 
\begin{equation}\label{eq:adjrate}
\mathbf{\Gamma}^{l}  = \beta_1\, \mathbf{A}\, ,
\end{equation}
where $\mathbf{A}$ is the adjacency matrix of the network. 
The subleading
transition rates likewise are determined by
currently incalculable quantities, so we use
\begin{equation}
\label{eq:gammasl}
\mathbf{\Gamma}^{sl}= \beta_2(\mathbf{J}-\mathbf{I})\,,
\end{equation}
where $\mathbf{J}$ and $\mathbf{I}$ are the all-one and identity matrices, respectively, and $\beta_2$ is a constant. 
Eq.~\eqref{eq:gammasl} simply indicates any geometry can tunnel to any other except itself. 

 We can understand the interplay between $\beta_1$ and $\beta_2$ by considering
 two limiting cases. Let $\beta_1 = 0, \beta_2 \neq 0$, so 
 the normalized late-time behavior is given by $\mathbf{p} =  \mathbf{1}/N$. This would give a delta-function-like spike in the distribution of $\mathbf{p}$, indicating no geometry selection, as one would expect from a universal tunneling rate; see the black lines in Figure~\ref{fig:facetree}. The effect of $\beta_2 \neq 0$ is to flatten the distribution
 of geometries, and would then indicate that the network of geometries is a complete graph, as every node is connected to every other node.
In the other limit with $\beta_2=0,\beta_1 \neq 0$, the late-time behavior of $\mathbf{p}$
is non-trivial, and is given by Eq.~\ref{eq:sumprob}. It is shown in~\cite{1475-7516-2006-01-017} that the entries of $\mathbf{p}$ are all positive.

For a general network, $\mathbf{p}$ is not expected to be uniform; therefore, $\beta_1 \neq 0$ provides a physical mechanism for vacuum selection. We assume $\beta_1 \gg  \beta_2$, so that nearby tunneling dominates over far-away tunneling effects.  In this case the matrix $\mathbf{R}$ in Eq.~\eqref{eq:M_matrix} takes the block form:
\begin{equation}
\mathbf{R} = -(\mathbf{L} + \mathbf{D}) \, ,
\end{equation}
where $\mathbf{L}$ is the graph Laplacian and $\mathbf{D}$ is the degree matrix of the graph, which contains node degrees along the diagonal and zeros elsewhere. We now turn to vacuum selection on our networks.

\subsection{The Tree Network}
\begin{figure}[!t]
\includegraphics[width=0.452\textwidth]{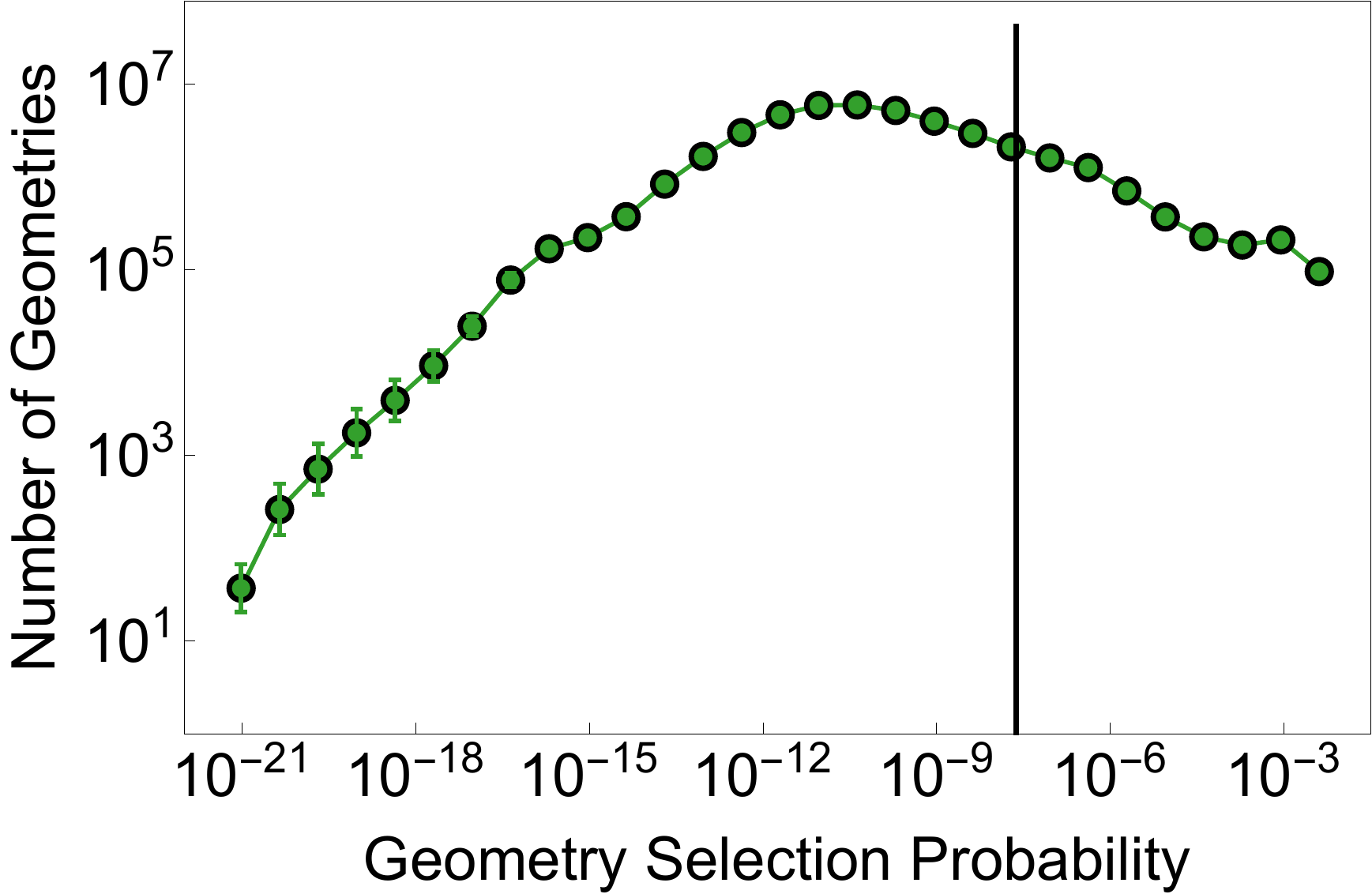}
\hspace{1.5cm}
\includegraphics[width=0.452\textwidth]{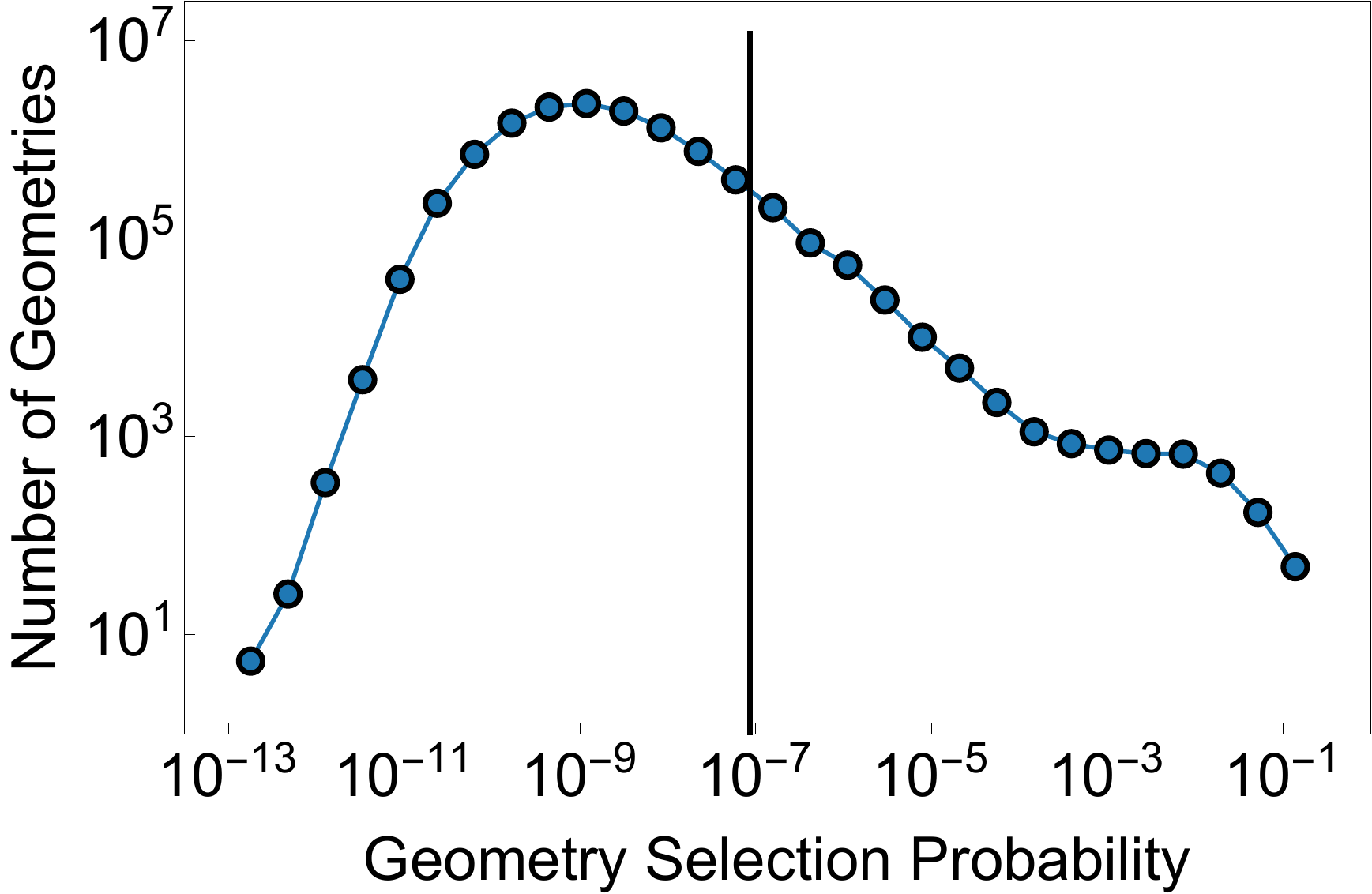}
\caption{\textbf{Geometry Selection in the Face Tree and Hypersurface Networks.} \emph{Left:}
The distribution of vacua $\mathbf{p}$ at $t\to\infty$ is shown for the face tree network $G_F$.
The largest entry is 0.007, while 98 percent of the entries are at least a factor of 1000 smaller, with selection strengths $\Xi=42.9, \Upsilon=18.4$, indicating strong vacuum selection. Note that instead of a single geometry being strongly selected, many geometries are preferred over the bulk.
\emph{Right:}
The same distribution is shown for the hypersurface network.
The largest entry is 0.17, while 99.9 percent of the entries are at least a factor of 1000 smaller, with selection strengths $\Xi=27.4, \Upsilon=18.6$, indicating a weaker, yet sharper selection. This distribution indicates that fewer geometries are selected over the bulk than in the face tree network. The vertical black line in each plot shows the trivial solution $\mathbf{p}=\mathbf{1}/N$, wherein each geometry is equally preferred, and no selection occurs. 
}
\label{fig:facetree}
\end{figure}
We first consider the network of toric trees $G_T$, which has the structure $G_T = G_E^{\Box 108} \Box G_F^{\Box 72}$. We analyze $G_T$ by analyzing $G_E$ and $G_F$ independently. Let us start with $G_F$. The probability distribution $\mathbf{p}$ is shown in Figure~\ref{fig:facetree} (left). The largest entry is 0.007, while 98 percent of entries are at least a factor of 1000 smaller, and the distribution is therefore highly skewed. The maximum selection strength in $G_F$ is $\Xi\equiv\ln(p_\mathrm{max}/p_\mathrm{min})\approx42.9$ while the typical selection strength is $\Upsilon\equiv\ln(p_\mathrm{max}/p^*)\approx18.4$, where $p^*$ is the selection probability for the typical (most probable) geometry. Recall that no vacuum selection corresponds to $\Xi =\Upsilon = 0$, and so these values of $\Xi$ and $\Upsilon$ indicate a highly nontrivial selection effect.

The structure of $G_E$ is simpler due to the smaller size of the network. The highest probability entry is 0.97, and the ratio of the largest probability to the smallest is $5 \times 10^4$. However, in the case of $G_E$ two geometries are preferred over the rest, by factors of $\sim 100$ and $\sim 20$, respectively.

Having analyzed $G_E$ and $G_F$ individually, we consider the Cartesian product $G_T$. The dominant eigenvector $\mathbf{s}_G$ of a Cartesian product $G = A \Box B$, with dominant eigenvectors $\mathbf{s}_A$ and  $\mathbf{s}_B$ is the tensor product 
$ \mathbf{s}_G = \mathbf{s}_A \otimes \mathbf{s}_B$. From this, it is simple to construct the probability distribution of the full $G_T$. We find the ratio of the largest to smallest probability is $\sim 10^{1555}$, i.e., $\Xi\sim 3580$. Note that this is a measure
of the maximal selection, not the typical selection $\Upsilon$, in the
network. It would be interesting to understand whether such
large selection effects are typical in the full landscape.

\subsection{The Hypersurface Network}
We next consider the network of $CY_3$ hypersurfaces. We ignore polytopes that are disconnected from the bulk; such a feature is due to the cutoff at 10 vertices and will disappear when more polytopes are included in the network. The probability distribution is shown in Figure~\ref{fig:facetree} (right). 
The largest eigenvector entry is 0.17. The maximum and typical selection strengths are respectively $\Xi=27.4$ and $\Upsilon=18.6$. It is interesting to compare the shapes of the two plots. The right tail of the distribution for the hypersurface network drops more rapidly than the right tail for $G_F$.  As in the tree network, there is a continuum of geometries with selection probability near the maximum, but 99.9 percent of the entries are at least a factor of 1000 smaller. We have thus demonstrated strong vacuum selection effects in both networks of geometries.

\section{Discussion} 
This work is the first step toward systematically under-
standing vacuum selection from cosmology on networks
of string vacua. Even in the absence of detailed knowledge of the microphysics governing bubble nucleation and quantum tunneling rates, it is possible to construct a semi-realistic model which permits interpolation between different cosmological paradigms. We found that if the network structure indicates preferred transitions, as opposed to universal quantum tunneling, then the vacuum probability distribution can be highly non-trivial, indicating a selection effect. This vacuum selection was explicitly realized on two separate networks of compactified geometries connected by topological transitions. 
In the future, it is of critical importance to add additional data, such as fluxes, to the networks to allow for the identification of gauge and cosmological sectors that contain the Standard Model and account for Cosmic Microwave Background data. This is plausible given the current knowledge of fluxes and branes, but is beyond current computational feasibility. In addition, allowing for a non-trivial distribution of the $H_\alpha$ would promote the graph of vacua to a weighted, directed graph, and the $\Gamma_{ij}$ would satisfy non-trivial relations as in \mbox{Eq.\ 2.5} of~\cite{Denef:2017cxt}.
However, it is natural to expect that non-trivial $H_\alpha$
should further aid in vacuum selection, i.e., it should not smooth
our $H_\alpha=1$ distributions into a flat distribution.

More broadly, the application of concepts and techniques commonly employed in network science promises to be fruitful in the study of the string theory landscape. Variations on the simple cosmological model presented herein can easily be incorporated by modifying the centrality measures used to study the network properties, by weighting the edges in appropriate ways, or by changing the governing equations to account for bubble collisions and decays. We anticipate such a network-centered approach will prove to be vital to making concrete, quantitative statements about vacuum selection in the string landscape.

\afterpage{\blankpage}

%\part{Useful Results and Conclusions}
\part{Conclusion}
\label{part:final}
\thispagestyle{empty}
\afterpage{\blankpage}
\chapter{Conclusion}
\label{chap:conclusion}
\thispagestyle{empty}

The high performance algorithms described in this dissertation have proven to be useful in a wide range of applications. Not only do they improve the performance of numerical experiments, but they also allow us to study areas of physics otherwise inaccessible. 

%\section{Summary}
We reviewed in Chapter~\ref{chap:parallel} how the CPU and GPU microarchitectures influence how we design algorithms which maximize instruction throughput, optimize memory access patterns, and distribute computations among multiple cores. After examining the physical components inside a CPU, we looked at several optimization techniques, including loop unrolling, branch elimination, and pipelined cache access (Algorithms~\ref{alg:unrolling}-\ref{alg:fetch}). We also considered how to distribute calculations across cores using OpenMP while avoiding read/write conflicts (Algorithms~\ref{alg:vecadd},~\ref{alg:wconf}), and then finally we considered how to use the Intel AVX library to vectorize certain mathematical operations (Algorithm~\ref{alg:avxadd}). \par

In Chapter~\ref{chap:sets}, we introduced and used the compact \texttt{FastBitset} data structure for sets. This binary representation allowed us to optimize the set operations~(\ref{eq:intersection}--\ref{eq:difference}), such as the (partial) set intersection (Algorithms~\ref{alg:intersection},~\ref{alg:partial_intersection}) and the bitcount (Algorithm~\ref{alg:popcnt}). We then combined these techniques using AVX to create a vectorized inner product (Algorithm~\ref{alg:vecprod}). In the final section of Chapter~\ref{chap:sets}, we considered partial order partitions into collections of chains and antichains (Algorithms~\ref{alg:chain},~\ref{alg:antichain}), while also introducing an efficient method to iterate through elements in an Alexandroff set (Algorithm~\ref{alg:bsf}). \par

Chapter~\ref{chap:graphs} extended these methods to random geometric graphs in Lorentzian spaces. These RGGs were generated by Poisson sprinkling elements into a compact spacetime region and then iterating over all possible pairwise relations (Algorithm~\ref{alg:nested_loop}) to construct the adjacency matrix. An extra speedup was found by instead using the GPU to construct a list of relations. The GPU algorithm written in CUDA used the GPU's shared memory (L1 cache) to efficiently read and write to the global GPU memory (Algorithms~\ref{alg:idxmap}-\ref{alg:wrrel}). These operations introduced speedup of a factor of $1000$ compared to the naive implementation, as demonstrated in Figure~\ref{fig:link_action_times} (left). \par

After introducing these data structures and algorithms for sets and graphs, we applied them in Part~\ref{part:qg} to causal set quantum gravity. In Chapter~\ref{chap:causets}, we examined the Benincasa-Dowker action, i.e., the discrete analogue to the Einstein-Hilbert action from general relativity. Since this quantity is a function of global graph structures (the inclusive-order-intervals), and its calculation is essential for research of causal set dynamics, we introduced a new highly efficient algorithm to find the inclusive-order-interval abundances (Algorithm~\ref{alg:cardinalities}) which is nearly $1000$ times faster than the naive calculation (Algorithm~\ref{alg:naive_cardinalities}). We also considered how to distribute this calculation across multiple computers using MPI (Sections~\ref{sec:mpi_static},~\ref{sec:mpi_balanced} and Figure~\ref{fig:mpi_flowchart}). The performance was studied in detail in Figures~\ref{fig:link_action_times} (right) and~\ref{fig:benchmarking}. \par

Chapter~\ref{chap:chull} then examined the causal set embedding problem, called the Hauptvermutung. Rather than solve the full embedding problem, we considered what information about extrinsic geometry we could extract using methods from computational geometry. In particular, since it is much easier to identify elements near spacelike boundaries compared to those near timelike ones, we introduced two new algorithms to measure timelike boundaries. The first detected elements believed to be ``close'' to a boundary (Algorithm~\ref{alg:candidates}), while the second took this set of candidates and constructed chains which covered and, therefore, measured the volume of that boundary (Algorithm~\ref{alg:timelike_measurement}). Three examples of usage were demonstrated in Section~\ref{sec:examples5}.

Chapter~\ref{chap:geodesics} in Part~\ref{part:geodesics} introduced new closed-form expressions for geodesics in Friedmann-Lema\^itre-Robertson-Walker manifolds. We found a general solution, Eqs.~(\ref{eq:geodesic4},~\ref{eq:distances_general_spatial},~\ref{eq:distances_general}), and also an efficient method to determine whether two points in a Lorentzian space can even be connected by a geodesic, Eqs.~(\ref{eq:conformal_time},~\ref{eq:max_time_constraint},~\ref{eq:geo_compl}). Useful numerical approximations for a spacetime with dark energy and dust matter, which is approximately the model for our physical universe, were discussed in Section~\ref{sec:num_approx}. \par

Part~\ref{part:apps} then studied some other applications of the methods developed in Part~\ref{part:foundations}. Using the solutions and numerical approximations from Part~\ref{part:geodesics}, we measured the navigability of random geometric graphs in Lorentzian spaces in Chapter~\ref{chap:navigation}. We constructed graphs in spacetimes containing dust matter, dark energy, and both, and then used a greedy routing algorithm (Algorithm~\ref{alg:routing}) to measure the success ratio and stretch for ensembles of graphs with constant sprinkling density or constant average degree (Figure~\ref{fig9-4}). Our results indicated that random geometric graphs in spacetimes with dark energy, i.e., those whose scale factors were asymptotically exponential, had a success ratio which tended toward $100\%$. \par

In Chapter~\ref{chap:multiverse}, we considered another application, this time to one of the branches of string theory called F-theory. We developed a graph model for bubble cosmology in the context of eternal inflation, ultimately to demonstrate vacuum selection in the string landscape, i.e., to show some vacua are preferred over others. Since vacua were said to be related by simple topological transitions called blowups, we constructed large graphs, $N\sim 10^7$, using Algorithm~\ref{alg:tree_construction} to model the structure of the string landscape. Then, we used two different networks of string geometries --- the tree network and the hypersurface network --- to measure the dominant eigenvector of $-(\mathbf{L}+\mathbf{D})$, where $\mathbf{L}$ is the network Laplacian and $\mathbf{D}$ the degree matrix. A non-uniform distribution of entries indicated an interesting vacuum selection effect (Figure~\ref{fig:facetree}).

The~\thealgorithm~set and graph algorithms presented in this dissertation have proven to be useful in a wide range of applications. We gave examples here in quantum gravity, network science, and string theory applied to eternal inflation in cosmology, and we found interesting results in general relativity along the way. These examples show that these algorithms have broad applicability to many systems modeled by sets or graphs, and they improve existing methods by reducing simulation runtimes by orders of magnitude.

%\afterpage{\blankpage}

\appendix
\twolinechapter
\chapter[Useful Expressions for Causal Sets]{\texorpdfstring{Useful Expressions\\[-0.8cm] for Causal Sets}{Useful Expressions for Causal Sets}}
\chaptermark{Useful Expressions}
\mainchapter
\label{chap:useful}
\thispagestyle{empty}

This appendix contains many unpublished yet useful results for causal sets. They are listed here for reference.

\section{Causal Set Sprinklings}
\label{sec:causet_sprinklings}
While it is common to use rejection sampling to sprinkle elements into a curved spacetime with nontrivial boundaries, in numerical experiments it is much more efficient to sample from the coordinate probability distributions directly, supposing they can be solved in closed form. The primary purpose of this section is to provide the yet unpublished equations used to construct causal sets in various regions of curved spacetimes. We begin with an overview of how such solutions are found in $(3+1)$ dimensions, and then report results for each spacetime studied over the course of this work.

\subsection{General Method}
To begin, we pick a particular manifold $\mathbb{M}$ with metric $g_{\mu\nu}$. Depending on the region we consider, we choose Cartesian coordinates $(t,x,y,z)$, spherical coordinates $(t,r,\theta,\phi)$, or spherical light cone coordinates $(u,v,\theta,\phi)$, where $u=(t+r)/\sqrt{2}$ and $v=(t-r)/\sqrt{2}$. In curved spacetimes, it can be particularly useful to work with the conformal time $\eta=\int\,dt^\prime/a(t^\prime)$, where $a(t)$ is the scale factor of a conformally flat FLRW spacetime. Using the metric tensor, we can write the volume form $dV=\sqrt{-|g_{\mu\nu}|}\,dx^0\,dx^1\,dx^2\,dx^3$. The volume of the region of interest is then $V=\int\,dV$, where the integration is performed over bounds which specify the region. \par

Coordinates are sampled using a Poisson point process with intensity $\nu$ so that the mean number of elements added to the region is $\bar{N}=\nu V$, and the true number for any realization is a Poisson random variable with mean $\bar{N}$. Coordinate distributions are extracted from the volume form when it is written like $dV=\rho(x^0,x^1,x^2,x^3)\,dx^0\,dx^1\,dx^2\,dx^3$. When $\rho$ is separable in all coordinates, one needs only to normalize the probability distributions $\rho(x^\mu)$, integrate them to find the cumulative probability distribution $C(x^\mu)$, and then invert $\fru=C(x^\mu)$ to find $x^\mu$ as a function of a uniform random variable $\fru\in[0,1)$. When $\rho$ is not separable, or it is partially separable, we must integrate the joint probability distribution $\rho_{X,Y}(x,y)$ to find the marginal distribution,
\begin{equation}
\rho_X(x)=\int\rho_{X,Y}(x,y)\,dy\,,
\end{equation}
and then use the two to find the conditional probability distribution,
\begin{equation}
\rho_{Y|X}(y|x^*)=\frac{\rho_{X,Y}(x^*,y)}{\rho_X(x^*)}\,,
\end{equation}
where $x^*$ is a variable sampled from the marginal distribution $\rho_X(x)$. \par

\begin{comment}
In certain cases, one can also analytically find the expected mean number of relations per element $\langle\bar{k}\rangle$, where $k$ is related to the causal set ordering fraction~\cite{myrheim1978statistical} by $\textswab{r}=k/(N-1)$. This expression is found by calculating the following:
%
\begin{equation}
\langle\bar{k}\rangle = \nu\int\left[V_p(x^\mu)+V_f(x^\mu)\right]\,dV\,,
\end{equation}
%
where $V_p(x^\mu)$ and $V_f(x^\mu)$ are the volumes of the past and future light cones, respectively, at point $x^\mu$. Thus, using these methods, we can derive the following expressions for various spacetimes.
\end{comment}

\subsection{Minkowski Spacetime}
\paragraph{1+1 Dimensional Square}\mbox{}\\*
Spacetime Interval: $ds^2=-dt^2+dx^2$\\*
Volume Form: $dV=dt\,dx$\\*
Region: $t\in[-t_0,t_0]\qquad x\in[-x_0,x_0]$\\*
Volume: $V(t_0,x_0)=4t_0x_0$\\*
Coordinate PDFs: $\rho(t)=\frac{1}{2t_0}\qquad\rho(x)=\frac{1}{2x_0}$\\*
Coordinates: $t^*=(2\fru-1)t_0\qquad x^*=(2\fru-1)x_0$
%Average Degree: $\avgk=\frac{2}{3\pi}Nt_0$ when $x_0\gg t_0$

\paragraph{1+1 Dimensional Cylinder}\mbox{}\\*
Spacetime Interval: $ds^2=-dt^2+d\theta^2$\\*
Volume Form: $dV=dt\,d\theta$\\*
Region: $t\in[-t_0,t_0]\qquad\theta\in[0,2\pi)$\\*
Volume: $V(t_0)=4\pi t_0$\\*
Coordinate PDFs: $\rho(t)=\frac{1}{2t_0}\qquad\rho(\theta)=\frac{1}{2\pi}$\\*
Coordinates: $t^*=(2\fru-1)t_0\qquad\theta^*=2\pi \fru$
%Average Degree: $\avgk=\frac{2}{3\pi}Nt_0$

\paragraph{1+1 Dimensional Diamond}\mbox{}\\*
Spacetime Interval: $ds^2=-2\,du\,dv$\\*
Volume Form: $dV=du\,dv$\\*
Region: $u\in[0,u_0]\qquad v\in[0,v_0]$\\*
Volume: $V(u_0,v_0)=u_0v_0$\\*
Coordinate PDFs: $\rho(u)=\frac{1}{u_0}\qquad\rho(v)=\frac{1}{v_0}$\\*
Coordinates: $u^*=\fru u_0\qquad v^*=\fru v_0$\\*
%Average Degree: $\avgk=N/2$\\*
Notes: $u_0=v_0\,;\quad\tau_0=\sqrt{2}u_0$ is the proper time for all diamonds

\paragraph{2+1 Dimensional Diamond}\mbox{}\\*
Spacetime Interval: $ds^2=-2\,du\,dv+\frac{1}{2}(u-v)^2\,d\theta^2$\\*
Volume Form: $dV=\frac{1}{\sqrt{2}}(u-v)\,du\,dv\,d\theta$\\*
Region: $u\in[0,u_0]\qquad v\in[0,u]\qquad \theta\in[0,2\pi)$\\*
Volume: $V=\frac{\pi}{3\sqrt{2}}u_0^3$\\*
Coordinate PDFs: $\rho(u,v)=\frac{6}{u_0^3}(u-v)\qquad\rho(\theta)=\frac{1}{2\pi}$\\*
Coordinates: $u^*=\fru^{1/3}u_0\qquad v^*=u^*\left(1-\sqrt{1-\fru}\right)\qquad\theta^*=2\pi\fru$

\paragraph{3+1 Dimensional Diamond}\mbox{}\\*
Spacetime Interval: $ds^2=-2\,du\,dv+\frac{1}{2}(u-v)^2\,d\Omega_2^2$\\*
Volume Form: $dV=\frac{1}{2}(u-v)^2\sin\theta\,du\,dv\,d\theta\,d\phi$\\*
Region: $u\in[0,u_0]\qquad v\in[0,u]\qquad\theta\in[0,\pi)\qquad\phi\in[0,2\pi)$\\*
Volume: $V=\frac{\pi}{6}u_0^4$\\*
Coordinate PDFs: $\rho(u,v)=\frac{12}{u_0^4}(u-v)^2\qquad\rho(\theta)=\frac{1}{2}\sin\theta\qquad\rho(\phi)=\frac{1}{2\pi}$\\*
Coordinates: $u^*=\fru^{1/4}u_0\qquad \fru=3\frac{v^*}{u^*}-3\left(\frac{v^*}{u^*}\right)^2+\left(\frac{v^*}{u^*}\right)^3\qquad\theta^*=\arccos(1-2\fru)\qquad\phi^*=2\pi\fru$\\*
Notes: $\fru(v^*)$ must be inverted numerically

\paragraph{4+1 Dimensional Diamond}\mbox{}\\*
Spacetime Interval: $ds^2=-2\,du\,dv+\frac{1}{2}(u-v)^2\,d\Omega_3^2$\\*
Volume Form: $dV=\frac{1}{2\sqrt{2}}(u-v)^3\sin^2\psi\sin\theta\,du\,dv\,d\psi\,d\theta\,d\phi$\\*
Region: $u\in[0,u_0]\qquad v\in[0,u]\qquad\psi,\theta\in[0,\pi)\qquad\phi\in[0,2\pi)$\\*
Volume: $V=\frac{\pi^2}{20\sqrt{2}}u_0^5$\\*
Coordinate PDFs: $\rho(u,v)=\frac{20}{u_0^5}(u-v)^3\qquad\rho(\psi)=\frac{2}{\pi}\sin^2\psi\qquad\rho(\theta)=\frac{1}{2}\sin\theta\qquad\rho(\phi)=\frac{1}{2\pi}$\\*
Coordinates: $u^*=\fru^{1/5}u_0\qquad v^*=u^*\left[1-(1-\fru)^{1/4}\right]$\\*
\phantom{Coordinates: }$\fru=(2\psi^*-\sin(2\psi^*))/2\pi\qquad\theta^*=\arccos(1-2\fru)\qquad\phi^*=2\pi\fru$\\*
Notes: $\fru(\psi^*)$ must be inverted numerically

\subsection{de Sitter Spacetime}
\paragraph{1+1 Dimensional Slab (Spherical Foliation)}\mbox{}\\*
Spacetime Interval: $ds^2=\lambda^2\sec^2\eta\left(-d\eta^2+d\theta^2\right)$\\*
Volume Form: $dV=\lambda^2\sec^2\eta\,d\eta\,d\theta$\\*
Region: $\eta\in[0,\eta_0]\qquad\theta\in[0,2\pi)$\\*
Volume: $V=2\pi\lambda^2\tan\eta_0$\\*
Coordinate PDFs: $\rho(\eta)=\frac{\sec^2\eta}{\tan\eta_0}\qquad\rho(\theta)=\frac{1}{2\pi}$\\*
Coordinates: $\eta^*=\arctan(\fru\tan\eta_0)\qquad\theta^*=2\pi \fru$\\*
%Average Degree: $\avgk=4\lambda^2\nu\left(\frac{\eta_0}{\tan\eta_0}+\ln\sec\eta_0-1\right)$
Notes: $\lambda$ is the de Sitter pseudo-radius.

\paragraph{1+1 Dimensional Diamond (Flat Foliation)}\mbox{}\\*
Spacetime Interval: $ds^2=-\frac{4\lambda^2}{(u+v)^2}\,du\,dv$\\*
Volume Form: $dV=\frac{2\lambda^2}{(u+v)^2}\,du\,dv$\\*
Region: $u\in[u_0,u_0+w]\qquad v\in[v_0,v_0+w]$\\*
Volume: $V(u_0,v_0,w)=2\lambda^2\ln\mu(u_0,v_0,w)$\\*
Coordinate PDFs: $\rho(u,v)=\frac{1}{(u+v)^2\ln\mu}$\\*
Coordinates: $u^*=\frac{\beta(v_0+w)-v_0}{1-\beta}\qquad v^*=\frac{\gamma u^*+v_0}{1-\gamma}$\\*
Notes: $u_0=v_0=-1/\sqrt{2}\,;\quad w\in(0,1/\sqrt{2})\,;\quad\lambda$ is the de Sitter pseudo-radius\\*
\phantom{Notes: }$\mu\equiv\frac{(u_0+v_0+w)^2}{(u_0+v_0)(u_0+v_0+2w)}\,;\quad\beta\equiv\frac{u_0+v_0}{u_0+v_0+w}\mu^\fru\,;\quad\gamma\equiv\frac{w\fru}{u^*+v_0+w}$

\paragraph{1+1 Dimensional Diamond (Spherical Foliation)}\mbox{}\\*
Spacetime Interval: $ds^2=-2\lambda\sec^2\left(\frac{u+v}{\sqrt{2}}\right)\,du\,dv$\\*
Volume Form: $dV=\lambda^2\sec^2\left(\frac{u+v}{\sqrt{2}}\right)\,du\,dv$\\*
Region: $u\in[0,w_0]\qquad v\in[0,w_0]$\\*
Volume: $V=2\lambda^2\ln\mu$\\*
Coordinate PDFs: $\rho(u,v)=\frac{1}{2\ln\mu}\sec^2\left(\frac{u+v}{\sqrt{2}}\right)$\\*
Coordinates: $u^*=\sqrt{2}\arctan\left[\left(1-\mu^{-\fru}\right)\cot\left(\frac{w_0}{\sqrt{2}}\right)\right]$\\*
\phantom{Coordinates: }$v^*=\sqrt{2}\arctan\left[\tan\left(\frac{u^*+w_0}{\sqrt{2}}\right)+(1-\fru)\tan\left(\frac{u^*}{\sqrt{2}}\right)\right]-u^*$\\*
Notes: $\mu\equiv\left(1+\sec\left(\sqrt{2}w_0\right)\right)/2$

\paragraph{3+1 Dimensional Slab (Flat Foliation)}\mbox{}\\*
Spacetime Interval: $ds^2=\left(\frac{\lambda}{\eta}\right)^2\left(-d\eta^2+dr^2+r^2\,d\Omega_2^2\right)$\\*
Volume Form: $dV=\left(\frac{\lambda}{\eta}\right)^4 r^2\,d\eta\,dr\,d\Omega_2^2$\\*
Region: $\eta\in[-1,\eta_0]\qquad r\in[0,r_0]\qquad\theta\in[0,\pi)\qquad\phi\in[0,2\pi)$\\*
Volume: $V=-\frac{4\pi}{9}\lambda^4r_0^3\left(1+\eta_0^{-3}\right)$\\*
Coordinate PDFs: $\rho(\eta)=-\frac{3\eta_0^3}{(\eta_0^3+1)\eta^4}\qquad\rho(r)=\frac{3r^2}{r_0^3}\qquad\rho(\theta)=\frac{1}{2}\sin\theta\qquad\rho(\phi)=\frac{1}{2\pi}$\\*
Coordinates: $\eta^*=\left[\fru\left(1+\eta_0^{-3}\right)-1\right]^{-1/3}\quad r^*=\fru^{1/3}r_0\quad\theta^*=\arccos(1-2\fru)\quad\phi^*=2\pi\fru$
%Average Degree: $\avgk=\frac{4\pi\lambda^4\nu}{9}\frac{\left(e^{-\tau_0}-1\right)\left(13-e^{-\tau_0}\left(14-13e^{-\tau_0}\right)\right)+6\tau_0\left(e^{-3\tau_0}+1\right)}{1-e^{-3\tau_0}}\,;\quad\tau_0=-\ln|\eta_0|$

\paragraph{3+1 Dimensional Slab (Spherical Foliation)}\mbox{}\\*
Spacetime Interval: $ds^2=\lambda^2\sec^2\eta\left(-d\eta^2+d\Omega_3^2\right)$\\*
Volume Form: $dV=\lambda^4\sec^4\eta\,d\eta\,d\Omega_3^2$\\*
Region: $\eta\in[0,\eta_0]\qquad\psi\in[0,\pi)\qquad\theta\in[0,\pi)\qquad\phi\in[0,2\pi)$\\*
Volume: $V=\frac{2\pi^2}{3}\lambda^4\left(2+\sec^2\eta_0\right)\tan\eta_0$\\*
Coordinate PDFs: $\rho(\eta)=\frac{3\sec^4\eta}{(2+\sec^2\eta_0)\tan\eta_0}\qquad\rho(\psi)=\frac{2}{\pi}\sin^2\psi\qquad\rho(\theta)=\frac{1}{2}\sin\theta\qquad\rho(\phi)=\frac{1}{2\pi}$\\*
Coordinates: $\eta^*=2\arctan x\quad\fru=(2\psi^*-\sin(2\psi^*))/2\pi\quad\theta^*=\arccos(1-2\fru)\quad\phi^*=2\pi\fru$\\*
%Average Degree: $\avgk=\frac{4\pi}{9}\frac{\lambda^4\nu}{2+\sec^2\eta_0}\left[12\left(\frac{\eta_0}{\tan\eta_0}+\ln\sec\eta_0\right)+\left(6\ln\sec\eta_0-5\right)\sec^2\eta_0-7\right]$\\*
Notes: $\mu\equiv\fru\left(\frac{2+\csc^2\zeta}{\tan\zeta}\right);\qquad\zeta\equiv\frac{\pi}{2}-\eta_0$\\*
\phantom{Notes: }$\mu=x(6+x(3\mu+x(-4+x(-3\mu+x(6+\mu x)))))$ must be solved for $x$ numerically\\*
\phantom{Notes: }$\fru(\psi^*)$ must be inverted numerically

\paragraph{3+1 Dimensional Diamond (Flat Foliation)}\mbox{}\\*
Spacetime Interval: $ds^2=\frac{2\lambda^2}{\left(u+v\right)^2}\left[-2\,du\,dv+\frac{1}{2}\left(u-v\right)^2\,d\Omega_2^2\right]$\\*
Volume Form: $dV=2\lambda^4\frac{(u-v)^2}{(u+v)^4}\sin\theta\,du\,dv\,d\theta\,d\phi$\\*
Region: $u\in[u_0,u_0+w]\qquad v\in[u_0,u]\qquad\theta\in[0,\pi)\qquad\phi\in[0,2\pi)$\\*
Volume: $V=\frac{4\pi}{3}\lambda^4\mu$\\*
Coordinate PDFs: $\rho(u,v)=\frac{6}{\mu}\frac{(u-v)^2}{(u+v)^4}\qquad\rho(\theta)=\frac{1}{2}\sin\theta\qquad\rho(\phi)=\frac{1}{2\pi}$\\*
Coordinates: $u^*=\frac{2u_0}{W_0(z)}\left[\sqrt{W_0(z)+1}-1\right]-u_0\qquad v^*=-\frac{1}{3\alpha}\left(\beta+C+\frac{\Delta_0}{C}\right)$\\*
\phantom{Coordinates: }$\theta^*=\arccos(1-2\fru)\qquad\phi^*=2\pi\fru$\\*
Notes: $w\in[0,1/\sqrt{2})\,;\qquad W_0(x)$ is the principal branch of the Lambert function\\*
\phantom{Notes: }$\mu\equiv\ln\left[\frac{(w+2u_0)^2}{4u_0(w+u_0)}\right]-\left(\frac{w}{w+2u_0}\right)^2\,;\quad z\equiv -e^{-(\mu \fru+1)}$\\*
\phantom{Notes: }$\alpha\equiv u^{*3}(\fru-2)-3u^{*2}\fru u_0+3u^*(\fru-2)u_0^2-\fru u_0^3$\\*
\phantom{Notes: }$\beta\equiv 3u^*\left[u^{*3}\fru-3u^{*2}(\fru-2)u_0+3u^*\fru u_0^2-(\fru-2)u_0^3\right]$\\*
\phantom{Notes: }$\gamma\equiv3u^{*2}\alpha\,;\quad\delta\equiv\frac{u^{*2}}{3}\beta\,;\quad\Delta_0\equiv\beta^2-(3u^*\alpha)^2\,;\quad\Delta_1\equiv2\beta\Delta_0$\\*
\phantom{Notes: }$C\equiv\left[(\beta+3u^*\alpha)^2(\beta-3u^*\alpha)\right]^{1/3}$

\paragraph{3+1 Dimensional Diamond (Spherical Foliation)}\mbox{}\\*
Spacetime Interval: $ds^2=\lambda^2\sec^2\left(\frac{u+v}{\sqrt{2}}\right)\left[-2du\,dv+\sin^2\left(\frac{u-v}{\sqrt{2}}\right)\,d\Omega_2^2\right]$\\*
Volume Form: $\lambda^4\sec^4\left(\frac{u+v}{\sqrt{2}}\right)\sin^2\left(\frac{u-v}{\sqrt{2}}\right)\sin\theta\,du\,dv\,d\theta\,d\phi$\\*
Region: $u\in[0,u_0]\qquad v\in[0,u]\qquad \theta\in[0,\pi)\qquad\phi\in[0,2\pi)$\\*
Volume: $dV=\frac{4\pi}{3}\lambda^4\mu$\\*
Coordinate PDFs: $\rho(u,v)=\frac{3}{\mu}\sec^4\left(\frac{u+v}{\sqrt{2}}\right)\sin^2\left(\frac{u-v}{\sqrt{2}}\right)\qquad\rho(\theta)=\frac{1}{2}\sin\theta\qquad\rho(\phi)=\frac{1}{2\pi}$\\*
Coordinates: $\fru=\frac{1}{\mu}\left[\ln\left(\frac{1}{2}\left(1+\sec\left(\sqrt{2}u^*\right)\right)\right)-\sec^2\left(\frac{u^*}{\sqrt{2}}\right)+1\right]$\\*
\phantom{Coordinates: }$\fru=\frac{1}{4}\cos\left(\sqrt{2}u^*\right)\csc\left(\frac{u^*}{\sqrt{2}}\right)\left[4\csc\left(\frac{u^*}{\sqrt{2}}\right)\left[1+3\cos\left(\sqrt{2}u^*\right)+\cos\left(2\sqrt{2}u^*\right)\right]+\right.$\\*
$\left.\qquad\qquad\qquad\qquad\cot^2\left(\frac{u^*}{\sqrt{2}}\right)\sec^3\left(\frac{u^*+v}{\sqrt{2}}\right)\left[\sin\left(\frac{2u^*-3v^*}{\sqrt{2}}\right)+3\sin\left(\frac{v^*}{\sqrt{2}}\right)+3\sin\left(\frac{v^*-2u^*}{\sqrt{2}}\right)-\right.\right.$\\*
$\left.\left.\qquad\qquad\qquad\qquad\sin\left(\frac{3(2u^*+v^*)}{\sqrt{2}}\right)-3\sin\left(\frac{4u^*+v^*}{\sqrt{2}}\right)+\sin\left(\frac{2u^*+3v^*}{\sqrt{2}}\right)\right]\right]$\\*
\phantom{Coordinates: }$\theta^*=\arccos\left(1-2\fru\right)\qquad\phi^*=2\pi\fru$\\*
Notes: $\fru(u^*)$ and $\fru(v^*)$ must be inverted numerically\\*
\phantom{Notes: }$\mu\equiv\ln\left[\frac{1}{2}\left(1+\sec\left(\sqrt{2}w_0\right)\right)\right]-\sec^2\left(\frac{w_0}{\sqrt{2}}\right)+1$

\subsection{Dust Spacetime}
\paragraph{3+1 Dimensional Slab}\mbox{}\\*
Spacetime Interval: $ds^2=-d\tau^2+\tilde\alpha^2\left(\frac{3\tau}{2}\right)^{4/3}\left(dr^2+r^2\,d\Omega_2^2\right)$\\*
Volume Form: $dV=\lambda\alpha^3\left(\frac{3\tau}{2}\right)^2r^2\,d\tau\,dr\,d\Omega_2^2$\\*
Region: $\tau\in[0,\tau_0]\qquad r\in[0,r_0]\qquad\theta\in[0,\pi)\qquad\phi\in[0,2\pi)$\\*
Volume: $V=\pi\lambda\left(\alpha\tau_0\right)^3$\\*
Coordinate PDFs: $\rho(\tau)=\frac{3\tau^3}{\tau_0^3}\qquad\rho(r)=\frac{3r^2}{r_0^3}\qquad\rho(\theta)=\frac{1}{2}\sin\theta\qquad\rho(\phi)=\frac{1}{2\pi}$\\*
Coordinates: $\tau^*=\tau_0\fru^{1/3}\qquad r^*=r_0\fru^{1/3}\qquad\theta^*=\arccos(1-2\fru)\qquad\phi^*=2\pi\fru$
%Average Degree: $\avgk=\frac{18\pi \lambda^4\nu}{385}\tau_0^4$

\paragraph{3+1 Dimensional Diamond}\mbox{}\\*
Spacetime Interval: $ds^2=\left(\frac{\alpha}{2}\right)^2\left[\frac{\tilde\alpha}{2}\left(u+v\right)\right]^4\left[-2\,du\,dv+\frac{1}{2}(u-v)^2\,d\Omega_2^2\right]$\\*
Volume Form: $dV=\frac{\lambda^4}{2}\left(\frac{\tilde\alpha}{2}\right)^{12}(u+v)^8(u-v)^2\sin\theta\,du\,dv\,d\theta\,d\phi$\\*
Region: $u\in[0,u_0]\qquad v\in[0,u]\qquad \theta\in[0,\pi)\qquad\phi\in[0,2\pi)$\\*
Volume: $V=\frac{1981\pi}{2970}\lambda^4\left(\frac{\tilde\alpha u_0}{2}\right)^{12}$\\*
Coordinate PDFs: $\rho(u,v)=\frac{5940}{1981 u_0^{12}}(u+v)^8(u-v)^2\qquad\rho(\theta)=\frac{1}{2}\sin\theta\qquad\rho(\phi)=\frac{1}{2\pi}$\\*
Coordinates: $u^*=u_0\fru^{1/12}\qquad \fru=\frac{495}{1981u^{*11}}\left[u^{*10}v^*+3u^{*9}v^{*2}+\frac{13}{3}u^{*8}v^{*3}+2u^{*7}v^{*4}-\right.$\\*
$\left.\qquad\qquad\qquad\qquad\qquad\qquad\qquad\qquad\frac{14}{5}u^{*6}v^{*5}-\frac{14}{3}u^{*5}v^{*6}-2u^{*4}v^{*7}+u^{*3}v^{*8}+\frac{13}{9}u^{*2}v^{*9}+\right.$\\*
$\left.\qquad\qquad\qquad\qquad\qquad\qquad\qquad\qquad\frac{3}{5}u^*v^{*10}+\frac{1}{11}v^{*11}\right]$\\*
\phantom{Coordinates: }$\theta^*=\arccos(1-2\fru)\qquad\phi^*=2\pi\fru$\\*
Notes: $\fru(v^*)$ must be inverted numerically; for details on $\tilde\alpha$ see Chapter~\ref{chap:navigation}.

\subsection{Dust and Dark Energy Spacetime}
\paragraph{3+1 Dimensional Slab}\mbox{}\\*
Spacetime Interval: $ds^2=-d\tau^2+\tilde\alpha^2\sinh^{4/3}\left(\frac{3\tau}{2}\right)\left(dr^2+r^2\,d\Omega_2^2\right)$\\*
Volume Form: $dV=\lambda\alpha^3\sinh^2\left(\frac{3\tau}{2}\right)r^2\,d\tau\,dr\,d\theta\,d\phi$\\*
Region: $\tau\in[0,\tau_0]\qquad r\in[0,r_0]\qquad\theta\in[0,\pi)\qquad\phi\in[0,2\pi)$\\*
Volume: $V=\frac{2\pi}{9}\lambda^4\tilde\alpha^3\left(\sinh(3\tau_0)-3\tau_0\right)$\\*
Coordinate PDFs: $\rho(\tau)=\frac{6\sinh^2\left(3\tau/2\right)}{\sinh(3\tau_0)-3\tau_0}\qquad\rho(r)=\frac{3r^2}{r_0^3}\qquad\rho(\theta)=\frac{1}{2}\sin\theta\qquad\rho(\phi)=\frac{1}{2\pi}$\\*
Coordinates: $\fru=\frac{\sinh(3\tau^*)-3\tau^*}{\sinh(3\tau_0)-3\tau_0}\qquad r^*=r_0\fru^{1/3}\qquad\theta^*=\arccos(1-2\fru)\qquad\phi^*=2\pi\fru$\\*
Notes: $\fru(\tau^*)$ must be inverted numerically
\clearpage

\section{Isolated Elements in de Sitter Spacetime}
%\sectionmark{Isolated Elements}
%%%%%%%%%%
The number of isolated elements, i.e., sprinkled elements with no relations in the resulting RGG, may be solved analytically for a closed $(1+1)$-dimensional de Sitter spacetime bounded by $\eta\in[0,\eta_0]$.  For a Poisson point process, the probability distribution has the form
%%%%%
\begin{equation}
P(x) = \frac{\left(\nu V\right)^x}{x!}e^{-\nu V}\,.
\end{equation}
%%%%%
\noindent The probability an element at conformal time $\eta$ is isolated is given by the above expression, where $x=0$ is the number of nodes in the sum of the light cone volumes $V=V_p\left(\eta\right)+V_f\left(\eta\right)$.
Manipulating this expression yields
%%%%%
\begin{equation}
\begin{split}
P\left(0\right) &= e^{-2\nu \lambda^2\left[\left(\eta_0-\eta\right)\tan\eta_0 + \ln\sec^2\eta - \ln\sec\eta_0\right]}\,, \\
  &= e^{-2\nu \lambda^2\left[\eta_0\tan\eta_0 - \ln\sec\eta_0\right]}e^{-2\nu \lambda^2\left[\ln\sec^2\eta - \eta\tan\eta_0\right]}\,, \\
  &=\xi e^{-2\nu \lambda^2\left[\ln\sec^2\eta-\eta\tan\eta_0\right]}\,.
\end{split}
\end{equation}
%%%%%
\noindent Then, the expected number of isolated nodes $N_0$ is given by
%%%%%
\begin{equation}
\begin{split}
\langle N_0\rangle &= N\int_0^{\eta_0}\!\rho\left(\eta\right)P\left(0\right)\,d\eta\,, \\
  &= \frac{N\xi}{\tan\eta_0}\int_0^{\eta_0}\!\sec^2\eta e^{-2\nu \lambda^2\left[\ln\sec^2\eta - \eta\tan\eta_0\right]}\,d\eta\,, \\
  &\qquad \Rightarrow e^{-2\nu \lambda^2\ln\sec^2\eta} = \left(\cos\eta\right)^{4\nu \lambda^2}\,, \\
  &= \frac{N\xi}{\tan\eta_0}\int_0^{\eta_0}\left(\cos\eta\right)^{4\nu \lambda^2 - 2} e^{\left(2\nu \lambda^2 \tan\eta_0\right)\eta}\,d\eta\,.
\end{split}
\end{equation}
%%%%%
\noindent It is possible to simplify this further by recognizing the following identity:
%%%%%
\begin{equation}
\begin{split}
\int_0^a\!\cos^bxe^{cx}\,dx = &\frac{\left(1+e^{2ia}\right)e^{ac}\cos^ba}{c-ib} {}_2F_1\left(1,\frac{b}{2}+1-i\frac{c}{2};1-\frac{b}{2}-i\frac{c}{2};-e^{2ia}\right) \\
& - \frac{2}{c-ib} {}_2F_1\left(1,\frac{b}{2}+1-i\frac{c}{2};1-\frac{b}{2}-i\frac{c}{2};-1\right)\,,
\end{split}
\end{equation}
%%%%%
\noindent so in the present case we find
%%%%%
\begin{equation}
\begin{split}
\int_0^{\eta_0}\!\left(\cos\eta\right)^{4\nu \lambda^2 - 2}&e^{2\nu \lambda^2\eta\tan\eta_0}\,d\eta = \left[\frac{\left(1+e^{2i\eta_0}\right)e^{2\nu \lambda^2\eta_0\tan\eta_0}\left(\cos\eta_0\right)^{4\nu \lambda^2 - 2}}{2\nu \lambda^2\tan\eta_0 - i\left(4\nu \lambda^2 - 2\right)}\right. \\
&\left.\times{}_2F_1\left(1,2\nu \lambda^2 - i\nu \lambda^2\tan\eta_0;2-2\nu \lambda^2 - i\nu \lambda^2\tan\eta_0;-e^{2i\eta_0}\right)\right] \\
& - \left[\frac{1}{\nu \lambda^2\tan\eta_0 - i\left(2\nu \lambda^2 - 1\right)}\right. \\
&\left.\times{}_2F_1\left(1,2\nu \lambda^2-i\nu \lambda^2\tan\eta_0;2-2\nu \lambda^2-i\nu \lambda^2\tan\eta_0;-1\right)\right]\,.
\end{split}
\end{equation}
%%%%%
\noindent Therefore, we find the approximate number of isolated elements can be written as
%%%%%
\begin{equation}
\begin{split}
\langle N_0\rangle &\approx N\frac{e^{-2\nu \lambda^2\eta_0\tan\eta_0}\left(\cos\eta_0\right)^{-2\nu \lambda^2}}{\tan\eta_0}\frac{\left(1+e^{2i\eta_0}\right)e^{2\nu \lambda^2\eta_0}\left(\cos\eta_0\right)^{4\nu \lambda^2-2}}{2\nu \lambda^2\tan\eta_0 - i\left(4\nu \lambda^2 - 2\right)} \\
& \quad\times {}_2F_1\left(1,2\nu \lambda^2 - i\nu \lambda^2\tan\eta_0;2-2\nu \lambda^2 - i\nu \lambda^2\tan\eta_0;-e^{2i\eta_0}\right)\,, \\
&\approx N\frac{\left(1+e^{2i\eta_0}\right)\left(\cos\eta_0\right)^{2\nu \lambda^2 - 2}}{2\nu \lambda^2\left(\tan\eta_0\right)^2} {}_2F_1\left(1,2\nu \lambda^2-i\nu \lambda^2\tan\eta_0;2-2\nu\lambda^2-i\nu \lambda^2\tan\eta_0;-e^{2i\eta_0}\right)\,, \\
&\approx N\frac{\left(1+e^{2i\eta_0}\right)\left(\cos\eta_0\right)^{2\nu \lambda^2}}{2\nu \lambda^2} {}_2F_1\left(1,2\nu \lambda^2-i\nu \lambda^2\tan\eta_0;2-2\nu\lambda^2-i\nu \lambda^2\tan\eta_0;-e^{2i\eta_0}\right)\,.
\end{split}
\end{equation}
%%%%%
\noindent This can be reduced by implementing the Pfaff transformation:
%%%%%
\begin{equation}
{}_2F_1\left(a,b;c;z\right) = \left(1-z\right)^{-a} {}_2F_1\left(a,c-b;c;\frac{z}{z-1}\right)\,,
\end{equation}
%%%%%
\noindent so that we find the relation
%%%%%
\begin{equation}
\begin{split}
\langle N_0\rangle \approx &N\frac{\left(1+e^{2i\eta_0}\right)\left(\cos\eta_0\right)^{2\nu \lambda^2}}{2\nu \lambda^2}\left(1+e^{2i\eta_0}\right)^{-1} \\
&\quad \times {}_2F_1\left(1,2-4\nu \lambda^2;2-2\nu \lambda^2-i\nu \lambda^2\tan\eta_0;\frac{e^{2i\eta_0}}{1+e^{2i\eta_0}}\right)\,,
\end{split}
\end{equation}
%%%%%
\noindent and by using the approximations (as $\eta_0\to\frac{\pi}{2}$)
%%%%%
\begin{equation}
e^{2i\eta_0}\to -1\,,\quad 1+e^{2i\eta_0}\to 2i\left(\frac{\pi}{2} - \eta_0\right)\,,\quad \tan\eta_0\to\frac{1}{\frac{\pi}{2}-\eta_0}\,,
\end{equation}
%%%%%
\noindent we arrive at the equation
%%%%%
\begin{equation}
\begin{split}
\langle N_0\rangle &\approx N\frac{\left(\cos\eta_0\right)^{2\nu \lambda^2}}{2\nu \lambda^2} {}_2F_1\left(1,2-4\nu \lambda^2;\frac{-i\nu \lambda^2}{\frac{\pi}{2}-\eta_0};\frac{-1}{2i\left(\frac{\pi}{2}-\eta_0\right)}\right)\,, \\
&\approx N\frac{\left(\cos\eta_0\right)^{2\nu \lambda^2}}{2\nu \lambda^2} {}_2F_1\left(1,2-4\nu \lambda^2;-i\nu \lambda^2\tan\eta_0;\frac{i}{2}\tan\eta_0\right)\,.
\end{split}
\end{equation}
%%%%%
\noindent Since the average number of isolated elements is known to be a real number, it makes sense to continue to reduce this expression to eliminate the complex quantities.
To do this, we need to apply four new relations:
%%%%%
\begin{enumerate}
  \item For $a - b\notin \mathbb{Z}$ and $x\notin\left(0, 1\right)$,
  \begin{equation}
    \begin{split}
    {}_2F_1\left(a,b;c;x\right) =& \frac{\Gamma\left(b-a\right)\Gamma\left(c\right)}{\Gamma\left(b\right)\Gamma\left(c-b\right)}\left(-x\right)^{-a} {}_2F_1\left(a,a-c+1;a-b+1;\frac{1}{x}\right) + \\
    & \frac{\Gamma\left(a-b\right)\Gamma\left(c\right)}{\Gamma\left(a\right)\Gamma\left(c-b\right)}\left(-x\right)^{-b} {}_2F_1\left(b,b-c+1;-a+b+1;\frac{1}{x}\right)\,,
    \end{split}
  \end{equation}
%%%%%
  \item The Kummer hypergeometric function is
  \begin{equation}
  {}_1F_1\left(a;c;bd\right) = \lim_{x\to\infty} {}_2F_1\left(a,bx;c;\frac{d}{x}\right)\,,
  \end{equation}
%%%%%
  \item
  \begin{equation}
  \lim_{x\to\infty}\frac{\Gamma\left(x\right)}{\Gamma\left(x-a\right)} = x^a\,,
  \end{equation}
%%%%%
  \item
  \begin{equation}
  \label{eq:1F1_limit}
  {}_1F_1\left(1;a;b\right) = \left(a-1\right)e^{b}b^{1-a}\left(\Gamma\left(a-1\right)-\Gamma\left(a-1,b\right)\right)\,.
  \end{equation}
\end{enumerate}
%%%%%
\noindent By using the definitions $\alpha\equiv 2-4\nu \lambda^2$, $\beta\equiv -i\nu \lambda^2$, $\gamma\equiv\frac{i}{2}$, and $x\equiv\tan\eta_0$ and the relations
%%%%%
\begin{equation}
\begin{split}
{}_2F_1\left(1,\alpha;\beta x;\gamma x\right) &= \frac{\Gamma\left(\alpha-1\right)\Gamma\left(\beta x\right)}{\Gamma\left(\alpha\right)\Gamma\left(\beta x-1\right)}\left(-\gamma x\right)^{-1} {}_2F_1\left(1,2-\beta x;2-\alpha;\frac{1}{\gamma x}\right) \\
&\quad + \frac{\Gamma\left(1-\alpha\right)\Gamma\left(\beta x\right)}{\Gamma\left(1\right)\Gamma\left(\beta x-\alpha\right)}\left(-\gamma x\right)^{-\alpha} {}_2F_1\left(\alpha,\alpha+1-\beta x;\alpha;\frac{1}{\gamma x}\right)\,, \\
&\rightarrow \frac{\beta x-1}{\left(\alpha-1\right)\left(-\gamma x\right)} {}_1F_1\left(1;2-\alpha;-\frac{\beta}{\gamma}\right) \\
&\quad + \frac{\Gamma\left(1-\alpha\right)\left(\beta x\right)^\alpha}{\left(-\gamma x\right)^\alpha} {}_1F_1\left(\alpha;\alpha;-\frac{\beta}{\gamma}\right)\,.
\end{split}
\end{equation}
\noindent as well as the limit with~\eqref{eq:1F1_limit}, we obtain
%%%%%
\begin{equation}
\lim_{x\to\infty} {}_2F_1\left(1,\alpha;\beta x;\gamma x\right) = e^{-\beta/\gamma}\left(-\frac{\beta}{\gamma}\right)^\alpha\Gamma\left(1-\alpha,-\frac{\beta}{\gamma}\right)\,,
\end{equation}
%%%%%
\noindent and finally arrive at the result
%%%%%
\begin{equation}
\langle N_0\rangle \approx N\left(e\cos\eta_0\right)^{2\nu \lambda^2}\left(2\nu \lambda^2\right)^{1-4\nu \lambda^2}\Gamma\left(4\nu \lambda^2-1,2\nu \lambda^2\right)\,.
\end{equation}
\afterpage{\blankpage}

%%%%%%%%%%%
\backmatter
%%%%%%%%%%%

\phantomsection
\addcontentsline{toc}{chapter}{References}
\renewcommand{\chapter}[2]{\clearpage%
\begin{center}
\vspace*{-1.2cm}\noindent\rule{\textwidth}{1pt}
\bibname\vspace*{-0.3cm}
\noindent\rule{\textwidth}{1pt}
\end{center}}
\thispagestyle{empty}
\pagestyle{refstyle}
\singlespacing
\bibliographystyle{apsrev4-1}
\bibliography{ms}
\clearpage

\phantomsection
\addcontentsline{toc}{chapter}{Academic Output}
\begin{center}
\vspace*{-1cm}\noindent\rule{\textwidth}{1pt}\vspace*{0.35cm}
{\Large \textbf{ACADEMIC OUTPUT}}
\noindent\rule{\textwidth}{1pt}\\[0.2cm]
\end{center}
\thispagestyle{empty}
\pagestyle{refstyle}
The following papers were published, or are in the process of being published, as a result of this dissertation:\\
\begin{enumerate}[label={[\arabic*]}]
\setcounter{enumi}{248}
\item W.\ Cunningham, K.\ Zuev \& D.\ Krioukov, \textit{Navigability of random geometric graphs in the universe and other spacetimes}, \href{https://doi.org/10.1038/s41598-017-08872-4}{{Sci.\ Rep.\ }{\bf 7}, 8699 (2017).}
\item W.\ J.\ Cunningham, D.\ Rideout, J.\ Halverson \& D.\ Krioukov, \textit{Exact geodesic distances in FLRW spacetimes}, \href{https://doi.org/10.1103/PhysRevD.96.103538}{{Phys.\ Rev.\ D} {\bf 96}, 103538 (2017).}
\item W.\ J.\ Cunningham, \textit{Inference of boundaries in causal sets}, \href{https://doi.org/10.1088/1361-6382/aaadc4}{{Class.\ Quant. Grav.\ }{\bf 35}, 094002 (2018).}
\item W.\ J.\ Cunningham \& D.\ Krioukov, \textit{Causal set generator and action computer}, {Submitted to Comput.\ Phys.\ Commun.\ } (2017) \href{https://arxiv.org/abs/1709.03013}{arXiv:1709.03013}.
\item J.\ Carifio, W.\ J.\ Cunningham, J.\ Halverson, D.\ Krioukov, C.\ Long \& B.\ Nelson, \textit{Vacuum selection from cosmology on networks of string geometries}, {Submitted to Phys.\ Rev.\ Lett.\ } (2017) \href{http://arxiv.org/abs/1711.06685}{arXiv:1711.06685}.
\end{enumerate}

\vspace*{1cm}
\noindent The algorithms described herein are available online in these libraries:\\
\begin{enumerate}[resume,label={[\arabic*]}]
\item W.\ J.\ Cunningham, \href{https://bitbucket.org/dk-lab/2015_code_fastmath}{\textit{FastMath}, Bitbucket Repository (2015).}
\item W.\ J.\ Cunningham, \href{https://bitbucket.org/dk-lab/causalsetgenerator}{\textit{Causal Set Generator}, Bitbucket Repository (2017).}
\end{enumerate}

\end{document}